\documentclass[a4paper,11pt]{memoir}

\pdfoutput=1

\usepackage[final]{pdfpages}

\usepackage[square,numbers,sort&compress]{natbib}

\usepackage{ragged2e}
\usepackage{geometry}

\newcommand\blankpage{\newpage\null\thispagestyle{empty}
\newpage}

\makepagestyle{MyPagestyleChapter}

\makepsmarks{MyPagestyleChapter}{%
\nouppercaseheads

\createmark{chapter}{left}{shownumber}{}{.\space}
\createmark{section}{right}{shownumber}{}{.\space}

\createplainmark{toc}{both}{\contentsname}
\createplainmark{lof}{both}{\listfigurename}
\createplainmark{lot}{both}{\listtablename}
\createplainmark{bib}{both}{\bibname}
\createplainmark{index}{both}{\indexname}
\createplainmark{glossary}{both}{\glossaryname}}

\makeevenhead{MyPagestyleChapter}{\normalfont\thepage}{}{\let\boldmath\relax\normalfont\itshape\relax\rightmark}
\makeoddhead{MyPagestyleChapter}{\let\boldmath\relax\normalfont\itshape\leftmark}{}{\normalfont\thepage}
\makeheadrule{MyPagestyleChapter}{\textwidth}{\normalrulethickness}

\makepagestyle{MyPagestyleAppendix}

\makepsmarks{MyPagestyleAppendix}{%
\nouppercaseheads

\createmark{chapter}{both}{shownumber}{Appendix\space}{.\space}
\createmark{section}{right}{shownumber}{}{.\space}

\createplainmark{toc}{both}{\contentsname}
\createplainmark{lof}{both}{\listfigurename}
\createplainmark{lot}{both}{\listtablename}
\createplainmark{bib}{both}{\bibname}
\createplainmark{index}{both}{\indexname}
\createplainmark{glossary}{both}{\glossaryname}}

\makeevenhead{MyPagestyleAppendix}{\normalfont\thepage}{}{\let\boldmath\relax\normalfont\itshape\relax\rightmark}
\makeoddhead{MyPagestyleAppendix}{\let\boldmath\relax\normalfont\itshape\leftmark}{}{\normalfont\thepage}
\makeheadrule{MyPagestyleAppendix}{\textwidth}{\normalrulethickness}

\usepackage{graphicx}
\usepackage[labelsep=period]{caption}
\usepackage{subcaption}
\usepackage{xcolor}
\usepackage{overpic}

\definecolor{complutense}{HTML}{a30100}
\definecolor{mydarkgrey}{HTML}{c1c1c1}
\definecolor{mylightgrey}{HTML}{e6e6e6}
\definecolor{mylightergrey}{HTML}{eaeaea}

\usepackage[hidelinks]{hyperref}
\hypersetup{colorlinks=true,linkcolor=complutense,citecolor=complutense,urlcolor=complutense}

\maxtocdepth{subsection}

\setsecnumdepth{subsubsection}

\counterwithout{footnote}{chapter}

\renewcommand{\contentsname}{Contents.}
\renewcommand{\listfigurename}{List of Figures.}
\renewcommand{\listtablename}{List of Tables.}
\renewcommand{\bibname}{Bibliography.}

\renewcommand\thesubsection{\arabic{chapter}.\arabic{section}.\arabic{subsection}}
\renewcommand\thesubsubsection{\arabic{chapter}.\arabic{section}.\arabic{subsection}.\arabic{subsubsection}}

\usepackage{titlesec}
\titlelabel{\thetitle.\enskip}

\usepackage[dotinlabels]{titletoc}
\usepackage{tocloft}

\renewcommand*{\partnamefont}{\huge\bfseries\scshape}

\renewcommand{\afterpartskip}{\vspace*{\fill}}

\titleformat{\section}
  {\normalfont\Large\bfseries\justifying}{\thesection .}{1em}{}
\titleformat{\subsection}
  {\normalfont\large\bfseries\justifying}{\thesubsection .}{1em}{}

\usepackage{calc,soul}
\makeatletter
\newlength\dlf@normtxtw
\newsavebox{\feline@chapter}
\newcommand\feline@chapter@marker[1][4cm]{%
\sbox\feline@chapter{%
\resizebox{!}{#1}{\fboxsep=1pt%
\colorbox{complutense}{\color{white}\bfseries\thechapter}%
}}%
\raisebox{\depthof{\usebox{\feline@chapter}}}{\usebox{\feline@chapter}}%
}
\newcommand\feline@chm[1][4cm]{%
\sbox\feline@chapter{\feline@chapter@marker[#1]}%
\makebox[0pt][l]{%
\makebox[2.5cm][r]{\usebox\feline@chapter}%
}}
\makechapterstyle{daleif1}{

\renewcommand\printchapternum{\null\hfill\feline@chm[2.5cm]\par}

}
\makeatother
\chapterstyle{daleif1}

\usepackage{enumitem}

\newlist{coloritemize}{itemize}{1}
\setlist[coloritemize]{label=\textcolor{complutense}{\textbullet}}

\usepackage{tabularx}
\newcolumntype{L}{>{\arraybackslash}X}

\usepackage{multirow}

\newcommand{\tabitemize}[1]{%
\begin{minipage}[t]{\linewidth}
\begin{coloritemize}[nosep,left=0pt]
#1
\end{coloritemize}
\end{minipage}}

\usepackage{amsmath}
\usepackage{amsfonts}
\usepackage{amssymb}

\usepackage{amsthm}

\usepackage{thmtools}

\theoremstyle{definition}
\declaretheorem[shaded={rulecolor=mylightergrey,rulewidth=3pt, bgcolor=mylightergrey}]{result}
\declaretheorem[numbered=no,shaded={rulecolor=mylightergrey,rulewidth=3pt, bgcolor=mylightergrey}]{corollary}

\makeatletter
\g@addto@macro\normalsize{%
  \setlength\abovedisplayshortskip{\abovedisplayskip}
  \setlength\belowdisplayshortskip{\belowdisplayskip}
}

\usepackage[utf8]{inputenc}
\usepackage[T1]{fontenc}
\usepackage[sc]{mathpazo}

\AtBeginDocument{
  \DeclareSymbolFont{AMSb}{U}{msb}{m}{n}
  \DeclareSymbolFontAlphabet{\mathbb}{AMSb}}

\usepackage{calrsfs}

\newcommand{\bigo}{\mathcal{O}}

\newcommand\titlemath[1]{\texorpdfstring{#1}\,}

\newcommand\titlebm[1]{\protect\boldmath\texorpdfstring{#1}\,}

\newcommand{\myskip}{\hspace{9pt}}
\newcommand{\mybigskip}{\hspace{18pt}}
\newcommand{\myhugeskip}{\hspace{27pt}}
\newcommand{\e}{\mathrm{e}}
\newcommand{\iu}{\mathrm{i}}
\newcommand{\dif}{\mathrm{d}}
\newcommand{\const}{\mathrm{const.}}
\newcommand{\diag}{\mathrm{diag}}
\newcommand{\tr}{\mathrm{tr}}
\DeclareMathOperator{\sign}{sign}
\DeclareMathOperator*{\arcsinh}{arcsinh}
\DeclareMathOperator*{\arccosh}{arccosh}

\DeclareMathOperator*{\Si}{Si}
\DeclareMathOperator*{\Ci}{Ci}
\DeclareMathOperator*{\Shi}{Shi}
\DeclareMathOperator*{\Chi}{Chi}
\newcommand{\eff}{\mathrm{eff}}
\newcommand{\Mp}{M_\mathrm{Pl}}
\newcommand{\matter}{\mathrm{matter}}

\newcommand{\PGG}{PGG}
\newcommand{\bspgt}{\mathrm{BSPGG}}
\newcommand{\BSPGT}{BSPGG}
\newcommand{\stress}{T}

\usepackage{accents}
\newcommand\MA[1]{\hat{#1}{}}
\newcommand\MAG{\MA{\Gamma}}
\newcommand\MAD{\MA{\nabla}}

\newcommand\MAR{\MA{R}}
\newcommand{\holst}{H}
\newcommand\LC[1]{\mathring{#1}{}}
\newcommand\LCG{\LC{\Gamma}}
\newcommand\LCD{\LC{\nabla}}
\newcommand\LCbox{\LC{\square}}
\newcommand\LCR{\LC{R}}
\newcommand\LCbar[1]{\mathring{\bar{#1}}{}}
\newcommand\LCRbar{\LCbar{R}}
\newcommand\LCEin{\LC{G}}
\newcommand\LCKre{\LC{\mathcal{K}}}
\newcommand\LCf{f(\LCR)}
\newcommand\LCfp{f'(\LCR)}
\newcommand\LCfpp{f''(\LCR)}
\newcommand\LCfppp{f'''(\LCR)}
\newcommand\LCten{\varepsilon}
\newcommand\axcurrent{J}
\newcommand\trcurrent{L}

\newcommand{\Left}{\mathopen{}\mathclose\bgroup\left}
\newcommand{\Right}{\aftergroup\egroup\right}

\newcommand\sSigma{\Sigma}
\newcommand\szero{{(0)}}
\newcommand\sone{{(1)}}
\newcommand\stwo{{(2)}}
\newcommand\sN{{(N)}}
\newcommand\sell{{(\ell)}}
\newcommand\sstar{{(*)}}
\newcommand\sGR{(\text{GR})}
\newcommand\salpha{(\alpha)}
\newcommand\sbeta{(\beta)}

\newcommand\jump[1]{[#1]^+_-}

\newcommand\udis[1]{\underline{#1}{}}
\newcommand\disLCD{\udis{\LCD}{}}

\newcommand\partialdis{\udis{\partial}{}}
\newcommand\regpart[1]{\varrho \Left[#1\Right]}
\newcommand\singpart[2]{\Delta^{#1} \Left[#2\Right]}
\newcommand\Regpart[1]{\varrho [#1]}
\newcommand\Singpart[2]{\Delta^{#1} [#2]}
\newcommand\regsym{\varrho}
\newcommand\singsym[1]{\Delta^{#1}}
\newcommand\ind{\chi_\sSigma}
\newcommand\del{\udis{\delta}^\sSigma}

\newcommand{\heavidis}[1]{\udis{\Theta}^{#1}}
\newcommand\tosigma[1]{\overline{#1}}

\newcommand*{\wasyfamily}{\fontencoding{U}\fontfamily{wasy}\selectfont}
\newcommand*{\sun}{{\odot}}
\newcommand*{\mercury}{{\text{\wasyfamily\char39}}}

\begin{document}

\title{Compact objects in modified gravity: junction conditions and other viability criteria}
\newcommand{\englishtitle}{Compact objects in modified gravity:}

\newcommand{\englishsubtitle}{junction conditions and other viability criteria}

\newcommand{\spanishtitle}{Objetos compactos en gravedad modificada:}

\newcommand{\spanishsubtitle}{condiciones de pegado y otros criterios de viabilidad}

\author{Adri\'{a}n Casado-Turri\'{o}n}
\newcommand{\spanishauthor}{Adrián Casado Turrión}

\newgeometry{
    top=1cm,
    bottom=1cm,
    outer=1cm,
    inner=1cm,
}

\clearpage\thispagestyle{empty}\addtocounter{page}{-1}
\vspace*{\fill}

\begin{center}

{\Large \textbf{Universidad Complutense de Madrid}}

{\Large Facultad de Ciencias Físicas}

\vspace{1cm}

{\Large \textbf{PhD THESIS}}

\vspace{1cm}

{\huge \textbf{\englishtitle}}

{\huge \englishsubtitle}

\noindent\makebox[\linewidth]{\rule{17.8cm}{1pt}}

\vspace{10pt}

{\huge \textbf{\spanishtitle}}

{\huge \spanishsubtitle}

\vspace{1cm}

{\large Dissertation submitted for the degree of Doctor of Philosophy in Physics by}

\smallskip

{\huge\textbf{\MakeUppercase{\theauthor}}}

\vspace{1cm}

\begin{overpic}[width=7cm]{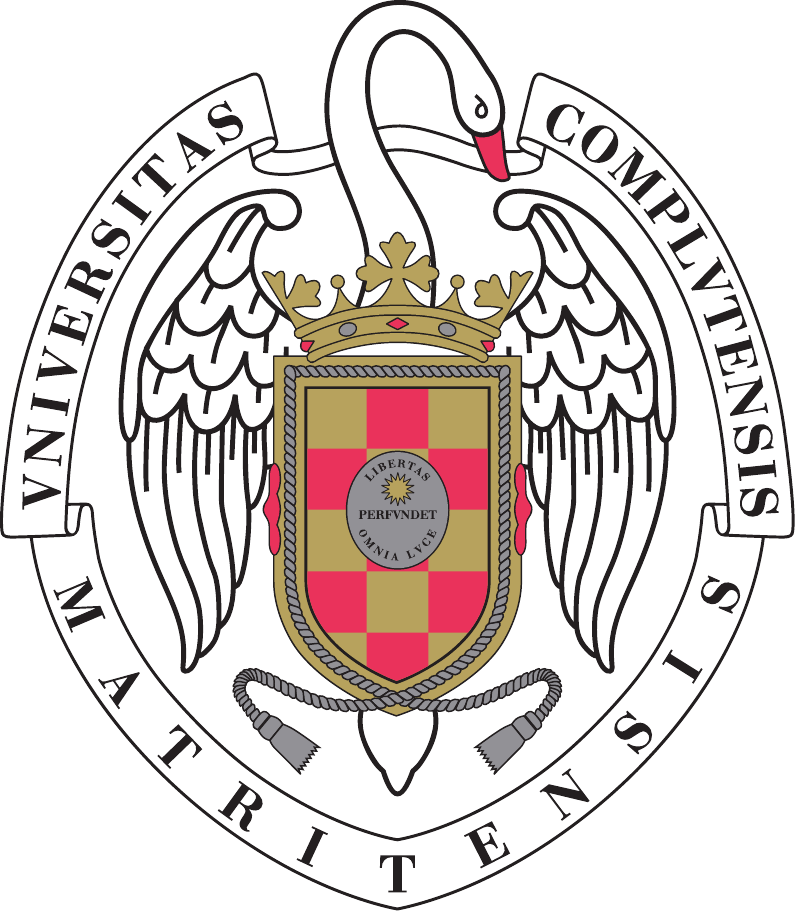}
\end{overpic}

\vspace{1cm}

{\large Supervisors}

\medskip

{\Large \textbf{Antonio Dobado}}

\smallskip

{\Large \textbf{Álvaro de la Cruz-Dombriz}}

\end{center}

\vfill

\clearpage
\blankpage

\newgeometry{
    top=2.5cm,
    bottom=2.5cm,
    outer=2.5cm,
    inner = 2.5cm
}

\clearpage\thispagestyle{empty}\addtocounter{page}{-1}
\vspace*{\fill}

\begin{center}

{\Large \textbf{Universidad Complutense de Madrid}}

{\Large Facultad de Ciencias Físicas}

{\Large Instituto de Física de Partículas y del Cosmos}

\smallskip

{\Large \textit{Programa de Doctorado en Física}}

\vspace{1cm}

{\Large \textbf{PhD THESIS}}

\vspace{1cm}

{\huge \textbf{\englishtitle}}

{\huge \englishsubtitle}

\noindent\makebox[\linewidth]{\rule{15cm}{1pt}}

\vspace{10pt}

{\huge \textbf{\spanishtitle}}

{\huge \spanishsubtitle}

\vspace{1cm}

{\large Dissertation submitted for the degree of Doctor of Philosophy in Physics by}

\smallskip

{\LARGE\textbf{\MakeUppercase{\theauthor}}}

\vspace{1cm}

{\large Thesis publicly defended on Wednesday, 8 November 2023}

\vspace{1cm}

{\large Supervisors}

\medskip

{\Large \textbf{Antonio Dobado}}

\smallskip

{\Large \textbf{Álvaro de la Cruz-Dombriz}}

\end{center}

\vfill

\blankpage

\newgeometry{
    top=2.5cm,
    bottom=2.5cm,
    outer=2.5cm,
    inner=3.5cm,
}

\thispagestyle{empty}
\vspace*{\fill}

\begin{center}
    \textit{A mis padres, Ana y Santi.}
\end{center}

\vfill

\blankpage

\thispagestyle{empty}
\vspace*{\fill}

\begin{center}
    \begin{tabular}{l}
        \textit{I cada cop que camino vaig més lluny,} \\
        \textit{coordenades d'un espai-temps que hem perdut.} \\
        \textit{Aquí les coses ja no cauen pel seu propi pes.} \\
        \textit{Relatius com els dies que hem viscut;} \\ 
        \textit{teories que s'escriuen i es fan fum.} \\
        \textit{Aquí les coses ja no pesen el seu propi pes...} \\ \\
        \textit{Houston, no hi ha cap problema;} \\
        \textit{aquí tot va bé, va tot sempre bé.} \\
        \textit{Houston, no hi ha cap estrella} \\
        \textit{quan ningú la veu, quan ningú la veu...} \\ \\ \\
        \hfill Blaumut, `Houston,' \\
        \hfill in \textit{Equilibri} (2017).
    \end{tabular}    
\end{center}

\vfill

\frontmatter

\newgeometry{
    top=3.5cm,
    bottom=2.5cm,
    outer=2.5cm,
    inner=3.5cm,
    footnotesep=1cm,
}

\chapter*{Acknowledgements.}
\addcontentsline{toc}{chapter}{Acknowledgements.}
\markboth{Acknowledgements.}{Acknowledgements.}

There is a number of people to whom I would like to express my most sincere gratitude for having accompanied me during these PhD years.

First and foremost, I would like to thank my supervisors, Antonio and Álvaro, for guiding me through the PhD with their invaluable help and advice. Álvaro, Antonio, you have given shape to the physicist I am today. You have always encouraged me to persist in the face of all difficulties. I also owe a sizeable portion of all I have learnt through these years to your always insightful comments and discussions.
Antonio, thank you for having been my advisor since my Bachelor's Thesis, and for having helped me navigate the abrupt transition from Quantum Field Theory in Curved Space-Times to Modified Gravity.
Álvaro, thank you for having embarked on this venture midway, with the pandemic still raging across the world, as well as for always helping me to collect my thoughts. I am very thankful for having the possibility of working with and learning from both of you.

I would also like to thank all the members of our Department, my fellow PhD students and office mates at Complutense (to name a few, Álvaro, Andrea, Arkaitz, Carlos, Clara, Dani, Eva, Felipe, Isi, José Alberto, Juanjo, Lucía, Luis, Merce, Patri, Rafa, Sergio, Teo, Valentín...), for all the wonderful moments we have shared along the years (the table talks after lunch, the traditional group dinners, the occasional \emph{seminarios-aperitivo}, \emph{viernes de juegos de mesa} and after-work activities; the multiple times we tidied up and redecorated our office, finding all sorts of long-forgotten treasures in the process...). I would like to specially thank Mercè Guerrero and Rita Neves, for their friendship and support during these years. Rita, Mercè, you have always been by my side during the whole PhD, in the good and bad moments, and you have made me a better person and a better scientist. I am very glad to have seen how we have evolved from first-year PhD students to actual theoretical physicists. I will never forget our excursions and travels, museum visits, photography tutorials, gastronomical experiences, \emph{gravidiscussions}, improvised musical sessions, and, of course, our long walks around Madrid. I am very fortunate to have met you both and to count on your friendship and help. Mercè, \emph{moltíssimes gràcies per haver-me ajudat a complir un dels meus somnis: aprendre a parlar català. Gràcies per la teva paciència, sobretot mentre estava aprenent les conjugacions verbals.} Rita, \emph{sei que não avancei muito com o português, mas prometo que vou continuar a tentar, e que um dia nós os dois finalmente podemos falar em português!}

My gratitude to Antonio Muñoz Sudupe, for introducing me to the art of lecturing, and for trusting the continuous evaluation of our students to me year after year. I would like to take the opportunity to thank all our first-year Algebra students as well, for their effort during the course, and for unavertedly helping me improve my teaching skills. I wish the best of luck to them, confident that they will all find success in their professional lives, be it in the academic world or the labour market.

I would like to express my deepest appreciation to all the members of the group at the University of Tartu (Aneta, Arpan, Daniel, Débora, Laur, Laxmipriya, Margus, María José, Sofía, Vaso...), 
for their help and hospitality during my three-month research visit in Estonia. Thank you for the immensely fruitful discussions and for making me feel like home for the entirety of my stay there. You are all exceptional, both at scientific and personal levels. I am very grateful for all the wonderful moments we shared at Physicum and outside work. A special thanks to Pepe Maldonado Torralba and Alejandro Jiménez Cano; without you, none of this would have been possible. Alejandro, Pepe, thank you for forming part of the \emph{junction-conditions team}. Working alongside you is an absolute pleasure; I can feel a flood of ideas coming to my brain every time we are close to a blackboard. Furthermore, on the personal side, I have found in you two marvellous, infinitely generous friends. I would also like to extend the appreciation to Jorge Gigante Valcárcel, who guided me in my first steps in Tartu, and would selflessly offer me his assistance whenever I needed it. Jorge, thank you for being such a kind person, and for having more trust in myself than I do. On a diffrent note, Vaso, although my Greek vocabulary and grammar are still on their alpha stages of development, \emph{$\varepsilon\upsilon\chi\alpha\rho\iota\sigma\tau\acute{\omega}$} for introducing me to such a fantasic language.

I would also like to thank all the people I have met in conferences, workshops or elsewhere in the academic world, such as Sebastián Bahamonde, Adrià Delhom, Ángel Murcia, Lucía Santamaría and all the PhD students and young postdocs attending the memorable EREP 2022 in Salamanca. Learning from you (and having fun together) has been a pleasure.

I wish to show my most sincere gratitude to Adrián, Alberto, Carlos, Coral, Cris, Joaquim, Laura, Natalia and Pablo for these ten years of true friendship. It never ceases to amaze me how we all belong together in spite of our very different characters and ideas. Ever since that chaotic first day of university, back in 2013, you have always been there, and you know you have become a fundamental pillar of my life. We have shared all sorts of moments, bright and dark; we have celebrated all our successes and offered each other our help, advice and affection in the face of disappointment. We have studied together, debated about the deepest questions, had fits of laughter, cooked actual \emph{paelles valencianes} and \emph{migas extremeñas}, travelled across Spain and Europe, got lost in the labyrinthine (and hilly) streets of Lisbon, hiked during heatwaves, risked our lives kayaking around the Medes islands, been on verge of missing planes and trains, been surprised by sudden storms and, in summary, had great fun together. So thank you for all these unforgettable experiences and for the many more to come in the future, for always being by my side, and---as I told Rita and Mercè before---for contributing to shape to the person I am today. There is nothing more wonderful than witnessing how, little by little, we are all starting to find our place in the world. By the way, Joaquim, \emph{a tu també et vull agrair el que m'hages ensenyat a parlar valencià, així com haver-me descobert "La insuportable lleugeresa de l'ésser" de Milan Kundera, el millor llibre mai escrit.}

I would like to extend the previous appreciation to the many more fantastic friends I have made during these ten years at Complutense, such as Bárbara, the very first person I met on my first day of university, and Jorge, my inseparable lab partner, among others. Thank you for being there all this time.

At this point, I would like to express my gratitude to my whole family---my parents, grandparents, aunts, uncles and cousins---for their unconditional love and support. I would like to specially thank my parents, Ana and Santi. Ever since I was born, you have devoted your lives to raising me, and have made great sacrifices to provide me with the best possible education. You have always encouraged me to be kind and altruistic, to dare to know, to be creative and to think critically. I owe you all my good qualities. I am not sure of whether I will ever be able to express how grateful I am for all you have selflessly given to me. I love you as much as you love me. \emph{Abuelos, Yaya, espero que estéis orgullosos de mí. El incansable esfuerzo que habéis realizado a lo largo de vuestras vidas ha sido para mí el mejor ejemplo. Os quiero muchísimo, y deseo compartir con vosotros muchos años más. Abuelo José Antonio, nunca llegué a conocerte, pero sabes que, para mí, eres mi referente.}

Last but not least, I would like to thank the External Evaluators and the members of the Thesis Committee (including the back-ups), for generously accepting the task and devoting part of their valuable time to assess this Thesis, helping to improve its quality.

The investigations condensed in this Thesis have been supported by a Universidad Complutense de Madrid-Banco Santander predoctoral contract CT63/19-CT64/19, as well as a Univesidad Complutense de Madrid short-term mobility grant EB25/22, which made my three-month research stay in Estonia possible. Further economic support was provided by project PID2019-108655GB-I00, funded by the Spanish Ministry of Science and Innovation (MICINN).

\blankpage

\newpage
\begin{KeepFromToc}
  \tableofcontents
\end{KeepFromToc}

\chapter*{List of Publications.}
\addcontentsline{toc}{chapter}{List of Publications.}

This Thesis is based on the following Publications:
\begin{enumerate}
  \item[\cite{Casado-Turrion:2022xkl}] \textbf{A.~Casado-Turri\'on}, \'A.~de la Cruz-Dombriz and A.~Dobado,

    \textit{Is gravitational collapse possible in $f(R)$ gravity?},

    Phys. Rev. D \textbf{105} (2022) no.~8, 084060,

    \href{https://arxiv.org/abs/2202.04439}{arXiv:2202.04439} [gr-qc].

    \item[\cite{SecondPaper}] \textbf{A.~Casado-Turri\'on}, \'A.~de la Cruz-Dombriz, A.~Jiménez-Cano and F.~J.~Maldonado Torralba,

    \textit{Junction conditions in bi-scalar Poincaré gauge gravity},

    JCAP \textbf{07} (2023) 023,

    \href{https://arxiv.org/abs/2303.01206}{arXiv:2303.01206} [gr-qc].

    \item[\cite{ThirdPaper}] \textbf{A.~Casado-Turri\'on}, \'A.~de la Cruz-Dombriz and A.~Dobado,

    \textit{Physical non-viability of a wide class of metric $f(R)$ models and their constant-curvature solutions},

    Phys. Rev. D \textbf{108} (2023) no.~6, 064006,

    \href{https://arxiv.org/abs/2303.02103}{arXiv:2303.02103} [gr-qc].
\end{enumerate}

\blankpage

\thispagestyle{empty}
\listoftheorems[ignoreall,show={result,corollary},swapnumber]
\label{List of Results}

\blankpage

\thispagestyle{empty}
\listoffigures

\blankpage

\thispagestyle{empty}
\listoftables

\chapter*{Abstract.}
\addcontentsline{toc}{chapter}{Abstract.}
\markboth{Abstract.}{Abstract.}

The purpose of this Thesis is to improve our understanding of the strong-field regime of gravitational interactions, where deviations from the theory of General Relativity (GR) are expected to be found on theoretical grounds. In particular, our investigations have been concerned with the formulation and application of junction conditions (which govern stellar collapse, black-hole formation and the dynamics of more exotic compact objects), as well as with other viability criteria necessary for modified gravity theories and their solutions to be physically admissible (such as stability or the avoidance of singularities). We will concentrate on two paradigmatic modifications of GR, namely, $f(R)$ theories (specially on the metric formulation) and the ghost-free sub-class of Poincaré Gauge Gravity.

After briefly summarising the postulates, predictive successes and shortcomings of GR, as well as motivating the need for postulating modified gravity theories (either higher-order or metric-affine), we will confront the problem of gluing two given solutions of a theory together across a time-like or space-like boundary. By making use of the language of tensor distributions and the theory of hypersurfaces, we will review the consistent way of formulating junction conditions, which allow some matched configurations to be proper solutions of the theory in question. We will provide a novel insight to the subject, first by introducing a convenient notation to deal with both regular and singular parts, and then by explicitly identifying and resolving most of the subtleties appearing when deriving junction conditions.

As our first application of the junction-condition formalism, we shall study gravitational collapse and black-hole formation within the framework of $f(R)$ gravity. By means of a systematic analysis of the relevant junction conditions, we will explain how to generalise the archetypal Oppenheimer-Snyder model of GR (which accounts for the simple scenario of a collapsing spherical, uniform-density dust star) so as to allow it to fit within metric $f(R)$ theories of gravity. We have been able to prove that the presence of more stringent junction conditions in these theories entails that the space-time outside the collapsing star must be highly non-trivial and substantially different from the Schwarzschild solution one has in GR. More precisely, we have found that such exterior space-time must be non-static, endowed with two non-trivial metric functions $g_{tt} g_{rr}\neq -1$ and lack a constant scalar curvature.  We have also shown that the Oppenheimer-Snyder construction is incompatible with the Palatini formulation of $f(R)$ gravity.

Subsequently, we will present, for the first time in the literature, the exhaustive derivation of the junction conditions in the ghost-free sub-class of quadratic Poincaré Gauge Gravity. We will show that these theories allow the matching interface to host surface spin densities, as well as for energy-momentum thin shells and double layers. This singular structure at the matching surface is richer than its counterparts in GR or $f(R)$ gravity, leading to possible interesting applications, as we shall discuss. We will also compare our results to those in the existing literature, as the theory reduces to well-known cases in some particular scenarios.

Finally, we will examine a special class of metric $f(R)$ models whose equations of motion are trivially solved by space-times with a certain constant scalar curvature. Some of these models had been been previously considered in the literature, for instance, due to their success in explaining cosmological observations. We will conclude that the vast majority of these $f(R)$ models suffer, in general, from several shortcomings rendering their physical viability extremely limited, even when not directly ruled out by current observational evidence. Among these deficiencies are instabilities (including previously unforeseen strong-coupling problems) and issues limiting the predictive power of the models. Furthermore, we will also show that novel, $f(R)$-exclusive constant-curvature solutions may also exhibit a variety of unphysical properties. Most notably, those having a non-zero Ricci scalar are generically unstable, whereas those with vanishing scalar curvature require a number of additional conditions to be satisfied in order to be at least metastable. Curvature singularities, geodesic completeness, horizons and regions where the metric signature becomes unphysical are also studied for some of the novel solutions, revealing that most of them are not physically-substantiated.

\chapter*{Resumen.}
\addcontentsline{toc}{chapter}{Resumen.}
\markboth{Resumen.}{Resumen.}

El propósito de esta Tesis es mejorar nuestro entendimiento del régimen de campo fuerte de las interacciones gravitacionales, donde, desde un punto de vista teórico, se espera encontrar desviaciones de la teoría de la Relatividad General (RG). En particular, nuestras investigaciones se han centrado en la formulación y aplicación de las condiciones de pegado (que gobiernan el colapso gravitacional, la formación de agujeros negros y la dinámica de objetos compactos más exóticos), así como en otros criterios de viabilidad necesarios para que las teorías de gravedad modificada y sus soluciones sean físicamente admisibles (como la estabilidad o la evasión de singularidades). Nos concentraremos en dos modificaciones paradigmáticas de la RG, a saber, las teorías $f(R)$ (especialmente en su formulación métrica) y en la subclase libre de \emph{ghosts} de la Gravedad Gauge Poincaré.

Tras resumir brevemente los postulados, éxitos predictivos y carencias de la RG, así como motivar la necesidad de postular teorías de gravedad modificada (o bien de orden superior o bien métrico-afines), haremos frente al problema del pegado de dos soluciones dadas de una teoría a lo largo de una interfaz de tipo tiempo o espacio. Haciendo uso del lenguaje de distribuciones tensoriales y de la teoría de hipersuperficies, repasaremos la forma consistente de formular las condiciones de pegado, que permiten que algunas configuraciones pegadas sean soluciones correctas de la teoría en cuestión. Ofreceremos un novedoso enfoque sobre la materia, primero introduciendo una notación práctica para lidiar tanto con las partes regulares como con las singulares, para después identificar explícitamente y resolver la mayor parte de las sutilezas que aparecen al derivar condiciones de pegado.

Como nuestra primera aplicación del formalismo de condiciones de pegado, estudiaremos el colapso gravitacional y la formación de agujeros negros en el marco de la gravedad $f(R)$. Mediante un análisis sistemático de las condiciones de pegado pertinentes, explicaremos cómo generalizar el arquetípico modelo de Oppenheimer-Snyder de RG (que describe el escenario simple de una estrella colapsante de polvo, esférica y con densidad uniforme), para que pueda encajar dentro de las teorías de gravedad $f(R)$. Hemos sido capaces de probar que la presencia de condiciones de pegado más restrictivas en estas teorías implica que el espacio-tiempo fuera de la estrella colapsante ha de ser altamente no trivial y sustancialmente diferente de la solución de Schwarzschild que uno tiene en RG. En concreto, hemos encontrado que tal espacio-tiempo exterior debe ser no estático, contar con dos funciones de la métrica no triviales $g_{tt} g_{rr}\neq -1$ y carecer de curvatura escalar constante. También hemos demostrado que la construcción de Oppenheimer-Snyder es incompatible con la formulación de Palatini de la gravedad $f(R)$.

Subsecuentemente, presentaremos, por primera vez en la literatura, la derivación exhaustiva de las condiciones de pegado en la subclase libre de \emph{ghosts} de la Gravedad Gauge Poincaré cuadrática. Mostraremos que dichas teorías permiten que la interfaz de pegado albergue densidades de espín superficiales, además de capas finas y dobles de energía-momento. Esta estructura singular en la superficie de pegado es más rica que sus homólogas en RG o gravedad $f(R)$, dando lugar a posibles aplicaciones de gran interés, como veremos. También compararemos nuestros resultados con los de la literatura existente, ya que la teoría se reduce a casos bien conocidos en algunos casos particulares.

Finalmente, examinaremos una clase especial de modelos $f(R)$ métricos cuyas ecuaciones de movimiento son trivialmente satisfechas por aquellos espacio-tiempos con una cierta curvatura escalar constante. Algunos de estos modelos se habían considerado previamente en la literatura, por ejemplo, debido a su éxito a la hora de explicar observaciones cosmológicas. Concluiremos que la amplia mayoría de estos modelos $f(R)$ sufre, en general, de algunas carencias que hacen que su viabilidad física sea extremadamente limitada, incluso cuando dichos modelos no están directamente descartados por la evidencia observacional actual. Entre dichas deficiencias se encuentran algunas inestabilidades (incluyendo problemas de acoplamiento fuerte previamente imprevistos) y problemas que limitan el poder predictivo de los modelos. Además, también mostraremos que las nuevas soluciones de curvatura constante exclusivas de $f(R)$ pueden exhibir también variadas propiedades no físicas. De forma notoria, aquellas que tienen una curvatura Ricci distinta de cero son genéricamente inestables, mientras que las que tienen curvatura escalar nula requieren que se satisfagan varias condiciones adicionales para ser al menos metaestables. También se estudiarán las singularidades de curvatura, la completitud geodésica, los horizontes y las regiones en las cuales la signatura de la métrica se convierte en no física para algunas de las nuevas soluciones, demostrando que la mayoría de ellas no están justificadas desde un punto de vista físico.

\chapter*{Notation and conventions.}
\addcontentsline{toc}{chapter}{Notation and conventions.}
\markboth{Notation and conventions.}{Notation and conventions.}

The ensuing notations and conventions shall be followed throughout this Thesis:
\begin{coloritemize}
    \item Regarding tensorial indices, our convention will be as follows:
    \begin{itemize}
        \item We will use Greek letters to denote four-dimensional, space-time indices in a coordinate basis, i.e.~$\mu,\nu,\rho,\sigma\ldots=0,1,2,3$.
        \item Lowercase Latin letters from the beginning of the alphabet shall denote three-dimensional indices associated to a coordinate system intrinsic to a given hypersurface $\Sigma$, i.e.~$a,b,c\ldots=0,1,2$ if $\Sigma$ is time-like and $a,b,c\ldots=1,2,3$ if $\Sigma$ is space-like.
        \item Uppercase Latin letters from the beginning of the alphabet ($A,B\ldots$) will denote arbitrary tensor indices.
        \item Uppercase Latin letters from the middle of the alphabet represent Lorentzian indices related to a frame (tetrad), i.e.~$I,J,K\ldots=0,1,2,3$.
    \end{itemize}
    Repeated indices imply summation over all allowed index values, as per the Einstein summation convention. For instance,
    $T^\mu{}_\mu=T^0{}_0+T^1{}_1+T^2{}_2+T^3{}_3$.
    
    \item The sign convention to be followed herein shall be the one denoted as $(+,+,+)$ by Misner, Thorne and Wheeler \cite{Misner:1973prb}. This entails that:
    \begin{itemize}
        \item The metric signature will be $(-,+,+,+)$; in other words, the components of the flat Minkowski metric in standard Cartesian-like coordinates $(t,x,y,z)$ shall be $(\eta_{\mu\nu})=\diag(-1,+1,+1,+1)$.
        \item The Riemann tensor of the Levi-Civita connection $\LCG^\rho{}_{\mu\nu}$ will be defined as
        \begin{equation*}
            \LCR^{\rho}{}_{\sigma\mu\nu}\equiv+2\left(\partial_{[\mu|}\LCG^{\rho}{}_{\sigma|\nu]}+\LCG^{\rho}{}_{\lambda[\mu|}\LCG^{\lambda}{}_{\sigma|\nu]}\right),
        \end{equation*}
        with the associated Ricci tensor being obtained by contracting its first and third indices, i.e.~$\LCR_{\mu\nu}\equiv\LCR^{\rho}{}_{\mu\rho\nu}$.
        \item Analogously, the Riemann and Ricci tensors associated to a generic affine connection $\MAG^\rho{}_{\mu\nu}$ shall be, respectively.
        \begin{equation*}
            \MAR^{\rho}{}_{\sigma\mu\nu}\equiv+2\left(\partial_{[\mu|}\MAG^{\rho}{}_{\sigma|\nu]}+\MAG^{\rho}{}_{\lambda[\mu|}\MAG^{\lambda}{}_{\sigma|\nu]}\right)\myskip\text{and}\myskip\MAR_{\mu\nu}\equiv\MAR^{\rho}{}_{\mu\rho\nu}.
        \end{equation*}
        \item Finally, given a theory with matter Lagrangian density $\mathcal{L}_\matter$, its corresponding stress-energy tensor will be defined as
        \begin{equation*}
            \stress_{\mu\nu}\equiv-\dfrac{2}{\sqrt{-g}}\dfrac{\delta(\sqrt{-g}\mathcal{L}_\matter)}{\delta g^{\mu\nu}},
        \end{equation*}
        meaning that the Einstein field equations of General Relativity read
        \begin{equation*}
            \LCEin_{\mu\nu}\equiv\LCR_{\mu\nu}-\dfrac{1}{2}g_{\mu\nu}\LCR=\kappa\stress_{\mu\nu},
        \end{equation*}
        where $\kappa\equiv 8\pi G$. Sometimes, it will be more convenient for us to employ the so-called \emph{Plank mass} $\Mp\equiv 1/\sqrt{\kappa}$ instead of $\kappa$ itself. Notice that we are making use of natural units ($c=\hbar=1$), as we will throughout the remainder of the Thesis.
    \end{itemize}
    \item Covariant derivatives taken using the Levi-Civita connection will bear an overring, i.e.~they will be denoted as $\LCD_\mu$. Conversely, covariant derivatives with respect to an arbitrary affine connection will be marked with a hat, i.e.~\smash{$\MAD_\mu$}. The covariant derivative of a vector $V^\mu$ with respect to a generic affine connection will be
    \begin{equation*}
        \MAD_\mu V^\rho=\partial_\mu V^\rho+\MAG^\rho{}_{\mu\nu} V^\nu,
    \end{equation*}
    whereas, for a one form, $\omega_\mu$, we will have
    \begin{equation*}
        \MAD_\mu \omega_\nu=\partial_\mu \omega_\nu-\MAG^\rho{}_{\mu\nu} \omega_\rho.
    \end{equation*}
    Notice that, in both cases, the index with respect to which we are differentiating is the first covariant index of the connection. This will be relevant for computations done using torsionful connections, i.e.~if $T^\rho{}_{\mu\nu}\equiv 2\MAG^\rho{}_{[\mu\nu]}\neq 0$. The aforementioned results generalise in a straightforward way to tensors of arbitrary rank, cf.~\eqref{eq:covariant derivative generic tensor}.
\end{coloritemize}

\mainmatter

\part{Introduction.}
\label{part:introduction}

\chapter*{Preface.}
\addcontentsline{toc}{chapter}{Preface.}
\markboth{Preface.}{Preface.}
\label{preface}

The importance of gravitational interactions in shaping the entire history of the Universe cannot be underestimated. Gravity is responsible for cosmological evolution and for the origin, dynamics and fate of most celestial bodies, from planets to stars, black holes and galaxies. General Relativity (GR), while formulated more than a century ago, has remained as the primary source of our understanding of gravitation up to these days, providing us with a variety of observationally-substantiated predictions (such as the existence of gravitational waves and black holes, the gravitational bending of light, the correct value for the perihelion precession of Mercury, etc.), as well as a piercing understanding of the relationship between gravity, geometry, space and time.

However, in spite of its success in describing gravitation and all its associated phenomena, GR exhibits some major flaws, including its apparent incompatibility with Quantum Mechanics (due to its non-renormalisability), the various tensions observed between cosmological data and predictions and the necessity of \emph{ad hoc} ingredients (dark matter, dark energy, and the inflaton) to produce the correct cosmological evolution. Furthermore, although GR has been extensively tested on the so-called \emph{weak-field regime}, it is still unknown whether its predictions continue to be accurate when the gravitational field becomes intense, as expected near ultracompact objects, such as neutron stars, or in very energetic and violent processes, such as black-hole mergers.

In order to address the aforementioned limitations of GR, or to make novel predictions concerning the yet-unprobed strong-field regime of gravity, a great profusion of alternative gravity theories and classes of models therein have been considered in the literature during the last decades. Some of the most successful modified gravity theories include simple, higher-order extensions of GR, such as $f(R)$ gravity, as well as more sophisticated theories where the Riemannian geometry of space-time characteristic of Einsteinian gravity is abandoned in favour of a more general, metric-affine construction.

The discovery of gravitational waves and the related technological advances have enabled experimentalists to delve into the study of the most compact objects in the Cosmos. More accurate results regarding the strong-field regime of gravity will certainly become available in the forthcoming decades, with the steady improvement of experimental capabilities. As such, in the era of gravitational-wave astronomy, it will be possible to set stringent constraints on modified gravity theories using measurements of compact objects. For this reason, it is instrumental, from a purely theoretical viewpoint, to be prepared for the observational revolution ahead. More precisely, we must ensure that, out of all possible modifications of GR, we select those providing the most mathematically- and observationally-consistent picture of gravitational interactions. This entails that, even before comparing their new predictions with experiments, the physical viability of the various modified gravity models must be assessed in first place, so as to guarantee the robustness of their foundational hypotheses and mathematical structure, which should also be physically well-motivated. In recent times, the previous considerations have fuelled interest in examining viability criteria such as stability and linearised spectra in modified gravity.

In what concerns compact objects specifically, junction conditions constitute yet another invaluable tool when evaluating the feasibility of a certain gravitational theory. Given that compact objects are most often isolated bodies, surrounded by vacuum, junction conditions become decisive in determining their evolution and properties, as they control the evolution of the boundary surface separating the object's interior and exterior. Furthermore, junction conditions ultimately delimit which kinds of compact objects can exist within a given gravitational framework. Because of this, it will be possible to rule out theories where existing compact objects cannot be correctly modelled or junction-condition-mediated processes (such as gravitational collapse) do not proceed as observed.

For all the reasons outlined above, a theoretical examination of the junction conditions and other viability criteria in modified gravity theories will be provided in this Thesis, in the hope that their application to physically-relevant scenarios will contribute to our understanding of said processes, as well as to eventually set bounds on the parameter space of the alternatives to GR. We will focus on some of the most promising and better motivated modified gravity theories, namely, $f(R)$ theories and Poincaré Gauge Gravity (\PGG). In particular, we shall study gravitational collapse, gravitational-wave spectra, the presence of instabilities and the physical properties of novel space-times, and even endeavour to obtain junction conditions from first principles.

The Thesis is divided into three Parts. First, the present Part \ref{part:introduction} contains a brief overview of modified gravity theories (Chapter \ref{chapter:Introduction: Modified Gravity}) and junction conditions (Chapter \ref{chapter:Introduction: JCs}). In each of these two introductory Chapters, we will present the fundamental geometrical structures and quantities that lead to a correct description of gravity and of matched solutions, respectively. We shall also concentrate on the physical reasonings and motivations leading to each of our choices. Afterwards, Part \ref{part:results} features the main findings of our research, encapsulated in the form of eleven fundamental Results (see the \hyperref[List of Results]{List of Results}). More precisely, we have studied gravitational collapse in $f(R)$ gravity (Chapter \ref{chapter:f(R) collapse}), the junction conditions in the ghost-free sub-class of \PGG, dubbed Bi-Scalar \PGG~(\BSPGT) theories (Chapter \ref{chapter:BSPGT JCs}), and the physical viability of the so-called `$\LCR_0$-degenerate' $f(\LCR)$ models and constant-curvature solutions (Chapter \ref{chapter:constant curvature}), including stability analyses and a thorough examination of the properties of said solutions. Finally, most detailed calculations have been relegated to the \hyperref[Part:Appendices]{Appendices} Part, where one may also find useful formulae and supplementary discussions related to some key issues covered in this Thesis.

\chapter{Beyond General Relativity: modified gravity theories.}
\label{chapter:Introduction: Modified Gravity}

In this Chapter, we will motivate the need to surpass the standard Einsteinian gravitational paradigm, synthesised in the theory of General Relativity (GR). Founded on both theoretical and observational grounds, modified gravity theories attempt to overcome the various shortcomings of GR, while preserving its wide range of experimentally-confirmed predictions, such as the existence of gravitational waves and black holes. We will also use this opportunity to introduce the alternative theories to be considered in the remainder of this Thesis, namely, $f(R)$ and Poincaré Gauge Gravity (\PGG) theories, both of which are paradigmatic examples for the reasons to be stated below.

It is important to remark that the term `modified gravity' almost exclusively refers to \emph{classical} alternatives to GR. In other words, we will not be concerned with the quantisation of gravity (even though the use of some of the theories treated on this Thesis might be incentivised by quantum-gravity arguments). This is because some of the deficiencies of GR arise even at the classical level, as will become apparent later. Hence, a non-quantum solution may be sought after, typically by dropping any of the initial hypotheses concerning the geometry of space-time and/or the number and kind of dynamical degrees of freedom associated to gravity. Nonetheless, many immanently quantum approaches to gravity have been pursued in the last decades, such as Loop Quantum Gravity, Ho\v{r}ava-Lifshitz Gravity or String Theory.

The structure of this introductory Chapter will be the following. To start, we shall briefly review the postulates and mathematical structure of GR in Section \ref{sec:intro:GR}. Therein, we will also discuss the particular problems one attempts to solve by modifying Einsteinian gravity. Afterwards, we shall begin to gradually modify GR, first by altering its action. This will lead us to metric $f(\LCR)$ gravity, discussed in Section \ref{sec:intro:metric f(R)}. Then, by requiring the connection to be independent from the metric, we will obtain Palatini $f(\MAR)$ gravity in Section \ref{sec:intro:Palatini f(R)}. After a brief stop on generic metric-affine theories in Section \ref{sec:intro:MAG}, we will move on to \PGG, a particular example thereof intended to bear a likeness to the field theories governing the remaining fundamental interactions found in Nature, in Section \ref{sec:intro:Poincaré Gauge Gravity}. More precisely, we shall concentrate on the quadratic, ghost-free subset of \PGG, dubbed Bi-Scalar \PGG~(\BSPGT), to which Subsection \ref{sec:bspgt} is devoted.

\section{General Relativity.} \label{sec:intro:GR}

Einstein laid the foundations of Special Relativity (SR) in two of his seminal works of 1905 \cite{Einstein:1905ve,Einstein:1905emc2}, his \emph{annus mirabilis}. SR successfully accomplished the task it was intended to fulfil: reconciling Electrodynamics with Classical Mechanics. Though incompatible with Galilean transformations, Maxwell's equations were covariant under the Lorentz transformations of SR, which, in turn, were devised to implement the experimentally-corroborated fact that the speed of light has a finite value which remains the same in all reference frames. Thus, the speed of light constitutes a fundamental upper bound on the velocity of any moving body, giving rise to the special-relativistic notion of causality. In spite of this, it soon became apparent that SR exhibited one major flaw, namely, that it could not be consistently employed to describe gravity (at the time, the only other known fundamental interaction, aside from electromagnetism). Hence, a decade-long, tireless endeavour to describe gravitational interactions in relativistic settings began, led by Einstein himself, which concluded with the formulation and subsequent observational validation of his theory of General Relativity (GR).

\subsection{Differential geometry and gravity: Einstein's road to GR.}
\label{sec:GR Geometry}

Einstein's most crucial observation in his quest for GR was the realisation that light---and, as a matter of fact, any form of energy---gravitates, as entailed by the so-called \emph{Einstein equivalence principle} (EEP), which derives directly from the earlier \emph{weak equivalence principle} (WEP) and the notion of \emph{mass-energy equivalence} posited in SR.
The WEP is the equality between the inertial and gravitational masses of any body,\footnote{
    In the context of Newtonian dynamics, the \emph{inertial mass} $m_\mathrm{i}$ of a body is defined through Newton's second law, $\mathbf{F}=m_\mathrm{i}\mathbf{a}$, and may thus be regarded as a measure of the body's resistance to acceleration. On the other hand, the \emph{gravitational mass} $m_\mathrm{g}$ of a body is the proportionality constant relating the gravitational force $\mathbf{F}_\mathrm{g}$ felt by the object with the gradient of the Newtonian gravitational potential $\Phi$; mathematically, $\mathbf{F}_\mathrm{g}=-m_\mathrm{g}\nabla\Phi$. Consequently, the equality $m_\mathrm{i}=m_\mathrm{g}$ (i.e.~the WEP) implies that all objects experience the same acceleration due to gravity, regardless of their mass or internal composition.
} known since the times of Galileo and experimentally confirmed today to extraordinarily high precision.\footnote{
    As of 2023, the latest experimental results available concerning the validity of the WEP are those of the MICROSCOPE Mission \cite{MICROSCOPE:2022doy}. According to these empirical findings, the WEP holds to a precision of one part in $10^{15}$. 
} Due to the WEP, uniform acceleration is indistinguishable from a uniform gravitational field. Since any gravitational field can always be approximated to be uniform \emph{locally} (i.e.~in small-enough space-time regions), there is a preferred class of observers who are unable to perceive any gravitational field whatsoever in theories satisfying the WEP. These observers are said to be \emph{locally-inertial} or \emph{freely falling}. The EEP then states that the results of all local, non-gravitational experiments carried out by freely-falling observers agree with the predictions made using special-relativistic dynamics, where the effects of gravity are ignored \cite{Einstein1907}. As such, \emph{locally}, space-time must resemble the Minkowski metric of SR in theories satisfying the EEP.

Having established the EEP, Einstein came up to the conclusion that light gravitates by means of the following thought experiment \cite{EinsteinBook}. He envisioned a freely falling rocket\footnote{
    Evidently, spacecrafts capable of carrying humans were science fiction in Einstein's days; here, we are adapting the thought experiment to contemporary audiences for pedagogical reasons.
} near the surface of the Earth (so that the gravitational field can be taken to be uniform). In accordance with the EEP, an experimenter turning a lantern on inside the rocket will observe that light rays follow straight trajectories, exactly as they would in Minkowski space-time. However, another observer at ground control would regard the spacecraft as descending towards Earth with acceleration $g_\oplus\simeq 9.8$ $\mathrm{m}/\mathrm{s}^{2}$, as predicted by Newton's law of gravitation. Therefore, for this second observer, the trajectories followed by light rays inside the rocket are no longer straight, but \emph{curved}, as they fall solidarily with the spaceship. Einstein was then able to extract three main conclusions from this simple reasoning:
\begin{coloritemize}
    \item First, photons see their trajectories changed by gravity; therefore, though massless, they interact gravitationally. This was to be expected from the equivalence between mass and energy; all forms of energy should gravitate.
    \item Second, in theories satisfying the EEP, gravitational phenomena must be entirely due to intrinsic properties of space-time, as implied by the universality of free fall and the fact that local experiments cannot reveal the presence of a non-trivial gravitational field.
    \item Last, but not least, the bending of freely-falling light rays hints at the possibility that space-time itself can be curved, since light-like trajectories entirely determine the causal structure of space-time (nothing can move faster than light in any reference frame). This intuition appears to be confirmed by another thought experiment \cite{Einstein1911}. If one imagines a light ray climbing up a weak gravitational field $\Phi(x,y,z)\ll 1$, one finds that it will be subject to a gravitationally-induced redshift
    \begin{equation}
        z=\dfrac{\lambda_\mathrm{obs}}{\lambda_\mathrm{emi}}-1\simeq\Delta\Phi,
    \end{equation}
    where $\lambda_\mathrm{emi}$ is the wavelength of the emitted photons, $\lambda_\mathrm{obs}$ is the wavelength of the detected light, and $\Delta\Phi$ is the difference between the Newtonian gravitational potentials at the locations of the source and the receptor. This is exactly the same result one would obtain if space-time were not given by the flat Minkowski metric, but by the slightly curved line element
    \begin{equation} \label{eq:weak field metric}
        \dif s^2=-(1+2\Phi)\dif t^2+(1-2\Phi)(\dif x^2+\dif y^2+\dif z^2).
    \end{equation}
\end{coloritemize}

Mentored by his former classmate and Mathematics professor Marcel Grossmann, who was an expert on differential geometry, Einstein concluded that \emph{(pseudo)-Riemannian geometry} was the appropriate mathematical framework to describe gravity in relativistic contexts, due to the large similarities between this mathematical construction and the physically-motivated insights explained above \cite{EinsteinGrossmann}. For instance, the EEP naturally emerges if one considers space-time to be a four-dimensional \emph{differentiable manifold} $\mathcal{M}$,\footnote{
    An $n$-dimensional differentiable manifold $\mathcal{M}$ is a topological Hausdorff space such that (i) every point $p\in\mathcal{M}$ has an open neighbourhood which is homeomorphic to an open subset of $\mathbb{R}^n$, and (ii) $\mathcal{M}$ possesses a $C^\infty$ \emph{atlas}, namely, a set of \emph{coordinate charts} (continuous homomorfisms relating open subsets $\mathcal{U}_i\in\mathcal{M}$ with open subsets of $\mathbb{R}^n$) which are \emph{compatible} (i.e.~the \emph{transition functions} relating overlapping charts are of class $C^\infty$) and cover all $\mathcal{M}$ (i.e.~$\cup_i\mathcal{U}_i=\mathcal{M}$).
} filled with \emph{tensor fields}\footnote{
    A tensor field of type $(p,q)$ on $\mathcal{M}$ might be understood as any object $F^{\mu_1\ldots\mu_p}{}_{\nu_1\ldots\nu_q}$ transforming as
    \begin{equation*}
        F^{\alpha_1\ldots\alpha_p}{}_{\beta_1\ldots\beta_q}=\frac{\partial x^{\alpha_1}}{\partial x^{\mu_1}}\ldots\frac{\partial x^{\alpha_p}}{\partial x^{\mu_p}} \frac{\partial x^{\nu_1}}{\partial x^{\beta_1}}\ldots\frac{\partial x^{\nu_q}}{\partial x^{\beta_q}} F^{\mu_1\ldots\mu_p}{}_{\nu_1\ldots\nu_q}\myhugeskip\text{[continues on the next page]}
    \end{equation*}
    under general coordinate transformations $x^\mu=x^\mu(x^\alpha)$. Tensors of type $(0,0)$ are known as \emph{scalars}, those of type $(1,0)$ are known as \emph{vectors} while those of the $(0,1)$ kind are known as \emph{one-forms}. A more rigorous definition of tensor (which we shall omit for the sake of brevity) might be found elsewhere, e.g.~\cite{Carroll:2004st}.
} representing physical quantities. When equipped with a metric tensor $g_{\mu\nu}$ of Lorentzian signature $(-,+,+,+)$, any sufficiently-small region in $\mathcal{M}$ may be well approximated by Minkowski space-time, as per the definition of manifold.\footnote{
    Note that Minkowski spacetime is just $\mathbb{R}^4$ endowed with a metric of Lorentzian signature.
}
Moreover, if, in analogy with SR, one requires free massive test particles on a given space-time $(\mathcal{M},g_{\mu\nu})$ to move along affinely-parametrised time-like paths $x^\mu(\lambda)$ extremising their proper time
\begin{equation}
    \tau=\int\dif\lambda\left|g_{\mu\nu}\dfrac{\dif x^\mu}{\dif\lambda}\dfrac{\dif x^\nu}{\dif\lambda}\right|^{1/2},
\end{equation}
then said particles may be shown to move according to the \emph{geodesic equation},
\begin{equation} \label{eq:LC geodesic eq}
    \dfrac{\dif^2 x^\rho}{\dif\lambda^2}+
    \dfrac{1}{2}g^{\rho\sigma}(\partial_\mu g_{\sigma\nu}+\partial_\nu g_{\mu\sigma}-\partial_\sigma g_{\mu\nu})
    \dfrac{\dif x^\mu}{\dif\lambda}\dfrac{\dif x^\nu}{\dif\lambda}=0.
\end{equation}
Similarly, instead of time-like geodesics, free massless particles are compelled to follow affinely-parametrised null paths $x^\mu(\lambda)$ satisfying \eqref{eq:LC geodesic eq}. In Minkowski space-time, all geodesics are straight lines. This will no longer be the case for a generic metric $g_{\mu\nu}$, due to the second term in the geodesic equation. However, in spite of its appearance, the aforementioned second term in \eqref{eq:LC geodesic eq} is \emph{not} a tensor. In fact, it can be shown that, associated to every point $p$ in $\mathcal{M}$, there exist coordinates (known as \emph{normal coordinates} at $p$) such that $\smash{g_{\mu\nu}|_p\overset{*}{=}\eta_{\mu\nu}}$ and $\smash{\partial_\rho g_{\mu\nu}|_p\overset{*}{=}0}$ (where the symbol `$\smash{\overset{*}{=}}$' implies that the equality only holds in this particular coordinate system). Thus, when using normal coordinates, the second term in \eqref{eq:LC geodesic eq} vanishes, and the solutions of \eqref{eq:LC geodesic eq}---the equation of motion of free test particles in $(\mathcal{M},g_{\mu\nu})$---reduce to straight lines, as in SR, in agreement with the EEP. Compare this situation with the case in which the particle is also subject to an electromagnetic force. In such scenario, we have to include a term corresponding to the electromagnetic Lorentz force on the right-hand side of \eqref{eq:LC geodesic eq}:
\begin{equation} \label{eq:LC plus EM}
    \dfrac{\dif^2 x^\rho}{\dif\lambda^2}+
    \dfrac{1}{2}g^{\rho\sigma}(\partial_\mu g_{\sigma\nu}+\partial_\nu g_{\mu\sigma}-\partial_\sigma g_{\mu\nu})
    \dfrac{\dif x^\mu}{\dif\lambda}\dfrac{\dif x^\nu}{\dif\lambda}=\dfrac{q}{m} F^\rho{}_\sigma\dfrac{\dif x^\sigma}{\dif\lambda},
\end{equation}
where $q$ is the particle's electric charge, $m\neq 0$ is its mass and $F_{\mu\nu}$ is the electromagnetic field tensor. If a tensor identically vanishes in one coordinate system, then it must necessarily do so in \emph{any} other one. Hence, unlike the second term on the left-hand side, the electromagnetic force is not a mere product of inertia, and cannot be removed with a particular choice of coordinates (notice as well that the Lorentz force depends on the internal properties of the particle, namely on its charge and mass, while gravity does not). In summary, in Einstein's gravity, the second term in \eqref{eq:LC geodesic eq} encapsulates the effect of gravitation on the trajectories of test particles, while normal coordinates constitute the mathematical embodiment of freely-falling frames in which gravity can be locally neglected (in contrast to the Lorentz force, which is necessarily non-zero in every frame provided that the particle is electrically charged).

Equation \eqref{eq:LC geodesic eq} also conceals the fact that, in the Einsteinian conception of gravity, gravitation might be attributed to \emph{curvature}, as suggested by the bending of light rays and the existence of gravitationally-induced redshift. In particular, in GR, gravity is due to the curvature of the \emph{Levi-Civita connection}. Hence, at this point, it becomes imperative for us to introduce the related concepts of \emph{covariant differentiation} and \emph{connection}. With the exception of Minkowski, partial derivatives in a generic space-time $(\mathcal{M},g_{\mu\nu})$ do \emph{not} transform as tensors under general coordinate transformations.\footnote{
    The ultimate reason for this is that tensor fields evaluated at different points in $\mathcal{M}$ actually belong to different spaces. Hence, one cannot compare tensors at two different space-time points unless one devises a procedure to \emph{transport} tensors between said points. In particular, one is interested in \emph{parallel transport} of vectors along curves. This requires the introduction of an \emph{affine connection} between space-time points (as we are about to do in the text). The term \emph{affine} refers precisely to the preservation of parallelism.
} However, one might introduce an object $\MAG^\rho{}_{\mu\nu}$, dubbed affine connection, which transforms in such a way that, if $F^{\mu_1\mu_2\ldots\mu_p}{}_{\nu_1\nu_2\ldots\nu_q}$ is a tensor of rank $(p,q)$, then
\begin{align}
    & \MAD_\rho F^{\mu_1\mu_2\ldots\mu_p}{}_{\nu_1\nu_2\ldots\nu_q}= \partial_\rho F^{\mu_1\mu_2\ldots\mu_p}{}_{\nu_1\nu_2\ldots\nu_q} \nonumber \\
    &\quad\quad\quad\quad\quad +\MAG^{\mu_1}{}_{\rho\sigma} F^{\sigma\mu_2\ldots\mu_p}{}_{\nu_1\nu_2\ldots\nu_q}+\MAG^{\mu_2}{}_{\rho\sigma} F^{\mu_1\sigma\ldots\mu_p}{}_{\nu_1\nu_2\ldots\nu_q}+\ldots+\MAG^{\mu_p}{}_{\rho\sigma} F^{\mu_1\mu_2\ldots\sigma}{}_{\nu_1\nu_2\ldots\nu_q} \nonumber \\
    &\quad\quad\quad\quad\quad -\MAG^{\sigma}{}_{\rho\nu_1} F^{\mu_1\mu_2\ldots\mu_p}{}_{\sigma\nu_2\ldots\nu_q}-\MAG^{\sigma}{}_{\rho\nu_2} F^{\mu_1\mu_2\ldots\mu_p}{}_{\nu_1\sigma\ldots\nu_q}-\ldots-\MAG^{\sigma}{}_{\rho\nu_q} F^{\mu_1\mu_2\ldots\mu_p}{}_{\nu_1\nu_2\ldots\sigma} \label{eq:covariant derivative generic tensor}
\end{align}
transforms as a $(p,q+1)$ tensor.\footnote{
    For this to happen, the connection must transform as
    \begin{equation*}
        \MAG^{\lambda}{}_{\alpha\beta}=\frac{\partial x^\lambda}{\partial x^\rho}\frac{\partial x^\mu}{\partial x^\alpha}\frac{\partial x^\nu}{\partial x^\beta}\MAG^\rho{}_{\mu\nu}-\frac{\partial^2 x^\lambda}{\partial x^\mu \partial x^\nu}\frac{\partial x^\mu}{\partial x^\alpha}\frac{\partial x^\nu}{\partial x^\beta}
    \end{equation*}
    under general coordinate transformations.
} From \eqref{eq:covariant derivative generic tensor} it is clear that $\MAD$ is a linear differential operator satisfying Leibniz's product rule, and hence constitutes a covariant derivative on $(\mathcal{M},g_{\mu\nu},\MAG^\rho{}_{\mu\nu})$.

\begin{figure}
     \centering
     \begin{subfigure}[b]{0.3\textwidth}
         \centering
         \includegraphics[width=\textwidth]{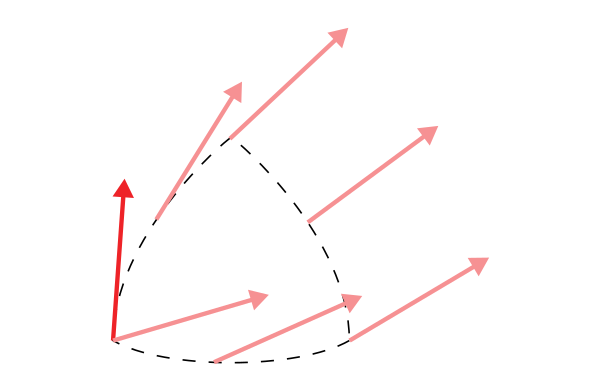}
         \caption{$\MAR^\rho{}_{\sigma\mu\nu}\neq 0$.}
         \label{fig:Curvature}
     \end{subfigure}
     \hfill
     \begin{subfigure}[b]{0.3\textwidth}
         \centering
         \includegraphics[width=\textwidth]{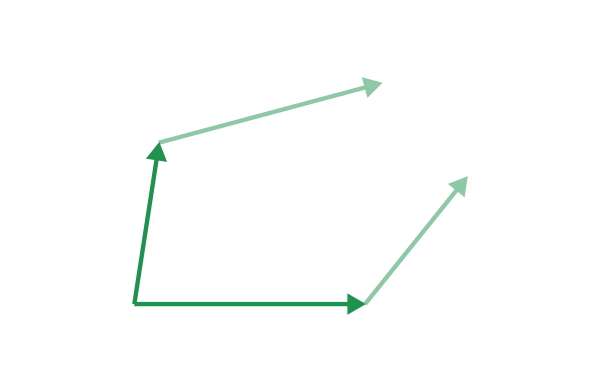}
         \caption{$T^\rho{}_{\mu\nu}\neq 0$.}
         \label{fig:Torsion}
     \end{subfigure}
     \hfill
     \begin{subfigure}[b]{0.3\textwidth}
         \centering
         \includegraphics[width=\textwidth]{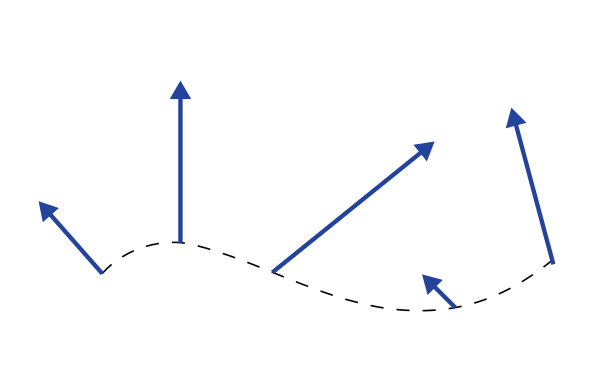}
         \caption{$Q_{\rho\mu\nu}\neq 0$.}
         \label{fig:NonMetricity}
     \end{subfigure}
        \caption{Pictorial representation of the effects of curvature, torsion and non-metricity in the parallel transport of vectors.}
        \label{fig:CurvatureTorsionNonMetricity}
\end{figure}

There is an infinite number of distinct connections one can define in a given space-time, with each of them giving rise to a different covariant derivative. In general, every connection in a manifold with metric is characterised by three tensorial quantities \cite{Bahamonde:2021gfp}:
\begin{coloritemize}
    \item \emph{Curvature}, mathematically expressed through the \emph{Riemann tensor},
    \begin{equation} \label{eq:def:MA Riemann}
        \MAR^{\rho}{}_{\sigma\mu\nu}\equiv 2\left(\partial_{[\mu|}\MAG^{\rho}{}_{\sigma|\nu]}+\MAG^{\rho}{}_{\lambda[\mu|}\MAG^{\lambda}{}_{\sigma|\nu]}\right).
    \end{equation}
    Geometrically, the presence of curvature entails that a vector's components will change when parallely transported along a closed curve.
    \item \emph{Torsion}, mathematically expressed through the \emph{torsion tensor},
    \begin{equation} \label{eq:def:torsion}
        T^\rho{}_{\mu\nu}\equiv 2\MAG^\rho{}_{[\mu\nu]},
    \end{equation}
    that is to say, the anti-symmetric part of the connection.\footnote{
        Notice that the difference between two connections transforms as a tensor.
    } Geometrically, infinitesimal parallelograms do not close in the presence of non-zero torsion.
    \item \emph{Non-metricity}, mathematically expressed though the \emph{non-metricity tensor},
    \begin{equation} \label{eq:def:non-metricity}
        Q_{\rho\mu\nu}\equiv\MAD_\rho g_{\mu\nu}.
    \end{equation}
    Geometrically, the consequence of a non-vanishing non-metricity is that vector transport changes lengths and angles.
\end{coloritemize}
A graphical depiction of the effects of curvature, torsion and non-metricity in the parallel transport of vectors is provided in Figure \ref{fig:CurvatureTorsionNonMetricity}.

It turns out that there exists only one connection which is simultaneously \emph{torsionless} ($T^\rho{}_{\mu\nu}=0$) and \emph{metric-compatible} ($Q_{\rho\mu\nu}=0$), the so-called \emph{Levi-Civita} or \emph{Riemannian connection}, whose components---often referred to as the \emph{Christoffel symbols}---are given in terms of the metric tensor through
\begin{equation} \label{eq:def:Christoffels}
    \LCG^{\rho}{}_{\mu\nu}=\dfrac{1}{2}g^{\rho\sigma}(\partial_\mu g_{\nu\sigma}+\partial_\nu g_{\mu\sigma}-\partial_\sigma g_{\mu\nu}).
\end{equation}
The Levi-Civita connection is automatically defined in every differentiable manifold endowed with a metric tensor (i.e.~in any space-time). In general, any other affine connection can be decomposed as
\begin{equation} \label{eq:MA connection decomposition}
    \MAG^\rho{}_{\mu\nu}= \LCG^\rho{}_{\mu\nu}+K^\rho{}_{\mu\nu}+L^\rho{}_{\mu\nu},
\end{equation}
where, apart from the Christoffel symbols \eqref{eq:def:Christoffels} associated to the Levi-Civita connection, we also have the \emph{contorsion tensor},
\begin{equation}
	K^\rho{}_{\mu\nu}\equiv-\dfrac{1}{2}g^{\rho\sigma}(T_{\mu\nu\sigma}+T_{\nu\mu\sigma}-T_{\sigma\mu\nu})
\end{equation}
and the \emph{disformation tensor},
\begin{equation}
	L^\rho{}_{\mu\nu}\equiv-\dfrac{1}{2}g^{\rho\sigma}(Q_{\mu\nu\sigma}+Q_{\nu\mu\sigma}-Q_{\sigma\mu\nu})
\end{equation}
Note the similarity between these expressions and \eqref{eq:def:Christoffels}.

Direct inspection of the geodesic equation \eqref{eq:LC geodesic eq} immediately reveals that it can be rewritten in terms of the Levi-Civita connection components $\LCG^{\rho}{}_{\mu\nu}$ as
\begin{equation}
    \dfrac{\dif^2 x^\rho}{\dif\lambda^2}+
    \LCG^\rho{}_{\mu\nu}
    \dfrac{\dif x^\mu}{\dif\lambda}\dfrac{\dif x^\nu}{\dif\lambda}=0.
\end{equation}
This led Einstein to believe that gravity was solely due to curvature, as stated before. The torsionless and metric-compatible geometry employed in GR is known as \emph{(pseudo)-Riemannian},\footnote{
    The prefix \emph{pseudo-} in \emph{pseudo-Riemannian} emphasises that the signature of space-time is Lorentzian instead of Euclidean; nonetheless, it is often dropped in the literature, and thus the geometry of space-time is referred to as simply as \emph{Riemannian}.
} after mathematician Bernhard Riemann, who pioneered its development starting in 1868 \cite{Riemann}.

The Riemann tensor associated to the Levi-Civita connection, which we shall denote
\begin{equation} \label{eq:def:LC Riemann}
    \LCR^{\rho}{}_{\sigma\mu\nu}\equiv 2\left(\partial_{[\mu|}\LCG^{\rho}{}_{\sigma|\nu]}+\LCG^{\rho}{}_{\lambda[\mu|}\LCG^{\lambda}{}_{\sigma|\nu]}\right),
\end{equation}
is also a function of the metric only, given that $\LCG^{\rho}{}_{\mu\nu}=\LCG^{\rho}{}_{\mu\nu}(g_{\alpha\beta})$. As a result, $\LCR^{\rho}{}_{\sigma\mu\nu}$ is said to represent the curvature of \emph{space-time itself}, whereas, for a non-Levi-Civita connection $\MAG^{\rho}{}_{\mu\nu}$, $\MAR^{\rho}{}_{\sigma\mu\nu}$ is the curvature \emph{of said connection}. Due to its symmetry properties, the Riemann tensor of the Levi-Civita connection has only one independent, non-vanishing contraction, the so-called \emph{Ricci-tensor},
\begin{equation} \label{eq:def:LC Ricci}
    \LCR_{\mu\nu}\equiv\LCR^{\rho}{}_{\mu\rho\nu}.
\end{equation}
Using the metric, one can further contract the Ricci tensor, obtaining the \emph{Ricci scalar},
\begin{equation} \label{eq:def:LC scalar}
    \LCR\equiv g^{\mu\nu}\LCR_{\mu\nu},
\end{equation}
also known as the \emph{scalar curvature}. It is clear, as per \eqref{eq:def:LC Riemann}, that $\LCR^{\rho}{}_{\sigma\mu\nu}$ and its contractions contain second-order derivatives of the metric tensor. Therefore, if $g_{\mu\nu}$ is taken to be the fundamental dynamical field of the theory of gravitation, then it is reasonable to assume that GR's equations of motion---which should be second-order differential equations for the metric---are constructed using the Levi-Civita Riemann tensor and its contractions. In particular postulating that the matter-energy content of space-time, i.e.~the source of gravity, is well described using the \emph{stress-energy tensor} $\stress_{\mu\nu}$, then the equations of motion for GR will be of the form $\LCEin_{\mu\nu}=\kappa\stress_{\mu\nu}$, where $\LCEin_{\mu\nu}$ is a symmetric, type-$(0,2)$ tensor built from $\LCR^{\rho}{}_{\sigma\mu\nu}$, and $\kappa$ is a constant with units of inverse mass squared. Considering that energy conservation ($\LCD_\mu\stress^{\mu\nu}=0$) is a fundamental physical tenet to be upheld in GR on the basis of empirical evidence, one is then forced to demand $\LCEin_{\mu\nu}$ to be covariantly conserved ($\LCD_\mu\LCEin^{\mu\nu}=0$). Through trial and error, Einstein noticed that
\begin{equation} \label{def:Einstein tensor}
    \LCEin_{\mu\nu}=\LCR_{\mu\nu}-\dfrac{1}{2}g_{\mu\nu}\LCR,
\end{equation}
nowadays known as the \emph{Einstein tensor}, is indeed a symmetric, covariantly-conserved, rank-$(0,2)$ tensor containing, at most, second-order derivatives of the metric, and thus fulfils all the desired requirements stated above. As such, he ended up postulating the following field equations for his theory of GR in 1915:
\begin{equation} \label{Einstein equations}
    \LCEin_{\mu\nu}=\kappa\stress_{\mu\nu}.
\end{equation}
Agreement with Newtonian gravity in the non-relativistic weak-field limit \eqref{eq:weak field metric} fixes $\kappa=8\pi G=\Mp^{-2}$ (in natural units where $c=\hbar=1$). Two years later, in 1917, Einstein modified his field equations \eqref{Einstein equations} through the inclusion of a new term,
\begin{equation} \label{GR + Lambda EOM}
    \LCEin_{\mu\nu}+ g_{\mu\nu} \Lambda=\kappa\stress_{\mu\nu}.
\end{equation}
where $\Lambda$ is the so-called \emph{cosmological constant}. The purpose of this new term was to prevent the Universe from expanding in cosmological scenarios \cite{EinsteinLambda}. Einstein eventually relinquished the cosmological-constant term---in his own words, his `biggest blunder'---after Edwin Hubble discovered that the Universe was, in fact, expanding, as suggested by empirical observations of a linear relationship between galaxy distances and recession speeds \cite{Hubble:1929ig}. Nowadays, the cosmological constant is an integral part of the consensus cosmological model ($\Lambda$CDM), as the simplest possible explanation of the acceleration of cosmic expansion, first observed in 1998 by the Supernova Search Team \cite{SupernovaSearchTeam:1998fmf} and the Supernova Cosmology Project \cite{SupernovaCosmologyProject:1998vns} collaborations.

\subsection{The predictive power of GR.} \label{sec:GR predictions}

Ever since 1915, the year Einstein presented his gravitational field equations \eqref{Einstein equations} before the Prussian Academy of Sciences, GR's predictions have been widely verified through numerous experiments. Some of the main achievements of GR are the following:
\begin{coloritemize}
    \item GR predicts the observed value for the perihelion precession of Mercury's orbit, which deviates from the Newtonian estimate by $(\Delta\varphi/\Delta t)_\mercury\simeq 43$ arcsec$/$century. This was one of Einstein's first computations using GR \cite{Einstein:1916vd}, and is considered one of the early tests of the theory leading to its widespread acceptance.
    \item The second classical test of GR is the prediction that the Sun is capable of deflecting light rays coming from distant stars by about $(\Delta\varphi)_\sun\simeq 1.74$ arcsec. Several expeditions---led by Sir Arthur Eddington, among others---corroborated this effect through observations of the solar eclipse of May 29, 1919 \cite{Dyson:1920cwa}.
    \item Gravitationally-induced redshift, another of the foundational predictions of GR, has been directly measured in a broad range of settings, such as the renowned Pound-Rebka experiment \cite{Pound:1960zz}. Nowadays, this effect has even become relevant in everyday life as well, as the proper functioning of GPS devices---which receive information from orbiting satellites---requires Earth's gravitational redshift to be taken into account \cite{Ashby:2003vja}.
    \item Other observed consequences of GR are the Shapiro time delay, occurring when radar pulses reflected by the inner planets pass near the Sun \cite{Shapiro:1964uw}, and gravitational lensing, which is the formation of multiple images corresponding to a single object when Earth-bound light rays pass near a massive body on their way to our planet. The Shapiro delay was first measured by Shapiro himself, along with other collaborators, in 1969 \cite{Shapiro:1964uw}, while the first gravitationally-lensed object---namely, quasar Q0957$+$561---was discovered by Walsh, Carswell and Weymann in 1979 \cite{Walsh:1979nx}.
    \item Very early on in the history of GR, Einstein was able to show that his theory admits gravitational-wave solutions \cite{Einstein:1916cc,Einstein:1918btx}, in analogy with the electromagnetic waves of Electrodynamics. In fact, relativistic sources, such as black-hole or neutron-star mergers, can emit gravitational radiation. The first indirect confirmation of the existence of gravitational-wave emission came from the observation of its effect on the period of the Hulse-Taylor binary pulsar \cite{Hulse:1974eb,Taylor:1979zz}. However, the weakness of gravitational-wave signals impeded its direct detection until the construction of high-precision laser interferometers became viable. The first direct observation of gravitational waves was carried out by the LIGO and VIRGO collaborations in 2015 \cite{LIGOScientific:2016aoc,LIGOScientific:2016lio}, paving the way towards a new era of \emph{Gravitational-Wave Astronomy}.
    \item Finally, GR predicts the existence of black holes, massive, compact objects whose gravitational pull cannot be avoided, not even by light. The first---and most renowned---black hole solution of GR was discovered by Karl Schwarzschild in 1915 \cite{Schwarzschild:1916uq}. The Schwarzschild black hole\footnote{
        As widely known, the Schwarzschild solution also represents the gravitational field outside any spherically-symmetric matter source (such as the Sun, to good approximation) in GR, by virtue of Birkhoff's theorem \cite{Birkhoff}. In fact, some of the classical predictions of GR presented above (such as perihelion precession or solar light-bending) are obtained by assuming a Schwarzschild background.
    } was subsequently generalised to include electric charge (Reissner-Nordström space-time) \cite{Reissner, Nordstrom}, rotation (Kerr space-time) \cite{Kerr:1963ud}, or both (Kerr-Newmann space-time) \cite{Newman:1965my}. While observations of stellar motions around galactic centres \cite{Kormendy:1995er} and gravitational-wave observations provide indirect confirmation of the existence of electrically-neutral, rotating black holes, the first direct observation was reported by the Event Horizon Telescope in 2019 \cite{EventHorizonTelescope:2019dse}.
\end{coloritemize}

In spite of the aforementioned successes, GR is not exempt from shortcomings, both on observational and theoretical grounds. Among others, the following compelling arguments suggest that generalisations of GR ought to be considered:
\begin{coloritemize}
    \item First, in order to successfully describe astrophysical and cosmological observations within the framework of GR, it is necessary to introduce \emph{ad hoc} components in the stress-energy tensor of the Universe, namely, \emph{dark matter}, to explain the measured rotation curves of galaxies \cite{Zwicky:1933gu}, and \emph{dark energy}, to explain the late-time expansion of the Universe \cite{Peebles:2002gy}. The standard cosmological model ($\Lambda$CDM) further demands a period of exponential expansion at early times, dubbed \emph{inflation} \cite{Guth:1980zm}. Accounting for inflation using GR requires positing the existence of additional fields capable of driving such period of rapid cosmic expansion. The impossibility of describing dark matter, dark energy or inflation using the Standard Model of Particle Physics and GR motivates the search of alternative gravity theories where these phenomena could arise naturally.
    \item Moreover, with the advent of more precise cosmological measurements, tensions within the available data are presently emerging \cite{Abdalla:2022yfr}, possibly hinting at the need of overcoming the consensus cosmological model or even GR in its entirety.
    \item It is also important to remark that GR has only been tested in the \emph{weak-field regime}, for instance, in the Solar System, as we have previously seen. Owing to the recent developments in gravitational-wave experiments, exploring the gravitational field generated by very compact objects (such as neutron stars) or by highly energetic processes (such as black-hole mergers) is starting to be feasible in our days and the decades to come. In these scenarios, higher-order corrections to GR (which would be suppressed in the weak-field regime) could potentially become relevant and leave imprints on observations. Hence, the possibility of probing the \emph{strong-gravitational-field} regime using gravitational waves could facilitate the discovery of deviations from GR---for a review, see \cite{Barack:2018yly}.
    \item From a more theoretical point of view, GR is capable of revealing its own domain of validity, as several theorems allow one to predict the appearance of space-time singularities \cite{Senovilla:1998oua}.\footnote{
        A space-time is said to contain singularities if it is \emph{geodesically incomplete}, that is, if some of its geodesics are inextendible past some fine value of the affine parameter. Before Penrose's pioneering work on singularities \cite{Penrose:1964wq}, where he introduced the notion of geodesic completeness and proved the first singularity theorem, a space-time was considered to be singular if it contained \emph{curvature singularities}, i.e.~points or regions where the curvature scalars blow up. However, curvature singularities can be considered to be admissible if no causal observer has access to them.
    } For instance, it is well known that black holes and the Friedmann-Lemaître-Robertson-Walker (FLRW) space-time employed in standard Cosmology feature such singularities \cite{Penrose:1964wq,Hawking:1966jv}. Examples of higher-order theories where gravitational (as well as electromagnetic) singularities can be resolved have been discussed in the literature; see, for instance, \cite[Ch. 2]{Gil:2022ubv} and references therein.
    \item In addition, as we shall discuss later, spinor fields---which model fundamental matter fields---are not compatible with GR unless additional structure, the so-called \emph{spin connection}, is introduced (see Sections \ref{sec:intro:MAG} and \ref{sec:intro:Poincaré Gauge Gravity}). More precisely, the presence of fermionic fields in the matter content forces the total connection to feature non-trivial torsion.
    \item Finally, GR is known to have a troubled relationship with Quantum Mechanics. While it is possible to quantise fields on curved backgrounds \cite{Birrell:1982ix}, GR itself is \emph{non-renormalisable} \cite{tHooft:1974toh} as a quantum field theory.\footnote{
        Quantum Field Theory in Curved Space-Times also leads to interesting predictions adding to the apparent incompatibility between GR and Quantum Mechanics, for instance, the discovery by Hawking that black holes radiate, and thus eventually evaporate \cite{Hawking:1974rv,Hawking:1975vcx}. Since nothing can escape from a black hole while it has not evanesced, the mere possibility that black holes can disappear entirely while leaving behind a mere thermal ensemble of particles leads to further tensions between GR and the unitary evolution found in standard Quantum Theory. This is the so-called \emph{black-hole information paradox}; for a review, see \cite{Unruh:2017uaw}.
    } Since Nature is quantum at a fundamental level, GR can only be regarded as a low-energy, effective description of gravity. One could devise modified gravity theories with improved renormalisability properties \cite{Stelle:1976gc}.
\end{coloritemize}

\subsection{Field-theoretical aspects of GR. Perspectives for generalisation.} \label{sec:GR field theory}

The weaknesses of GR we have just summarised suggest that the Einsteinian paradigm needs to be surpassed by a more general theory capable of resolving most of them, if not all, while preserving the numerous successful predictions of the theory. As mentioned before, in this Thesis, we shall be only be concerned with classical modifications of GR, also known as `modified gravity' theories. Thus, having already understood the physical arguments leading to Einsteinian gravity, as well as its fundamental building blocks, the first step towards generalising GR consists in understanding its main features as a classical, relativistic field theory. This will allow us to better discern what can and cannot be changed in any possible generalisation.

Similarly to any other field theory, the Einstein field equations \eqref{Einstein equations} can be obtained from an action principle. As is evident from the discussion in Section \ref{sec:GR Geometry}, the fundamental field of the theory shall be the metric tensor $g_{\mu\nu}$, from which all all relevant geometrical quantities---namely, the Levi-Civita connection and its associated curvature tensors---emanate. In 1915, Hilbert \cite{Hilbert:1915tx} proposed an action functional of the form
\begin{equation} \label{eq:EH action}
    S_\mathrm{EH}[g_{\mu\nu}]=\dfrac{1}{2\kappa}\int_\mathcal{M}\dif^4 x\,\sqrt{-g}\,\LCR
\end{equation}
which is is known today as the \emph{Einstein-Hilbert action}. Action \eqref{eq:EH action} turns out to be incomplete, and several other terms must be added to it in order to recover Einstein's equations \eqref{Einstein equations} in a mathematically-consistent way. First, in order to describe non-vacuum scenarios, the Einstein-Hilbert action needs to be supplemented by
\begin{equation} \label{matter action}
    S_\matter[g_{\mu\nu},\Psi]=\int_\mathcal{M}\dif^4 x\,\sqrt{-g}\,\mathcal{L}_\matter(g_{\mu\nu},\Psi),
\end{equation}
i.e.~the action for the matter fields on $\mathcal{M}$, collectively denoted as $\Psi$. Here, $\mathcal{L}_\matter(g_{\mu\nu},\Psi)$ represents the matter Lagrangian density. Requiring the full action $S_\mathrm{EH}+S_\matter$ to be stationary under variations of the inverse metric $g^{\mu\nu}$, one finds\footnote{
    The detailed computation might be found in any of the standard references on GR, e.g.~\cite{poisson2004relativist}.
}
\begin{align}
    0&=\delta(S_\mathrm{EH}+S_\matter)=\dfrac{1}{2\kappa}\int_\mathcal{M}\dif^4 x\,\sqrt{-g}\,\left[\LCEin_{\mu\nu}+\dfrac{2\kappa}{\sqrt{-g}}\dfrac{\delta(\sqrt{-g}\mathcal{L}_\matter)}{\delta g^{\mu\nu}}\right]\delta g^{\mu\nu} \nonumber \\
    &\quad +\dfrac{1}{2\kappa}\oint_{\partial\mathcal{M}}\dif^3 y\,|h|^{1/2}\,\epsilon n_\mu(g^{\rho\sigma}\delta\LCG^\mu{}_{\rho\sigma}-g^{\rho\mu}\delta\LCG^\sigma{}_{\rho\sigma}), \label{eq:EH + matter variation}
\end{align}
where $\partial\mathcal{M}$ is the boundary of $\mathcal{M}$ (which we shall assume to be closed and non-null), $h_{\mu\nu}$ is the induced metric on $\partial\mathcal{M}$, $h\equiv\det(h_{\mu\nu})$, $n_\mu$ is the unit normal to $\partial\mathcal{M}$ and $\epsilon\equiv n^\mu n_\mu=\pm 1$ depending on whether $\partial\mathcal{M}$ is time-like or space-like, respectively.\footnote{
    The aforementioned quantities characterise the geometry of hypersurface $\partial\mathcal{M}$. For more details on said geometrical quantities in any time-like or space-like (but otherwise arbitrary) hypersurface $\Sigma\subset\mathcal{M}$, we refer the interested reader to Section \ref{sec:geometry of Sigma} (where surface $\Sigma$ is a boundary layer along which two distinct space-times are glued together) or to external sources, such as \cite{poisson2004relativist}. 
} Since $\delta g^{\mu\nu}$ is only required to vanish on $\partial\mathcal{M}$, the term in brackets inside the bulk integral must be identically equal to zero on classical solutions. The resulting tensorial equations are, as expected, the Einstein field equations \eqref{Einstein equations}, provided that one \emph{defines}
\begin{equation} \label{eq:stress-energy tensor}
    \stress_{\mu\nu}\equiv-\dfrac{2}{\sqrt{-g}}\dfrac{\delta(\sqrt{-g}\mathcal{L}_\matter)}{\delta g^{\mu\nu}}.
\end{equation}
Nonetheless, even though $\delta g^{\mu\nu}|_{\partial\mathcal{M}}=0$, the boundary integral on \eqref{eq:EH + matter variation} remains non-vanishing, since
\begin{equation}
    \left.n_\mu(g^{\rho\sigma}\delta\LCG^\mu{}_{\rho\sigma}-g^{\rho\mu}\delta\LCG^\sigma{}_{\rho\sigma})\right|_{\partial\mathcal{M}}=-h^{\rho\sigma} n^\mu\delta(\partial_\mu g_{\rho\sigma})|_{\partial\mathcal{M}}
\end{equation}
and $\delta(\partial_\mu g_{\rho\sigma})$ does not need to equal zero on $\partial\mathcal{M}$; this entails that the variational problem is not well-posed. Hence, one must introduce a counterterm whose variation cancels this non-vanishing boundary contribution. This is the so-called Gibbons-Hawking-York counterterm \cite{York:1972sj,Gibbons:1976ue}, which can be compactly expressed in terms of the trace $K$ of the extrinsic curvature of $\partial\mathcal{M}$ as\footnote{
    Again, we refer to the reader to Section \ref{sec:geometry of Sigma} or to references such as \cite{poisson2004relativist} for more details about the extrinsic curvature of an embedded submanifold on $\mathcal{M}$.
}
\begin{equation} \label{eq:GHY term}
    S_\mathrm{GHY}[g_{\mu\nu}]=\dfrac{1}{\kappa}\oint_{\partial\mathcal{M}}\dif^3 y\,|h|^{1/2}\,\epsilon\,K.
\end{equation}
Finally, one might introduce an additional, non-dynamical term of the form
\begin{equation} \label{eq:non-dynamical term}
    S_0=\dfrac{1}{\kappa}\oint_{\partial\mathcal{M}}\dif^3 y\,|h|^{1/2}\,\epsilon\,K_0,
\end{equation}
where $K_0$ is the extrinsic curvature of $\partial\mathcal{M}$ when embedded in flat Minkowski space-time. This term will not contribute to the field equations, but will render the complete gravitational action of GR finite when evaluated on Minkowski space-time \cite{poisson2004relativist}. Despite having no effect on the equations of motion of GR, boundary terms \eqref{eq:GHY term} and \eqref{eq:non-dynamical term} are crucial for the mathematical consistency of the theory. Furthermore, they play a crucial role in the Hamiltonian formulation of GR and the definition of mass \cite{Arnowitt:1962hi}, as well as in black-hole thermodynamics \cite{Hawking:1978jz}.

All in all, the correct action for the purely gravitational sector of GR reads
\begin{equation} \label{eq:GR action}
    S_\mathrm{GR}[g_{\mu\nu}]=S_\mathrm{EH}[g_{\mu\nu}]+S_\mathrm{GHY}[g_{\mu\nu}]+S_0,
\end{equation}
which leads to a well-defined field theory for $g_{\mu\nu}$, whose second-order equations of motion have a well-posed initial-value problem \cite{Choquet-Bruhat:1969ywq}, meaning that the theory is predictive once a suitable set of initial conditions is given. Moreover, a theorem due to Lovelock \cite{Lovelock:1971yv} guarantees that, in four-dimensional space-time, the only symmetric, covariantly-conserved, rank-$(0,2)$ tensors containing, at most, second-order derivatives of the metric are (i) the Einstein tensor $\LCEin_{\mu\nu}$ and (ii) the metric itself. As such, GR $+$ $\Lambda$\footnote{
    If one intends to recover GR plus a cosmological constant $\Lambda$, i.e.~field equations \eqref{GR + Lambda EOM}, it suffices to replace $\LCR\rightarrow\LCR-2\Lambda$ in $S_\mathrm{EH}$, as given by \eqref{eq:EH action}.
} is the most general field theory for gravity---in four space-time dimensions---where the geometry of space-time is built solely from the metric tensor. At linearised level \emph{in vacuo}, GR propagates only one massless and traceless graviton, i.e.~a spin-2 field with two polarisations, which has been observed \cite{LIGOScientific:2016lio}. The graviton is free from ghost instabilities, possessing a kinetic term with the correct sign. Also, it turns out that full GR is the only self-consistent theory for such a spin-2 field which is also invariant under general coordinate transformations---for a derivation, see, for instance \cite[Ch.~3]{Ortin:2015hya}.

Thus, the possibilities for extending GR---in an entirely classical way, i.e.~without invoking quantum-mechanical effects---are clear from the aforementioned results:
\begin{coloritemize}
    \item Increasing the order of the field equations. This could potentially lead to the appearance of instabilities, as per Ostrogradski's theorem \cite{Ostrogradsky:1850fid}. However, higher-order corrections to GR emerge naturally in the low-energy, effective action for gravity, due to quantum corrections \cite{Buchbinder:1992rb}. Furthermore, the inclusion of higher-order curvature terms may improve the renormalisability of the theory, as shown by Stelle \cite{Stelle:1976gc}. Therefore, higher-order gravity theories are well-motivated, natural extensions of GR, but their viability is inexorably tied to the absence of instabilities.
    \item Increasing the number of fields associated to gravity in the linearised spectrum of the theory. While an effectively-massless graviton must always be present in the spectrum of the theory so as to account for gravitational-wave observations \cite{Ezquiaga:2017ekz}, other gravitational fields of different spins---scalars, vectors---might be introduced.
    \item Giving up Riemannian geometry in favour of metric-affine geometry. In this case, tensions with the observationally-supported arguments in favour of preferring Riemannian geometry---reviewed back in Section \ref{sec:GR Geometry}---could arise. Metric-affine theories often propagate additional fields \emph{in vacuo} and might feature instabilities.
    \item Modifying the dimensionality of space-time or breaking the invariance under diffeomorphisms (general coordinate transformations). Other possibilities include violating the EEP by breaking local Lorentz invariance, or non-locality. We will not pursue any of these options in this Thesis.
\end{coloritemize}
As the reader might imagine, the aforementioned avenues for enhancing GR are general enough to give rise to plethora of modified gravity theories---for a review, see, for example, \cite{Faraoni:2010pgm,Capozziello:2011et,CANTATA:2021ktz}. It should be clear, however, that any deviation from the well-grounded fundamental postulates of GR requires significant substantiation. Because the observational validity of Einsteinian gravity \emph{within its regime of application} remains entirely out of question, successful modified gravity theories would be those providing an explanation to phenomena which cannot be explained within the GR paradigm, while leaving the weak-field regime (where Einstein's theory is experimentally well-tested) unaltered.

\section[Metric \titlemath{$\LCf$} gravity.]{Metric \titlebm{$\LCf$} gravity.} \label{sec:intro:metric f(R)}

The first modified gravity theory we are going to consider in this Thesis---$f(\LCR)$ gravity in its metric formulation \cite{Sotiriou:2008rp,DeFelice:2010aj}---is, perhaps, the most straightforward deviation from GR one could possibly conceive. Furthermore, it constitutes a versatile alternative to Einsteinian gravity, remaining observationally viable in a wide variety of scenarios, hence its privileged status among all modified gravity theories.

In metric $f(\LCR)$ gravity, one simply replaces the (Levi-Civita) Ricci scalar $\LCR$ appearing on the Einstein-Hilbert action \eqref{eq:EH action} by an arbitrary function $f(\LCR)$, with each particular $f(\LCR)$ model corresponding to a certain choice for $f$. This has the effect of increasing the order of the field equations; however, a unique feature of metric $f(\LCR)$ theories is that they are capable of avoiding Ostrogradski instabilities \cite{Woodard:2006nt}. Such a surprising result is related to the fact that $f(\LCR)$ gravities can be understood as GR plus an extra, interacting scalar field, as we will become evident soon. For this reason, the linearised spectrum of most $f(\LCR)$ models contains both the massless and traceless graviton of GR plus the (still unobservable) scalar, and hence these theories are compatible with gravitational-wave observations in all but some pathological cases---see Section \ref{Section: Pathologies of the models} in Chapter \ref{chapter:constant curvature}, where this issue is dealt with in detail, and references therein.

Historically, $f(\LCR)$ gravity theories were first introduced by Buchdahl in 1970 \cite{Buchdahl:1970ynr}. However, they did not receive much attention until it was discovered that some $f(\LCR)$ models successfully explain early-universe inflation \cite{Starobinsky:1980te} and late-time accelerated expansion without the need of introducing \emph{ad-hoc} inflaton and dark-energy fields \cite{Nojiri:2008nt}. In fact, there exist $f(\LCR)$ models from which the entire evolutionary history of the universe can be recovered, including a period of matter-domination followed by acceleration \cite{Hu:2007nk,Nojiri:2006gh,Nojiri:2006be,Evans:2007ch}. In all previous cases, the accelerated expansion is due to the presence of the extra scalar field, which provides the violation of the energy conditions necessary to trigger the various periods of cosmic expansion. Some $f(\LCR)$ models also reproduce the observed rotation curves of galaxies without resorting to dark matter \cite{Capozziello:2006ph,Salucci:2014oka,Naik:2019moz}---although some of these models might not be physically viable, cf.~our results in Chapter \ref{chapter:constant curvature}.

Metric $f(\LCR)$ gravities exhibit even more desirable properties. The underlying geometry of space-time in these models is still Riemannian, as in GR, and thus $f(\LCR)$ theories comply with the EPP. Furthermore, a large number of solutions of Einsteinian gravity also solve the modified equations of motion of $f(\LCR)$. For example, the all-important Minkowski and Schwarzschild space-times are vacuum solutions of $f(\LCR)$ models satisfying $f(0)=0$ \cite{delaCruz-Dombriz:2009pzc,Nzioki:2009av}. Therefore, most $f(\LCR)$ models reducing to GR in the appropriate limit comply with the theoretical and observational requirements described in Sections \ref{sec:GR Geometry} and \ref{sec:GR predictions}. At the same time, metric $f(\LCR)$ theories remain sufficiently general and distinct from GR for them to be able to produce rich, novel phenomenology \cite{delaCruz-Dombriz:2021fuw}. A case in point would be the fact that neutron stars are allowed to have larger masses and smaller radii in $f(\LCR)$ in comparison with GR \cite{AparicioResco:2016xcm}, a prediction which could eventually be contrasted with observations. Another remarkable consequence of $f(\LCR)$ gravities is the fact that their spherically-symmetric vacuum solutions do not need to be static, in contrast with GR \cite{Birkhoff}. This permits the existence of new, time-dependent black-hole solutions \cite{Clifton:2006ug}.

Notwithstanding these virtues, there are issues in metric $f(\LCR)$ theories of gravity which should be addressed. On the one hand, progress on the study of compact objects in $f(\LCR)$ gravity has been hampered by the higher complexity of their field equations and junction conditions with respect to those of GR \cite{Casado-Turrion:2022xkl,Senovilla:2013vra}. Junction conditions play an essential role for compact-object dynamics, as isolated matter sources are often described using two solutions: one for the interior and one for the exterior. Examples would include regular stars surrounded by vacuum, or more exotic compact objects, such as \emph{thin-shell wormholes} \cite{Visser:1989kg}. Understanding the conditions under which two space-times can be smoothly glued together is thus fundamental. Moreover, most studies of black-hole and singularity formation through simple models of collapse rely heavily on the junction-condition formalism \cite{Misner:1973prb}, which completely determines the evolution of the stellar surface (and, hence, whether the star truly collapses or not). We will delve into the topic of junction conditions and gravitational collapse in $f(\LCR)$ gravity in Chapter \ref{chapter:f(R) collapse}.

On the other hand, unlike in GR, instabilities associated to the fundamental fields propagated by $f(\LCR)$ gravity may arise. For instance, some previously-unnoticed strong-coupling instabilities may appear for some pathological choices of function $f$; this shall be precisely one of the key points of the investigations contained in Chapter \ref{chapter:constant curvature}. Moreover, the dynamical, scalar degree of freedom characteristic of metric $f(\LCR)$ gravities may become a ghost should $f''(\LCR)<0$. This is related to the widely known result that $f(\LCR)$ models fulfilling $f''(\LCR)<0$ can develop a certain matter instability, known as the Dolgov-Kawasaki instability \cite{Dolgov:2003px,Faraoni:2006sy,Seifert:2007fr}, whereby the presence of matter sources might trigger an infinite feedback loop allowing small curvature pertubations to grow without bounds. This behaviour is clearly unphysical, and thus $f(\LCR)$ models featuring it should be discarded. Another viability concern raised very early on in the history of $f(\LCR)$ gravity was the apparent finding that these theories were falsified by Solar-System experiments \cite{Chiba:2003ir}. This result was obtained by exploiting the dynamical equivalence between $f(\LCR)$ gravities and Brans-Dicke theories with $\omega_0=0$, and using the then-available observational bounds on parameter $\omega_0$, namely, $|\omega_0|\gtrsim 40000$ \cite{Bertotti:2003rm}. However, with the introduction of the so-called `Chameleon mechanisms' \cite{Khoury:2003rn,Khoury:2003aq}, it became evident that it is always possible to construct $f(\LCR)$ models meeting all viability and stability criteria, settling the issue.

\subsection[Field-theoretical aspects of metric \titlemath{$f(\LCR)$} models.]{Field-theoretical aspects of metric \titlebm{$f(\LCR)$} models.}

Having discussed the main features, accomplishments and limitations of metric $f(\LCR)$ gravity models, let us comment on their most relevant field-theoretical aspects, as done with GR in Section \ref{sec:GR field theory}, in order to fix the notation and conventions for the remainder of the Thesis.

As can be deduced from the discussion above, metric $f(\LCR)$ models provide a geometric description of gravity based on the Levi-Civita connection and related curvature tensors, exactly as in GR. Thus, the fundamental dynamical field of the theory shall be, once again, the metric tensor $g_{\mu\nu}$, out of which all relevant geometrical quantities will be built. Moreover, the additional scalar degree of freedom shall also turn out to be a function of the metric. The initial-value problem in $f(\LCR)$ is well posed in vacuum or in the presence of reasonable matter sources \cite{Salgado:2005hx,Faraoni:2006sy,Salgado:2008xh}, similarly to GR; hence, the theory is fully predictive.

Replacing the $\LCR$ in the Einstein-Hilbert Lagrangian by $f(\LCR)$ entails that the boundary and non-dynamical sectors of the action need to be changed accordingly, in order for the theory to have a well-posed variational problem under variations of $g^{\mu\nu}$ vanishing on the boundary of space-time $\partial\mathcal{M}$. The correct action for metric $f(\LCR)$ theories---including matter sources---ends up being \cite{Sotiriou:2008rp,DeFelice:2010aj}
\begin{align} \label{f(R) action}
    S_{f(\LCR)}[g_{\mu\nu},\Psi]&=\dfrac{1}{2\kappa}\int_\mathcal{M}\dif^4 x\,\sqrt{-g}\,f(\LCR)
    +\dfrac{1}{\kappa}\oint_{\partial\mathcal{M}}\dif^3 y\,|h|^{1/2}\,\epsilon\,f'(\LCR)\,(K-K_0) \nonumber \\
    &\quad +S_\matter[g_{\mu\nu},\Psi]. \vphantom{\dfrac{0}{0}}
\end{align}
Notice that the action of metric $f(\LCR)$ gravity includes a Gibbons-Hawking-York-like term, whose presence is strictly necessary for the variational problem to be well-posed.

It may be readily checked that imposing the stationarity of $S_{f(\LCR)}[g_{\mu\nu},\Psi]$ under variations of the inverse metric produces the following fourth-order equations of motion:
\begin{equation} \label{f(R) EOM}
    \LCfp\LCR_{\mu\nu}-\dfrac{\LCf}{2}g_{\mu\nu}-(\LCD_\mu\LCD_\nu-g_{\mu\nu}\LCbox)\LCfp=\kappa\stress_{\mu\nu},
\end{equation}
where $\stress_{\mu\nu}$ is defined as in \eqref{eq:stress-energy tensor}. Contracting $\eqref{f(R) EOM}$ with the metric, one obtains the trace of the equations of motion,
\begin{equation} \label{f(R) EOM trace}
    \LCfp\LCR-2\LCf+3\LCbox\LCfp=\kappa\stress,
\end{equation}
where $\stress\equiv g^{\mu\nu}\stress_{\mu\nu}$. Hence, in generic $f(\LCR)$ gravity models, $\LCR$ and $\stress$ are related through a differential equation. In GR, said relation is algebraic, namely, $\LCR=-\kappa\stress$. Thus, in GR, $\stress=0$ implies $\LCR=0$ and $\LCR_{\mu\nu}=0$, but this is no longer the case in metric $f(\LCR)$ gravity.

The fourth-order equations of motion \eqref{f(R) EOM} of $f(\LCR)$ gravity can be disentangled into two second-order equations by means of a field redefinition; more precisely, through a conformal transformation of the form \cite{Sotiriou:2008rp,DeFelice:2010aj} 
\begin{equation} \label{Einstein-frame metric}
    \bar{g}_{\mu\nu}=f'(\LCR)\,g_{\mu\nu}.
\end{equation}
If, in addition, one introduces the so-called \emph{scalaron} field
\begin{equation} \label{scalaron}
    \phi(\LCR)=\sqrt{\dfrac{3}{2\kappa}}\ln \LCfp,
\end{equation}
action \eqref{f(R) action} transforms into
\begin{align} \label{f(R) action Einstein frame}
    S_{f(\LCR)}[\bar{g}_{\mu\nu},\phi,\Psi]&=\dfrac{1}{2\kappa}\int_{\bar{\mathcal{M}}}\dif^4 x\,\sqrt{-\bar{g}}\,\LCRbar
    +\dfrac{1}{\kappa}\oint_{\partial\bar{\mathcal{M}}}\dif^3 y\,|\bar{h}|^{1/2}\,(\bar{K}-\bar{K}_0) \nonumber \\
    &\quad -\int_{\bar{\mathcal{M}}}\dif^4 x\,\sqrt{-\bar{g}}\,\left[\dfrac{1}{2}\bar{g}^{\mu\nu}\partial_\mu\phi\partial_\nu\phi+V(\phi)\right] \nonumber \\
    &\quad +\int_{\bar{\mathcal{M}}}\dif^4 x\,\sqrt{-\bar{g}}\,\mathcal{L}_\matter\Left(\e^{\sqrt{2\kappa/3}\,\phi}\bar{g}_{\mu\nu},\Psi\Right),
\end{align}
where barred quantities are computed using $\bar{g}_{\mu\nu}$ and the scalaron potential is given by
\begin{equation} \label{scalaron potential phi}
    V(\phi)=\dfrac{f'(\LCR) \LCR-f(\LCR)}{2\kappa f'^2(\LCR)},
\end{equation}
where it is understood that $\LCR=\LCR(\phi)$, i.e.~that relationship \eqref{scalaron} can be inverted. Action \eqref{f(R) action Einstein frame} demonstrates that metric $f(\LCR)$ gravity is dynamically equivalent to GR plus a scalar field, the scalaron $\phi$, which self-interacts through the $f(\LCR)$-dependent potential \eqref{scalaron potential phi} and is minimally coupled to $\bar{g}_{\mu\nu}$ (but not to the matter fields $\Psi$). For this reason, the conformally-transformed action \eqref{f(R) action Einstein frame} is often referred to as the \emph{Einstein-frame representation} of $f(\LCR)$ gravity, whereas the original action \eqref{f(R) action} is known as the \emph{Jordan-frame representation}. Computations in any of the two frames provide equally valid results; hence, in cases where the dynamical equivalence between \eqref{f(R) action} and \eqref{f(R) action Einstein frame} is not broken due to the peculiarities of $f$, one might choose to work in the most suitable frame, as long as the final outcomes are always transformed to the frame considered to be \emph{physical}. It is a matter of convention to decide which frame represents the actual physical theory, and which one is a mere computational tool. In this Thesis, we shall consider that the Jordan frame is to be regarded as the \emph{defining representation} of the theory.

\section[Palatini \titlemath{$f(\MAR)$} gravity.]{Palatini \titlebm{$f(\MAR)$} gravity.} \label{sec:intro:Palatini f(R)}

In the last Section, we pursued a modification of GR where gravitational dynamics were altered by the inclusion of higher-order terms in the action, whereas the Riemannian character of space-time geometry remained unchanged. However, as discussed back in Section \ref{sec:GR Geometry}, the preference for Riemannian geometry is a choice based on physical arguments (in particular, by the EEP); from a purely mathematical point of view, nothing forces the metric and affine structures in a manifold to be intertwined. Hence, one could adopt the so-called \emph{Palatini formalism}, in which both metric and connection are treated as independent dynamical variables, but matter fields are assumed not to couple to the independent connection (for test particles to still follow geodesics of the Levi-Civita connection, in agreement with the EEP). For instance, the Palatini version of GR is obtained by replacing $\LCR\rightarrow\MAR\equiv g^{\mu\nu}\MAR_{\mu\nu}$ in the Einstein-Hilbert action, while keeping the matter action unmodified:
\begin{equation} \label{eq:Palatini GR}
    S_\mathrm{Palatini}[g_{\mu\nu},\MAG^\rho{}_{\mu\nu},\Psi]=\dfrac{1}{2\kappa}\int_\mathcal{M}\dif^4 x\,\sqrt{-g}\,\MAR+S_\matter[g_{\mu\nu},\Psi].
\end{equation}
In the particular case of Palatini GR, this effort turns out to be all for naught, as the equation for motion resulting from varying \eqref{eq:Palatini GR} with respect to $\MAG^\rho{}_{\mu\nu}$ is
\begin{equation}
    \MAD_\rho(\sqrt{-g}\,g^{\mu\nu})=0,
\end{equation}
which inevitably entails that $\smash{\MAD_\rho g_{\mu\nu}=0}$, i.e.~that the connection is actually Levi-Civita. Hence, the theory reduces to standard GR, albeit \emph{dynamically}.

Nonetheless, if one applies the Palatini formalism to a gravitational action containing an arbitrary function $f(\MAR)$, i.e. \cite{Buchdahl:1970ynr}
\begin{equation} \label{eq:Palatini f(R) action}
    S_{f(\MAR)}[g_{\mu\nu},\MAG^\rho{}_{\mu\nu},\Psi]=\dfrac{1}{2\kappa}\int_\mathcal{M}\dif^4 x\,\sqrt{-g}\,f(\MAR)+S_\matter[g_{\mu\nu},\Psi],
\end{equation}
one finds that the connection remains non-trivial even on-shell. As such, Palatini $f(\MAR)$ gravity, as a theory, is different from metric $f(\LCR)$, although both reduce to GR in the appropriate limits, i.e.~$f(\MAR)=\MAR$ and $f(\LCR)=\LCR$. The Palatini formulation of $f(\MAR)$ gravity exhibits some advantages with respect to its metric counterpart. For instance, its equations of motion are of second-order, the theory is free from ghost and Dolgov-Kawasaki instabilities, and vacuum solutions are the same as in GR. This is because Palatini gravity is equivalent to a scalar-tensor theory with a non-dynamical scalaron, and thus the only relevant degrees of freedom of the theory are those corresponding to the massless and traceless graviton \cite{Olmo:2011fh}. These findings are not surprising, in the light of the well-known result that Palatini $f(\MAR)$ gravity---along a larger class of similar theories dubbed \emph{Ricci-based gravities}, or RBGs---can be transformed into pure GR through field redefinitions, at the expense of having modified matter sources \cite{Afonso:2018bpv,Delhom:2021bvq}. Other shortcomings affecting Palatini $f(\MAR)$ gravity include doubts about the well-posedness of its initial-value problem in the presence of matter sources \cite{Salgado:2005hx,Sotiriou:2008rp} and possible conflicts with the Standard Model \cite{Flanagan:2000nx}. Another early concern was that polytropic stars in Palatini $f(\MAR)$ appeared to develop curvature singularities at their surfaces \cite{Barausse:2007pn}; this issue was resolved upon the formulation of the junction conditions of the theory \cite{Olmo:2020fri}.

In this Thesis, we shall concentrate on metric $f(\LCR)$ gravity rather than in Palatini $f(\MAR)$ gravity; we will only briefly comment on the situation of analytic collapse studies in the latter in Chapter \ref{chapter:f(R) collapse}, Section \ref{sec:OS Palatini}. The present introductory Section has been included here mainly for completeness and for pedagogic reasons, as Palatini $f(\MAR)$ gravity is an intermediate step in passing from metric theories to fully metric-affine gravities.

\section{Metric-affine theories of gravity.} \label{sec:intro:MAG}

Palatini $f(\MAR)$ is among the simplest possible \emph{metric-affine theories} of gravity, deviating minimally from GR, as we have just seen. In general, we shall say that a metric-affine theory is one where gravity can be explained in terms of dynamical non-Riemannian geometry. The theory's action will then contain scalars built from both $g_{\mu\nu}$ and $\MAG^\rho{}_{\mu\nu}$.

One natural question to ask when considering metric-affine theories is whether the move to non-Riemannian geometries is completely justified, given the observationally-motivated physical reasonings that led Einstein to believe that the geometry of space-time was indeed Riemannian. For instance, as discussed back on Section \ref{sec:GR Geometry}, if test particles are required to move along geodesics (that is to say, paths extremising proper time in the massive case), then the connection naturally describing such motion is the Levi-Civita one.

Contrary to Einstein's first intuition, it turns out that there is a class of metric-affine theories which are dynamically equivalent to GR, but whose geometry is non-Riemannian. The simplest examples are the \emph{Teleparallel Equivalent of GR} (TEGR) \cite{Hessenberg} and the \emph{Symmetric Teleparallel Equivalent of GR} (STEGR) \cite{Nester:1998mp}.\footnote{
    In fact, Einstein himself was one of the pioneers of teleparallelism, approximately one decade after the completion of GR \cite{EinsteinTeleparallel}.
} In the former, the connection is assumed to be flat and metric-compatible (i.e.~$\MAR^\rho{}_{\sigma\mu\nu}=0$ and $Q_{\rho\mu\nu}=0$, but $T^\rho{}_{\mu\nu}\neq 0$), and the action is chosen to be of the form
\begin{equation} \label{eq:TEGR action}
	S_\mathrm{TEGR}[g_{\mu\nu},\MAG^\rho{}_{\mu\nu},\Psi]=-\int_\mathcal{M}\dif^4 x\,\sqrt{-g}\,\LC{T}+S_\matter[g_{\mu\nu},\Psi],
\end{equation}
where $\LC{T}$ is the so-called \emph{torsion scalar},
\begin{equation}
	\LC{T}\equiv \dfrac{1}{4}T_{\rho\mu\nu}T^{\rho\mu\nu}+\dfrac{1}{2}T_{\mu\nu\rho}T^{\nu\mu\rho}-T_\mu T^\mu,
\end{equation}
with $T_\mu\equiv T^\rho{}_{\mu\rho}$ the \emph{torsion (trace) vector}. It can be shown that, in the absence of both total curvature and non-metricity, the Levi-Civita Ricci scalar \smash{$\LCR$} and \smash{$\LC{T}$} differ solely on a total derivative, in particular
\begin{equation}
	\LCR=-\LC{T}+2\LCD_\mu T^\mu.
\end{equation}
As a result, the GR action \eqref{eq:GR action} and the TEGR action \eqref{eq:TEGR action} both produce Einstein's equations \eqref{Einstein equations} when performing variations of the inverse metric.\footnote{
	It is customary to use the \emph{tetrad formalism} instead, where the metric tensor is replaced by the tetrad field $e^I{}_\mu$; this formalism shall be discussed in Section \ref{sec:intro:Poincaré Gauge Gravity}. We also note that the equations of motion of TEGR do not determine the independent connection. A similar situation occurs in STEGR.
} Moreover, the counting of degrees of freedom in TEGR reveals that it propagates just two, exactly as GR with its massless and traceless graviton. Hence, the two theories can be considered to be dynamically equivalent, as previously anticipated.

Similarly, in STEGR, the affine structure of space-time is flat and torsion-free, but not metric compatible (i.e.~$\MAR^\rho{}_{\sigma\mu\nu}=0$ and $T^\rho{}_{\mu\nu}=0$, but $Q_{\rho\mu\nu}\neq 0$), while the gravitational dynamics are provided by
\begin{equation} \label{eq:STEGR action}
	S_\mathrm{STEGR}[g_{\mu\nu},\MAG^\rho{}_{\mu\nu},\Psi]=\int_\mathcal{M}\dif^4 x\,\sqrt{-g}\,\LC{Q}+S_\matter[g_{\mu\nu},\Psi],
\end{equation}
where $\LC{Q}$ is now the \emph{non-metricity scalar}, defined as
\begin{equation}
	\LC{Q}\equiv-\dfrac{1}{4}Q_{\rho\mu\nu}Q^{\rho\mu\nu}+\dfrac{1}{2}Q_{\mu\nu\rho}Q^{\nu\mu\rho}+\dfrac{1}{4}Q_\mu Q^\mu-\dfrac{1}{2}Q_\mu\tilde{Q}^\mu,
\end{equation}
with $Q_\mu\equiv Q_{\mu\rho}{}^\rho$ and $\tilde{Q}_\mu\equiv Q^\rho{}_{\rho\mu}$ the \emph{non-metricity vectors}. Again, in the absence of both total curvature and torsion,
\begin{equation}
	\LCR=\LC{Q}-\LCD_\mu(Q^\mu-\tilde{Q}^\mu),
\end{equation}
and thus STEGR is equivalent to both GR and TEGR; this is confirmed by a counting of propagating degrees of freedom. As mentioned before, TEGR and STEGR are not the only metric-affine theories which produce the same gravitational dynamics as in GR. The most \emph{General Teleparallel Equivalent of GR} (GTEGR) was found in \cite{BeltranJimenez:2019odq}; TEGR and STEGR result from making particular gauge choices in GTEGR.

As previously mentioned, metric-affine theories are further motivated by the impossibility of defining spinor fields in GR without introducing a \emph{spin connection}, which is independent of the metric.\footnote{
	The construction of the spin connection will be reviewed in Section \ref{sec:intro:Poincaré Gauge Gravity}, in the context of gauging the Poincaré group.
} Hence, in any dynamical theory with fermionic fields, said spinors couple to a dynamical, independent connection, i.e.~the theory is metric-affine. In particular, torsion plays a fundamental role on spinor dynamics \cite{Cembranos:2018ipn}.

Given the large number of independent scalars which can be constructed from $\MAR^\rho{}_{\sigma\mu\nu}$, $T^\rho{}_{\mu\nu}$ and $Q_{\rho\mu\nu}$,\footnote{
	Note that, in non-Riemannian settings, $\MAR^\rho{}_{\sigma\mu\nu}$ possesses fewer symmetries than its Levi-Civita counterpart $\LCR^\rho{}_{\sigma\mu\nu}$. This allows for the existence of novel, non-trivial contractions, such as the \emph{Holst scalar} $\holst\equiv\LCten^{\rho\sigma\mu\nu}\MAR_{\rho\sigma\mu\nu}$ or the \emph{homothetic curvature} $\tilde{R}_{\mu\nu}\equiv\MAR^{\rho}{}_{\rho\mu\nu}$, which vanish identically in Riemannian geometry.
} there is a plethora of possible metric-affine theories, each with its own distinctive set of field equations for the metric and the independent connection. For instance, some of the better known examples---apart from TEGR, STEGR and GTEGR and Palatini $f(\MAR)$---are New General Relativity (NGR), $f(\LC{T})$, $f(\LC{Q})$, Eddington-inspired Born-Infeld gravity (EiBI) and other Ricci-Based theories (RBGs), as well as Poincaré Gauge Gravity (\PGG). We will discuss the last one separately, given that it will be considered in the Results Part \ref{part:results} (more precisely, on Chapter \ref{chapter:BSPGT JCs}).

For this reason, it becomes apparent that additional criteria are necessary so as to discern whether these theories are actually physically viable or not. From a purely theoretical perspective, stability criteria appear to be the most effective tool for discarding action terms. For example, it has been shown that $f(\LC{T})$ gravity suffers from a so-called strong-coupling instability in Minkowski \cite{BeltranJimenez:2020fvy} and cosmological backgrounds \cite{Golovnev:2018wbh,Sahlu:2019bug,Bahamonde:2022ohm}.

Apart from purely theoretical considerations, experimental searches for additional torsion- and non-metricity-related degrees of freedom can be considered \cite{Lammerzahl:1997wk}. It has been shown that some of these degrees of freedom can manifest in Nature in the form of dark-matter mimickers \cite{delaCruzDombriz:2021nrg}. As was the case with metric $f(\LCR)$ and Palatini  $f(\MAR)$ gravities, well-founded metric-affine theories could leave observable imprints on cosmological dynamics \cite{Shie:2008ms}, gravitational-wave spectra \cite{Hohmann:2018jso,Hohmann:2018wxu,Jimenez-Cano:2020lea}, and stellar physics \cite{Guerrero:2021fnz}, paving the way for searches of indirect evidence of non-Riemannian geometry.

Another possible way to experimentally assess the Riemannian character of space-time geometry would be the discovery of violations of the EEP. As previously mentioned, in Palatini $f(\MAR)$ gravity, as well as in TEGR and STEGR, matter fields do not couple to the independent connection, in order to avoid violating the EEP. But one could perfectly go one step further and allow the matter fields to couple to the independent connection; in such case, test particles would no longer follow geodesics. Some authors have proposed \cite{Ponomariev,Kleinert:1998cz} that test particles would instead move along \emph{autoparallels} of the independent connection $\MAG^\rho{}_{\mu\nu}$, i.e.~curves whose tangent vector is paralelly transported along itself. The autoparallel equation turns out to be
\begin{equation} \label{eq:autoparallel}
	\dfrac{\dif^2 x^\rho}{\dif\lambda^2}+\MAG^\rho{}_{\mu\nu}\dfrac{\dif x^\mu}{\dif \lambda}\dfrac{\dif x^\nu}{\dif x^\nu}=0,
\end{equation}
where $\lambda$ is the affine parameter chosen for the autoparallel.\footnote{
    Observe that geodesics are autoparallels of the Levi-Civita connection, as per \eqref{eq:LC geodesic eq} and \eqref{eq:autoparallel}.
} Using \eqref{eq:MA connection decomposition}, \eqref{eq:autoparallel} can be decomposed in terms of a Levi-Civita part and a purely tensorial part,
\begin{equation}
    \dfrac{\dif^2 x^\rho}{\dif\lambda^2}+\LCG^\rho{}_{\mu\nu}\dfrac{\dif x^\mu}{\dif \lambda}\dfrac{\dif x^\nu}{\dif\lambda}=-(K^\rho{}_{\mu\nu}+L^\rho{}_{\mu\nu})\dfrac{\dif x^\mu}{\dif \lambda}\dfrac{\dif x^\nu}{\dif\lambda},
\end{equation}
As any other tensor, the right-hand side of this expression cannot be made to vanish in any coordinate system. Thus, it gives rise to a true \emph{gravitational force}; gravity and inertia are not indistinguishable in these theories.

\section{Poincaré Gauge Gravity.} \label{sec:intro:Poincaré Gauge Gravity}

Up to this point, all modifications of GR we have considered share one common trait: none of them appears to bear any resemblance to the \emph{gauge theories} describing all other known fundamental interactions, namely, Electrodynamics, Chromodynamics and weak interactions. Non-gravitational gauge theories are characterised by the following:
\begin{coloritemize}
    \item They are all physically grounded on the empirical observation of certain \emph{conserved charges}. According to Noether's theorem \cite{Noether:1918zz}, the existence of conserved charges entails that the theory's action is invariant under a certain continuous \emph{symmetry group}, also known as \emph{Lie group}.\footnote{
        A Lie group is a group whose internal group multiplication operation is continuous and differentiable. Lie groups may be alternatively characterised as being differential manifolds satisfying the group axioms. 
    } For instance, in the case of Electrodynamics, the conserved charge is the usual electric charge whereas the symmetry group preserving the action is $\mathrm{U}(1)$. The symmetry groups associated to weak and strong interactions are $\mathrm{SU}(2)$ and $\mathrm{SU}(3)$, respectively.
    \item Symmetry transformations in gauge theories are \emph{local}, meaning that one requires the action to be invariant under group transformations which depend on the space-time point. It is thus necessary to introduce a \emph{gauge connection} so as to compare fields at different space-time points, in particular, when performing derivatives.\footnote{
        Actually, this the only explicit similarity between gauge theories and geometric descriptions of gravity.
        }
    \item Gauge theories whose local symmetry group is $\mathrm{SU}(N)$ are known as \emph{Yang-Mills theories}, after Yang and Mills, who were the first to employ an $\mathrm{SU}(2)$-based gauge theory to describe strong interactions in terms of isospin symmetry \cite{Yang:1954ek}. In Yang-Mills theories, the gauge connection is constructed using a one-form \emph{gauge field} $A_\mu$,\footnote{
        Notice that the gauge field $A_\mu$ takes values on the group's \emph{Lie algebra}, which consists of all group elements connected to the identity. For more information about Lie groups and their representations and their applications in Physics, we refer the reader to the standard references on the topic, e.g.~\cite{Georgi:1999wka,Nakahara:2003nw}.
    } which is the fundamental dynamical object of the theory (together with the charged matter fields).
    \item The kinetic and self-interaction terms of Yang-Mills theories in curved space-time are provided by the following action:
    \begin{equation} \label{eq:YM action}
        S_\mathrm{YM}[A_\mu]=\dfrac{1}{4}\int\dif^4 x\,\sqrt{-g}\,\tr( F_{\mu\nu} F^{\mu\nu}),
    \end{equation}
    where $F_{\mu\nu}$ is the so-called \emph{field-strength tensor} or \emph{gauge curvature}, which is a two-form field with components
    \begin{equation}
        F_{\mu\nu}=2\left(\partial_{[\mu} A_{\nu]}-\iu g A_{[\mu}A_{\nu]}\right),
    \end{equation}
    with $g$ being the theory's \emph{coupling constant}, while $\tr$ denotes the usual matrix trace operation.\footnote{
        Since $A_\mu$ (and, hence, $F_{\mu\nu}$) take values on the Lie algebra $\mathfrak{su}(N)$ of the gauge group $\mathrm{SU}(N)$, they can be understood as being matrices of dimension $N\times N$.
    }
\end{coloritemize}
Thus, in order to construct a gauge theory of gravitational interactions, one would have to find a suitable continuous, local symmetry of space-time, and use its associated gauge connection to construct an action quadratic in its field strength. Furthermore, this symmetry should be intimately related to the geometric properties of space-time, since gravity is correctly described by a geometric theory (GR) at least in the weak-field limit. While the task of encountering such a symmetry might seem to be unfeasible, it turns out that any description of space-time in terms of differentiable manifolds actually provides us with a natural candidate: Poincaré symmetry.

Recalling that, being a differentiable manifold, any space-time is locally homeomorphic to the Minkowski metric (in compliance with the EEP), one has that attached to every point in space-time there is a local copy of the group of symmetries of SR, which comprises both translations and Lorentz transformations. This is the so-called \emph{Poincaré group}, mathematically expressed as $\mathbb{R}^{1+3}\times\mathrm{SO}(1,3)$, with the first factor corresponding to translations, and second to the Lorentz group. The two conserved quantities associated to the Poincaré group are mass and spin. The ideas behind a gauge theory for gravity built out of local Poincaré symmetry were first laid out by Sciama \cite{Sciama1963} and Kibble \cite{Kibble1961}. More mathematically-consistent treatments of the theory were presented thereafter, first by Hayashi \cite{Hayashi:1968hc} and later by Hehl and collaborators \cite{Hehl:1976kj}. For a review of Poincaré Gauge Gravity (\PGG) theories, we refer the reader to \cite{Blagojevic:2013xpa,Blagojevic2001,Obukhov2006b}.

Let us now briefly review why \PGG~theories are indeed metric-affine theories \cite{Blagojevic:2003cg}. To that end, we start by noting that the locally-Minkowskian structure of any space-time is unravelled when using \emph{tetrad fields} $e^I{}_\mu$, also known as \emph{frames}, which satisfy
\begin{equation} \label{eq:tetrad}
    g_{\mu\nu}=\eta_{IJ}\,e^I{}_\mu\,e^J{}_{\nu},
\end{equation} 
where $\eta_{IJ}$ is the Minkoski metric, and $I$, $J$...
are indices associated to the local Poincaré symmetry. Conversely, one introduces the inverse tetrads $e_I{}^{\mu}$, which satisfy
\begin{equation} \label{eq:inverse tetrad}
    g^{\mu\nu}=\eta^{IJ}\,e_I{}^\mu\,e_J{}^{\nu}.
\end{equation}
The components of a vector $V^\mu$ in a local Lorentz frame are given by $V^I=e^I{}_\mu V^\mu$, where $V^I$ behaves as a Lorentz vector, and similarly for tensor fields of other ranks. We already know that parallel transport defines an affine structure---i.e.~a connection $\smash{\MAG^\rho{}_{\mu\nu}}$ and a covariant derivative $\MAD_\mu$---when using coordinate indices, since the action of $\partial_\mu$ on non-scalar tensor fields---for instance, $\partial_\mu V^\nu$---does not behave as a tensor under global coordinate transformations. Analogously, when using frames, one needs to introduce the so-called \emph{spin connection} $\omega^I{}_{J\mu}$ and the \emph{spin covariant derivative} $D_\mu$, because, again, the action of $\partial_\mu$ on non-scalar Lorentzian tensors---for instance, $\partial_\mu V^I$---is not a Lorentzian tensor itself. More precisely, if a field $\psi$ transforms under a representation of the Lorentz group with generators $\Sigma_{IJ}$, then
\begin{equation} \label{eq:spin derivative}
    D_\mu\psi=\left(\partial_\mu+\dfrac{1}{2}\omega^{IJ}{}_{\mu}\Sigma_{IJ}\right)\psi.
\end{equation}
Notice that $\omega^{IJ}{}_{\mu}=-\omega^{JI}{}_{\mu}$ for $\eta_{IJ}$
to remain constant under parallel transport.

The aforementioned two notions of parallel transport coincide if and only if the spin connection $\omega^I{}_{J\mu}$ is related to the affine connection $\MAG^\rho{}_{\mu\nu}$ through the so-called \emph{tetrad postulate},
\begin{equation} \label{eq:tetrad postulate}
    \MAG^\rho{}_{\mu\nu}=e^\rho{}_I(\partial_\mu e^I{}_\nu+\omega^I{}_{J\mu} e^J{}_\nu).
\end{equation}
In combination with \eqref{eq:spin derivative}, the tetrad postulate \eqref{eq:tetrad postulate} entails that the \emph{total} covariant derivative of the tetrad vanishes, i.e.~$\MAD_\mu e^I{}_\nu=D_\mu e^I{}_\nu-\MAG^\rho{}_{\mu\nu} e^I{}_\rho=0$, and thus the total connection is metric-compatible: $\smash{Q_{\rho\mu\nu}=\MAD_\rho g_{\mu\nu}=0}$. However, it can also be shown that
\begin{align}
    T^\rho{}_{\mu\nu} &= 2e_I{}^\rho D_{[\mu} e^I{}_{\nu]}, \label{eq:PG torsion} \\
    \MAR^\rho{}_{\sigma\mu\nu} &= 2e_I{}^\rho e^J{}_\sigma\left(\partial_{[\mu|}\omega^I{}_{J|\nu]}+\omega^I{}_{K[\mu|}\omega^K{}_{J|\nu]}\right). \label{eq:PG Riemann}
\end{align}
Therefore, the total connection remains torsionful and curved. Hence, \PGG~theories are metric-affine, as previously stated.

Almost without noticing, we have already built the Lorentz-covariant derivative associated to Poincaré invariance. We have also identified the gauge field associated to the theory: the antisymmetric spin connection, or, equivalently, the metric-compatible affine connection. Exploting the equivalence between these two affine structures, it is possible to find that the field strength associated to translations is the torsion tensor $T^\rho{}_{\mu\nu}$, as given by \eqref{eq:PG torsion}, whereas the field strength corresponding to Lorentz transformations is the Riemann tensor \smash{$\MAR^\rho{}_{\sigma\mu\nu}$}, as given by \eqref{eq:PG Riemann}. Taking this into account, and in the same spirit as in YM theories, one then postulates that the appropriate generalisation of \eqref{eq:YM action} in \PGG~shall be the most general action up to quadratic order in field strengths \smash{$\MAR^\rho{}_{\sigma\mu\nu}$} and $T^\rho{}_{\mu\nu}$, namely,
\begin{align}
    S_\mathrm{PG}[g_{\mu\nu},T^\rho{}_{\mu\nu}]&=\dfrac{1}{2\kappa}\int_\mathcal{M}\dif^4 x\,\sqrt{-g}\,\left(\MAR+c_1\,\MAR^2
    +c_2\,\MAR_{\mu\nu}\MAR^{\mu\nu}
    +c_3\,\MAR_{\mu\nu}\MAR^{\nu\mu}\right. \nonumber \\
    &\quad +c_4\,\MAR_{\mu\nu\rho\sigma}\MAR^{\mu\nu\rho\sigma}
    +c_5\,\MAR_{\mu\nu\rho\sigma}\MAR^{\rho\sigma\mu\nu}
    +c_6\,\MAR_{\mu\nu\rho\sigma}\MAR^{\mu\rho\nu\sigma} \vphantom{\dfrac{1}{2\kappa}} \nonumber \\
    &\quad+\left.c_7\,T_{\rho\mu\nu}T^{\rho\mu\nu}+
    c_8\,T_{\mu\nu\rho}T^{\nu\mu\rho}
    +c_9\,T_\mu T^\mu\vphantom{\MAR^{\mu\rho\nu\sigma}}\right), \vphantom{\dfrac{1}{2\kappa}} \label{eq:pgg action}
\end{align}
for some dimensionless parameters $\{c_i\}_{i=1}^9$. Notice that we are also taking into account the requirement that the independent connection must be metric-compatible but torsionful and curved, as can be seen in the functional dependency of $S_\mathrm{PG}$.

\subsection{Bi-Scalar Poincaré Gauge Gravity.} \label{sec:bspgt}

Action \eqref{eq:pgg action} can give rise to instabilities. For this reason, in this Thesis we shall focus on the sub-class of PGG theories which can be stable. As shown by Beltrán Jiménez and Maldonado Torralba, ghost-free, quadratic \PGG s propagate, at most, a scalar and a pseudoscalar field in addition to the usual graviton of GR \cite{BeltranJimenez:2019hrm}. Hence, they are known as \emph{Bi-Scalar PGG} (\BSPGT).

The only dynamical fields in \BSPGT~are the metric $g_{\mu\nu}$ together with the irreducible pieces of $T^\rho{}_{\mu\nu}$ when decomposed under the pseudo-orthogonal group:
\begin{equation}
    T^\rho{}_{\mu\nu} = t^\rho{}_{\mu\nu} + \dfrac{2}{3}T_{[\mu}\delta^\rho_{\nu]} + \dfrac{1}{6} \LCten^\rho{}_{\mu\nu\sigma}S^\sigma.
\end{equation}
$T_\mu\equiv T^\rho{}_{\mu\rho}$ is the \emph{torsion trace vector}, $S_\mu= \LCten_{\mu\nu\sigma\lambda}T^{\nu\sigma\lambda}$ is the \emph{torsion axial vector} and $t^\sigma{}_{\mu\nu}$ is the \emph{tensor part of torsion}, which is traceless and has vanishing totally-antisymmetric part. In terms of these fields, the full action of \BSPGT~coupled to matter is
\begin{equation} 
    S_\bspgt[g_{\mu\nu}, T_\mu, S_\mu, t^\rho{}_{\mu\nu},\Psi]=S_\mathrm{GFPG}[g_{\mu\nu}, T_\mu, S_\mu, t^\rho{}_{\mu\nu}] + S_\matter[g_{\mu\nu}, T_\mu, S_\mu, \Psi],\label{eq:fullaction}
\end{equation}
with the ghost-free, purely-gravitational sector being
\begin{align}
    S_\mathrm{GFPG}[g_{\mu\nu}, T_\mu, S_\mu, t^\rho{}_{\mu\nu}]&=\int\dif^4x \sqrt{-g}\left(\frac{\Mp^2}{2}\MAR+\beta\MAR^2+\alpha\holst^2 \right. \nonumber \\
    &\quad \left.+\frac{m_T^2}{2}T_\mu T^\mu+\frac{m_S^2}{2}S_\mu S^\mu+\frac{m_t^2}{2}t_{\mu\nu\rho}t^{\mu\nu\rho}\right). \label{eq:gravaction}
\end{align}
where $m_T$, $m_S$ and $m_t$ are mass parameters, $\alpha$ and $\beta$ are dimensionless coupling constants, and the Ricci scalar ($\MAR\equiv g^{\mu\nu}\MAR_{\mu\nu}$) and Holst pseudoscalar ($\holst\equiv \LCten^{\mu\nu\rho\sigma}\MAR_{\mu\nu\rho\sigma}$) admit the following post-Riemannian decompositions using the irreducible parts of torsion:
\begin{align}
    \MAR &=\LCR+2\LCD_\mu T^\mu+\frac{1}{24}S_\mu S^\mu-\frac{2}{3}T_\mu T^\mu +\frac{1}{2}t_{\mu\nu\rho}t^{\mu\nu\rho}, \label{eq:RicciPostRiem} \\
    \holst &= \frac{2}{3}T_\mu S^\mu-\LCD_\mu S^\mu+\frac{1}{2}\LCten_{\mu\nu\rho\sigma}t^{\lambda\mu\nu}t_\lambda{}^{\rho\sigma}. \label{eq:HolstPostRiem}
\end{align}
As usual, the equations of motion of \BSPGT~are obtained by requiring action \eqref{eq:fullaction} to be stationary under variations of $g_{\mu\nu}$, $T_\mu$, $S_\mu$ and $t^\rho{}_{\mu\nu}$. The left-hand sides of said equations of motion---obtained when varying the purely-gravitational sector \eqref{eq:gravaction}---are encapsulated on Appendix \ref{app:BSPGT variations}. The corresponding right-hand sides---as well as the equations for the non-gravitational fields $\Psi$---are then obtained by varying the matter action $S_\matter$. Observe that we are assuming that the matter sector is coupled to only two of the dynamical pieces of torsion, $T_\mu$ and $S_\mu$. This is because, as shown in Appendix \ref{app:EoMt}, the purely-tensorial part of torsion $t^\rho{}_{\mu\nu}$ is non-dynamical in \BSPGT.\footnote{
     Of course, one could consider scenarios where $t^\rho{}_{\mu\nu}$ becomes dynamical by introducing \emph{ad hoc} non-minimal derivative couplings to matter, but we shall not consider such cases in this Thesis.
} Even under this restriction, the matter action in \eqref{eq:fullaction} is sufficiently general to cover most relevant cases, such as the usual Lagrangians for scalars or $\mathrm{U}(1)$ gauge fields (which are, in fact, independent of torsion), as well as the Dirac Lagrangian minimally-coupled to the torsionful connection (which depends solely on $S_\mu$). At this point, we highlight that, strictly speaking, the Dirac Lagrangian would not fit in the description provided by \eqref{eq:fullaction}, as it requires the use of the tetrad formulation discussed before. However, the application of our results in Chapter \ref{chapter:BSPGT JCs} to the Dirac case is straightforward (as will be explained in Section \ref{sec:Dirac}).

As stated above, the equation of motion of $t^\rho{}_{\mu\nu}$ trivialises for the reasons explained in Appendix \ref{app:EoMt}. Because of this, in what follows, all objects depending on $t^\rho{}_{\mu\nu}$ will be assumed to be evaluated for $t^\rho{}_{\mu\nu}=0$ . For instance,
\begin{align}
    \MAR &=\LCR-\frac{2}{3}T_\mu T^\mu+\dfrac{1}{24}S_\mu S^\mu+2\LCD_\mu T^\mu, \label{eq:RicciPostRiem t=0} \\
    \holst &= \dfrac{2}{3}S_\mu T^\mu-\LCD_\mu S^\mu. \label{eq:HolstPostRiem t=0}
\end{align}
In consequence, the remaining non-trivial equations of motion in Appendix \ref{app:BSPGT variations}, namely the equations of motion of $T_\mu$, $S_\mu$ and $g_{\mu\nu}$, are respectively given by
\begin{align}
    \dfrac{4\alpha}{3}\holst S_\mu +\left(M_T^2-\dfrac{8\beta}{3}\MAR\right)T_\mu - 4\beta\LCD_\mu \MAR &= \trcurrent_\mu, \label{eq:EoM1} \\
     \dfrac{4\alpha}{3}\holst T_\mu +\left(M_S^2+\dfrac{\beta}{6}\MAR\right)S_\mu +2\alpha\LCD_\mu \holst&=  \axcurrent_\mu, \label{eq:EoM2} \\
   E_{\mu\nu}-\dfrac{1}{2}g_{\mu\nu}E-4\kappa\beta(\LCD_{\mu}\LCD_{\nu}- g_{\mu\nu}\LCbox)\MAR &= \kappa \stress_{\mu\nu}. \label{eq:EoM3}
\end{align}
Here, we have introduced the tensor $E_{\mu\nu}$, defined in \eqref{eq:deftensorE}, its trace $E\equiv g^{\mu\nu}E_{\mu\nu}$, the effective masses
\begin{equation} \label{eq:defMSMT}
    M_T^2\equiv m_T^2-\dfrac{2\Mp^2}{3},\mybigskip M_S^2\equiv m_S^2 + \dfrac{\Mp^2}{24},
\end{equation}
and the matter currents
\begin{equation} \label{eq:defLJT}
    \trcurrent_\mu\equiv-\dfrac{1}{\sqrt{-g}}\dfrac{\delta(\sqrt{-g}\mathcal{L}_\matter)}{\delta T^\mu},\myhugeskip \axcurrent_\mu\equiv-\dfrac{1}{\sqrt{-g}}\dfrac{\delta(\sqrt{-g}\mathcal{L}_\matter)}{\delta S^\mu},
\end{equation}
which are, respectively, the \emph{vector spin density} and the \emph{axial vector spin density}. Observe that the axial current $\axcurrent_\mu$ is indeed a pseudo-vector, i.e.~it gets an extra sign under a reflection of all the coordinates, as can be easily deduced by consistency of the equation of motion of $S_\mu$, \eqref{eq:EoM2}. The stress-energy tensor $T_{\mu\nu}$ is once again defined as in \eqref{eq:stress-energy tensor}.

\chapter{Junction conditions and tensor distributions.}
\label{chapter:Introduction: JCs}

It is not difficult to conceive physically-relevant scenarios in which two solutions of a given theory are separated by a boundary. A classical example in gravitational physics would be to consider an isolated, static, perfect-fluid star surrounded by vacuum; in GR, this is the celebrated Tolman-Oppenheimer-Volkov construction \cite{Tolman:1939jz,Oppenheimer:1939ne}, which constitutes the basis of our understanding of stellar objects of various kinds, such as neutron stars. While it is often straightforward to solve the field equations of a given theory on either side of the boundary, it is not immediately clear whether the \emph{combined} solution resulting from \emph{gluing} the two parts together also constitutes a mathematically-consistent solution of the theory in question. The conditions ensuring that a glued pair of solutions of a certain theory gives rise to a properly-matched solution of such theory are known as \emph{junction conditions}. These junction conditions also govern the nature and dynamics of boundary sources, that is to say, matter distributions and/or field configurations confined within the separating surface.

As might be inferred from the previous paragraph, knowledge of junction conditions is of paramount importance when modelling realistic physical settings. In the context of modern gravitation, the junction conditions of GR were first studied by Lanczos \cite{https://doi.org/10.1002/andp.19243791403}. Subsequent studies provided a more mathematically precise formulation depending on the character (time-like, space-like, null or general) of the matching hypersurface \cite{Darmois,Lichnerowicz:107002,o1952jump,Israel:1966rt,bel1967conditions,taub1980space,bonnor1981junction,clarke1987junction,barrabes1989singular,barrabes1991thin,Mars:1993mj}. Of these investigations, the first to consider the possibility of having matter distributions on the boundary surface (the so-called \emph{thin shells}) were those by Darmois \cite{Darmois} and Israel \cite{Israel:1966rt}; for this reason, the junction conditions of GR are often known as the \emph{Darmois-Israel conditions}. Nonetheless, Darmois' and Israel's results are only valid for time-like matching interfaces; the generalisation to null and arbitrary hypersurfaces was developed later on by Barrab\`{e}s and Israel \cite{barrabes1991thin} and by Mars and Senovilla \cite{Mars:1993mj}.

Regarding gravitational theories surpassing standard GR, junction conditions have been studied within the framework of $f(R)$ gravity (in the metric \cite{Deruelle:2007pt,Senovilla:2013vra}, Palatini \cite{Vignolo:2018eco,Olmo:2020fri}, and hybrid \cite{Rosa:2021mln} formalisms), quadratic gravity \cite{Reina:2015gxa}, generalised scalar-tensor theories \cite{Padilla:2012ze}, extended teleparallel gravity \cite{delaCruz-Dombriz:2014zaa}, Einstein-Cartan theory \cite{Arkuszewski:1975fz}, and metric-affine gravity \cite{Macias:2002sr}, among other alternative gravity theories.
    It is evident that modifying the gravitational theory (i.e.~the junction conditions) may alter the allowed matchings. Hence, it is instrumental to study such variations thoroughly, specially in cases where the space-times to be glued are solutions of both GR and the alternative gravity theory. A paradigmatic example of this situation would be the so-called \emph{Oppenheimer-Snyder construction} \cite{Oppenheimer:1939ue}, which is a simple (yet illustrative) account of gravitational collapse in standard GR. As we shall see later, the Oppenheimer-Snyder model relies heavily on the use of the Darmois-Israel junction conditions. Because of this, the model cannot be immediately carried over to other theories, such as metric $f(\LCR)$ gravity, without substantial modifications to the solutions inside and (most importantly) outside the star \cite{Casado-Turrion:2022xkl}.

In this Chapter, we intend to present a general account of junction conditions, as well of the mathematical framework in which they are naturally and consistently formulated, namely that of \emph{tensor distributions}. Our approach will parallel that in Publication \cite{SecondPaper}, where we offered novel insights on the formalism of junction conditions, while extracting the main ideas from \cite{Reina:2015gxa} and, to a lesser extent, \cite{poisson2004relativist}. The chapter shall be organised as follows. First, in Section \ref{sec:matching}, we will formalise the intuitive concept of space-time gluing and define of the matching hypersurface, $\Sigma$. With this purpose, we shall provide a mathematically rigorous construction of the full, matched space-time from its two constituent parts in Subsection \ref{sec:construction of M}. Afterwards, in Subsection \ref{sec:geometry of Sigma}, we will focus on the intrinsic and extrinsic geometry of $\Sigma$, while some useful notation shall be introduced in Subsection \ref{sec:jumps etc.}. Tensor distributions will be introduced in Section \ref{sec:tensordistributions}. The definition and basic properties of both \emph{regular} and \emph{singular} distributions will be given in Subsections \ref{sec:defregtensor} and \ref{sec:singparts}, respectively. Subsequently, in Section \ref{sec:procedure}, we will expound on the generic procedure by means of which one may obtain the junction conditions of any field theory in curved space-time. As part of the discussion therein, we shall introduce the all important conceptual distinction among \emph{consistency}, \emph{singular-layer} and \emph{smooth-matching} conditions. Finally, in Section \ref{sec:JCs in gravity}, we will particularise to metric theories of gravity. After promoting the various geometrical quantities to distributions in Subsection \ref{sec:metricdistrib}, we will review the junction conditions in GR and metric $f(\LCR)$ gravity in Subsections \ref{sec:GR JCs} and \ref{sec:f(R) JCs}, respectively.

\section{The matching hypersurface.} \label{sec:matching}

\subsection[Construction of the matched space-time \titlemath{$\mathcal{M}$}.]{Construction of the matched space-time \titlebm{$\mathcal{M}$}.} \label{sec:construction of M}

\begin{figure}[t]
    \centering
    \begin{overpic}[width=\textwidth]{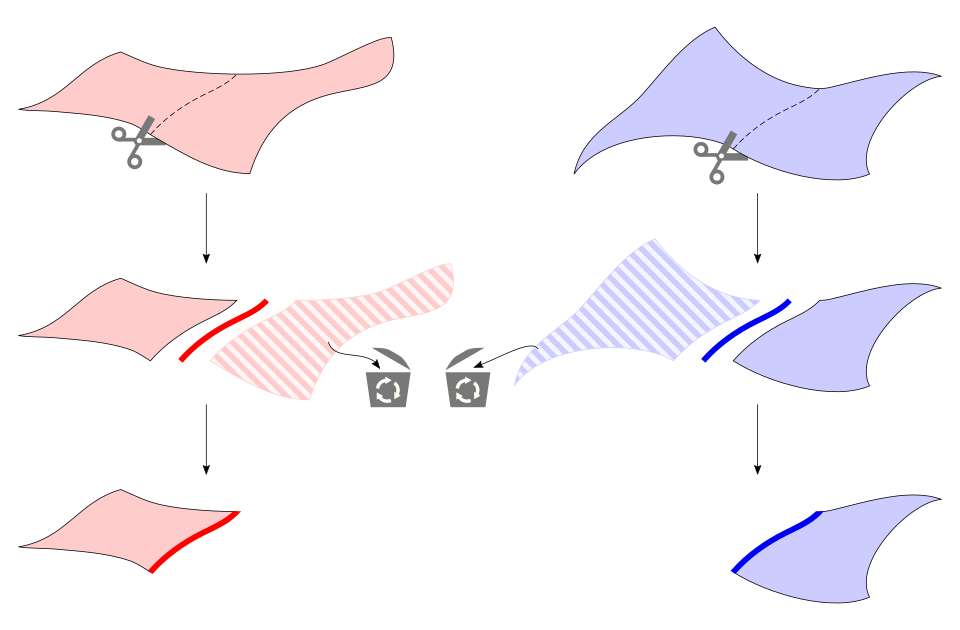}
            \put(21.5,64){\makebox(0,0){\textcolor{red}{$\mathcal{V}^+$}}}
            \put(14,56.2){\makebox(0,0){$\mathcal{M}^+$}}
            \put(24.5,53.5){\makebox(0,0){$\mathcal{D}^+$}}
            \put(25.5,60){\makebox(0,0){$\Sigma^+$}}

            \put(80,64){\makebox(0,0){\textcolor{blue}{$\mathcal{V}^-$}}}
            \put(86.15,52){\makebox(0,0){$\mathcal{M}^-$}}
            \put(74,56){\makebox(0,0){$\mathcal{D}^-$}}
            \put(86,58.5){\makebox(0,0){$\Sigma^-$}}

            \put(14,32.5){\makebox(0,0){$\mathcal{M}^+$}}
            \put(31,29.8){\makebox(0,0){$\mathcal{D}^+$}}
            \put(29,36.3){\makebox(0,0){$\Sigma^+$}}

            \put(86.15,30){\makebox(0,0){$\mathcal{M}^-$}}
            \put(68,34){\makebox(0,0){$\mathcal{D}^-$}}
            \put(83,36.5){\makebox(0,0){$\Sigma^-$}}

            \put(14,10.6){\makebox(0,0){$\mathcal{M}^+$}}
            \put(25.5,14.4){\makebox(0,0){$\Sigma^+$}}

            \put(86.15,8){\makebox(0,0){$\mathcal{M}^-$}}
            \put(86,14.5){\makebox(0,0){$\Sigma^-$}}
        \end{overpic}
        \caption[\emph{Cutting} initial spacetimes $\mathcal{V}^\pm$ along surfaces $\Sigma^\pm$ (respectively).]{\emph{Cutting} initial spacetimes $\mathcal{V}^\pm$ along surfaces $\Sigma^\pm$ (respectively) to separate the parts to be eventually glued together, $\mathcal{M}^\pm$, from the physically irrelevant ones, $\mathcal{D}^\pm$.}
        \label{fig:cutting spacetimes}
\end{figure}
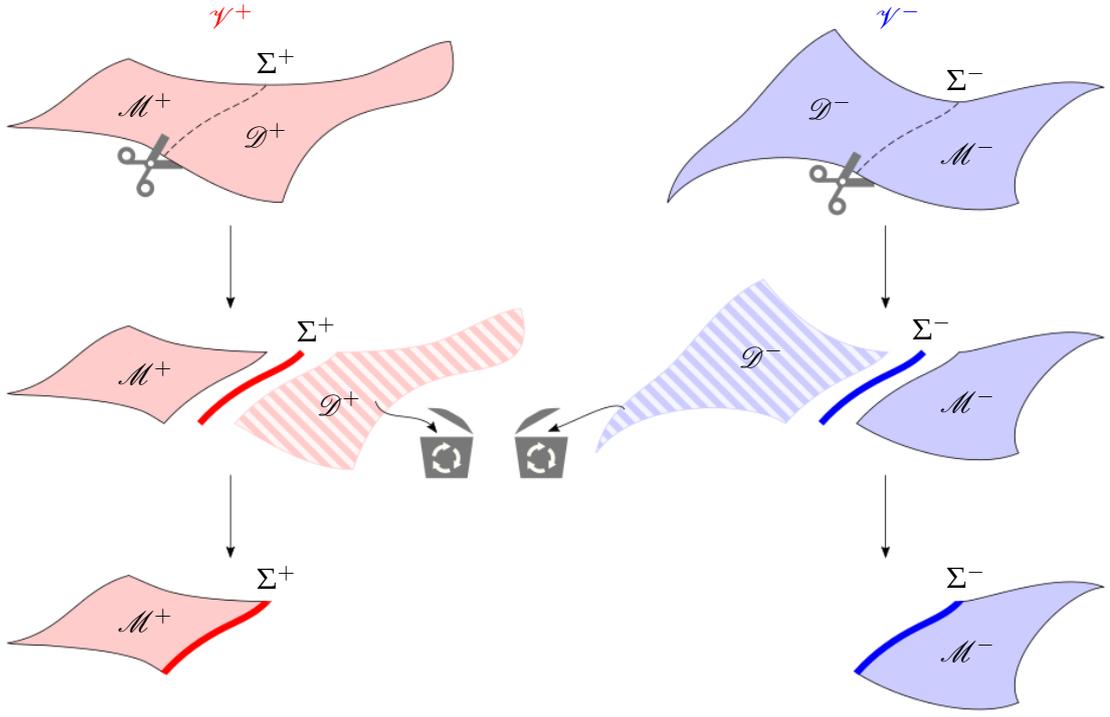

Let $\mathcal{V}^\pm$ denote the two space-times that we intend to glue together. As any other space-times, both $\mathcal{V}^+$ and $\mathcal{V}^-$ will be pseudo-Riemannian manifolds of dimension four, each endowed with (and represented by) a certain metric tensor, $g_{\mu\nu}^\pm$. The first step in our construction is to select the respective matching hypersurfaces $\Sigma^\pm\subset\mathcal{V}^\pm$, which will be either time-like or space-like\footnote{
    The null case, which requires a dedicated formalism and a much more involved treatment, shall not be considered in what follows; the formalism presented in this Thesis will nonetheless cover most relevant physical applications (for instance, stars).}
co-dimension one (i.e.~three-dimensional) embedded sub-manifolds within $\mathcal{V}^\pm$. Since one is only interested in matching just a portion of $\mathcal{V}^\pm$ (for example, the region of Schwarzschild spacetime $r>r_*$ outside some stellar radius $r_*$), hypersurfaces $\Sigma^\pm$ must be such that they divide $\mathcal{V}^+$ and $\mathcal{V}^-$ in two pieces each: regions $\mathcal{M}^\pm$, which are the ones to be eventually glued together (i.e.~$\mathcal{M}^\pm$ represent the actual physical solutions at either side of the matching surface), and regions $\mathcal{D}^\pm$, which comprise the parts of $\mathcal{V}^\pm$ that we do not want to preserve (on physical grounds) after the matching. The aforementioned set-up may be visualised in Figure \ref{fig:cutting spacetimes}. It is important to stress that we are defining $\mathcal{M}^\pm$ and $\mathcal{D}^\pm$ in such a way that $\Sigma^\pm\not\subset\mathcal{M}^\pm,\mathcal{D}^\pm$.

\begin{figure}[t]
    \centering
    \begin{overpic}[width=\textwidth]{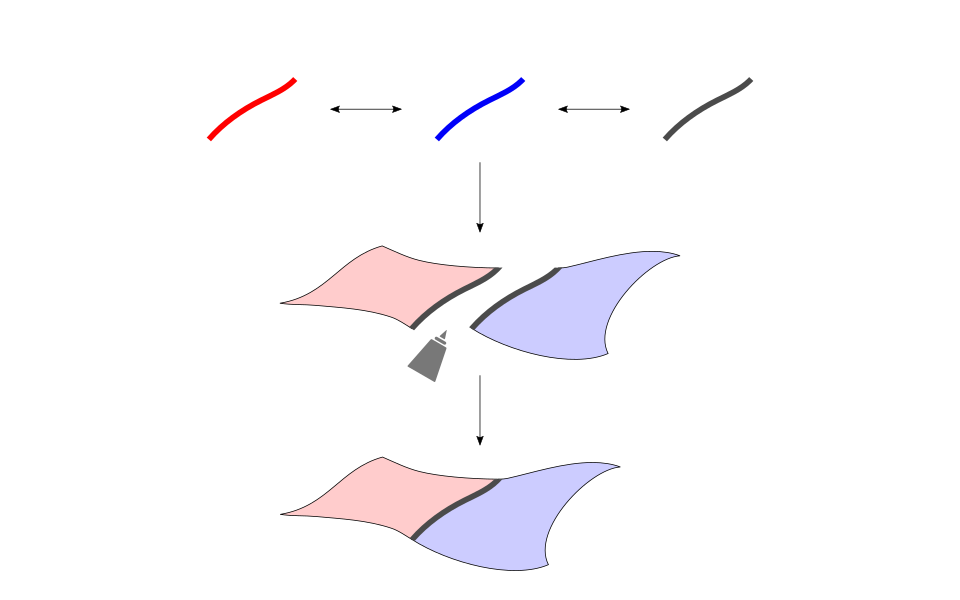}
            \put(24.5,54){\makebox(0,0){$\Sigma^+$}}
            \put(49,54){\makebox(0,0){$\Sigma^-$}}
            \put(73.5,54){\makebox(0,0){$\Sigma$}}
            \put(38.2,54){\makebox(0,0){diff.}}
            \put(61.75,54){\makebox(0,0){def.}}

            \put(41,33){\makebox(0,0){$\mathcal{M}^+$}}
            \put(59,31){\makebox(0,0){$\mathcal{M}^-$}}
            \put(52,36.5){\makebox(0,0){$\Sigma$}}
            \put(58,36.5){\makebox(0,0){$\Sigma$}}

            \put(41,11){\makebox(0,0){$\mathcal{M}^+$}}
            \put(53,9){\makebox(0,0){$\mathcal{M}^-$}}
            \put(52,14.5){\makebox(0,0){$\Sigma$}}
    \end{overpic}
    \caption{Identifying $\Sigma^+$ and $\Sigma^-$ as \emph{the} (unique) matching surface $\Sigma$, and \emph{gluing} $\mathcal{M}^+$ and $\mathcal{M}^-$ together along $\Sigma$ to form the total space-time, $\mathcal{M}=\mathcal{M}^+\sqcup\,\Sigma\,\sqcup\mathcal{M}^-$.}
    \label{fig:gluing spacetimes}
\end{figure}
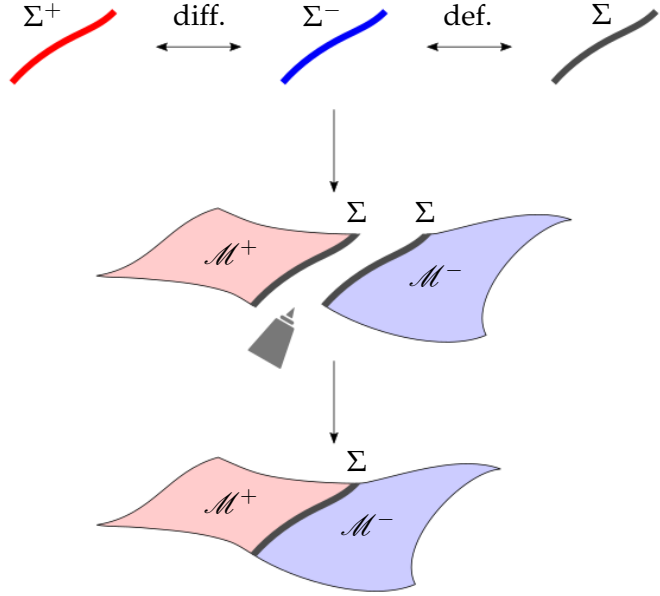

Once $\mathcal{V}^\pm$ have been \emph{cut} along $\Sigma^\pm$ (as described above), and the physically meaningless regions $\mathcal{D}^\pm$ have been separated and discarded, one may proceed to \emph{glue} the two desired physical sub-manifolds $\mathcal{M}^+$ and $\mathcal{M}^-$. However, it should be noted that, in order to do so, one must be able to identify both $\Sigma^+$ and $\Sigma^-$ with \emph{the} (unique) matching hypersurface $\Sigma$ of the resulting, matched space-time, $\mathcal{M}$, which is then constructed as
\begin{equation} \label{glued M}
    \mathcal{M}=\mathcal{M}^+\sqcup\,\Sigma\,\sqcup\mathcal{M}^-,
\end{equation}
where $\sqcup$ denotes the disjoint union of sets. The requirement that $\Sigma^+$ and $\Sigma^-$ must be identified with each other (as well as with $\Sigma\subset\mathcal{M}$) imposes the additional condition that said hypersurfaces must be related by a diffeomorphism. In principle, one assumes the particular diffeomorphism enabling the matching to occur to be known. The formalisation of the matching problem is completed once this last technical issue is overcome. We refer the reader to Figure \ref{fig:gluing spacetimes} for a graphical depiction of the gluing process.

To better illustrate the previous formal construction, let us consider once again the paradigmatic case of an isolated star in a metric theory of gravity of choice. The solution of the field equations sourced by the stress-energy tensor corresponding to the star will consist of a space-time $(\mathcal{V}^-,g^-)$ which is physically meaningless outside the (closed) stellar surface $\Sigma^-\subset\mathcal{V}^-$. Therefore, one needs to \emph{cut} $\mathcal{V}^-$ along $\Sigma^-$ and keep only the the part $\mathcal{M}^-\subset\mathcal{V}^-$ contained inside the stellar surface, which will comprise the actual interior space-time to be matched to exterior solution. Analogously, one solves the vacuum equations of motion of the theory, obtaining a space-time $(\mathcal{V}^+,g^+)$ which is physically meaningless within the stellar surface $\Sigma^+\subset\mathcal{V}^+$. Once more, one \emph{cuts} $\mathcal{V}^+$ along $\Sigma^+$ and disposes of the part located within said surface, thus obtaining the true exterior space-time $\mathcal{M}^+$. Upon identification of $\Sigma^+$ and $\Sigma^-$ as \emph{the} stellar surface $\Sigma$, one finally glues $\mathcal{M}^-$ and $\mathcal{M}^+$ together along $\Sigma$ to obtain the complete, matched space-time $\mathcal{M}$, as given by \eqref{glued M}.

At this point, it is important to notice that, in the previous discussion, we have adopted an \emph{active} viewpoint whereby we are \emph{generating} space-time $\mathcal{M}$ starting from two other space-times, $\mathcal{V}^\pm$. One may alternatively understand the problem from a \emph{passive} perspective, in other words, one could as well consider a space-time $\mathcal{M}$ to be divided in two regions $\mathcal{M}^\pm$ by hypersurface $\Sigma$ (perhaps because the energy contents at either side of $\Sigma$ are different). Even though the passive viewpoint is the one commonly assumed in the literature \cite{SecondPaper,Reina:2015gxa}, we have decided to adopt and present the active perspective because, in most practical situations (such as those to be considered in Part \ref{part:results} of this Thesis), one is not directly given the final space-time $\mathcal{M}$, and is instead forced to deal with two initial solutions $\mathcal{V}^\pm$ which need to be \textit{cut} and then matched, as explained before. Thus, adopting the active perspective right from the start appears to be more appropriate for our purposes. Needless to say, both procedures yield exactly the same results, and should be considered as compatible for all intents and purposes.

Let us make a last technical remark of relative importance. Since the full, matched space-time $\mathcal{M}$ is a smooth manifold, we can always take a chart around any given point $p$ in the matching hypersurface $\Sigma$. Such a chart, which intersects (at least partially) both sides $\mathcal{M}^+$ and $\mathcal{M}^-$, will be denoted $\{x^\mu\}$, and, hereafter, all tensor components will be referred to it. This simplifies the formalism, although, in practice, one normally starts with different coordinate charts $\{x_\pm^\mu\}$ on both sides, and thus some preliminary work is required to connect them via parametric equations; see e.g.~\cite[Sec.~3.8]{poisson2004relativist}.

\subsection[Geometry of the matching surface \titlemath{$\Sigma$}.]{Geometry of the matching surface \titlebm{$\Sigma$}.} \label{sec:geometry of Sigma}

\begin{figure}[t]
    \centering
    \vspace{5pt}
    \begin{overpic}[width=0.6\textwidth]{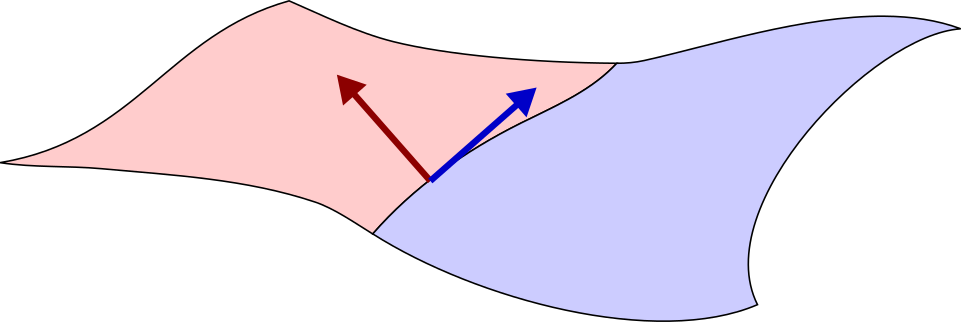}
            \put(21,19){\makebox(0,0){$\mathcal{M}^+$, $\{x_+^\mu\}$}}
            \put(66,5){\makebox(0,0){$\mathcal{M}^-$, $\{x_-^\mu\}$}}
            \put(55,13){\makebox(0,0){$\Sigma$, $\{y^a\}$}}
            \put(33,27){\makebox(0,0){$n^\mu$}}
            \put(66,25){\makebox(0,0){$e^\mu_a=\dfrac{\partial x^\mu}{\partial y^a}$}}
            \put(-7,0){$\mathcal{M}=\textcolor{red}{\mathcal{M}^+}\sqcup\,\Sigma\,\sqcup\textcolor{blue}{\mathcal{M}^-}, \{x^\mu\}$}
    \end{overpic}
    \vspace{5pt}
    \caption[Geometry of the matching surface $\Sigma$ and the various coordinate systems employed in any matching problem.]{Geometry of the matching surface $\Sigma$ and the various coordinate systems employed in any matching problem. Coordinates $\{y^a\}$ are intrinsic to $\Sigma$, while $\{x^\mu\}$ cover, at leats, the entirety $\Sigma$ and a neighbourhood of $\Sigma$ including points in both $\mathcal{M}^+$ and $\mathcal{M}^-$. Finally, coordinates $\{x^\mu_\pm\}$ cover a portion of $\mathcal{M}^\pm$ containing $\Sigma$.}
    \label{fig:sigma geometry}
\end{figure}
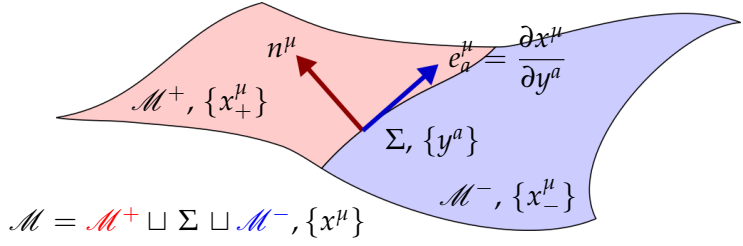

Once the construction of the glued space-time $\mathcal{M}$ is complete, one may readily use the theory of sub-manifolds to study the matching surface $\Sigma$. A pictorial representation of the geometrical quantities we are about to introduce can be found on Figure \ref{fig:sigma geometry}. We will work with a generic chart $\{y^a\}$ intrinsic to $\Sigma$ (index $a$ thus runs from 0 to 2 or from 1 to 3 depending on whether $\Sigma$ is time-like or space-like, respectively), often adapted to the specific physical problem in question. Thus, one way of characterising $\Sigma$ would be by specifying a set of parametric equations of the form $x^\mu=x^\mu(y^a)$. The tangent space at any given point in $\Sigma$ will be spanned by the basis of vectors $\{e_a^\mu, n^\mu\}$, where
\begin{equation}
    e_a^\mu\equiv\dfrac{\partial x^\mu}{\partial y^a}
\end{equation}
are tangent to $\Sigma$, while $n^\mu$ is the \emph{unit normal} to $\Sigma$, chosen to point from $\mathcal{M}^-$ to $\mathcal{M}^+$. For instance, if $\Sigma$ is implicitly given through a relation of the form
\begin{equation} \label{eq:Sigma Phi = 0}
    \Phi(x^\mu)=0,
\end{equation}
with $\Phi$ increasing in the direction of $\mathcal{M}^+$,\footnote{
    This implies that $n^\mu\partial_\mu\Phi>0$, for the unit normal is to point from $\mathcal{M}^-$ to $\mathcal{M}^+$, as stated above.}
then the components of the unit normal one-form $n_\mu$ are univocally fixed to be
\begin{equation}
    n_\mu=\dfrac{\epsilon\,\partial_\mu\Phi}{|g^{\mu\nu}\,\partial_\mu\Phi\,\partial_\nu\Phi|},
\end{equation}
where we have defined
\begin{equation} \label{eq:defepsilon}
    \epsilon\equiv n^\mu n_\mu
    =\begin{cases}+1 & \text{if $\Sigma$ is time-like}, \\
    -1 & \text{if $\Sigma$ is space-like}.
    \end{cases}
\end{equation}
Although there is no actual distinction between $\{e_a^{+\mu}, n^\mu_+\}$ and $\{e_a^{-\mu}, n^\mu_-\}$, because the unit normal $n^\mu$ and tangent vectors $e_a^\mu$ are unique, in practical situations it is often convenient to work with
\begin{equation}
    e_a^{\pm\mu}\equiv\dfrac{\partial x_\pm^\mu}{\partial y^a},\myhugeskip n_\mu^\pm\equiv\dfrac{\epsilon\,\partial_\mu\Phi^\pm}{|g_\pm^{\mu\nu}\,\partial_\mu\Phi^\pm\,\partial_\nu\Phi^\pm|},
\end{equation}
where, as seen from either side $\mathcal{M}^\pm$, $\Sigma$ is given implicitly by relations of the form $\Phi^\pm(x_\pm^\mu)=0$ such that $n_\pm^\mu\partial_\mu\Phi^\pm>0$.

Two fundamental geometrical quantities defined only on $\Sigma$ are the \emph{induced metric} (or \emph{first fundamental form}) and the \emph{extrinsic curvature} (or \emph{second fundamental form}). Let us first introduce the former. Both $g_{\mu\nu}^\pm$ induce corresponding intrinsic metrics on $\Sigma$, namely
\begin{equation} \label{eq:h_ab+-}
    h^\pm_{ab}\equiv\dfrac{\partial x_\pm^\mu}{\partial y^a}\dfrac{\partial x_\pm^\nu}{\partial y^b} g^\pm_{\mu\nu}.
\end{equation}
The agreement between these induced metrics, in other words, the existence of a \emph{unique} first fundamental form $h_{ab}\equiv h^+_{ab}=h^-_{ab}$ in $\Sigma$, requires the metric to be continuous, i.e.~$g^+_{\mu\nu}=g^-_{\mu\nu}$ in all $\Sigma$. As we will comment later on Section \ref{sec:metricdistrib}, the continuity of the metric across $\Sigma$ is an indispensable requirement in any geometrical theory of gravity, since it is necessary in order to have a well-defined curvature in the distributional sense. As such, we shall assume that, hereafter, the metric (and, hence, the induced metric $h_{ab}$) is continuous, so we are allowed to drop the $\pm$ from the symbol and consider only\footnote{
    Notice that, in practice, it is often very convenient to work with $h_{ab}^\pm$ as defined in \eqref{eq:h_ab+-}, as was the case with $e_a^{\pm\mu}$ and $n_\mu^\pm$.}
\begin{equation} \label{eq:h_ab def}
    h_{ab}=\dfrac{\partial x^\mu}{\partial y^a}\dfrac{\partial x^\nu}{\partial y^b} g_{\mu\nu},
\end{equation}
or its corresponding four-dimensional tensor equivalent,
\begin{equation} \label{eq:h, g, n}
    h_{\mu\nu}=g_{\mu\nu}-\epsilon n_\mu n_\nu,
\end{equation}
which acts as a \emph{projector} onto $\Sigma$, given that it is purely transverse, i.e.~$n^\mu h_{\mu\nu}=0$. Notice that, by construction, both $h_{ab}$ and $h_{\mu\nu}$ are symmetric under permutation of their indices.

On the other hand, the second fundamental forms, or extrinsic curvatures of $\Sigma$ as embedded in $\mathcal{M}^\pm$ are defined as
\begin{equation}
    K^\pm_{\mu\nu}\equiv h^\rho{}_\mu h^\sigma{}_\nu \LCD^\pm_\rho n_\sigma,  
\end{equation}
i.e.~as the projections onto $\Sigma$ of the (Levi-Civita) covariant derivatives of the normal vector. Notice that the notation $\LCD^\pm$ is needed because, even though we have assumed the metric to be continuous across $\Sigma$, the corresponding Levi-Civita connection $\LCG\sim g^{-1}\partial g$ may have a non-vanishing discontinuity coming from the discontinuity of $\partial_\rho g_{\mu\nu}$, which is necessarily directed along the normal to $\Sigma$. The discontinuity of the connection turns to be related to the difference between $K_{\mu\nu}^+$ and $K_{\mu\nu}^-$.\footnote{
    In particular, denoting the discontinuity of a quantity $Q$ as $\jump{Q}$, as we shall do in the following (see Subsection \ref{sec:jumps etc.}), one then has the following relationship:
    \begin{equation*} 
        \jump{K_{\mu\nu}}=-n_\lambda h^\rho{}_\mu h^\sigma{}_\nu \jump{\LCG^\lambda{}_{\rho\sigma}}.
    \end{equation*}
}
Thus, the discontinuity of the extrinsic curvature is related to the discontinuity of the Riemann and Ricci tensors. This is not surprising, as $K_{\mu\nu}^\pm$ are also \emph{curvature tensors} in the sense that they intuitively describe how $\Sigma$ changes in the direction of $\mathcal{M}^\pm$ (i.e.~the direction of $n^\mu$); in other words, how $\Sigma$ bends when embedded within $\mathcal{M}^\pm$.

Similarly to the first fundamental form, the extrinsic curvatures $K_{\mu\nu}^\pm$ are also symmetric and transverse four-dimensional tensors (i.e.~$K_{\mu\nu}^\pm=K_{\nu\mu}^\pm$ and $n^\mu K_{\mu\nu}^\pm=0$) admitting intrinsic three-tensor equivalents
\begin{equation}
    K_{ab}^\pm=\dfrac{\partial x^\mu}{\partial y^a}\dfrac{\partial x^\nu}{\partial y^b} \LCD^\pm_\mu n_\nu.
\end{equation}
The previous expression may be readily transformed into
\begin{equation} \label{eq:K_ab+-}
    K_{ab}^\pm=-n_\mu^\pm\left(\dfrac{\partial^2 x_\pm^\mu}{\partial y^a \partial y^b}+\LCG^{\pm\mu}{}_{\rho\sigma}\dfrac{\partial x_\pm^\rho}{\partial y^a}\dfrac{\partial x_\pm^\sigma}{\partial y^b}\right)
\end{equation}
by applying Leibniz's rule, while taking into account that $e^\mu_a n_\mu=0$. Equation \eqref{eq:K_ab+-} is particularly well suited for explicit computations of the components $K_{ab}^\pm$ of the extrinsic curvatures. In fact, we will systematically make use of it in all the practical cases considered in Chapter \ref{chapter:f(R) collapse} and its companion Appendix \ref{app:regsing}.

Finally, the traces of the two extrinsic curvatures are given by
\begin{equation}
    K^\pm\equiv g^{\mu\nu} K_{\mu\nu}^\pm=h^{ab} K_{ab}^\pm.
\end{equation}
Both $K^\pm$ will be recurring quantities in the remainder of this Thesis, as they will consistently appear in junction conditions of theories where the metric is a dynamical field.

\subsection[Discontinuities, averages and decompositions on $\Sigma$.]{Discontinuities, averages and decompositions on $\boldsymbol{\Sigma}$.} \label{sec:jumps etc.}

The purpose of this brief subsection is to introduce several definitions and notations which will prove to be of great utility in the remainder of this Chapter.

Consider any tensorial quantity $T_A$ on $\mathcal{M}$, where subindex $A$ is an abbreviation for arbitrary tensor indices. For regions $\mathcal{M}^+$ and $\mathcal{M}^-$ are physically dissimilar, i.e.~the matter and field contents are different at each side of $\Sigma$, physical quantities could potentially be discontinuous across $\Sigma$. As such, for any point $p$ in the matching surface $\Sigma$, it is convenient to define the \emph{discontinuity} (also known as `jump') of $T_A$ as
\begin{equation} \label{eq:defjump}
    \jump{T_A}(p) \equiv T^+_A(p^+)- T^-_A(p^-),
\end{equation}
where $T^\pm_A(p^\pm)$ represent the limits of $T^\pm_A$ as point $p$ is approached from $\mathcal{M}^\pm$. Similarly, it is also helpful to introduce the \emph{average value} of $T_A$ on $\Sigma$:
\begin{equation} \label{eq:defonsigma}
    T^\sSigma_A(p)\equiv\dfrac{1}{2}\left[T^+_A(p^+)+ T^-_A(p^-)\right].
\end{equation}
We highlight that quantities $\jump{T_A}$ and $T^\sSigma_A$ are only defined on $\Sigma$, but they are not well defined for objects which are not defined outside $\Sigma$, such as $n^\mu$, $h^\pm_{\mu\nu}$ or $K^\pm_{\mu\nu}$.\footnote{
    Notice that one can make these tensors compatible with definitions \eqref{eq:defjump} and \eqref{eq:defonsigma} by extending them outside $\Sigma$ in an appropriate way. For instance, the problem might be circumvented by introducing a foliation of $\mathcal{M}$ given by the level sets of a certain scalar function $\Phi$, such that $\Sigma=\{p\in\mathcal{M}|\Phi=0\}$, i.e.~expressing $\Sigma$ implicitly, as in Equation \eqref{eq:Sigma Phi = 0}; cf.~\cite[Sec.~3.7]{poisson2004relativist}.}
In any case, it is always \emph{convenient} to introduce, on practical grounds, the \emph{notation}
\begin{align}
    \jump{K_{\mu\nu}} &\equiv K^+_{\mu\nu}-K^-_{\mu\nu},\\
    K^\sSigma_{\mu\nu} &\equiv\dfrac{1}{2}(K^+_{\mu\nu}+K^-_{\mu\nu}),
\end{align}
and similarly for $n^\mu$, $h_{\mu\nu}$, etc. Observe that, by construction,
\begin{equation}
    \jump{n^\mu}=0 \mybigskip \text{and}\mybigskip \jump{g_{\mu\nu}}=0~\Leftrightarrow~\jump{h_{ab}}=0.
\end{equation}
It is also important to note that the average $T_A^\sSigma$ of $T_A$ on $\Sigma$ does not always equal the direct \emph{evaluation} of $T_A$ on $\Sigma$, which we shall denote $T_A|_\sSigma$. 

Finally, we conclude this subsection by reminding the reader that any 1-form field $V_\mu$ on $\Sigma$ can be decomposed into normal and tangential components as
\begin{equation} \label{eq:Vdecomp}
    V_\mu=\tosigma{V}_\mu+\epsilon n_\mu V_\perp,
\end{equation}
where we have introduced the notation $\tosigma{V}_\mu\equiv h^\rho{}_\mu V_\rho$ and $V_\perp\equiv n^\mu V_\mu$.

\section{Tensor distributions.} \label{sec:tensordistributions}

The equations of motion of any classical field theory (including gravitational theories) are expressions involving functions of the theory's \emph{basic fields}\footnote{
    In other words, the fundamental, dynamical fields appearing on the theory's action.}
and their derivatives. As mentioned before, the full, matched spacetime $\mathcal{M}$ is divided in two regions by the matching hypersurface $\Sigma$, implying that all physical (tensorial) quantities will be given by piecewise functions. This renders their derivatives undefined within the realm of ordinary functions; instead, the equations of motion of the theory can only be well-defined in a \emph{distributional} sense. As a result, the use of distributions (rather than mere functions) is pivotal in any physical problem involving matched solutions.

We must remind the reader that, as widely known, it is not possible to define a product of distributions. This is one of the main obstacles and the source of many ambiguities when developing a framework for junction conditions. As we shall see in Section \ref{sec:procedure}, we will bypass this problem by imposing appropriate consistency conditions and by choosing a suitable prescription when promoting products of tensors to the distributional framework. 

In general, a tensor distribution $\udis{F}_A$ is a linear and continuous functional acting on \emph{test tensors} $\varphi^A$, i.e.~sufficiently differentiable tensor fields on $\mathcal{M}$ of compact support.\footnote{
    For a fully rigorous mathematical development of the theory of tensor distributions, we refer the interested reader to \cite[App.~A]{Reina:2015gxa} and references therein. Here, we shall only concentrate on the definitions and nomenclature that are relevant for our purposes.}
Such action will be denoted using angle brackets: $\langle\udis{F}_A,\varphi^A\rangle$. Notice that the action of $\udis{F}_A$ on $\varphi^A$ is a \emph{scalar} relation in the sense that there are no free tensor indices in it; if $\udis{F}_A$ is a $p$-covariant and $q$-contravariant tensor distribution, then it acts on $q$-covariant and $p$-contravariant test tensors $\varphi^A$. We also note that, throughout this Thesis, all distributions will be underlined, in order to be distinguished from ordinary functions.

Because integrable tensor fields on $\mathcal{M}$ are necessarily of compact support, a tensor distribution $\udis{F}_A$ can be associated to any locally integrable tensor field $F_A$ as follows:
\begin{equation} \label{eq:tensor distribution action}
    \langle\udis{F}_A,\varphi^A\rangle\equiv\int_\mathcal{M}\dif^4 x\,\sqrt{-g}\,F_A\varphi^A.
\end{equation}
For instance, one of the fundamental distributions to be considered in this Thesis will be the \emph{Heaviside distributions} $\heavidis{\pm}$ associated to the Heaviside functions \cite{abramowitz1972handbook} for $\mathcal{M}^\pm$,
\begin{equation} \label{def:heaviside functions}
    \Theta^\pm(p)\equiv\begin{cases} 1 & \text{if }p\in\mathcal{M}^\pm, \\
    1/2 & \text{if }p\in\Sigma, \\
    0 & \text{if }p\in \mathcal{M}^\mp.
    \end{cases}
\end{equation}
Heaviside functions $\Theta^\pm$ implement the restriction of physical quantities to regions $\mathcal{M}^\pm$; therefore, piecewise functions with different expressions at either side of $\Sigma$ are naturally expressed in terms of them. As locally integrable scalar functions on $\mathcal{M}$, there exist associated scalar distributions $\heavidis{\pm}$ whose action of $\heavidis{\pm}$ on test scalar functions $\varphi$ may be readily deduced from \eqref{eq:tensor distribution action}:
\begin{equation} \label{eq:Heaviside action}
    \langle\heavidis{\pm},\varphi\rangle=\int_{\mathcal{M}}\dif^4 x\,\sqrt{-g}\,\Theta^\pm\varphi=\int_{\mathcal{M}^\pm}\dif^4 x\,\sqrt{-g}\,\varphi.
\end{equation}
Heaviside distributions $\heavidis{\pm}$ thus allow one to express the distributions associated to picewise functions on $\mathcal{M}$, as we shall explain in detail in Subsection \ref{sec:defregtensor}.

As mentioned before, distributions are introduced, among other reasons to have well defined derivatives for piecewise quantities. Given \emph{any} tensor-valued distribution $\udis{F}_A$ (even in cases where $\udis{F}_A$ is unrelated to an ordinary tensor $F_A$), its \emph{distributional covariant derivative} associated to the Levi-Civita connection $\LCD$ is another distribution $\disLCD_\mu\udis{F}_A$ whose action on a test tensor $\varphi^{\mu A}$ is defined to be
\begin{equation} \label{eq:defdisD}
    \langle\disLCD_\mu\udis{F}_A,\varphi^{\mu A}\rangle\equiv\langle\udis{F}_A,-\LCD_\mu\varphi^{\mu A}\rangle.
\end{equation}
In particular, combining \eqref{eq:Heaviside action}, \eqref{eq:defdisD} and Gauss-Stokes theorem \cite{poisson2004relativist}, one finds:
\begin{equation} \label{eq:heaviside derivative}
    \langle\disLCD_\mu\heavidis{\pm},\varphi^{\mu}\rangle=-\int_{\mathcal{M}^\pm}\dif^4 x\,\sqrt{-g}\,\LCD_\mu\varphi^{\mu}=\pm\int_\Sigma\dif^3 y\,|h|^{1/2}\,\epsilon n_\mu\varphi^\mu.
\end{equation}
The covariant derivative of $\heavidis{\pm}$ is related to another scalar distribution, the \emph{(Dirac) delta distribution}, or $\del$, which shall also turn out to be instrumental for our purposes, as we will discuss later. This distribution is \emph{not} associated to any ordinary tensor on $\mathcal{M}$,\footnote{
    Nonetheless, we still choose use an underline to make the distributional character of $\del$ explicit.}
and its action on scalar test functions defined at least on $\Sigma$ is given by
\begin{equation} \label{eq:delta action}
    \langle\del,\varphi\rangle\equiv\int_\Sigma\dif^3 y\,|h|^{1/2}\,\varphi.
\end{equation}
The delta distribution can be multiplied by any locally integrable tensor field defined at least on $\Sigma$, such as $n_\mu$. Hence, expressions \eqref{eq:heaviside derivative} and \eqref{eq:delta action} together entail
\begin{equation}
    \disLCD_\mu\heavidis{\pm}=\pm\epsilon n_\mu\del.
\end{equation}
Higher-order derivatives of the Heaviside distribution need to be considered in theories whose equations of motion include higher-order derivatives. For instance \cite{SecondPaper,Reina:2015gxa},
\begin{equation}
    \disLCD_\mu\disLCD_\nu\heavidis{\pm}=\pm\left[\disLCD_\rho(n_\mu n_\nu n^\rho \del)+(K_{\mu\nu}^\sSigma-\epsilon n_\mu n_\nu K^\sSigma)\del\right].
\end{equation}
Distributional derivatives of $\del$ are not always well-defined and must be handled with care \cite{Reina:2015gxa}. The physical significance of such kind of distributions shall be discussed in Subection \ref{sec:singparts}.

\subsection{Regular tensor fields and distributions.} \label{sec:defregtensor}

A tensor field $T_A$ on $\mathcal{M}$ is said to be \emph{regular} if it satisfies the following requirements:
\begin{enumerate}[label=(\roman*)]
    \item It is smooth in $\mathcal{M}^{\pm}$.
    \item Both $T_A$ and its successive (Levi-Civita) covariant derivatives have well-defined lateral limits on the matching hypersurface $\Sigma$ from both sides $\mathcal{M}^\pm$.
\end{enumerate}
As mentioned before, we notice that such limits could be non-coincident, meaning that the field and/or its derivatives could be discontinuous. Therefore, regular tensors describe physical quantities with possibly differing values on $\mathcal{M}^\pm$. For this reason, basic tensor fields will inevitably either be regular or contain, at least, a \emph{regular part}, as we will explain in the upcoming Section \ref{sec:singparts}.

Observe that, when doing a matching in physical situations, there is an ambiguity in the value of the regular field precisely on the matching hypersurface $\Sigma$. By convention, we shall \emph{choose} said value to be the average of the lateral limits from both sides of $\Sigma$, as defined back in \eqref{eq:defonsigma}. This entails that we can always express any regular tensor $T_A$ in terms of Heaviside functions \eqref{def:heaviside functions}:
\begin{equation} \label{eq:def:regular tensor}
    T_A = T^{+}_A\,\Theta^+ +T^{-}_A\,\Theta^-=\begin{cases} T_A^+ & \text{in }\mathcal{M}^+, \\
    T_A^\sSigma & \text{in }\Sigma, \\
    T_A^- & \text{in }\mathcal{M}^-.
    \end{cases}
\end{equation}
Notice that setting $T_A|_\sSigma\equiv T_A^\sSigma$ is a choice only for \emph{basic} tensor fields. Once we start combining those basic fields to form \emph{compound} tensors (for instance, through multiplication), the property that the value on $\Sigma$ coincides with the average \eqref{eq:defonsigma} is not maintained. For example, consider two regular tensor fields $A$ and $B$ (we omit the arbitrary tensor indices for simplicity) fulfilling $A|_\sSigma = A^\sSigma$ and $B|_\sSigma=B^\sSigma$. One may easily prove that
\begin{equation} \label{eq:ssigmaproduct}
    (AB)^\sSigma=A^\sSigma B^\sSigma+\dfrac{1}{4}\jump{A}\jump{B},
\end{equation}
which does not coincide with the restriction of their product to $\Sigma$,
\begin{equation} \label{eq:sigma evaluation product}
    (AB)|_\sSigma=A|_\sSigma B|_\sSigma=A^\sSigma B^\sSigma,
\end{equation}
unless either $A$ or $B$ is continuous. In computing junction conditions, it is crucial to take the aforementioned subtlety into account; otherwise, the results coming from terms involving products of regular tensors might be incorrect. Notice as well that \eqref{eq:ssigmaproduct} is compatible with the standard multiplication rule for Heaviside functions \eqref{def:heaviside functions},
\begin{equation} \label{eq:heaviside fun. prod.}
    \Theta^{\pm}\Theta^{\pm}=\Theta^{\pm}+\dfrac{\ind}{2} \mybigskip\text{and}\mybigskip \Theta^{\pm}\Theta^{\mp}=\dfrac{\ind}{4},
\end{equation}
where $\ind$ is the indicator function associated to $\Sigma$, i.e.~$\ind(p)=1$ if $p\in\Sigma$ and $\ind(p)=0$ otherwise. This implies that the product of two regular tensors is:
\begin{equation} \label{eq:def:regular tensor product}
    AB=(AB)^{+}\,\Theta^+ +(AB)^{-}\,\Theta^-+\dfrac{\ind}{4}\jump{A}\jump{B}.
\end{equation}

Promoting the Heaviside functions $\Theta^\pm$ in \eqref{eq:def:regular tensor} to distributions $\heavidis{\pm}$, it is possible to associate a so-called \emph{regular distribution}
\begin{equation}
    \udis{T}_A\equiv T^{+}_A\,\heavidis{+}+T^{-}_A\,\heavidis{-}
\end{equation}
to any regular tensor $T_A$. One of the most remarkable features of regular distributions is that their product is well-defined, in stark contrast with most other distributions. The product of regular distributions is similar to that of regular tensors, \eqref{eq:def:regular tensor product}; in particular, if, in analogy with relations \eqref{eq:heaviside fun. prod.}, one \emph{defines}
\begin{equation} \label{eq:heaviside distrib. product}
    \heavidis{\pm}\heavidis{\pm}\equiv\heavidis{\pm} \mybigskip\text{and}\mybigskip \heavidis{\pm}\heavidis{\mp}\equiv 0,
\end{equation}
then product of two regular distributions $\udis{A}$ and $\udis{B}$ is also a regular distribution:
\begin{equation} \label{eq:def:regular distrib. product}
    \udis{A}\udis{B}=A^+ B^+ \,\heavidis{+} + A^-B^-\,\heavidis{-}.
\end{equation}
Notice that the terms proportional to $\ind$ in \eqref{eq:heaviside fun. prod.} and \eqref{eq:def:regular tensor product} have no distributional counterpart in \eqref{eq:heaviside distrib. product} and \eqref{eq:def:regular distrib. product}, respectively.

It should be noted that other distributional products involving at least one regular distribution, such as $\udis{\Theta}\,\del$, are \emph{not} well-defined. Nonetheless, in certain matching problems, one might need to \emph{make sense} of these products, because they arise naturally when promoting the equations of motion to distributional equalities. The way in which one might treat the product $\udis{\Theta}\,\del$ is explained in Section \ref{sec:procedure}, with the subtleties arising in the process being discussed Appendix \ref{app:subtleties}; we nonetheless recommend reading Sections \ref{sec:singparts} and \ref{sec:procedure} before consulting said Appendix.

\subsection{Singular tensor distributions.}\label{sec:singparts}

As we will justify below, all distributions appearing on any set of equations of motion will be, in general, of the form
\begin{equation} \label{eq:decompF}
    \udis{F}_A=\regpart{\udis{F}_A}+\singpart{\sstar}{\udis{F}_A},
\end{equation}
where $\regpart{\udis{F}_A}$ represents the \emph{regular part} of $\udis{F}_A$, while $\singpart{\sstar}{\udis{F}_A}$ is its so-called \emph{singular part}. As its name suggests, the regular part of $\udis{F}_A$ is a regular distribution corresponding to a certain regular tensor field $F_A$:
\begin{equation}
    \regpart{\udis{F}_A}=F^{+}_A\,\heavidis{+}+F^{-}_A\,\heavidis{-}.
\end{equation}
The singular part of $\udis{F}_A$, on the other hand, consists in a sum of distributions
\begin{equation} \label{eq:singpart def}
    \singpart{\sstar}{\udis{F}_A}=\sum_{\ell=0}^N\singpart{\sell}{\udis{F}_A}
\end{equation}
for some finite $N$. The \smash{$\{\singpart{\sell}{\udis{F}_A}\}_{\ell=0}^N$} in \eqref{eq:singpart def} all follow the same structure:
\begin{align}
   \singpart{\szero}{\udis{F}_A} &= F^\szero_A\del, \label{eq:defD0} \\
   \singpart{\sone}{\udis{F}_A} &= \disLCD_\mu\left(F^\sone_A n^\mu \del\right), \label{eq:defD1}\\
   \singpart{\stwo}{\udis{F}_A} &= \disLCD_\mu\disLCD_\nu\left(F^\stwo_A n^\mu n^\nu \del\right),\\
   \vdots\nonumber \\
   \singpart{\sN}{\udis{F}_A} &= \disLCD_{\mu_1}\ldots\disLCD_{\mu_N}\left(F^\sN_A n^{\mu_1}\ldots n^{\mu_N} \del\right). \label{eq:defDN}
\end{align}
Here, \smash{$\{F^\sell_A\}^N_{\ell=0}$} are tensors defined on the matching surface $\Sigma$, which can be interpreted as multipole densities of order $\ell$ \cite{Senovilla:2014yea}. In other words, \smash{$F^\szero_A$} symbolises a monopolar field on $\Sigma$, $F^\sone_A$ represents a surface dipole, $F^\stwo_A$ describes a quadrupolar contribution, and so forth. To be more precise, \smash{$\{F^\sell_A\}^N_{\ell=0}$} are just certain regular tensor fields which vanish outside $\Sigma$ but take some finite value on it. For example, in practical situations, the \smash{$F^\sell_A$}s will be typically comprised of discontinuities, including $\jump{F_A}$ or $\jump{\LCD_\mu F_A}$; averages on $\Sigma$, such as \smash{$F_A^\sSigma$}; or combinations thereof. The reader may find useful expressions for particular cases in Appendix \ref{app:regsing}.

As expressions \eqref{eq:defD0}--\eqref{eq:defDN} suggest, singular parts of higher orders are generated when taking successive (Levi-Civita) distributional derivatives of regular distributions. Indeed, let us consider a continuous regular distribution with discontinuous first derivative. If we differentiate the original continuous distribution twice, we will generate a monopolar piece $\singsym{\szero}$. A dipolar term $\singsym{\sone}$ will emerge after the third distributional differentiation, and so on. Schematically:
\begin{equation} \label{eq:schemeD}
    \regsym\,\text{(continuous)}\myskip\overset{\disLCD}{\longrightarrow}\myskip
    \regsym\myskip\overset{\disLCD}{\longrightarrow}\myskip
    \regsym+\singsym{\szero}\myskip\overset{\disLCD}{\longrightarrow}\myskip
    \regsym+\singsym{\szero}+\singsym{\sone}\myskip \overset{\disLCD}{\longrightarrow}\myskip\ldots 
\end{equation}

With these ideas in mind, it is not difficult to understand why every possible (distributional) equation of motion will contain terms of the general form \eqref{eq:decompF}. As mentioned before in Section \ref{sec:defregtensor}, the basic fields of the theory shall be assumed to have, at least, a regular part indicating the value of the field on $\mathcal{M}^\pm$ and $\Sigma$. The equations of motion contain derivatives of the basic fields; hence, according to \eqref{eq:schemeD}, only distributions of the type \eqref{eq:decompF} will be present.

It is important to stress at this point that products involving singular distributions cannot be defined in general; as explained before, one can only formulate a mere prescription for the particular case in which a regular distribution multiplies a purely monopolar one (see Section \ref{sec:procedure} and Appendix \ref{app:subtleties}). For this reason, basic tensor fields will not be allowed to have a singular part when the equations of motion involve their products. As the reader might infer, this is almost always the case. Preventing these ill-defined products from appearing on the distributional field equations is the ultimate reason why junction conditions are necessary.

\section{Junction conditions in field theories: the general procedure.} \label{sec:procedure}

Having introduced the preceding formalism, we are now in a position to obtain the junction conditions in any field theory (involving the metric and other dynamical fields) defined in a matched space-time, constructed as stipulated in Section~\ref{sec:matching}. In essence, such junction conditions have three distinct purposes:
\begin{coloritemize}
    \item Guaranteeing that all quantities appearing on the equations of motion can be promoted to well-defined distributions, i.e.~providing the basic \emph{consistency conditions} under which the matching of the two space-times and their field contents are mathematically well-posed problems.
    \item Providing said distributions' singular parts with a physical interpretation. In particular, interpreting singular contributions as describing \emph{surface sources} on $\Sigma$ (as these terms are only defined on the matching surface).
    \item Determining the circumstances under which the matching is \emph{smooth}, in other words, that no singular parts (surface sources) are allowed.
\end{coloritemize}
In general, junction conditions can be obtained by following the ensuing protocol: 
\begin{enumerate}[label=(\roman*)]
    \item Compute the equations of motion of the theory.
    \item Write the equations of motion explicitly in terms of the basic fields---i.e.~those appearing in the functional dependency of the theory's action---and their Levi-Civita covariant derivatives. In the case of the metric, one will obtain the associated Riemann tensor, its traces and potential Levi-Civita covariant derivatives of them.
    \item Expand all the derivatives; in other words, use the Leibniz rule of the Levi-Civita derivative. This step is introduced in order to avoid facing distributional derivatives of products of distributions.
    \item Promote the metric and the rest of the fields to distributions of the type \eqref{eq:decompF}. Similarly, promote covariant derivatives to distributional derivatives, i.e.~$\LCD \to \disLCD$.
    \item Find a set of conditions (as minimal as possible) to prevent ill-defined distributions, such as $\del\,\del$, from appearing in the equations of motion, the so-called \emph{consistency conditions} of the theory in question.

    This is indeed a key stage even when the equations of motion do not contain terms which are quadratic in singular parts. The reason is that, since the equations are usually non-linear, they could still feature products of the type $\udis{\Theta}\,\del$. An example of this is the product of $n$ regular fields, $X_i$, and an additional one, $Y$, which contains both a regular and a monopolar part. (For our purposes, only the cases $n=1$ and $n=2$ will be relevant, but we compute the general case for the sake of completeness.) If we naively promote the aforementioned product to the distributional framework, we find 
    \begin{equation} \label{eq:naive product promotion}
        X_1\ldots X_n Y\myskip\overset{\text{naively}}{\longrightarrow}\myskip\regpart{X_1} \ldots  \regpart{X_n}\left(\regpart{Y}+\singpart{\szero}{Y}\right).
    \end{equation}
    Although the product of distributions is not well-defined in general, it is possible to make sense of the $\udis{\Theta}\,\del$ terms arising from \eqref{eq:naive product promotion} by introducing a consistent prescription. In particular, we will \emph{define} the promotion of these terms to the distributional framework as follows:
    \begin{align}
        X_1\ldots X_n Y \myskip \overset{\text{def.}}{\longrightarrow}  \myskip \udis{X}_1 \ldots\udis{X}_n\udis{Y} &\equiv X^+_1\ldots X^+_n Y^+ \udis{\Theta}^+  + X^-_1\ldots X^-_nY^-\udis{\Theta}^-  \nonumber\\
        &\quad+ X^\sSigma_1 \ldots X^\sSigma_n \singpart{\szero}{Y}. \label{eq:prescTheDel}
    \end{align}
     This is the same result one would obtain through naive promotion, provided some appropriate identifications (needed to fix a certain ambiguity) are considered. A more detailed discussion on these subtleties can be found in Appendix \ref{app:subtleties}.
    
    \item Finally, under the restrictions pointed out in the previous step, decompose the equations of motion into regular and singular parts, as in \eqref{eq:decompF}, and find the conditions required to obtain the desired kind of matching, either smooth (with no surface sources) or non-smooth (including surface sources). Thus, on a purely terminological level, we will distinguish between \emph{consistency}, \emph{singular-layer} and \emph{smooth-matching} junction conditions.
\end{enumerate}
This procedure, as outlined above, is general for any classical relativistic field theory, even those not including dynamical gravity. In the sections to come, we will apply this formalism to theories including GR and metric $\LCf$ gravity. What is more, one of the main results in this Thesis is the derivation of the junction conditions in \BSPGT, which may be found in Chapter \ref{chapter:BSPGT JCs} of Part \ref{part:results}.

\subsection{A warm-up example: junction conditions in Klein-Gordon theory in flat space-time.}

As an enlightening first example, devoid of all the complications we will encounter in more sophisticated theories such as GR, let us consider the case of a self-interacting scalar field $\phi$ of mass $m$ in flat Minkowski space-time (for simplicty, we neglect gravitational interactions). The field is sourced by a classical (scalar) current $j$. Then action of such theory is thereby given by
\begin{equation}
    S=-\int\dif^4 x \left[\dfrac{1}{2}\eta^{\mu\nu}\partial_\mu\phi\partial_\nu\phi+\dfrac{1}{2}m^2\phi^2+V(\phi)+j\phi\right],
\end{equation}
where $V(\phi)$ is the potential providing the self-interactions for $\phi$. Without loss of generality, we shall assume $V(\phi)$ to be an analytic function of $\phi$, i.e.
\begin{equation}
    V(\phi)=\sum_{n=3}^\infty c_n \phi^n=c_3\phi^3+c_4\phi^4+\ldots
\end{equation}
for some self-coupling constants $c_n$. The corresponding equation of motion---step (i) of the procedure outlined above---is the non-homogeneous Klein-Gordon equation,
\begin{equation} \label{eq:example:Klein-Gordon}
    \eta^{\mu\nu}\partial_\mu\partial_\nu\phi-m^2\phi-V'(\phi)=j.
\end{equation}
Since points (ii) and (iii) are automatically fulfilled, we move on directly to step (iv), in which we promote the scalar field $\phi$ and source $j$ to distributions
\begin{gather}
    \udis{\phi}
        =\phi^+\,\heavidis{+}+\phi^-\,\heavidis{-}+\Singpart{\sstar}{\udis{\phi}}, \\
    \udis{j}
        =j^+\,\heavidis{+}+j^-\,\heavidis{-}+\Singpart{\sstar}{\udis{j}},
\end{gather}
in accordance with the decomposition of Minkowski space-time in two regions with different values of $\phi$ and $j$. Promoting ordinary derivatives to distributional ones, the distributional generalisation of the Klein-Gordon equation \eqref{eq:example:Klein-Gordon} then reads
\begin{equation} \label{eq:example:distributional Klein-Gordon}
    \eta^{\mu\nu}\partialdis_\mu\partialdis_\nu\udis{\phi}-m^2\udis{\phi}-V'(\udis{\phi})=\udis{j}.
\end{equation}
One immediately notices---step (v)---that \eqref{eq:example:distributional Klein-Gordon} is ill-defined in the distributional sense unless consistency condition $\Singpart{\sstar}{\udis{\phi}}=0$ is imposed. This is because $V'(\udis{\phi})$ contains powers of $\udis{\phi}$, such as $\udis{\phi}^2$, $\udis{\phi}^3$... Ill-defined distributional products (such as $\del\,\del$) will not arise if and only if $\udis{\phi}$ is regular, and thus this requirement is unavoidable. One may also check that no further conditions are necessary for the other terms on the left-hand side of \eqref{eq:example:distributional Klein-Gordon} to be properly-defined distributions.

In fact, we observe that the regularity of $\udis{\phi}$ entails that the potential term $V'(\udis{\phi})$ in \eqref{eq:example:distributional Klein-Gordon} is automatically regular as well. Thus, the only singular contributions to $\udis{j}$---step (vi)---are the ones arising from \smash{$\eta^{\mu\nu}\partialdis_\mu\partialdis_\nu\udis{\phi}$}. In other words,
\begin{equation}
   \Singpart{\sstar}{\udis{j}}=\Singpart{\sstar}{\eta^{\mu\nu}\partialdis_\mu\partialdis_\nu\udis{\phi}}\myskip\Longleftrightarrow\myskip\Singpart{\sell}{\udis{j}}=\Singpart{\sell}{\eta^{\mu\nu}\partialdis_\mu\partialdis_\nu\udis{\phi}}\text{ for all }\ell.
\end{equation}
According to formulae \eqref{eq:decomDDReg2} and \eqref{eq:decomDDReg3} in Appendix \ref{app:regsing}, the singular part of the scalar current $\udis{j}$ decomposes into
\begin{equation}
    \Singpart{\sstar}{\udis{j}}=\Singpart{\szero}{\udis{j}}+\Singpart{\sone}{\udis{j}},
\end{equation}
with the monopolar contribution being
\begin{equation} \label{eq:example:KG monopole}
    \Singpart{\szero}{\udis{j}}
    =\epsilon n^\mu\jump{\partial_\mu\phi}\,\del,
\end{equation}
while the dipolar part reads
\begin{equation} \label{eq:example:KG dipole}
    \Singpart{\sone}{\udis{j}}
    =\partialdis_\mu\Left(\epsilon n^\mu\jump{\phi}\, \del\Right).
\end{equation}
Alternatively, singular parts \eqref{eq:example:KG monopole} and \eqref{eq:example:KG dipole} may be expressed in terms of ordinary tensor kernels $j^\sell$, defined as in equations \eqref{eq:defD0}--\eqref{eq:defDN}:
\begin{equation}
    \Singpart{\sstar}{\udis{j}}=j^\szero\,\del+\partialdis_\mu\Left(n^\mu j^\sone \del\Right),
\end{equation}
where, as is evident from expressions \eqref{eq:example:KG monopole} and \eqref{eq:example:KG dipole},
\begin{equation}
    j^\szero=\epsilon n^\mu\jump{\partial_\mu\phi},\myhugeskip j^\sone=\epsilon\jump{\phi}.
\end{equation}
This notation turns out to be quite convenient, since the conditions ensuring a smooth matching---$\Singpart{\sstar}{\udis{j}}=0$---turn out to be simply $j^\szero=0$ and $j^\sone=0$. This way of expressing singular parts will also prove to be of great utility in more complex theories, where singular parts are given by more convoluted expressions.

Having found the expressions for the singular parts of $\udis{j}$, we can finally summarise the junction conditions in a self-interacting Klein-Gordon theory in flat space-time as follows:
\begin{coloritemize}
    \item No consistency condition (apart from the regularity of $\udis{\phi}$) is required for the Klein-Gordon equation \eqref{eq:example:distributional Klein-Gordon} to be well-defined in the distributional sense.
    \item If the scalar field $\phi$ is discontinuous at the matching surface, a dipolar source of the form \eqref{eq:example:KG dipole} is induced on $\Sigma$.
    \item Moreover, if $n^\mu\partial_\mu\phi$ is also discontinuous at the matching surface, then a monopolar contribution to $\udis{j}$ given by \eqref{eq:example:KG monopole} emerges on $\Sigma$.
    \item Therefore, in smooth-matching scenarios (physical situations in which surface sources, i.e.~singular contributions to $\udis{j}$, are not allowed), the continuity of the scalar field and its normal derivative on $\Sigma$ are unavoidable. Mathematically, these smooth-matching conditions read $\jump{\phi}=0$ and $n^\mu\jump{\partial_\mu\phi}=0$.
\end{coloritemize}

\section{Junction conditions in metric theories of gravity.} \label{sec:JCs in gravity}

\subsection{Geometrical quantities as distributions.} \label{sec:metricdistrib}

As explained in Chapter \ref{chapter:Introduction: Modified Gravity}, in any metric theory of gravity, all relevant geometrical quantities (connection, Riemmann and Ricci tensors and scalar curvature) are built solely upon the metric, which is the only dynamical field responsible for gravitational interactions. Hence, any study of junction conditions in metric theories must necessarily start by obtaining the proper distributional generalisations of the Levi-Civita connection and curvature tensors. It will turn out that these geometrical objects (the basic building blocks of the equations of motion) will only be well-defined as distributions provided that the metric is regular and continuous across $\Sigma$. Furthermore, we will also see that the singular parts of the Riemann and Ricci curvature distributions are related to the extrinsic curvature of the matching surface.

In principle, one should promote the metric to a distribution of the form \eqref{eq:decompF}, as stipulated in the generic procedure outlined in Section \ref{sec:procedure}. Thus, we start by postulating
\begin{equation}
    \udis{g}_{\mu\nu}=
    g_{\mu\nu}^+\,\heavidis{+}+g_{\mu\nu}^-\,\heavidis{-}
    +\Singpart{\sstar}{\udis{g}_{\mu\nu}}.
\end{equation}
Recalling the definition \eqref{eq:def:Christoffels} of the Christoffel symbols $\LCG^{\rho}{}_{\mu\nu}$, we immediately realise that their distributional counterparts $\udis{\LCG}^{\rho}{}_{\mu\nu}$ will be quadratic in singular pieces, and thus ill-defined, if the metric is allowed to have a singular part. As a result, mathematical consistency of the Christoffel symbols requires the metric to be regular, i.e.
\begin{equation} \label{eq:regularity of the metric}
    \Singpart{\sstar}{\udis{g}_{\mu\nu}}=0.
\end{equation}
Now, if the metric is regular but discontinuous, then its partial derivatives are singular; more precisely, they have a monopolar contribution proportional to the discontinuity:
\begin{equation}
    \partial_\rho\udis{g}_{\mu\nu}=\partial_\rho g_{\mu\nu}^+\,\heavidis{+}+\partial_\rho g_{\mu\nu}^-\,\heavidis{-}+\epsilon n_\rho \jump{g_{\mu\nu}}\,\del.
\end{equation}
For this reason, if the metric were allowed to be discontinuous across $\Sigma$, the distributional Riemann tensor $\udis{\LCR}^{\rho}{}_{\sigma\mu\nu}$---which, as per \eqref{eq:def:LC Riemann}, is quadratic in the Christoffel symbols, and thus quadratic in derivatives of the metric---would contain ill-defined products of delta distributions. Thus, the mathematical consistency of any metric field theory of gravity inevitably entails that, when matching two space-times together,
\begin{equation} \label{1st JC g}
    \jump{g_{\mu\nu}}=0.
\end{equation}
Observe that this condition can be recast in two equivalent ways. First, remembering that the induced metric on $\Sigma$ is related to the total metric through equation \eqref{eq:h, g, n}, and that the normal vector $n_\mu$ is trivially continuous across $\Sigma$, then \eqref{1st JC g} is equivalent to
\begin{equation} \label{1st JC h_mu nu}
    \jump{h_{\mu\nu}}=0.
\end{equation}
Second, using equation \eqref{eq:h_ab def}, the transversality of $h_{\mu\nu}$ and the fact that coordinates $\{x^\mu\}$ are continuous across $\Sigma$, one has that \eqref{1st JC h_mu nu} may also be expressed as
\begin{equation} \label{1st JC h_ab}
    \jump{h_{ab}}=0.
\end{equation}
This last expression has the advantage of being entirely \emph{intrinsic} in nature, i.e.~independent of the particular coordinate system $\{x^\mu\}$ in which we are expressing all four-dimensional tensors.

Conditions \eqref{eq:regularity of the metric} and \eqref{1st JC g}--\eqref{1st JC h_ab} enable us to treat the metric as an \emph{ordinary function} $g_{\mu\nu}$ instead of a distribution $\udis{g}_{\mu\nu}$ in all situations, even though the Riemann tensor (as an intrinsically quadratic object containing derivatives in its definition) will still involve products of distributions of the types $\udis{\Theta}\,\udis{\Theta}$ and $\udis{\Theta}\,\del$. The reason is that, by proceeding in this way, all computations will be consistent with our prescription for products, as given by equation \eqref{eq:prescTheDel}.\footnote{Notice as well that $\disLCD$ is also metric-compatible, in the sense that $g_{\mu\nu} \disLCD_\rho \udis{F}_A =  \disLCD_\rho (g_{\mu\nu}\udis{F}_A)$, as can be derived from \eqref{eq:defdisD}. Thus, one may freely \emph{diffuse} the metric tensor across distributional derivatives.}

Having made the foregoing clarifications, we are now entitled to introduce the following connection and curvature distributions associated to a continuous metric:
\begin{align}
    \udis{\LCG}^\rho{}_{\mu\nu} & \equiv\LCG^{+\rho}{}_{\mu\nu}\,\heavidis{+}+\LCG^{-\rho}{}_{\mu\nu}\,\heavidis{-}, \label{eq:distribGamma} \\
    \udis{\LCR}^{\rho}{}_{\sigma\mu\nu} & \equiv 
    \LCR^{+\rho}{}_{\sigma\mu\nu}\,\heavidis{+}+\LCR^{-\rho}{}_{\sigma\mu\nu}\heavidis{-}+\LCR^{\szero\rho}{}_{\sigma\mu\nu}\,\del, \label{eq:distribRiemann}
\end{align}
where, using the notation in \eqref{eq:defD0}, we have introduced $\Singpart{\szero}{\udis{\LCR}^{\rho}{}_{\sigma\mu\nu}}\equiv\LCR^{\szero\rho}{}_{\sigma\mu\nu}\,\del$, with
\begin{equation}
    \LCR^{\szero\rho}{}_{\sigma\mu\nu}=-2\epsilon\jump{\LCG^\rho{}_{\sigma[\mu}}\, n_{\nu]}.
\end{equation}
Indeed, \eqref{eq:distribRiemann} is what we would have obtained from the usual definition of the Riemann tensor \eqref{eq:def:LC Riemann} after substituting $\LCG\rightarrow\udis{\LCG}$---as given in \eqref{eq:distribGamma}---and the partial derivatives by partial distributional derivatives. After some calculations, it is possible to establish a relationship between the discontinuity of the Christoffel symbols and the discontinuity of the extrinsic curvature \cite{Reina:2015gxa,poisson2004relativist}. Because of this, the singular part of the Riemann tensor turns out to be given by
\begin{equation} \label{eq:singpart0Riem}
    \LCR^{\szero\rho}{}_{\sigma\mu\nu}=4g^{\rho\lambda} n_{[\lambda}\jump{K_{\sigma][\mu}}\, n_{\nu]}.
\end{equation}
Using this expression, one may readily find that the singular parts of the Ricci tensor distribution $\udis{\LCR}_{\mu\nu}\equiv \udis{\LCR}^\rho{}_{\mu\rho\nu}$ and the Ricci scalar distribution $\udis{\LCR}\equiv g^{\mu\nu}\udis{\LCR}_{\mu\nu}$ (both of which are purely monopolar) are
\begin{align}
    \LCR^\szero_{\mu\nu} & = -n_\mu n_\nu\jump{K}-\epsilon\jump{K_{\mu\nu}}, \label{eq:ricci tensor singular} \\
    \LCR^\szero & =-2\epsilon\jump{K}. \label{eq:ricci scalar singular}
\end{align}

Before closing this Section, let us list the expressions for the first and second distributional derivatives of the Ricci scalar, which we will encounter several times further on when discussing metric $\LCf$ theories or bi-scalar Poincaré Gauge Gravity. In general, the first distributional derivative of the Ricci scalar consists of a regular part alongside with monopolar and dipolar singular contributions \cite{Senovilla:2013vra,SecondPaper}. However, for reasons that will become clear later, we will only need to compute derivatives of the Ricci scalar in scenarios where $\jump{K}=0$, in which the expression for $\disLCD_\mu\udis{\LCR}$ simplifies considerably. In particular, the dipolar contribution vanishes, while the remaining parts read
\begin{equation} \label{eq:derivative of ricci scalar continuous K}
    \left.\disLCD_\mu\udis{\LCR}\right|_{\jump{K}=0}=(\LCD_\mu\LCR)^{+}\,\heavidis{+}+(\LCD_\mu\LCR)^{-}\,\heavidis{-}+\epsilon n_\mu\jump{\LCR}\,\del,
\end{equation}
since $\udis{\LCR}$ becomes regular when $\jump{K}=0$, cf.~\eqref{eq:ricci scalar singular}. As for the second distributional derivative of $\udis{\LCR}$, it is given by
\begin{align}
    \left.\disLCD_\mu\disLCD_\nu\udis{\LCR}\right|_{\jump{K}=0}&=(\LCD_\mu\LCD_\nu\LCR)^{+}\,\heavidis{+}+(\LCD_\mu\LCD_\nu\LCR)^{-}\,\heavidis{-} \nonumber \\
    &\quad+ \left.\Singpart{\szero}{\disLCD_\mu\disLCD_\nu\udis{\LCR}}\right|_{\jump{K}=0}
    +\left.\Singpart{\sone}{\disLCD_\mu\disLCD_\nu\udis{\LCR}}\right|_{\jump{K}=0},
\end{align}
where
\begin{align}
   \left.\Singpart{\szero}{\disLCD_\mu\disLCD_\nu\udis{\LCR}}\right|_{\jump{K}=0}&=
   \epsilon\left[
   n_{(\mu} h^\rho{}_{\nu)}\jump{\LCD_\rho\LCR}
   +\epsilon n_\mu n_\nu n^\rho\jump{\LCD_\rho\LCR}\right. \\
   &\quad+\left.(K_{\mu\nu}^\sSigma-\epsilon n_\mu n_\nu K^\sSigma)\jump{\LCR}\right]\del,  \label{eq:Ricci second derivative cont. K monopole} \\
   \left.\Singpart{\sone}{\disLCD_\mu\disLCD_\nu\udis{\LCR}}\right|_{\jump{K}=0}&=\disLCD_\rho \left(\jump{\LCR} n_\mu n_\nu n^\rho\del\right). \label{eq:Ricci second derivative cont. K quadrupole}
\end{align}

\subsection{An example: junction conditions in GR.} \label{sec:GR JCs}

The previous results may be forthrightly applied to the simplest metric theory of gravity, GR. Assuming consistency condition \eqref{1st JC g} holds, the distributional version of the Einstein equations \eqref{Einstein equations} turns out to be free of ill-defined terms. More precisely,
\begin{equation}
    \udis{\LCEin}_{\mu\nu}=\LCEin_{\mu\nu}^+\,\heavidis{+}+\LCEin_{\mu\nu}^-\,\heavidis{-}+\LCEin_{\mu\nu}^\szero\,\del=\kappa\udis{\stress}_{\mu\nu},
\end{equation}
where
\begin{align}
    \kappa\stress_{\mu\nu}^\pm &= \LCEin_{\mu\nu}^\pm=\LCR_{\mu\nu}^\pm-\dfrac{1}{2}g_{\mu\nu}^\pm\LCR^\pm, \\
    \kappa\stress_{\mu\nu}^\szero &= \LCEin_{\mu\nu}^\szero=\LCR_{\mu\nu}^\szero-\dfrac{1}{2}h_{\mu\nu}\LCR^\szero.
\end{align}
As per equations \eqref{eq:h, g, n}, \eqref{eq:ricci tensor singular} and \eqref{eq:ricci scalar singular}, the singular part of the stress energy tensor (which is purely monopolar) is
\begin{equation} \label{eq:GR thin shell}
    \stress_{\mu\nu}^\szero=-\dfrac{\epsilon}{\kappa}\left(\jump{K_{\mu\nu}}-h_{\mu\nu}\jump{K}\right).
\end{equation}
Consequently, the renowned Darmois-Israel junction conditions of GR may be summarised as follows:
\begin{coloritemize}
    \item The Einstein field equations of GR are only well-defined in the distributional sense provided that the metric is regular and continuous across $\Sigma$, i.e.~$\jump{g_{\mu\nu}}=0$ $\Leftrightarrow$ $\jump{h_{\mu\nu}}=0$ $\Leftrightarrow$ $\jump{h_{ab}}=0$.
    \item The matching is smooth if and only if $\jump{K_{\mu\nu}}=0$ $\Leftrightarrow$ $\jump{K_{ab}}=0$. In cases where the extrinsic curvature is discontinuous, a monopolar stress-energy tensor given by \eqref{eq:GR thin shell} is induced on $\Sigma$.
\end{coloritemize}
Two terminological comments are pertinent at this point. On the one hand, the two junction conditions stated above are commonly referred to as the \emph{first} and \emph{second} Darmois-Israel conditions, respectively. We will also use this nomenclature in the rest of the Thesis. On the other hand, the singular matter source induced on $\Sigma$ when $\jump{K_{\mu\nu}}\neq 0$ is often known in the literature as the \emph{thin shell}; indeed, a monopolar stress-energy tensor at $\Sigma$ describes an infinitesimally narrow matter distribution confined to $\Sigma$.

\subsection[Another example: junction conditions in metric \titlemath{$\LCf$} gravity.]{Another example: junction conditions in metric \titlebm{$\LCf$} gravity.} \label{sec:f(R) JCs}

Another relevant case we cannot avoid examining in certain detail is that of metric $\LCf$ gravity. The junction conditions in these theories were independently derived by Deruelle, Sasaki and Sendouda \cite{Deruelle:2007pt}, on the one hand, and by Senovilla \cite{Senovilla:2013vra}, on the other. As always, the derivation starts with the equations of motion of the theory, which in this case are given by \eqref{f(R) EOM}. Following the protocol prescribed in Section \ref{sec:procedure}, we first have to expand all covariant derivatives. This results in
\begin{align}
    \kappa\stress_{\mu\nu}&=\LCfp\LCR_{\mu\nu}-\dfrac{\LCf}{2}g_{\mu\nu} \nonumber \\
    &\quad-\LCfppp(\LCD_\mu\LCR\LCD_\nu\LCR-g_{\mu\nu}g^{\rho\sigma}\LCD_\rho\LCR\LCD_\sigma\LCR) \vphantom{\dfrac{1}{2}} \nonumber \\
    &\quad-\LCfpp(\LCD_\mu\LCD_\nu\LCR-g_{\mu\nu}\LCbox\LCR). \vphantom{\dfrac{1}{2}} \label{expanded f(R) EOM}
\end{align}
When promoting the previous expression to the distributional realm, one finds that there are two further necessary consistency conditions which must inexorably hold, apart from \eqref{1st JC g}. First, it is clear that $f(\udis{\LCR})$, $f'(\udis{\LCR})$, etc.~will typically contain powers of $\LCR$ (for instance, consider an analytic $f$, which is always expandable in power series). Because of this, it is indispensable to demand the Ricci scalar to be regular; owing to \eqref{eq:ricci scalar singular}, this requirement is tantamount to $\jump{K}=0$. Second, as Equation \eqref{eq:derivative of ricci scalar continuous K} evinces, $\disLCD_\mu\udis{\LCR}$ is a singular distribution unless $\jump{\LCR}=0$. This could be potentially problematic, as the distributional version of \eqref{expanded f(R) EOM} contains products such as $\disLCD_\mu\udis{\LCR}\disLCD_\nu\udis{\LCR}$ (which could give rise to ill-defined terms of the form $\del\,\del$) unless $\LCfppp=0$ for all $\LCR$. Thus, the analysis branches into two distinct sub-cases, depending on whether $\LCfppp$ is vanishing or not for all $\LCR$:
\begin{coloritemize}
    \item In models satisfying $\LCfppp\neq 0$ (the generic case), consistency condition $\jump{\LCR}=0$ is unavoidable. Equations \eqref{eq:derivative of ricci scalar continuous K}--\eqref{eq:Ricci second derivative cont. K quadrupole} supplying the singular parts of $\disLCD_\mu\udis{\LCR}$ and $\disLCD_\mu\disLCD_\nu\udis{\LCR}$ simplify considerably, and after some additional computations one finds that the singular stress-energy tensor at $\Sigma$ is purely monopolar, i.e.
    \begin{equation}
        \udis{\stress}_{\mu\nu}=\stress_{\mu\nu}^+\,\heavidis{+}+\stress_{\mu\nu}^-\,\heavidis{-}+\stress_{\mu\nu}^\szero\del,
    \end{equation}
    with the thin-shell contribution being\footnote{It might seem that Equation \eqref{eq:generic f(R) thin shell} does not reduce to \eqref{eq:GR thin shell} in the GR limit $\LCf=\LCR$. Nonetheless, this is not in fact the case, for one has to remember that the $\LCf$-exclusive consistency condition $\jump{K}=0$ has been applied in deriving \eqref{eq:generic f(R) thin shell}.}
    \begin{equation} \label{eq:generic f(R) thin shell}
    \stress_{\mu\nu}^\szero=-\dfrac{\epsilon}{\kappa}\left[f'(\LCR^\sSigma) \jump{K_{\mu\nu}}-f''(\LCR^\sSigma) h_{\mu\nu} n^\rho\jump{\LCD_\rho\LCR}\right].
\end{equation}
    \item On the contrary, in the quadratic case $\LCf=\LCR-2\Lambda+\alpha\LCR^2$ (for some constants $\Lambda$ and $\alpha$), in which $\LCfppp=0$ for all $\LCR$, the Ricci scalar does not need to be continuous on $\Sigma$. This gives rise to a stress-energy tensor with both monopolar and dipolar singular parts. Mathematically,
    \begin{equation}
        \udis{\stress}_{\mu\nu}=\stress_{\mu\nu}^+\,\heavidis{+}+\stress_{\mu\nu}^-\,\heavidis{-}+\stress_{\mu\nu}^\szero\del+\disLCD_\rho\left(\stress_{\mu\nu}^\sone n^\rho \del\right),
    \end{equation}
    where the dipolar stress-energy tensor on $\Sigma$, often known in the literature as the \emph{double-layer} contribution, is given by
    \begin{equation} \label{eq:quadratic f(R) double layer}
        \stress^\sone_{\mu\nu}=\dfrac{2\alpha\epsilon}{\kappa}\jump{\LCR}\,h_{\mu\nu},
    \end{equation}
    while the thin-shell (monopolar) term decomposes as
    \begin{equation} \label{eq:quadratic f(R) thin shell}
        \stress^\szero_{\mu\nu}=\tau_{\mu\nu}+2n_{(\mu}\tau_{\nu)}+\tau n_\mu n_\nu ,
    \end{equation}
    where, following the notation in \cite{Senovilla:2013vra,Reina:2015gxa}, we have introduced
    \begin{equation} \label{eq:quadratic f(R) thin shell decomposition}
        \tau_{\mu\nu}\equiv h^\rho{}_\mu h^\sigma{}_\nu T^\szero_{\rho\sigma},\mybigskip
        \tau_\mu\equiv\epsilon h^\rho{}_\mu n^\sigma T^\szero_{\rho\sigma},\mybigskip
        \tau\equiv n^\mu n^\nu T^\szero_{\mu\nu},
    \end{equation}
    with the various pieces in \eqref{eq:quadratic f(R) thin shell decomposition} being given by
    \begin{align}
        \tau_{\mu\nu} &= -\dfrac{\epsilon}{\kappa}\left[(1+2\alpha\LCR^\sSigma)\jump{K_{\mu\nu}}
        +2\alpha(K^\sSigma_{\mu\nu}\jump{\LCR}-h_{\mu\nu} n^\rho\jump{\LCD_\rho\LCR})\right], \label{eq:quadratic f(R) thin shell piece 1} \\
        \tau_\mu &= -\dfrac{2\alpha\epsilon}{\kappa}
        h^\rho{}_\mu\jump{\LCD_\rho\LCR}, 
        \label{eq:quadratic f(R) thin shell piece 2} \\
        \tau &= \dfrac{2\alpha}{\kappa}K^\sSigma\jump{\LCR}. \label{eq:quadratic f(R) thin shell piece 3}
    \end{align}
    The aforementioned quantities have a straightforward physical interpretation: $\tau_{\mu\nu}$ is sometimes known the \emph{surface energy-momentum tensor} (even though we will most often reserve this nomenclature for the full stress-energy tensor, $\udis{\stress}_{\mu\nu}$), $\tau_\mu$ represents the \emph{external momentum flux}, while $\tau$ is the \emph{tension scalar}. Furthermore, we note that, by definition, $\tau_{\mu\nu}$ and $\tau_\mu$ are tangential objects, i.e.~$n^\mu\tau_{\mu\nu}=0$ and $n^\mu\tau_\mu=0$.
\end{coloritemize}

All in all, the junction conditions in metric $\LCf$ gravity may be summarised as follows:
\begin{coloritemize}
    \item For every choice of function $f$, consistency conditions $\jump{g_{\mu\nu}}=0$ and $\jump{K}=0$ are necessary to guarantee that the equations of motion are well-defined in the distributional sense.
    \item If $\LCfppp\neq 0$ for all $\LCR$, one must also impose $\jump{\LCR}=0$ for the sake of consistency. In this case, the surface stress-energy tensor induced at $\Sigma$ is purely monopolar, with the corresponding thin-shell contribution being found on \eqref{eq:generic f(R) thin shell}.
    \item If $\LCfppp=0$ for all $\LCR$, i.e.~$\LCf=\LCR-2\Lambda+\alpha\LCR^2$, it is no longer necessary to have a continuous Ricci scalar, and the induced stress-energy tensor at $\Sigma$ contains both a dipolar (double-layer) term, \eqref{eq:quadratic f(R) double layer}, and a monopolar (thin-shell) part, given by \eqref{eq:quadratic f(R) thin shell}--\eqref{eq:quadratic f(R) thin shell piece 3}.
    \item A smooth matching at $\Sigma$ is achieved provided that $\jump{K_{\mu\nu}}=0$ $\Leftrightarrow$ $\jump{K_{ab}}=0$, $f''(\LCR^\sSigma)\jump{\LCR}=0$ and $f''(\LCR^\sSigma)\,n^\mu\jump{\LCD_\mu\LCR}=0$, for any choice of $f$.
\end{coloritemize}
Let us make three final observations. First, it is straightforward to check that the junction conditions outlined above reduce to the Darmois-Israel conditions in GR after setting $\LCf=\LCR$ and lifting the $\LCf$-exclusive consistency condition $\jump{K}=0$, as expected. The Darmois-Israel conditions are also obtained for $f(\LCR)=\LCR-2\Lambda$, in which case $f''(\LCR)=0$ for all $\LCR$. The junction conditions of GR are also recovered in very specific matchings where $\LCR^\sSigma$ is such that $f''(\LCR^\sSigma)=0$.

Second, we note that, in generic scenarios,\footnote{
    That is to say, in cases where $f(\LCR)\neq\LCR-2\Lambda+\alpha\LCR^2$ (for any real $\Lambda$ and $\alpha$) and $\LCR^\sSigma$ is such that $f''(\LCR^\sSigma)\neq 0$.
} the smooth-matching conditions in $\LCf$ gravity are the same as in GR (continuity of the first and second fundamental forms across $\Sigma$) together with the continuity of the Ricci scalar and its normal derivative. This is the reason why the latter two conditions are often referred to as the \emph{third} and \emph{fourth} junction conditions in the literature focusing on smooth-matching problems in $\LCf$ theories of gravity.

Third, the appearance of these two additional junction conditions in metric $\LCf$ gravity is not coincidental; they are connected to the fact that these theories propagate an additional degree of freedom, the scalaron. As explained in Section \ref{sec:intro:metric f(R)}, in the Einstein frame, the scalaron is given by \eqref{scalaron}. Because the natural logarithm is a continuous function, requiring $\LCR$ to be continuous across $\Sigma$ is equivalent to requiring the scalaron to be continuous, and the same is true for their respective derivatives. These naive results are precisely the ones that are rigorously obtained by applying the formalism we have developed in this Section to the Einstein-frame version of $\LCf$ gravity \cite{Deruelle:2007pt}.

\part{Main results.}
\label{part:results}

\chapter{Collapsing stars in \titlebm{$f(R)$} theories of gravity.}
\label{chapter:f(R) collapse}

Despite their success in explaining cosmological observations, $f(R)$ theories of gravity\footnote{
    Throughout this Chapter, we will use the symbol `$f(R)$'  to refer collectively to Palatini $f(\MAR)$ and metric $f(\LCR)$ gravity theories.
} have not proven to be equally fruitful when one attempts to describe compact-object dynamics. So far, only static stellar configurations have been studied within the context of $f(R)$, both in the relativistic \cite{AparicioResco:2016xcm,Astashenok:2017dpo} and non-relativistic cases---for a review, see \cite{Olmo:2019flu}. What is more, gravitational collapse is still poorly understood in the context of $f(R)$ gravity, conversely, exact collapsing solutions in GR have been known from very early on. For example, the Oppenheimer-Snyder model \cite{Oppenheimer:1939ue}, describing the gravitational collapse of a uniform-density dust star, was conceived as early as 1939. The Oppenheimer-Snyder construction is particularly insightful in the sense that it is the simplest possible model of gravitational collapse. More realistic descriptions are all qualitatively similar to the Oppenheimer-Snyder picture, hence its importance.

When one attempts to describe the gravitational collapse of uniform-density dust stars in $f(R)$ gravity, most difficulties arise because the junction conditions of these theories differ from the renowned Darmois-Israel junction conditions of GR discussed in Section \ref{sec:GR JCs}. In particular, junction conditions in $f(R)$ gravity put tighter bounds than those of GR, both in the Palatini \cite{Olmo:2020fri} and metric \cite{Deruelle:2007pt,Senovilla:2013vra} formalisms. On the one hand, any construction involving a dust-star interior and a vacuum exterior is impossible on Palatini $f(\MAR)$ gravity, as per its junction conditions.\footnote{
    To the best of our knowledge, the fact that dust stars are incompatible with the junction conditions of Palatini $f(\MAR)$ gravity presented in \cite{Olmo:2020fri} has not been pointed out on any previous works, even though it is a straightforward consequence of such junction conditions. We will derive this result in Section \ref{sec:OS Palatini}.
} On the other hand, dust stars are not \emph{a priori} incompatible with the junction conditions of metric $f(\LCR)$ gravity; however, progress towards a simple account of their collapse within the metric formalism has been further hindered by the necessary non-triviality of the exterior space-time, which remains unknown \cite{Bueno:2017sui}. Because of this, previous works in the literature did not take into account the junction conditions nor the exterior, and instead focused on determining the evolution of the interior space-time for various equations of state \cite{Cembranos:2012fd,Astashenok:2018bol}.

For all these reasons, in this Chapter---which is based on Publication \cite{Casado-Turrion:2022xkl}---we shall endeavour to take a first step towards shedding some light on the issue of stellar collapse in $f(R)$ gravity. We will consider a spherically-symmetric, uniform-density dust star collapsing under its own gravitational pull, with the purpose of extracting as much information as possible from the relevant junction conditions. In particular, we will able to obtain some no-go results which severely constrain the form of the exterior metric in the metric formalism. Furthermore, we will also show that the Oppenheimer-Snyder model of gravitational collapse is not possible in Palatini $f(\MAR)$ gravity.

Our work is further motivated by the existing connection between gravitational collapse in metric $f(\LCR)$ gravity and the black-hole no-hair theorems (NHTs) \cite{Bueno:2017sui}. As mentioned back in Section \ref{sec:intro:metric f(R)}, metric $f(\LCR)$ gravity is dynamically equivalent to a scalar-tensor theory. It is well known that the NHTs hold in $f(\LCR)$ theories whose additional scalar degree of freedom satisfies some technical (but still general and easily achievable) conditions \cite{Sotiriou:2011dz}. The NHTs guarantee that the only stationary, linearly-stable black holes resulting from gravitational collapse are the same as in GR, i.e.~those belonging to the Kerr family. In particular, this implies that a non-rotating star should collapse into a Schwarzschild black hole in theories satisfying the NHTs. However, the resulting Schwarzschild space-time would have trivial (i.e.~constant) scalar hair, while the space-time outside the collapsing star is hairy---as originally pointed out in \cite{Bueno:2017sui}, and discussed in Section \ref{intandext}. Consequently, the scalar field should disappear dynamically as the star collapses; a detailed account of the process of gravitational collapse in metric $f(\LCR)$ gravity could thus shed light on the mechanism by which a star would dispose of its originally non-trivial scalar hair.

The Chapter shall be organised as follows. First, in Section, \ref{sec:Oppenheimer-Snyder}, we will review the Oppenheimer-Snyder model of gravitational collapse in GR. Afterwards, in Section \ref{Section2}, we shall present the rudiments for the collapse of spherical dust configurations in the context of metric $f(\LCR)$ models of gravity. Therein, in Section \ref{OS incompatible with f(R)}, we shall revisit the incompatibility of the Oppenheimer-Snyder model with these theories. In Section \ref{intandext}, we will provide the interior space-time---i.e.~a Friedmann-Lema\^itre-Robertson-Walker (FLRW)-like metric---as well as the hypotheses for the exterior metric. In \ref{SystematicApproach}, we shall sketch the systematic approach to follow in the upcoming Sections. Subsequently, in Section \ref{junction conditions f(R)}, we shall make the specific form of junctions conditions explicit when the smooth matching of the interior and exterior space-times under consideration is imposed. Sections \ref{ruling out} and \ref{sec:OS Palatini} then constitute the core of the investigation condensed in this Chapter. In \ref{ruling out}, we present a series of five crucial no-go results---plus one corollary---severely constraining the viable exterior space-times which can be smoothly connected with dust FLRW interiors in metric $f(\LCR)$. Proofs of such results appear in  Subsection \ref{Proofs Results 1 2 Corollary} and Subection \ref{Proofs Results 3 4 5}. Section \ref{sec:OS Palatini} is then devoted to proving that the Oppenheimer-Snyder model of gravitational collapse is unfeasible in Palatini $f(\MAR)$ gravity. Finally, we conclude the Chapter with Section \ref{Conclusions}, where conclusions and future work prospects are summarised. Most detailed calculations related to this Chapter have been relegated to Appendix \ref{Appendix:stellar collapse formulae}. Therein, Appendix \ref{ArealRadiusJCAppendix} resorts to the so-called `areal-radius' coordinates to present the $f(\LCR)$ junction conditions in this system of coordinates in detail, whereas Appendix \ref{NOTArealRadiusJCAppendix} introduces such conditions without turning to `areal-radius' coordinates.

Hereafter, we will consistently refer to $f(\LCR)$ theories satisfying $f''(\LCR)\neq 0$ as `generic $f(\LCR)$ gravity,' so as to differentiate them from GR plus a cosmological constant. This is because, in generic $f(\LCR)$ models, all four junction conditions described in Section \ref{sec:f(R) JCs} remain non-trivial (i.e.~their junction conditions are not merely the Darmois-Israel ones).

\section{Oppenheimer-Snyder collapse.} 
\label{sec:Oppenheimer-Snyder}

\begin{figure}[t]
    \centering
    \begin{overpic}[width=0.85\textwidth]{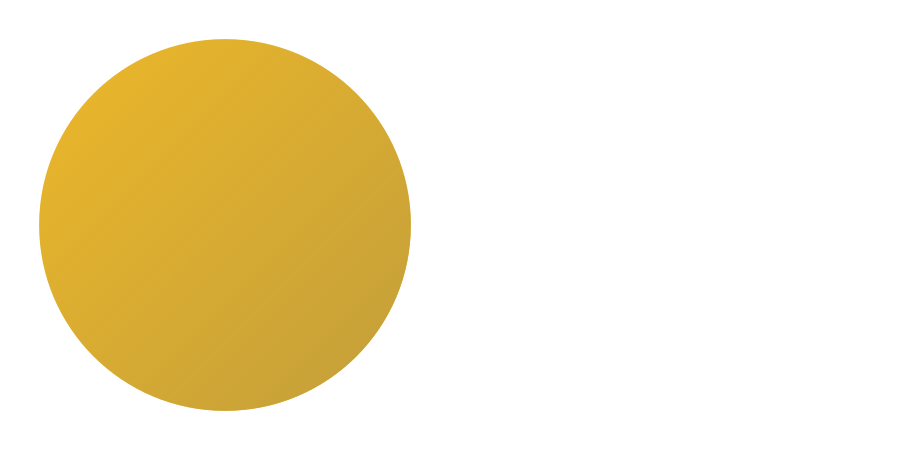}
        \put(25,25){\makebox(0,0){$\mathcal{M^-},\myskip\rho\neq\rho(\mathbf{x}),\myskip p=0$}}
        \put(40,40){$\leftarrow\Sigma_*\equiv\text{stellar surface}$}
        \put(75,25){\makebox(0,0){$\mathcal{M^+},\myskip\rho=0,\myskip p=0$}}
    \end{overpic}
    \caption{Schematic representation of a spatially-uniform dust star surrounded by vacuum.}
    \label{fig:dust star}
\end{figure}

In their seminal 1939 article \cite{Oppenheimer:1939ue}, Oppenheimer and Snyder considered a spatially-uniform dust star---in other words, a star made of a pressureless perfect fluid---collapsing under its own gravitational pull within the framework of GR (a pictorial representation of the Oppenheimer-Snyder construction can be found on Figure \ref{fig:dust star}). Although unrealistic and purely academic in nature, this construction is the simplest possible description of gravitational collapse and black-hole formation one can possibly devise, hence being an excellent starting point for any study on the subject. The absence of pressure inside the star implies that no other interaction aside from gravity is present. Therefore, the model is able to capture the essential features of gravitational collapse while retaining computational simplicity.

Because the matter that makes up the star is only subject to gravity, any fluid element falls freely, that is to say, following time-like geodesics. This allows one to introduce a coordinate system $x_-^\mu=(\tau,\chi,\theta,\varphi)$ adapted to such motion on the interior space-time. Time coordinate $\tau$ represents the proper time along the geodesics, while $\chi$ is a comoving radial coordinate, i.e.~each infinitesimal fluid element is associated to a single, fixed value of $\chi$, which remains unchanged during the entire process of collapse. This implies that the stellar surface---which we shall denote $\Sigma_*$---is always located at
\begin{equation} \label{intstellarsurface}
    \chi=\chi_*=\const.
\end{equation}

In GR, the line elements inside and outside the collapsing star are determined by solving Einstein's equations \eqref{Einstein equations} with the corresponding matter sources: dust for the interior region and vacuum for the exterior. Hence, $\stress^-_{\mu\nu}=\rho\,u_\mu u_\nu$ and $\stress^+_{\mu\nu}=0$, where $\rho=\rho(\tau)$ is the star's energy density, while $u_\mu$ is the four-velocity of any fluid element (thus, in interior, comoving coordinates \smash{$x^\mu_-$}, its components are $u_\mu=\delta^\tau_\mu$).
On the one hand, the metric inside the star turns out to be \cite{Weinberg:1972kfs} a closed ($k>0$) Friedmann-Lema\^{i}tre-Robertson-Walker (FLRW) space-time,
\begin{equation} \label{FLRW}
    \dif s_-^2=-\dif\tau^2+a^2(\tau)\left(\dfrac{\dif\chi^2}{1-k\chi^2}+\chi^2\dif\Omega^2\right),
\end{equation}
whose scale factor $a(\tau)$ satisfies the cycloid equation
\begin{equation} \label{cycloid}
    \dot{a}^2(\tau)=k\left[\dfrac{1}{a(\tau)}-1\right],
\end{equation}
where we are denoting $\dot{\,}\equiv\dif/\dif\tau$. The scale factor $a(\tau)$ is subject to the initial condition $a(0)=1$, which, in turn, implies that $\dot{a}(0)=0$. As per the Friedmann equation, constant $k>0$ is related to the initial energy density $\rho_0\equiv\rho(0)$ of the star; in particular,
\begin{equation} \label{k OS}
    k=\dfrac{\kappa\rho_0}{3}.
\end{equation}
Furthermore, the conservation equation requires the energy density to evolve in $\tau$ as $\rho(\tau)=\rho_0/a^3(\tau)$, as expected.

On the other hand, in GR, the only possible exterior solution is, by virtue of Birkhoff's theorem \cite{Birkhoff}, the Schwarzschild metric,
\begin{equation} \label{Schwarzschild}
    \dif s_+^2=-\left(1-\dfrac{2GM}{r}\right)\dif t^2+\left(1-\dfrac{2GM}{r}\right)^{-1}\dif r^2+r^2\dif\Omega^2,
\end{equation}
where $M$ is the total ADM mass of the space-time, as measured by an observer at spatial infinity \cite{Arnowitt:1962hi}. Coordinates $x_+^\mu=(t,r,\theta,\varphi)$---hereafter, the `exterior' coordinates---are not comoving with the matter within the star. As a result, $\Sigma_*$ becomes $\tau$-dependent when expressed in exterior coordinates:
\begin{equation} \label{stellarsurface out r}
    t=t_*(\tau),\myhugeskip r=r_*(\tau).
\end{equation}

The space-time resulting from smoothly-matching the interior and exterior metrics \eqref{FLRW} and \eqref{Schwarzschild} at $\Sigma_*$ will be a properly-glued solution of the Einstein field equations \eqref{Einstein equations} provided that the the two Darmois-Israel junction conditions of GR---reviewed in Section \ref{sec:GR JCs}---are satisfied:
\begin{coloritemize}
    \item \emph{First junction condition}: continuity of the induced metric $h_{ab}$ across the matching surface $\Sigma_*$, in other words, $\jump{h_{ab}}=0$.
    \item \emph{Second junction condition}: continuity of the extrinsic curvature $K_{ab}$ of $\Sigma_*$, that is to say, $\jump{K_{ab}}=0$.
\end{coloritemize}
We have decided to employ the three-dimensional formulation of the aforementioned junction conditions due to the fact that intrinsic coordinates $y^a=(\tau,\theta,\varphi)$ are ideally-suited to study the problem. In particular, owing to \eqref{stellarsurface out r}, one may readily compute \smash{$h^\pm_{\mu\nu}$} and \smash{$K^\pm_{\mu\nu}$} using said $y^a$ and expressions \eqref{eq:h_ab+-} and \eqref{eq:K_ab+-}, respectively.

For illustrative purposes, and to introduce notation, we shall briefly review in the following how the junction conditions of GR give shape to the Oppenheimer-Snyder model of gravitational collapse, without detailing the computations that lead to the matching equations. For further reference (and completeness), the exhaustive derivation---which closely follows the treatment of the problem given in reference \cite{Poisson:2009pwt}---may be found in Appendix \ref{ArealRadiusJCAppendix}, more precisely in its first three Subsections.

If the junction conditions are satisfied, they must provide us with two crucial pieces of information:
\begin{coloritemize}
    \item How the matching surface $\Sigma_*$ evolves, in particular, the functional dependence of $t_*$ and $r_*$ on proper time $\tau$.
    \item Whether there is a relationship between the parameters of the interior and exterior space-times, which in the case of Oppenheimer-Snyder collapse are $k\propto\rho_0$ and $M$, respectively.\footnote{
        We must stress that $\chi_*$, i.e.~the star's comoving radius, should not be regarded as a \emph{parameter} of the interior metric but an \emph{initial condition} for the matching. Nonetheless, $\chi_*$ will still appear in the equation relating $k\propto\rho_0$ and $M$, as one would intuitively expect.
    }
\end{coloritemize}
Hence, in order to understand not only \emph{how}, but also \emph{why} the dust-star interior matches the Schwarzschild exterior, we first need to decipher which of these two roles each junction condition plays in the gluing of both space-times.

As shown in Appendix \ref{ArealRadiusJCAppendix}, the first junction condition---the matching the induced metrics at $\Sigma_*$---will yield only two independent equations. The first of these is
\begin{equation}
    r_*(\tau)=a(\tau)\,\chi_*, \label{1.1 OS int}
\end{equation}
and simply states that $r_*(\tau)$ is proportional to the scale factor $a(\tau)$ of the interior metric,\footnote{
    We note that we do not write `$a_*$' in the junction conditions because $a$ is a function of $\tau$ only, and thus $a_*(\tau)=a(\tau)$.
} which evolves in $\tau$ according to equation \eqref{cycloid}. Because $a(0)=1$, the stellar radius decreases from its initial value $r_*(0)=\chi_*$ (i.e.~the star's comoving radius) to zero in finite time, following cycloid curve \eqref{cycloid}. Let us also remark that, upon differentiation, \eqref{1.1 OS int} supplies all derivatives of $r_*$ with respect to $\tau$, in particular $\dot{r}_*$. This is important because the second contraint coming from requiring $\jump{h_{ab}}=0$ is
\begin{equation}
    \dot{t}_*(\tau)=\dfrac{\sqrt{\dot{r}_*^2(\tau)+A_*(\tau)}}{A_*(\tau)}, \label{1.2 OS int}
\end{equation}
where $A(r)=1-2GM/r$ and $A_*(\tau)\equiv A(r_*(\tau))$. The only unknown function in this expression is $t_*(\tau)$, since one may readily obtain the dependence of $r_*$ and $\dot{r}_*$ on $\tau$ from \eqref{1.1 OS int}. As such, \eqref{1.2 OS int} is merely an ordinary, first-order differential equation for $t_*(\tau)$, which always has a unique solution given some initial condition. Consequently, the physical interpretation of the two equations coming from the first junction condition is clear: \eqref{1.1 OS int} and \eqref{1.2 OS int} allow for a complete determination of the evolution of the stellar surface, which is characterised by functions $r_*(\tau)$ and $t_*(\tau)$.

Having considered the first junction condition, we must also impose the second one, namely, the continuity of the extrinsic curvatures of $\Sigma_*$. As previously mentioned (and shown in Appendix \ref{Oppenheimer-Snyder Appendix}), the second junction condition only provides one additional equation, which is
\begin{equation} \label{rdot OS}
    \dot{r}_*^2(\tau)=1-k\chi_*^2-A_*(\tau).
\end{equation}
Replacing $r_*(\tau)$ and $\dot{r}_*(\tau)$ in favour of $a(\tau)$ and $\dot{a}(\tau)$ using the first junction condition \eqref{1.1 OS int}, Equation \eqref{rdot OS} becomes
\begin{equation}
    \dot{a}^2(\tau)=-k+\dfrac{2GM}{a(\tau)\,\chi_*^3}.
\end{equation}
Meanwhile, $\dot{a}(\tau)$ may be expressed in terms of $a(\tau)$ using the cycloid equation \eqref{cycloid}. This leads to
\begin{equation}
    k\left[\dfrac{1}{a(\tau)}-1\right]=-k+\dfrac{2GM}{a(\tau)\,\chi_*^3}.
\end{equation}
We immediately notice that the $a$s cancel; hence, all the dependence in $\tau$ disappears from the previous expression. As a result, one finally obtains an algebraic relation between parameters $M$, $k\propto\rho_0$ and $\chi_*$, specifically
\begin{equation} \label{M OS}
    M=\dfrac{k}{2G}\chi_*^3=\dfrac{4\pi}{3}\rho_0\chi_*^3,
\end{equation}
where in the last step we have made use of \eqref{k OS}. We thus see that, in GR, the dust-star interior and the Schwarzschild exterior can only be smoothly matched provided that the stellar mass $M$ equals the product of the star's initial volume, $4\pi\chi_*^3/3$, and its initial energy density, $\rho_0$. This completes the Oppenheimer-Snyder construction in pure GR.

With minimal modifications, the Oppenheimer-Snyder model of gravitational collapse is also a properly matched solution of GR plus a cosmological constant $\Lambda$ \cite{Markovic:1999di}, which may be understood as being the simple metric $f(\LCR)$ model $f(\LCR)=\LCR-2\Lambda$. As discussed at the end of Section \ref{sec:f(R) JCs}, this $f(\LCR)$ model is unique since it satisfies $f''(\LCR)=0$ for any $\LCR$, which entails that its junction conditions are exactly the same as in GR without a cosmological constant \cite{Deruelle:2007pt,Senovilla:2013vra}.

\begin{figure}[h!]
     \centering
     \caption[Two possible outcomes for the collapse of a uniform-density dust star in metric $f(\LCR)$ gravity.]{Two possible outcomes for the collapse of a uniform-density dust star in metric $f(\LCR)$ gravity. As we shall see in the following Sections, the evolution of the stellar surface parallels the behaviour of the interior FLRW spacetime's scale factor.}
     \begin{subfigure}[b]{\textwidth}
         \centering
         \vspace{0.25cm}
         \includegraphics[height=2.9cm]{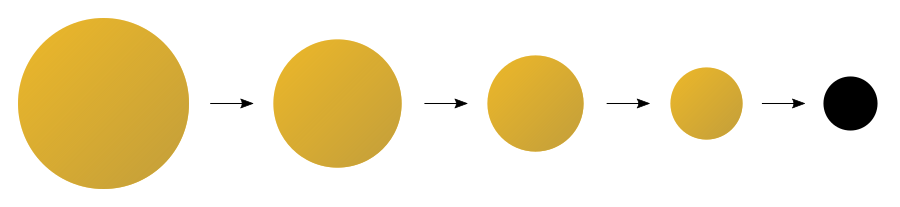}
         \vspace{0.25cm}
         \caption{The star collapses into a black hole, as in the Oppenheimer-Snyder model.}
         \label{fig:collapse outcome 1}
     \end{subfigure}
     \vspace{0.25cm}
     \begin{subfigure}[b]{\textwidth}
         \centering
         \vspace{0.25cm}
         \includegraphics[height=2.9cm]{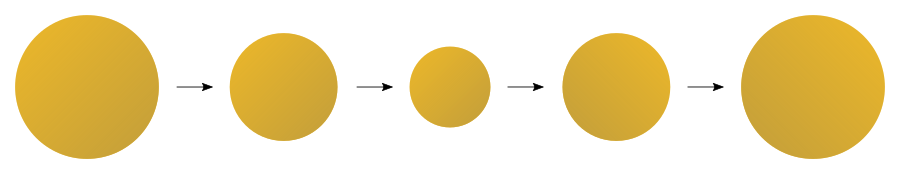}
         \vspace{0.05cm}
         \caption{The star \emph{bounces}, i.e.~contracts until it reaches a certain radius, and then re-expands again.}
         \label{fig:collapse outcome 2}
    \end{subfigure}
    \label{fig:collapse outcomes}
\end{figure}

If a cosmological constant $\Lambda$ is included in the gravitational action, the exterior space-time is necessarily the Schwarzschild-(Anti-)de Sitter metric, also known as the Kottler space-time:
\begin{equation} \label{Kottler}
    \dif s_+^2=-\left(1-\dfrac{2GM}{r}-\dfrac{\Lambda r^2}{3}\right)\dif t^2+\left(1-\dfrac{2GM}{r}-\dfrac{\Lambda r^2}{3}\right)^{-1}\dif r^2+r^2\,\dif\Omega^2,
\end{equation}
As a result, equations \eqref{1.2 OS int} and \eqref{rdot OS} are also valid in this case, but now with $A(r)=1-2GM/r-\Lambda r^2/3$ instead of just $A(r)=1-2GM/r$. The equation for the scale factor, however, gets modified if a cosmological constant is present, and becomes
\begin{equation}
    \dot{a}^2(\tau)=\dfrac{\kappa\rho_0}{3a(\tau)}+\dfrac{\Lambda a^2(\tau)}{3}-k.
\end{equation}
Thus, the value of $k$ also changes in this model:
\begin{equation}
    k=\dfrac{\kappa\rho_0+\Lambda}{3}.
\end{equation}
Combining these expressions with junction conditions \eqref{1.2 OS int} and \eqref{rdot OS}, one finds that the matching is possible for any $\Lambda$ provided that the relation between $M$, $k$ and $\chi_*$ is given, once again, by \eqref{M OS}. However, in the presence of a cosmological constant, the star may either collapse or bounce depending on the specific values of $\Lambda$ and $M$. Moreover, if $GM>\sqrt{\Lambda}/3$, the exterior space-time cannot avoid contracting into a `big-crunch' singularity as well, dragged by the gravitational pull of the collapsing dust star. It is clear that more complicated choices of function $f$ might produce similar effects. In particular, a modification of the scale factor dynamics and of the expression for $k$ is always to be expected in any $f(\LCR)$ theory. As we have seen, the modified dynamics could potentially lead to a complete evasion of gravitational collapse (as illustrated on Figure \ref{fig:collapse outcomes}), or even to the formation of singularities. Therefore, these possible issues must always be carefully considered.

\section[Collapsing dust stars in generic metric \titlemath{$f(\LCR)$} models.]{Collapsing dust stars in generic metric \titlebm{$f(\LCR)$} models.}
\label{Section2}

\subsection[Revisiting the incompatibility of Oppenheimer-Snyder collapse with generic metric \titlemath{$f(\LCR)$} models.]{Revisiting the incompatibility between Oppenheimer-Snyder collapse and generic metric \titlebm{$f(\LCR)$} models.} \label{OS incompatible with f(R)}

Given its importance for our understanding of gravitational collapse, it is natural to wonder whether the Oppenheimer-Snyder construction, as described in Section \ref{sec:Oppenheimer-Snyder}, is a properly-matched solution of generic $f(\LCR)$ models in the metric formalism. The answer is that it is \emph{not}, as carefully proven in \cite{Senovilla:2013vra}, because the junction conditions of metric $f(\LCR)$ gravity---reviewed back in Section \ref{sec:f(R) JCs}---differ from those of GR. The modified junction conditions also imply that a larger class of collapse models are incompatible with metric $f(\LCR)$ gravity as well \cite{Senovilla:2013vra,Goswami:2014lxa}. The more stringent junction conditions entail that the Oppenheimer-Snyder construction must be modified in $f(\LCR)$ gravity; in particular, as we shall see now, the exterior cannot be Schwarzschild.

Besides the two renowned Darmois-Israel junction conditions of GR, generic metric $f(\LCR)$ gravity models require two additional constraints to be satisfied so as to make a given smooth matching possible. In particular, as shown in Section \ref{sec:f(R) JCs}, the two supplementary junction conditions in metric $f(\LCR)$ gravity are:
\begin{coloritemize}
    \item \emph{Third junction condition}: continuity of the Ricci scalar across $\Sigma_*$, that is to say, $\jump{\LCR}=0$.
    \item \emph{Fourth junction condition}: continuity of the normal derivative of the Ricci scalar across $\Sigma_*$, in other words, $n^\mu\jump{\LCD_\mu\LCR}=0$.
\end{coloritemize}
Let us now comment on how the incompatibility between Oppenheimer-Snyder collapse and these two novel junction conditions arises.\footnote{
    In the original proof of the incompatibility \cite{Senovilla:2013vra}, the author considers several known glued solutions of GR (including Oppenheimer-Snyder), and then determines whether they are solutions of $f(\LCR)$ gravity as well. His derivation consists of a proof by \emph{reductio ad absurdum}: the Oppenheimer-Snyder construction is assumed to be a solution of $f(\LCR)$ gravity, and then it is shown that this implies that the dust star has vanishing energy density everywhere. Herein, we shall follow a different, although equivalent, approach.
} It is well known that both the interior FLRW metric \eqref{FLRW} and the exterior Schwarzschild space-time \eqref{Schwarzschild} are solutions of $f(\LCR)$ gravity. In the case of FLRW, only the dependence of $a(\tau)$ in $\tau$ gets modified \cite{Cembranos:2012fd}, as we shall see later, while Schwarzschild is a solution of every $f(\LCR)$ model satisfying $f(0)=0$ \cite{delaCruz-Dombriz:2009pzc}. If one attempted to glue these two space-times at the stellar surface, the third junction condition $\jump{\LCR}=0$ would then require
\begin{equation} \label{3 OS f(R)}
    \LCR^-_*=6\left[\dfrac{\dot{a}^2(\tau)+k}{a^2(\tau)}+\dfrac{\ddot{a}(\tau)}{a(\tau)}\right]=\LCR^+_*=0.
\end{equation}
This is an ordinary differential equation for the scale factor $a(\tau)$ which may be integrated using the standard initial conditions $a(0)=1$ and $\dot{a}(0)=0$, yielding
\begin{equation} \label{constantRa}
    a(\tau)=\sqrt{1-k\tau^2}.
\end{equation}
Therefore, the third junction condition $\jump{\LCR}=0$---which is exclusive to generic metric $f(\LCR)$ gravity---\emph{fixes} the scale factor of the interior space-time to be given by \eqref{constantRa}, instead of the cycloid equation \eqref{cycloid} one has in GR. Moreover, as explained in Section \ref{sec:Oppenheimer-Snyder}, fixing $a(\tau)$ amounts to fixing the evolution of $r_*(\tau)$ and $t_*(\tau)$, i.e.~of the stellar surface \eqref{stellarsurface out r}, by virtue of the first junction condition.

At first sight, \eqref{constantRa} seems to be an appropriate replacement of the cycloid-like scale factor obtained from expression \eqref{cycloid}; at least, it shares a number of its most distinguishing features. Indeed, an interior FLRW space-time with a scale factor given by expression \eqref{constantRa} would still be dynamic (that is to say, dependent on $\tau$), despite having a vanishing Ricci scalar. Moreover, if $k>0$, the star would collapse to zero proper volume in finite proper time, just as in the Oppenheimer-Snyder model. And, remarkably, \emph{all four} junction conditions would be satisfied by construction.\footnote{
    If $\LCR^+=\LCR^-=0$, then $n^\mu\jump{\LCD_\mu\LCR}=0$, and the fourth junction condition is satisfied automatically.
} However, we must bear in mind that \eqref{3 OS f(R)} is simply an additional \emph{constraint}, in the sense that it is a relationship between metrics. This relationship may or may not be compatible with the equations of motion of $f(\LCR)$ gravity. In other words, the third junction condition requires the scale factor to be given by \eqref{constantRa}; this, in turn, implies that the matching between the interior FLRW space-time and the exterior Schwarzschild metric can only occur in those $f(\LCR)$ theories whose equations of motion give rise to a scale factor which evolves in $\tau$ according to expression \eqref{constantRa}. As we shall prove in Section \ref{ruling out} (more precisely, in Subsection \ref{Proofs Results 1 2 Corollary}), an interior dust FLRW space-time whose scale factor is given by \eqref{constantRa} does not solve the equations of motion of $f(\LCR)$ gravity for any choice of function $f$. 

In consequence, Oppenheimer-Snyder collapse is not possible in metric $f(\LCR)$ gravity, as anticipated. Furthermore, we clearly see that the incompatibility arises because we have insisted that the exterior space-time be Schwarzschild. Thus, we are led to conclude that, in generic $f(\LCR)$ gravity, the exterior must be a different, more general space-time.

\subsection{Interior and exterior metrics.} \label{intandext}

Since the Oppenheimer-Snyder model of collapse is no longer a valid matched solution of generic $f(\LCR)$ gravity in the metric formalism, one needs to reconsider whether metrics \eqref{FLRW} and \eqref{Schwarzschild} correctly describe the interior and the exterior of the collapsing uniform-density star (respectively) in these theories.\footnote{
    We must remark at this point that we assume that the interior stress-energy tensor corresponds to dust in the Jordan frame representation of the theory only, and not in the conformally-related Einstein frame.
}

As shown in \cite{Cembranos:2012fd}, the space-time corresponding to a spherically-symmetric, uniform-density distribution of dust in any $f(\LCR)$ theory of gravity is still a portion of FLRW space-time \eqref{FLRW}, which in our case stretches out up until the stellar surface $\chi=\chi_*$. Thus, the $f(\LCR)$ field equations do not change the interior metric; however, they \emph{do} alter the scale-factor dynamics. More precisely, the equation for $a(\tau)$ now becomes
\begin{equation} \label{ap2 f(R)}
    \dfrac{\dot{a}^2(\tau)+k}{a^2(\tau)}=\dfrac{1}{2f'(\LCR^-)}\left[\dfrac{\kappa\rho_0}{3a^3(\tau)}+\ddot{f}'(\LCR^-)+\dfrac{\dot{a}(\tau)}{a(\tau)}\,\dot{f}'(\LCR^-)+\dfrac{f(\LCR^-)}{3}\right],
\end{equation}
which reduces to the cycloid equation \eqref{cycloid} of GR when $f(\LCR)=\LCR$, as expected. Assuming the usual initial conditions $a(0)=1$ and $\dot{a}(0)=0$, the expression for $k$ is modified accordingly,
\begin{equation} \label{k f(R)}
    k=\dfrac{\kappa\rho_0+3\ddot{f}'(\LCR_0^-) +f(\LCR_0^-)}{6f(\LCR_0^-)},
\end{equation}
where $\LCR_0^-\equiv\LCR^-(0)$.

\begin{figure}[t]
    \centering
    \begin{overpic}[width=0.8\textwidth]{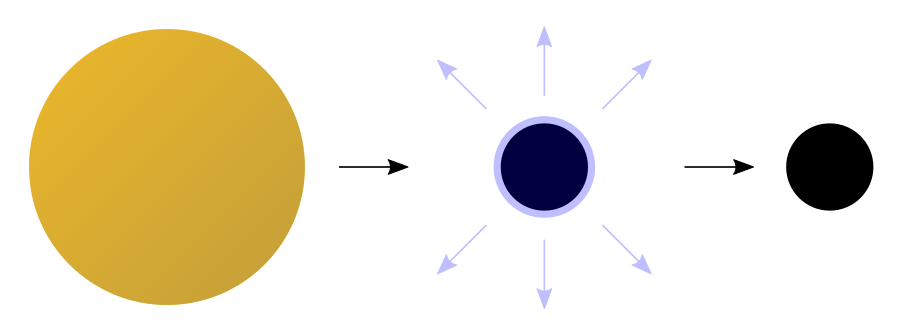}
        \put(18.5,-2.5){\makebox(0,0){(a)}}
        \put(60.5,-2.5){\makebox(0,0){(b)}}
        \put(91.75,-2.5){\makebox(0,0){(c)}}
    \end{overpic}
    \vspace{0.6cm}
    \caption[Formation of a Schwarzschild black hole through gravitational collapse in metric $f(\LCR)$ gravity. Expected stages.]{Formation of a Schwarzschild black hole through gravitational collapse in metric $\smash{f(\LCR)}$ gravity. Expected stages. (a) The collapsing star is surrounded by a non-trivial exterior. (b) Eventually, a horizon forms, and the solution is now a non-Schwarzschild black hole. From an Einstein-frame perspective, this black hole possesses scalar hair, which needs to be radiated away to infinity in order to comply with the NHTs. (c) Once the black hole has disposed of all its scalar hair, it becomes an ordinary Schwarzschild solution, as required by the NHTs.}
    \label{fig:bh descalarisation}
\end{figure}

As mentioned in Section \ref{OS incompatible with f(R)}, the space-time outside the collapsing dust star cannot be Schwarzschild. Consequently, it must be a non-trivial, possibly time-dependent, spherically-symmetric, vacuum solution of $f(\LCR)$ gravity. These assertions are further supported by a theorem in Reference \cite{Bueno:2017sui}, which states that only in theories propagating just a traceless and massless graviton \emph{in vacuo} the gravitational field outside a spherically-symmetric mass distribution can be represented by metrics of the form
\begin{equation} \label{singlefunction}
    \dif s_+^2=-A(r)\,\dif t^2+A^{-1}(r)\,\dif r^2+r^2\,\dif\Omega^2,
\end{equation}
of which Schwarzschild is a particular example. As discussed back in Section \ref{sec:intro:metric f(R)}, apart from the usual massless and traceless graviton, generic metric $f(\LCR)$ theories propagate an additional scalar degree of freedom, known as the scalaron, given by \eqref{scalaron} in the Einstein-frame representation of the theory. Thus, the theorem in \cite{Bueno:2017sui} guarantees that, in generic metric $f(\LCR)$ gravity, even though Schwarzschild remains as a vacuum solution of the field equations, it can only exist as a black hole, not as an exterior space-time smoothly matching \emph{any} interior matter distribution whatsoever. One can intuitively understand this result by noticing that the scalaron \eqref{scalaron} will---as widely known---couple to the trace of the stress-energy tensor. Hence, it will be excited provided that there is matter somewhere in the space-time. Nonetheless, equation \eqref{scalaron} reveals that Schwarzschild space-time only supports a trivial (i.e.~constant) scalar field.\footnote{
    For the time being, we are assuming that $f'(0)\neq 0$, so that the scalaron remains finite and constant for solutions with $\LCR=0$. in Chapter \ref{chapter:constant curvature}, we will prove that $f(\LCR)$ models such that $f(0)=0$ and $f'(0)=0$ give rise to instabilities and other pathologies which put their physical viability into question. 
} The fact that the original black hole is necessarily hairy entails that the scalaron must be eventually radiated away in the last stages of collapse (see Figure \ref{fig:bh descalarisation}), so as to comply with the NHTs for $f(\LCR)$ gravity \cite{Sotiriou:2011dz}.

\subsection{A systematic approach for junction conditions.}
\label{SystematicApproach}

Given that we do not know \emph{a priori} what the space-time outside a collapsing uniform-density dust star is in $f(\LCR)$ gravity (in fact, we can only assume that it is spherically symmetric, as the stellar surface), there are essentially two ways of approaching the problem:
\begin{coloritemize}
    \item On the one hand, one could \emph{verify} whether a known vacuum solution of $f(\LCR)$ gravity matches the interior FLRW spacetime, using the junction conditions, by explicitly checking that all four junction conditions are satisfied simultaneously.
    \item On the other, one may attempt to \emph{infer properties} of the exterior metric from the junction conditions, choosing a suitable spherically-symmetric \emph{Ansatz} for the exterior metric.
\end{coloritemize}
None of these two options are exempt from difficulties. For instance, the absence of a theorem analogous to Birkhoff's entails that generic $f(\LCR)$ models admit a variety of exterior vacuum solutions.\footnote{
    The strongest result in this respect is that the Schwarzschild space-time is the only static, spherically-symmetric solution \emph{with vanishing Ricci scalar} in $f(\LCR)$ theories satisfying $f(0)=0$ and $f'(0)\neq 0$, cf.~\cite{Nzioki:2009av}. However, this does not exclude the possibility that there exist other exterior vacuum solutions with a non-constant Ricci scalar even in these models.
} Virtually, \emph{any} spherically-symmetric, vacuum metric could be the one to match the interior uniform-density dust-star solution. Hence, without further guidance, it might seem that the pursuing the first approach consists in `looking for a needle in a haystack.' Therefore, the first way of proceeding must be inevitably supplemented by the second: it is necessary to use the junction conditions of $f(\LCR)$ gravity delimit the set of spherically-symmetric space-times which could potentially be the exterior to the collapsing dust star. In fact, as we shall show in Section \ref{ruling out}, the junction conditions of $f(\LCR)$ gravity turn out to impose strong constraints on the exterior space-time. Moreover, these conditions are so restricting that very few known vacuum solutions of metric $f(\LCR)$ gravity comply with them, as far as we are aware.

Notwithstanding this, junction conditions must be handled with special care when extracting information about the exterior space-time from them. As should have become evident from the explanations in Chapter \ref{chapter:Introduction: JCs}, the junction-condition formalism is plagued with subtleties which need to be resolved, often by resorting to physically-motivated arguments. Indeed, in spite of the simple form the matching equations might turn out to take, the most intuitive line of reasoning may present some loopholes, and apparent exceptions become possible, as we will see in the following.

Let us consider, for example, the assertion made in \cite{Goswami:2014lxa} claiming that, for any generic $f(\LCR)$ model, a dynamic, homogeneous space-time with non-constant Ricci scalar cannot be matched with a static space-time across a fixed boundary. Therein, authors argued that this result follows from the third junction condition $\jump{\LCR}=0$, because $\LCR^-_*$ would be a function of $\tau$ only, while $\LCR^+_*$ would be $\tau$-independent. Thus, equating both quantities for all $\tau$ would be impossible.

However, one can exploit a hole in this reasoning in order to find a static exterior solution which would still satisfy the third junction condition. The key is to realise that, for $\jump{\LCR}=0$ to hold when the interior solution has Ricci scalar $\LCR^-=\LCR^-(\tau)\neq\const$, the exterior solution cannot be static \emph{with respect to time coordinate} $\tau$---i.e.~it cannot have a time-like Killing vector field $\partial_\tau$. However, this does not prevent the exterior from being static with respect to a different time coordinate $t$---i.e.~it could have a different time-like Killing vector field $\partial_t$. In order to illustrate this, let us consider the interior FLRW space-time \eqref{FLRW} and an exterior space-time which is static with respect to some time coordinate $t$. Then it is always possible to choose `areal-radius' coordinates $(t,r,\theta,\varphi)$ in which the latter line element takes the form
\begin{equation}
\label{MetricStaticAB}
    \dif s_+^2=-A(r)\,\dif t^2+B(r)\,\dif r^2+r^2\,\dif\Omega^2.
\end{equation}
Let us also assume that the exterior metric has a non-constant Ricci scalar $\LCR^+=\LCR^+(r)$.\footnote{
    This is perfectly reasonable even for a vacuum solution of generic $f(\LCR)$ gravity because, \emph{in vacuo}, the trace \eqref{f(R) EOM trace} of the $f(\LCR)$ equations of motion remains a differential equation for the scalar curvature, rather than algebraic relation as in GR, where $\stress^+=0$ necessarily implies $\LCR^+=0$.
} As shown in the case of Schwarzschild in Section \ref{sec:Oppenheimer-Snyder}, the radial coordinate of this space-time will become a function of $\tau$ upon changing to interior coordinates $(\tau,\chi,\theta,\varphi)$---see, for instance, Equation \eqref{stellarsurface out r}. This reasoning can be shown to carry over to line element \eqref{MetricStaticAB}. Hence, its Ricci scalar $\LCR^+_*$ will generically depend upon $\tau$ even though the exterior metric is static, and thus the matching with a $\tau$-dependent interior scalar curvature (such as that of FLRW space-time) could still be possible. Nonetheless, we shall prove in Section \ref{ruling out} that a thorough analysis of the remaining junction conditions (including their compatibility with the field equations) rules out all static exteriors, and thus the conclusions in \cite{Goswami:2014lxa} remain valid, although for reasons different from those given in that Reference.

Finally, another important point to be considered is that one must always make sure that the constraints imposed by the junction conditions are compatible with the $f(\LCR)$ equations of motion. Recall, for instance, the discussion in Section \ref{OS incompatible with f(R)}. As explained therein, the third junction condition led to a result---equation \eqref{constantRa}---which was incompatible with the equations of motion of $f(\LCR)$ gravity. We thus concluded that the matching was impossible. Analogous scenarios will appear when considering generic exterior space-times in Section \ref{ruling out} below. In summary, after finding a family of metrics which are are not ruled out by the junction conditions, one must still verify whether the equations of motion of the theory hold for said solutions.\footnote{
    To be even more precise, junction conditions are relations between \emph{metrics}, as pointed out before. One may perfectly input \emph{any} desired metric on the junction conditions of a given theory, regardless of whether said space-time is a solution of the field equations.
}

\section[Junction conditions in generic metric \titlemath{$f(\LCR)$} gravity models.]{Junction conditions in generic metric \titlebm{$f(\LCR)$} gravity models.} \label{junction conditions f(R)}

As per the discussion in the previous Sections, the only assumption one can make on the exterior space-time is that it must be spherically symmetric. Consequently, our starting point will be the most general spherically-symmetric line element. We aim at determining which conditions such a space-time should satisfy so as to match an interior FLRW-like metric, whose line element is \eqref{FLRW} in interior coordinates $x^\mu_-=(\tau,\chi,\theta,\varphi)$. Once again, $y^a=(\tau,\theta,\varphi)$ will serve as the natural, induced coordinates on $\Sigma_*$.

In order to establish the junction conditions, one needs to choose a suitable coordinate system to express the exterior metric. For spherically-symmetric exteriors, the most natural one is the coordinate system $(t,r,\theta,\varphi)$ in which the line element is of the form
\begin{equation} \label{sphericallysymmetric}
    \dif s^2_+=-A(t,r)\,\dif t^2+B(t,r)\,\dif r^2+r^2\,\dif\Omega^2,
\end{equation}
i.e.~$r$ is the `areal radius,' in the sense that spherical time-like hypersurfaces centred at the origin have proper area $4\pi r^2$. `Areal-radius' coordinates $(t,r,\theta,\varphi)$ are specially convenient, for various reasons. First, the stellar surface is still given by \eqref{stellarsurface out r} when using these coordinates. Furthermore, the junction conditions remain similar in appearance to those of Oppenheimer-Snyder collapse. In fact, each of the four junction conditions essentially retains its physical interpretation, as given in Section \ref{sec:Oppenheimer-Snyder}. In what follows, we will merely present the equations resulting from imposing the four junction conditions. For the thorough derivation of the junctions conditions appearing immediately below, we refer the reader to Appendix \ref{ArealRadiusJCAppendix}.

Using `areal-radius' coordinates, there are six independent equations coming from the four junction conditions. Two of these equations,
\begin{gather}
    r_*(\tau)=a(\tau)\,\chi_*, \label{1.1 f(R)} \\
    A_*(\tau)\,\dot{t}_*^2(\tau)-B_*(\tau)\,\dot{r}_*^2(\tau)=1, \label{1.2 f(R)}
\end{gather}
are obtained by imposing the first junction condition, $\jump{h_{ab}}=0$. Similarly to Section \ref{sec:Oppenheimer-Snyder}, overdots denote differentiation with respect to $\tau$, while $A_*(\tau)\equiv A(t_*(\tau),r_*(\tau))$ and $B_*(\tau)\equiv B(t_*(\tau),r_*(\tau))$. Equations \eqref{1.1 f(R)} and \eqref{1.2 f(R)} univocally determine the evolution of the matching surface $\Sigma_*$, as in Oppenheimer-Snyder collapse. On the one hand, \eqref{1.1 f(R)} reveals that the stellar areal radius $r_*(\tau)$ is simply proportional to $a(\tau)$; it is exactly the same equation as \eqref{1.1 OS int}. On the other hand, \eqref{1.2 f(R)} is a first-order ordinary differential equation for $t_*(\tau)$.\footnote{
    Since $r_*(\tau)\propto a(\tau)$ as per the first junction condition \eqref{1.1 f(R)}, the $n$-th derivative of $r_*$ with respect to $\tau$ is proportional to the $n$-th derivative of $a$. One can determine $a(\tau)$ by solving \eqref{ap2 f(R)}; once $a(\tau)$ is known, all of its derivatives can be readily obtained. Therefore, amongst all the functions of $\tau$ appearing on \eqref{1.2 f(R)}, the only one which remains unknown---before solving \eqref{1.2 f(R)}---is $t_*(\tau)$. Therefore, junction condition \eqref{1.2 f(R)} is always a differential equation for $t_*(\tau)$, as stated in the text.
} Thus, its solution is unique given an initial condition, in the same way as \eqref{1.2 OS int}. As such, we clearly see that \eqref{1.1 f(R)} and \eqref{1.2 f(R)} allow us to know whether the star either collapses, expands or bounces, depending on the specific dynamics for the scale factor $a(\tau)$ associated to a given choice of $f(\LCR)$. Therefore, the interpretation of the first junction condition in $f(\LCR)$ gravity remains unchanged with respect to GR.

The second pair of equations,
\begin{gather}
    \dot{\beta}(\tau)=\dfrac{A_{t*}(\tau)\,\dot{t}_*^2(\tau)-B_{t*}(\tau)\,\dot{r}_*^2(\tau)}{2}, \label{2.1 f(R)} \\
    \beta(\tau)=\beta_0\,\sqrt{A_*(\tau)\,B_*(\tau)}, \label{2.2 f(R)}
\end{gather}
is obtained by requiring the second junction condition, $\jump{K_{ab}}=0$, to hold. Here (and in what remains of this Chapter), subscript $t$ or $r$ denote differentiation with respect to those coordinates, i.e.~$A_t(t,r)\equiv\partial A(t,r)/\partial t$, $A_{t*}(\tau)\equiv A_t(t_*(\tau),r_*(\tau))$, etcetera. Moreover, we have also introduced 
\begin{equation} \label{betadeff(R)}
    \beta(\tau)\equiv A_*(\tau)\,\dot{t}_*(\tau)=\sqrt{A_*(\tau)+A_*(\tau)\,B_*(\tau)\,\dot{r}_*^2(\tau)},
\end{equation}
and
\begin{equation} \label{beta0 f(R)}
    \beta_0\equiv\sqrt{1-k\chi_*^2}=\const
\end{equation}
In Oppenheimer-Snyder collapse, where $B_*(\tau)=1/A_*(\tau)$ and $A_{t*}(\tau)=B_{t*}(\tau)=0$, expressions \eqref{2.1 f(R)} and \eqref{2.2 f(R)} are satisfied simultaneously, and thus they actually provide only one condition. In the general case, however, \eqref{2.1 f(R)} and \eqref{2.2 f(R)} are distinct equations. What remains unchanged with respect to GR is the interpretation of the second junction condition, namely, that equations \eqref{2.1 f(R)} and \eqref{2.2 f(R)} should provide a relationship between the parameters of the interior and exterior space-times, once the constraints imposed by the other junction conditions are also taken into account.

Finally, the remaining two junction conditions---those exclusive to $f(\LCR)$ gravity---are
\begin{gather}
    \LCR^+_*(\tau)=6\left[\dfrac{\dot{r}_*^2(\tau)+k\chi_*^2}{r_*^2(\tau)}+\dfrac{\ddot{r}_*(\tau)}{r_*(\tau)}\right], \label{3 f(R)} \\
    \dfrac{\dot{r}_*(\tau)}{A_*(\tau)}\LCR_{r*}^+(\tau)+\dfrac{\dot{t}_*(\tau)}{B_*(\tau)}\LCR_{t*}^+(\tau)=0. \label{4 f(R)}
\end{gather}
These two additional constraints arise from imposing $\jump{\LCR}=0$ and $n^\mu\jump{\LCD_\mu\LCR}=0$, respectively. Notice that we have made use of the first junction condition \eqref{1.1 f(R)} to express $a(\tau)$ in terms of $r_*(\tau)$ in \eqref{3 f(R)}. Given that the behaviour of the stellar surface is already fixed by the first junction condition alone, expressions \eqref{3 f(R)} and \eqref{4 f(R)} should, in principle, contribute only to parameter determination, in the same manner as \eqref{2.1 f(R)} and \eqref{2.2 f(R)}.

To sum up, the relevant junction conditions between the interior FLRW space-time \eqref{FLRW} and the spherically-symmetric exterior \eqref{sphericallysymmetric} in generic $f(\LCR)$ gravity are equations \eqref{1.1 f(R)}--\eqref{2.2 f(R)}, \eqref{3 f(R)} and \eqref{4 f(R)}, together with the definitions of $\beta$ and $\beta_0$ given by \eqref{betadeff(R)} and \eqref{beta0 f(R)}, respectively.

Before closing this Section, it is important to stress that the procedure by which matching equations are obtained is fully coordinate-dependent. Accordingly, any coordinate change in expression \eqref{sphericallysymmetric} would require the junction conditions to be derived again. When the exterior space-time is non-static, this introduces some additional difficulties. For example, if the exterior space-time is naturally given in a coordinate system different than $(t,r,\theta,\varphi)$, the change to `areal-radius' coordinates might be very difficult---if not impossible---to perform analytically. Moreover, the junction conditions may become harder to implement if they are derived in a coordinate system different from $(t,r,\theta,\varphi)$. For example, it is always possible to choose coordinates $(\eta,\xi,\theta,\varphi)$ in which the spherically-symmetric exterior line element can be expressed as
\begin{equation}
    \dif s^2_+=-C(\eta,\xi)\,\dif \eta^2+D(\eta,\xi)\,\dif\xi^2+r^2(\eta,\xi)\,\dif\Omega^2.
\end{equation}
Some of the most renowned non-trivial vacuum solutions of metric $f(\LCR)$ gravity, such as the so-called Clifton space-time \cite{Clifton:2006ug} are naturally given in this form, hence its importance. The derivation of the junction conditions using these coordinates, as well as a discussion on their convoluted interpretation, is presented in Appendix \ref{NOTArealRadiusJCAppendix}.

\section{Exterior metrics forbidden by the junction conditions.} \label{ruling out}

In this Section, we intend to obtain constraints on the exterior space-time by means of the junction conditions of $f(\LCR)$ gravity, that is to say, equations \eqref{1.1 f(R)}--\eqref{2.2 f(R)}, \eqref{3 f(R)} and \eqref{4 f(R)} presented in the previous Section. The specific form of these conditions is valid provided that the exterior line element is given by expression \eqref{sphericallysymmetric}, i.e.~when one uses the ideally suited `areal-radius' coordinates $(t,r,\theta,\varphi)$. Thus, in this Section, we aim at building a system of differential equations for functions $A(t,r)$ and $B(t,r)$ in \eqref{sphericallysymmetric} out of the junction conditions. As mentioned back in Section \ref{SystematicApproach}, it is of paramount importance to devise a systematic procedure for manipulating the matching equations in order to draw conclusions drawn.

One can immediately notice that Equations \eqref{1.1 f(R)}--\eqref{2.2 f(R)}, \eqref{3 f(R)} and \eqref{4 f(R)} are not given exclusively in terms $A_*$ and $B_*$; they also depend on $t_*$, $r_*$ and their $\tau$-derivatives. Accordingly, the first step will then be to express $\dot{t}_*$ and $\dot{r}_*$ in terms of $A_*$ and $B_*$. This can be easily achieved by combining expressions \eqref{1.2 f(R)}, \eqref{2.2 f(R)} and \eqref{betadeff(R)}. We then obtain
\begin{gather}
    \dot{t}_*(\tau)=\beta_0\,\sqrt{\dfrac{B_*(\tau)}{A_*(\tau)}}, \label{tp} \\
    \dot{r}_*(\tau)=\sqrt{\beta_0^2-\dfrac{1}{B_*(\tau)}} \label{rp}.
\end{gather}
Expressions \eqref{tp} and \eqref{rp} may now be substituted back in equations \eqref{2.1 f(R)}, \eqref{2.2 f(R)}, \eqref{3 f(R)} and \eqref{4 f(R)}. For example, condition \eqref{2.1 f(R)} becomes
\begin{equation}
    \dot{\beta}(\tau)=\dfrac{1}{2}\left[\beta_0^2\,A_{t*}(\tau)\dfrac{B_*(\tau)}{A_*(\tau)}-B_{t*}(\tau) \left(\beta_0^2-\dfrac{1}{B_*(\tau)}\right)\right],
\end{equation}
while equation \eqref{4 f(R)} turns into
\begin{equation} \label{inserted 4 f(R)}
    \sqrt{\beta_0^2-\dfrac{1}{B_*(\tau)}}\,\LCR_{r*}^+(\tau)+\beta_0\,\sqrt{\dfrac{A_*(\tau)}{B_*(\tau)}}\,\LCR_{t*}^+(\tau)=0,
\end{equation}
and so forth. Proceeding in a similar fashion, the whole system of equations can be re-expressed in such a way that only $A_*$, $B_*$ and their derivatives with respect to $t$ and $r$ are present.
These equations hold on the stellar surface only, i.e.~on $t=t_*(\tau)$ and $r=r_*(\tau)$. However, as we will argue in what follows, it is reasonable to require the system to be satisfied for all $(t,r)$, in which case it would become a set of differential equations involving solely functions $A(t,r)$ and $B(t,r)$. For example, instead of \eqref{inserted 4 f(R)}, we could demand
\begin{equation} \label{all t and r 4 f(R)}
    \sqrt{\beta_0^2-\dfrac{1}{B(t,r)}}\,\LCR_{r}^+(t,r)+\beta_0\,\sqrt{\dfrac{A(t,r)}{B(t,r)}}\,\LCR_{t}^+(t,r)=0
\end{equation}
to be satisfied. Certainly, should this equation hold, then \eqref{inserted 4 f(R)} will be satisfied as well (and likewise for the remaining junction conditions). Let us thus explain why requiring the junction conditions to be satisfied for all $t$ and all $r$ is indeed a reasonable assumption. If the star ends up collapsing, then $r_*(\tau)$ will evolve continuously from its initial value $r_*(0)=\chi_*$---which is arbitrary\footnote{
    In general, the surface of a star is chosen to be any spherical surface where the stellar pressure $p$ vanishes. Hence, in realistic stars (whose interior pressure is non-zero), there might be upper or lower bounds on the radius, depending on the dynamics of $p$. However, in dust stars, $p$ vanishes everywhere, and \emph{any} spherical surface can be the initial stellar surface; $r_*(0)=\chi_*$ is thus arbitrary.\label{chistar arbitrary}
}---to zero. Similarly, $t_*(0)$ is also arbitrary, and one could expect that the black hole resulting from gravitational collapse requires infinite exterior time $t$ to form---this is the case, for example, in Oppenheimer-Snyder collapse \cite{Weinberg:1972kfs}. As a result, we conclude that, in the particular case of collapsing uniform-density dust stars, it is a well-motivated hypothesis to assume that $t_*(\tau)$ and $r_*(\tau)$ may respectively take all allowed values for coordinates $t$ and $r$. Therefore, it is sensible to require the junction conditions to hold for all possible $t$ and $r$.

Demanding that junction conditions \eqref{1.1 f(R)}--\eqref{2.2 f(R)}, \eqref{3 f(R)} and \eqref{4 f(R)} hold for every $t$ and $r$, we are able to construct a system of differential equations, which includes \eqref{all t and r 4 f(R)},\footnote{
    Equation \eqref{all t and r 4 f(R)} is the most compact equation in the aforementioned system; the remaining ones are long and convoluted, and their inclusion would provide no additional insight.
} to be satisfied by the exterior space-time if it is to match the interior one, \eqref{FLRW}.
The intrinsic difficulty of solving this system for 
$A(t,r)$ and $B(t,r)$---or even of extracting any kind of information on the exterior space-time from it---renders the approach of specifying certain simpler \emph{Ans\"{a}tze} for the exterior metric pragmatic. By proceeding this way, we have been able to establish the following no-go results:

\begin{result}[name={[no constant-curvature exterior in metric $f(\LCR)$ gravity]}] \label{theoconstcurv}
    No exterior space-time---either static or not---with constant scalar curvature can be smoothly matched to a uniform-density dust-star interior in generic $f(\LCR)$ gravity.
\end{result}

\begin{result}[name={[no static exterior in metric $f(\LCR)$ gravity]}] \label{theostatic}
    No static exterior space-time can be smoothly matched to a uniform-density dust-star interior in generic $f(\LCR)$ gravity.
\end{result}

\begin{result}[name={[no `single-function' exterior in metric $f(\LCR)$ gravity]}] \label{theosinglefunction}
    No exterior metric of the form
    \begin{equation} \label{singlefunctionmetric}
        \dif s^2_+=-A(t,r)\,\dif t^2+A^{-1}(t,r)\,\dif r^2+r^2\,\dif\Omega^2
    \end{equation}
    can be smoothly matched to a uniform-density dust-star interior in generic $f(\LCR)$ gravity.
\end{result}

\begin{result}[name={[no `single-function-plus-an-anomalous-redshift' exterior in metric $f(\LCR)$ gravity]}] \label{theoalmostANV}
    No exterior metric of the form
    \begin{equation}
        \dif s^2_+=-\e^{2\Phi(t,r)} A(r)\,\dif t^2+A^{-1}(r)\,\dif r^2+r^2\,\dif\Omega^2 \label{eqnResult4}
    \end{equation}
    can be smoothly matched to a uniform-density dust-star interior in generic $f(\LCR)$ gravity.
\end{result}

\begin{result}[name={[another other forbidden exterior in metric $f(\LCR)$ gravity]}] \label{theoanothernonstatic}
    No exterior metric of the form
    \begin{equation}
        \dif s^2_+=-U(r)\,\dif t^2+V(t)\,U^{-1}(r)\,\dif r^2+r^2\,\dif\Omega^2 \label{eqnResult5}
    \end{equation}
    can be smoothly matched to a uniform-density dust-star interior in generic $f(\LCR)$ gravity.
\end{result}

Result \ref{theoconstcurv} above together with the fourth junction condition also implies the following:

\begin{corollary}[name={[to Result \ref{theoconstcurv}]to Result \ref{theoconstcurv}}]
    No exterior space-time of the form \eqref{sphericallysymmetric} can be smoothly matched to a uniform-density dust-star interior in generic $f(\LCR)$ gravity if its Ricci scalar $R^+$ is either a function of $t$ only or of $r$ only.
\end{corollary}

In the following, we will offer a proof of each of these statements.

\subsection{Proof of Result \ref{theoconstcurv}, its Corollary, and Result \ref{theostatic}.}
\label{Proofs Results 1 2 Corollary}

Let us first consider a static exterior space-time of the form \eqref{MetricStaticAB}: the starting hypothesis of Result \ref{theostatic}.\footnote{
    The reason why we have chosen to start with Result \ref{theostatic} (and not with Result \ref{theoconstcurv}) will soon become apparent.
} With this particular choice for functions $A$ and $B$, junction condition \eqref{2.1 f(R)} reads $\dot{\beta}=0$, and thus we must require the $\tau$ derivative of \eqref{2.2 f(R)} to vanish. This yields
\begin{equation} \label{static beta dot}
    \sqrt{\beta_0^2-\dfrac{1}{B(r)}}\,\left[A_r(r) B(r)+A(r) B_r(r)\right]=0,
\end{equation}
which is satisfied provided that
\begin{equation}
\label{Eq29}
    \text{either}\myskip B(r)=\dfrac{1}{\beta_0^2}=\const\myskip\text{or}\myskip A(r) B(r)=\const,
\end{equation}
depending on whether we demand the square root or the parenthesis in \eqref{static beta dot} to be vanishing, respectively. The former choice is inconsistent with gravitational collapse, since it implies that $\dot{r}_*(\tau)=0$, as per equation \eqref{rp}. Therefore, we shall only concentrate on the latter,\footnote{
    Nonetheless, by choosing the first option, $B(r)=1/\beta_0^2=\const$, we have been able to obtain a novel static, non-collapsing solution of $f(\LCR)$ gravity, as we shall prove on Appendix \ref{staticsol}. The properties of this new solution will be thoroughly analysed on Chapter \ref{chapter:constant curvature}, as it will turn out to be a so-called `$(\LCR_0=0)$-degenerate' constant-curvature solution of $f(\LCR)$ gravity.
} i.e.~$A(r) B(r)=\const$, where the constant can always be set to 1, since it can be removed from line element \eqref{MetricStaticAB} through a suitable redefinition of time coordinate $t$ (in particular, $\dif t\rightarrow\dif t/\sqrt{\const}$). Junction condition \eqref{all t and r 4 f(R)} then becomes
\begin{equation}
    \sqrt{\beta_0^2-A(r)}\,\LCR_{r}^+(r)=0.
\end{equation}
This equality is satisfied if either $A(r)=1/B(r)=\beta_0^2$ or $\LCR^+=\const$ The former case must be discarded once again as explained above; therefore, the assumption of $\LCR^+=\const$ renders junction condition \eqref{3 f(R)} as follows:
\begin{equation} \label{JC static constant curvature single function}
    \LCR^+=\dfrac{6}{r^2}\left[1-A(r)-\dfrac{r A_r(r)}{2}\right],
\end{equation}
whose the right-hand side has been obtained by substituting \eqref{rp} on the right-hand side of \eqref{3 f(R)}. The general solution of \eqref{JC static constant curvature single function} is
\begin{equation} \label{mRN(A)dS}
    A(r)=1+\dfrac{Q^2}{r^2}-\dfrac{R^+ r^2}{12},
\end{equation}
$Q$ being an integration constant. This is the massless\footnote{
    We would like to highlight that a mass term $-2GM/r$ is present in the general solution of $\LCR^+=\const$ for a static, spherically-symmetric, `single-function' space-time with $A(t,r)=1/B(t,r)=A(r)$. However, compliance with the third junction condition $\jump{\LCR}=0$ explicitly requires $M\neq0$. Therefore, the mass term, which was of paramount importance in standard Oppenheimer collapse, is absent from \eqref{mRN(A)dS} due to one of the novel junction conditions of generic metric $f(\LCR)$ gravity. This is indeed a remarkable fact.
} Reissner-Nordstr\"{o}m (Anti-)de Sitter space-time, which is known to be a solution of any $f(\LCR)$ theory minimally coupled to the usual electromagnetic field \cite{delaCruz-Dombriz:2009pzc}.

Therefore, we have found that the only static and spherically-symmetric solution of $f(\LCR)$ gravity which satisfies junction conditions \eqref{2.1 f(R)}, \eqref{2.2 f(R)} and \eqref{3 f(R)}---and also \eqref{4 f(R)}---is \eqref{mRN(A)dS}. This solution possesses another crucial property: its Ricci scalar is constant. Therefore, if we are able to prove that constant-curvature solutions are incompatible with gravitational collapse in generic $f(\LCR)$ gravity (Result \ref{theoconstcurv}), then will have also proven that the exterior cannot be static (Result \ref{theostatic}).

Consequently, let us now prove Result \ref{theoconstcurv}. Consider a spherically-symmetric exterior space-time with $R^+=\const$ This space-time could either be static, such as \eqref{mRN(A)dS}, or non-static; our proof covers both situations. The third junction condition $\jump{\LCR}=0$ would imply that the Ricci scalar of the interior FLRW space-time must also be constant:
\begin{equation} \label{constantR+a}
   \LCR^-= 6\left[\dfrac{\dot{a}^2(\tau)+k}{a^2(\tau)}+\dfrac{\ddot{a}(\tau)}{a(\tau)}\right]=\LCR^+=\const
\end{equation}
Observe that this situation is analogous to the one we encountered back on Section \ref{Section2}; we will now demonstrate that no $f(\LCR)$ model can give rise to a collapsing, uniform-density dust star having constant Ricci scalar.

In generic metric $f(\LCR)$ gravity, it can be shown \cite{Cembranos:2012fd} that, for constant $\LCR^-=\LCR^+$,
\begin{equation}
    \dfrac{\ddot{a}(\tau)}{a(\tau)}=-\dfrac{2[\dot{a}^2(\tau)+k]}{a^2(\tau)}+\dfrac{f(\LCR^+)}{2f'(\LCR^+)}.
\end{equation}
Substituting this expression in \eqref{constantR+a}, we find that
\begin{equation}
    -\dfrac{\dot{a}^2(\tau)+k}{a^2(\tau)}+\dfrac{f(\LCR^+)}{2f'(\LCR^+)}=\dfrac{\LCR^+}{6}.
\end{equation}
For constant $\LCR^-=\LCR^+$, \eqref{ap2 f(R)} would also require
\begin{equation}
    \dfrac{\dot{a}^2(\tau)+k}{a^2(\tau)}=\dfrac{1}{f'(\LCR^+)}\left[\dfrac{\kappa\rho_0}{6a^3(\tau)}+\dfrac{f(\LCR^+)}{6}\right],
\end{equation}
and we thus finally have that
\begin{equation}
    -\dfrac{\kappa\rho_0}{a^3(\tau)}+2f(\LCR^+)=\LCR^+ f'(\LCR^+),
\end{equation}
which can be reformulated as
\begin{equation} \label{constantaconstantR+}
    a^3(\tau)=\dfrac{\kappa\rho_0}{2f(\LCR^+)-\LCR^+ f'(\LCR^+)}=\const
\end{equation}
This result is incompatible with gravitational collapse, as per the first junction condition \eqref{1.1 f(R)}. As a result, we have proven Result \ref{theoconstcurv}, i.e.~that no constant-curvature exterior solution---either static or not---can be matched to a collapsing dust-star interior.\footnote{
    The result anticipated in Section \ref{OS incompatible with f(R)} follows from evaluating \eqref{constantaconstantR+} for $\LCR^+=0$. The resulting expression is incompatible with \eqref{constantRa}, as previously stated without proof.
} \hfill $\blacksquare$

Since \eqref{mRN(A)dS}---which is the only static solution satisfying the second, third and fourth junction conditions---happens to have constant scalar curvature, it cannot be smoothly matched to the collapsing dust-star interior \eqref{FLRW} due to Result \ref{theoconstcurv}. Consequently, we have also proven Result \ref{theostatic}. \hfill $\blacksquare$

We must stress that Result \ref{theostatic} is in agreement with the theorem in \cite{Bueno:2017sui} we discussed back in Section \ref{intandext}, because \eqref{mRN(A)dS} is also a `single-function' space-time, i.e.~of the form \eqref{singlefunction}.

Finally, the Corollary issued from  Result \ref{theoconstcurv} follows almost immediately from the fourth junction condition: if either $R^+_t$ or $R^+_r$ vanish, then equation \eqref{all t and r 4 f(R)} forces the other derivative---$R^+_r$ or $R^+_t$, respectively---to vanish as well. Consequently, the exterior solution would have a constant Ricci scalar. This is forbidden by Result \ref{theoconstcurv}. \hfill $\blacksquare$

To sum up, throughout Section \ref{Proofs Results 1 2 Corollary} we have found that, if there exists a spherically-symmetric exterior solution smoothly matching a dust-star interior in $f(\LCR)$ gravity, then such space-time must be non-static and have a non-constant Ricci scalar. Furthermore, its scalar curvature $R^+(t,r)$ cannot depend only on either $t$ or $r$ when expressed in `areal-radius' coordinates.

\subsection{Proof of Results \ref{theosinglefunction}, \ref{theoalmostANV} and \ref{theoanothernonstatic}.}
\label{Proofs Results 3 4 5}

Having discarded the possibility of having a static exterior space-time, we shall now analyse one of the simplest non-static \emph{Ans\"{a}tze} one can possibly conceive: that given by \eqref{singlefunctionmetric}. In analogy with the Schwarzschild exterior of GR, \eqref{singlefunctionmetric} is also a `single-function' space-time, albeit a time-dependent one.

It is almost immediate to show that the exterior metric cannot be of the form \eqref{singlefunctionmetric}. In fact, to prove this, one only needs to realise that junction condition \eqref{2.2 f(R)} implies that $\beta=\beta_0=\const$ Accordingly, \eqref{2.1 f(R)} becomes
\begin{equation} \label{ansantz2 eq}
    A_{t*}(\tau)\left[2\beta_0^2-A_*(\tau)\right]=0.
\end{equation}
The first option, $A_{t*}(\tau)=0$, is to be discarded as per Result \ref{theostatic}, while the second one, $A_*(\tau)=2\beta_0^2=\const$, leads to the inconsistent result $\dot{r}_*^2(\tau)=-\beta_0^2<0$ when one resorts to \eqref{rp}. As a result, we conclude that an exterior metric of the form \eqref{singlefunctionmetric} is not compatible with gravitational collapse in $f(\LCR)$ theories of gravity with $f''(\LCR)\neq 0$. Result \ref{theosinglefunction} is thus proven: no `single-function' space-time, either static or not, matches the dust-star interior smoothly in generic $f(\LCR)$ gravity. \hfill $\blacksquare$

Result \ref{theosinglefunction} conveys a generalisation of the theorem in \cite{Bueno:2017sui} stating that no \emph{static} `single-function' space-time can be a exterior space-time in generic metric $f(\LCR)$ gravity. Nonetheless, Result \ref{theosinglefunction} only holds provided that the interior space-time is a dust-star FLRW metric, while the theorem in \cite{Bueno:2017sui} applies regardless of the interior matter source.

At this stage, another simple, non-static ansatz satisfying the necessary condition $B(t,r)\neq 1/A(t,r)$ would be of the form \eqref{eqnResult4}, i.e.~a generalisation of static `single-function' space-times in which we have included a $t$- and $r$-dependent anomalous redshift $\Phi$. It is not difficult to show that this ansatz is also unsatisfactory: the combination of junction condition \eqref{2.1 f(R)} with the $\tau$ derivative of \eqref{2.2 f(R)} implies
\begin{equation}
    \Phi_{r}(t,r)=0\myskip\Rightarrow\myskip \Phi(t,r)=\Phi(t).
\end{equation}
In consequence, metric \eqref{eqnResult4} becomes static, since $\Phi(t)$ can always be absorbed in the differential of $t$ through a coordinate transformation, namely, $\e^{\Phi(t)} \dif t\rightarrow\dif t$. Since static exteriors are ruled out by Result \ref{theostatic}, we have thus shown the validity of Result \ref{theoalmostANV}. \hfill $\blacksquare$

Another simple ansatz for time-dependent generalisations of the exterior metric would be as given in \eqref{eqnResult5}. However, this ansatz does not work either; equating \eqref{2.1 f(R)} with the $\tau$ derivative of \eqref{2.2 f(R)} one obtains
\begin{equation}
    \left[2\beta_0^2-\dfrac{U(r)}{V(t)}\right]V_{t}(t)=0\,.
\end{equation}
This equation is satisfied if either $V(t)=\const$ or if $U(r)/V(t)=2\beta_0^2=\const$ $\Rightarrow$ $U(r)=\const$ and $V(t)=\const$; in both cases, the metric becomes static. As a result, by virtue of Result \ref{theostatic}, line elements of the form \eqref{eqnResult5} do not satisfy the junction conditions of $f(\LCR)$ gravity. This is precisely the content of Result \ref{theoanothernonstatic}. \hfill $\blacksquare$

To conclude, let us mention that, throughout this Section, we have been capable of imposing restrictive constraints on the exterior space-time. In particular, our results indicate that, in generic metric $f(\LCR)$ gravity, the space-time outside a collapsing uniform-density dust star---if it exists---must be of the form \eqref{sphericallysymmetric}, with highly non-trivial---and probably non-separable---functions $A(t,r)$ and $B(t,r)\neq 1/A(t,r)$. Thus, the exterior in these theories seems to be substantially different from the Schwarzschild metric appearing in GR, yet the former should somehow reduce to the latter in the appropriate limits. Given the fact that most renowned solutions of generic metric $f(\LCR)$ gravity are either constant-curvature or static, Results \ref{theoconstcurv} and \ref{theostatic} rule out such exterior space-times as viable for matching FLRW-like, spatially-uniform dust-star interiors. For example, one of the most promising candidates, the Clifton-Barrow space-time \cite{Clifton:2005aj,Clifton:2006ug} fails to comply with the junction conditions because it is static.

\section[Oppenheimer-Snyder collapse in Palatini \titlemath{$f(\MAR)$} gravity.]{Oppenheimer-Snyder collapse in Palatini \titlebm{$f(\MAR)$} gravity.}
\label{sec:OS Palatini}

The junction conditions of Palatini $f(\MAR)$ gravity---where $\MAR$ is now the Ricci scalar of the independent connection---are different from those of GR and of metric $f(\LCR)$ gravity. More precisely, according to \cite{Olmo:2020fri}, the relevant constraints in the case allowing for thin shells to be present are
\begin{gather}
    \jump{h_{\mu\nu}}=0, \label{Palatini JC 1} \\
    \jump{\stress}=0, \label{Palatini JC 2} \\
    f'(\MAR^\sSigma) \left(\jump{K_{\mu\nu}}-\dfrac{1}{3}h_{\mu\nu}\jump{K}\right)=\kappa \stress^\szero_{\mu\nu}, \label{Palatini JC 3} \\
    \stress^\szero=0, \label{Palatini JC 4}
\end{gather}
where we recall that $\stress^\szero_{\mu\nu}$ is the thin shell's stress-energy tensor (i.e.~the monopolar part of the stress-energy tensor), while $\stress^\szero\equiv h^{ab}\stress^\szero_{\mu\nu}$ is its trace.

The case in which there are no thin shells at the matching surface is recovered by setting $\stress^\szero_{\mu\nu}=0$. It is then immediate to see that \eqref{Palatini JC 4} holds automatically, while condition \eqref{Palatini JC 3} becomes
\begin{equation}
    f'(\MAR^\sSigma)\left(K_{\mu\nu}-\dfrac{1}{3}h_{\mu\nu}\jump{K}\right)=0,
\end{equation}
whose trace is satisfied automatically for all $K_{\mu\nu}$. Conditions \eqref{Palatini JC 1} and \eqref{Palatini JC 2} are unaffected by the choice $\stress^\szero_{\mu\nu}=0$.

Condition \eqref{Palatini JC 2} suffices to show that Oppenheimer-Snyder collapse is impossible in Palatini $f(\MAR)$ gravity. A dust star has an energy-momentum tensor whose trace is $T^-=\rho\neq 0$, while the exterior vacuum solution has $T^+=0$. Therefore, the trace of the stress-energy tensor cannot be continuous at the matching surface unless $\rho=0$, and thus the gluing with any dust-star interior is impossible as per the junction conditions of Palatini $f(\MAR)$ gravity derived in  \cite{Olmo:2020fri}. What is more, if the interior stress-energy tensor is that of a perfect fluid, then the matching is only possible provided that $\rho_*=3p_*$, i.e.~that the equation of state evaluated at the stellar surface is that of radiation. This is of course true if the fluid \emph{is} radiation, but also if one requires the equation of state to be such that $\rho_*=0$ entails $p_*=0$. This is typically the case in more realistic stars, but not in those made of dust, in which $p=0$ everywhere and the star can end abruptly at any given radius. The generic incompatibility of dust-star interiors with junction condition \eqref{Palatini JC 2} shows that the Oppenheimer-Snyder collapse model is not viable within Palatini $f(\MAR)$ gravity. This leads us to state the following Result:

\begin{result}[name={[impossibility of Oppenheimer-Snyder collapse in Palatini $f(\MAR)$ gravity]}] \label{result:OS Palatini}
    Isolated bodies made of pressureless matter (i.e.~dust) are incompatible with the junction conditions of Palatini $f(\MAR)$ gravity as presented in \cite{Olmo:2020fri}. Therefore, the Oppenheimer-Snyder model of gravitational collapse is also incompatible with such junction conditions.
\end{result}

Nonetheless, we must stress that one can still study gravitational collapse in Palatini $f(\MAR)$; however, a different equation of state for the star is required by the junction conditions \eqref{Palatini JC 1}--\eqref{Palatini JC 4}. For example, non-dust perfect fluids and polytropic stars are still allowed by \eqref{Palatini JC 2}. It seems reasonable to expect, though, that a change in the equation of state might significantly complicate the mathematical treatment of the problem.

\section{Conclusions and future prospects.}
\label{Conclusions}

Aware of the fact that the junction conditions in $f(R)$ gravity are much more restrictive than their Einsteinian counterparts, the investigation presented in this Chapter has paved the way for a better understanding gravitational collapse in these theories. Most of the Chapter has been devoted to finding the appropriate generalisation of Oppenheimer-Snyder collapse in metric $f(\LCR)$ gravity. Such generalisation is not only interesting in itself; via transformation to the Einstein frame, it would shed light on the issue of the de-scalarization of matter on scalar-tensor theories, i.e.~on the precise mechanism by which collapsing solutions could get rid of its scalar hair so as to form a hairless black hole, as required by the no-hair theorems.

Herein, we have developed the general formalism allowing one to determine whether a spherically-symmetric exterior solution matches a dust star FLRW-like interior in $f(\LCR)$ gravity, regardless of the particular function $f$ chosen. There are two ways of tackling the problem: either one employs these junction conditions \eqref{1.1 f(R)}--\eqref{4 f(R)}, \eqref{tp} and \eqref{rp} directly in order to infer information about the exterior metric, or one simply inserts a known vacuum solution of $f(\LCR)$ gravity into the junction conditions and checks whether such conditions are satisfied. Neither procedure is exempt from difficulties: analytic computations can be hard or even impossible to perform, while the way in which the problem could be tackled numerically remains unclear to us. Furthermore, there are very few known exact vacuum solutions of metric $f(\LCR)$ gravity \cite{Faraoni:2021nhi}, a fact hampering the research on the topic.

Notwithstanding these shortcomings, the foundations of both approaches have been presented herein. We have also ruled out several classes of exterior solutions for the uniform-density dust star using the aforementioned junction conditions in Section \ref{ruling out}. In particular, we have proven that no constant-curvature exterior space-time---either static or dynamic---can be smoothly matched to the dust-star interior \eqref{FLRW} in generic $f(\LCR)$ gravity (Result \ref{theoconstcurv}). Furthermore, we have also offered a rigorous proof of the result that no static exterior space-time can be glued to the FLRW interior in generic $f(\LCR)$ gravity (Result \ref{theostatic}).

The power of our formalism has allowed us to extend the known result that static space-times satisfying $g_{tt}g_{rr}=-1$ cannot be the exterior of any matter source in $f(\LCR)$ gravity. Herein, we have shown that, at least in the case of a dust-star interior, even a non-static exterior satisfying $g_{tt}g_{rr}=-1$ cannot match the interior FLRW-like space-time (Result \ref{theosinglefunction}). We have also been able to establish further constraints on the exterior metric (Results \ref{theoalmostANV}, \ref{theoanothernonstatic} and the Corollary to Result \ref{theoconstcurv}). Finally, to the best of our knowledge, we have found for the first time in the literature that the junction conditions in Palatini $f(\LCR)$ gravity presented in \cite{Olmo:2020fri} are incompatible with Oppenheimer-Snyder collapse (Result \ref{result:OS Palatini}).

One of the most surprising consequences of our Results \ref{theoconstcurv}--\ref{theoanothernonstatic} is that the space-time outside a collapsing, uniform-density dust star in metric $f(\LCR)$ gravity must differ substantially from the Schwarzschild solution one necessarily has in GR. As we have seen, in metric $f(\LCR)$ gravity, the exterior cannot be static, constant-curvature or `single-function,' all of which are some of the most characteristic features of the Schwarzschild solution. These findings appear to be even more remarkable when one takes into account that the novel exterior should reduce to Schwarzschild in the appropriate limit.

Is gravitational collapse of a uniform-density dust star possible in $f(R)$ theories of gravity? Even though this simple and illustrative model of gravitational collapse is not feasible in the Palatini version of the theory, there are reasons to expect that it is still possible in the metric formulation. However, we have shown that the mathematical description of the process must be highly non-trivial, convoluted, and radically different from the Oppenheimer-Snyder picture. Hence, there is still plenty of work to be done so as to fully understand gravitational collapse in metric $f(\LCR)$ gravity. Since the catalogue of possible suitable exteriors is meagre, looking for novel time-dependent solutions---of the forms not ruled out by our results---would certainly contribute to a better understanding of gravitational collapse in generic metric $f(\LCR)$ gravity, as well as to a better understanding of the no-hair theorems for black-holes.

It is possible that the no-hair theorems could allow one to rule out exteriors by simple inspection, at least in principle. If the no-hair theorems hold, the exterior space-time must \emph{dynamically} become Schwarzschild at some point. Therefore, space-times which do not reduce to Schwarzschild after some time\footnote{
    Recall that, depending on the coordinate system, this time could be infinite. For example, it is well-known that, in Oppenheimer-Snyder collapse, the Schwarzschild black hole takes infinite time to form as seen from the exterior; see, for instance, \cite{Weinberg:1972kfs}.
} cannot describe gravity outside a collapsing non-rotating uniform-density dust star in  metric $f(\LCR)$ models satisfying the no-hair theorems.

Furthermore, in $f(\LCR)$ models which do not satisfy the no-hair theorems, such as the so-called `power-of-GR' models $f(\LCR) \propto \LCR^{1+\delta}$, one does not know \emph{a priori} whether a given solution describes the gravitational field outside the star, the outcome of collapse, neither, or both. Coincidentally, the only known solution which \emph{could be} compatible with all of our results is---as far as we are aware---the so-called Clifton space-time \cite{Clifton:2006ug}, which is a solution of these `power-of-GR' theories. The Clifton solution is highly non-trivial \cite{Faraoni:2009xb}, and the analysis of the junction conditions following the lines of Appendices \ref{ArealRadiusJCAppendix} and \ref{NOTArealRadiusJCAppendix} seems to be impossible to perform analytically. Furthermore, as we shall see in Chapter \ref{chapter:constant curvature}, the `power-of-GR' models are highly pathological in nature, exhibiting strong-coupling instabilities and other unphysical traits. Hence, it seems not to be appropriate to focus on the Clifton solution as a viable candidate.

For all the reasons stated above, further (numerical) studies are required in order to determine whether any space-time can be the exterior space-time corresponding to a collapsing dust star in metric $f(\LCR)$ gravity. These future works will likely require a change in the way in which the problem is dealt with; for example, novel numerical techniques might be required. A number of additional considerations could be the key to solving the problem of collapse in $f(R)$ gravity, both in the metric and Palatini formalisms, and both in numerical and analytical investigations. For instance, assuming that the star is spatially-homogeneous made of dust might constitute a significant oversimplification of the problem; perhaps in more realistic scenarios---such as having an anisotropic interior or choosing a different equation of state for the star, amongst others---some of the difficulties posited herein might be alleviated or even entirely resolved.

\chapter{Junction conditions in bi-scalar Poincaré Gauge Gravity.} \label{chapter:BSPGT JCs}

In this Chapter, we shall derive and present the junction conditions in the ghost-free subclass of quadratic Poincaré Gauge Gravity, which, as mentioned back in Section \ref{sec:intro:Poincaré Gauge Gravity}, propagates two additional degrees of freedom (one scalar and one pseudo-scalar) aside from the usual spin-2 graviton of GR, hence being known as bi-scalar Poincaré gravity or \BSPGT~for short. With this purpose, we will make full use of the mathematical framework expounded on Chapter \ref{chapter:Introduction: JCs}. We will show that, generically, \BSPGT~allows the matching interface $\Sigma$ to host surface spin densities, as well as for energy-momentum thin shells and double layers. We shall also discuss the relevance of our results and their possible practical applications in a wide variety of physical scenarios. The contents of this Chapter are based on Publication \cite{SecondPaper}.

The Chapter shall be organised as follows. First, in Section \ref{sec:consistency}, we will derive the minimum consistency requirements needed for the equations of motion of \BSPGT~to be well-defined in the distributional sense. After that, in Section \ref{sec:decomp}, we will supply an exhaustive derivation of the singular spin-density and stress-energy sources induced in $\Sigma$ when matching two solutions of \BSPGT. Next, Section \ref{sec:partcases} will contain a summary of the junction conditions in \BSPGT~and some of its most paradigmatic sub-cases (depending on the values of parameters $\alpha$ and $\beta$ in the action). Given that \BSPGT~reduces to other theories---such as metric $\LCf$---in some of these particular instances, we will be able to compare our results with the ones already found in the literature. Finally, after briefly commenting on some subtleties related to spinor-field sources in Section \ref{sec:Dirac}, we will close the Chapter with a brief summary and discussion of our findings, conclusions and future prospects, in Section \ref{sec:JCsBSPGT conclusions}.

\section{Consistency conditions in \BSPGT.}
\label{sec:consistency}

Following the generic procedure outlined in Section \ref{sec:procedure}, the first step towards finding the junction conditions in any field theory is to find the requisites under which the equations of motion are properly defined when the basic fields of the theory are promoted to distributions. In the case of \BSPGT, these basic fields are $g_{\mu\nu}$, $T_\mu$ and $S_\mu$. Given the inherent complexity of the \BSPGT~field equations \eqref{eq:EoM1}--\eqref{eq:EoM3}, we will divide the derivation in three steps. First, in Subsection \ref{sec:preconsistency}, we will detail the consistency conditions which may be inferred straight away by simple inspection of the equations of motion. Then, in Subsection \ref{sec:torsionconsistency}, we will investigate in more detail the torsion equations \eqref{eq:EoM1} and \eqref{eq:EoM2}; similarly, we will delve into the structure of the equation of the metric \eqref{eq:EoM3} in Subsection \ref{sec:metriconsistency}. Our findings regarding consistency conditions in \BSPGT~will be eventually summarised in Subsection \ref{cons:cond}. 

\subsection{Preliminary consistency conditions.}
\label{sec:preconsistency}

The Ricci scalar $\MAR$ of the torsionful connection \smash{$\MAG^\rho{}_{\mu\nu}$} of \BSPGT~can always be expressed in terms of the Levi-Civita scalar curvature \smash{$\LCR$}, cf.~\eqref{eq:RicciPostRiem t=0}. Thus, as argued in Section \ref{sec:metricdistrib}, we must assume the metric to be regular and continuous across $\Sigma$; otherwise, the various curvature tensors appearing in the equations of motion \eqref{eq:EoM1}--\eqref{eq:EoM3} would not be well-defined in the distributional sense.

Regarding the two vectors $T_\mu$ and $S_\mu$, one could in principle consider them to possess arbitrary regular and singular parts. Nonetheless, the distributional equations of motion are plagued by products involving both vectors, such as $\udis{T}_\mu\udis{T}^\mu$, $\udis{S}_\mu\udis{S}^\mu$, $\udis{T}_\mu\udis{S}^\mu$, etc. Therefore, it is imperative to assume from the beginning that the two vector parts of torsion are regular, so as to avoid both products of Dirac deltas and higher-order singular contributions. In other words,
\begin{align}
    \udis{T}_\mu &= T^{+}_\mu\,\heavidis{+}+T^{-}_\mu\,\heavidis{-}, \\
    \udis{S}_\mu &= S^{+}_\mu\,\heavidis{+}+S^{-}_\mu\,\heavidis{-}.
\end{align}
Notice, however, that $T_\mu$ and $S_\mu$ could in principle be discontinuous across $\Sigma$. Determining whether it is indeed possible to have discontinuous vectors requires a more careful examination of the field equations, which we shall carry on in the ensuing Subsections.

\subsection{Ill-defined combinations from torsion equations.}
\label{sec:torsionconsistency}

In this Subsection, we shall prove that no consistency condition is obtained from the torsion equations \eqref{eq:EoM1} and \eqref{eq:EoM2}. In order to understand why this is the case, let us represent by $V_\mu$ any of the regular fields $T_\mu$ or $S_\mu$. One may readily verify that all the terms in \eqref{eq:EoM1} and \eqref{eq:EoM2} have one of the following schematic structures (we drop coefficients as well as indices for simplicity):
\begin{align}
    \LCD\MAR &\sim \LCD\LCR+\LCD\LCD V+V\LCD V, \\
    \LCD\holst &\sim \LCD\LCD V+V\LCD V, \\
    V\holst &\sim  V\LCD V+VVV,\\
    V\MAR &\sim V\LCR+V\LCD V+VVV,\\
    V. &
\end{align}
When promoted to distributions using formuale \eqref{eq:LCR derivative}--\eqref{eq:Delta1DR} and \eqref{eq:decompDV}--\eqref{eq:Delta1DDV} in Appendix \ref{app:regsing} and prescription \eqref{eq:prescTheDel}, it is clear that none of the terms above gives rise to ill-defined products involving singular parts. Therefore, we conclude that torsion equations \eqref{eq:EoM1} and \eqref{eq:EoM2} are well-defined in the distributional sense under the preliminary conditions stated in Subsection \ref{sec:preconsistency}.

\subsection{Ill-defined combinations from the equation of the metric.}
\label{sec:metriconsistency}

As done for Equations \eqref{eq:EoM1} and \eqref{eq:EoM2}, we now present the schematic form of all the terms in the left-hand side of the metric equation \eqref{eq:EoM3}, once expanded in terms of the torsion vectors and the Levi-Civita Ricci curvature (again, irrelevant indices and coefficients are dropped out):
\begin{align}
    \LCR_{\mu\nu}\MAR &\sim \LCR_{\mu\nu} \LCR + \LCR_{\mu\nu}\LCD V + \LCR_{\mu\nu}VV, \label{eq:problems1}\\
    \LCD\LCD\MAR &\sim \LCD\LCD \LCR + \LCD\LCD\LCD V + \LCD V \LCD V +  V \LCD\LCD V , \label{eq:problems2}\\
    \MAR^2 &\sim 
        \LCR\LCR + \LCR\LCD V +\LCD V \LCD V+ \LCR VV + V V\LCD V +VVVV  ,\\
    \holst^2&\sim 
        \LCD V \LCD V+ \LCR VV + V V\LCD V +VVVV , \label{eq:problems4}\\
    V\LCD\MAR &\sim V\LCD\LCD \LCR + V\LCD\LCD V + VV \LCD V  ,\\
    V\LCD \holst &\sim V\LCD\LCD V + VV\LCD V ,\\
    VV\holst &\sim  VV\LCD V + VVVV,\\
    VV\MAR &\sim VV\LCR + VV\LCD V + VVVV,\\
    VV, & \\
    \LCR_{\mu\nu}.&
\end{align}
It is not difficult to realise that only the first four expressions above---\eqref{eq:problems1}--\eqref{eq:problems4}---could yield quadratic terms in the Dirac delta. In particular, the problematic combinations are $\LCR_{\mu\nu} \LCR$, $\LCR\LCR$, $\LCD V\LCD V$, $\LCR_{\mu\nu} \LCD V$ and $\LCR \LCD V$. Let us thus analyse them in full detail, so as to find the corresponding consistency conditions.

We first focus on terms $\beta\MAR^2$ and $\alpha \holst^2$ in Equation \eqref{eq:EoM3}. After expanding them in terms of the basic fields $g_{\mu\nu}$, $T_\mu$ and $S_\mu$, one finds
\begin{align}
    \beta\MAR^2 &= \beta \LCR^2+4\beta (\LCD_\mu T^{\mu})^2+\ldots\\
    \alpha\holst^2 &= \alpha (\LCD_\mu S^{\mu})^2+\ldots
\end{align}
We clearly see that these two quantities are well-defined in the sense of distributions provided that $\beta\udis{\LCR}$, $\beta\disLCD_\mu \udis{T}^{\mu}$ and $\alpha\disLCD_\mu \udis{S}^{\mu}$ are regular distributions. Due to the preliminary consistency conditions in Section \ref{sec:preconsistency}, the singular part of the aforementioned three distributions is of the lowest order, i.e.~proportional to $\del$. The nullity of each of these pieces leads us to impose the following conditions, respectively:
\begin{align}
        \beta\jump{K}=0, \label{eq:continuity of beta K} \\
        \beta n^\mu\jump{T_\mu}=0, \label{eq:continuity of beta T perp} \\
        \alpha n^\mu\jump{S_\mu}=0. \label{eq:continuity of alpha S}
\end{align}
Notice that we included the parameters here to avoid branching our analysis in different subcases depending on whether $\alpha$ and/or $\beta$ vanish or not (we will comment on these particular scenarios in Section \ref{sec:partcases}). 

Let us now focus on the two remaining problematic structures, namely \eqref{eq:problems1} and \eqref{eq:problems2}. Owing to \eqref{eq:continuity of beta K}--\eqref{eq:continuity of alpha S}, term \eqref{eq:problems1} is well-defined in the sense of distributions, so we are only left with \eqref{eq:problems2}. In this case, we find that the sole potentially problematic combinations are those of the form $\LCD V\LCD V$: 
\begin{equation}
    \beta \LCD_\mu \LCD_\nu\MAR = \dfrac{\beta}{12} \LCD_\mu S^\rho \LCD_\nu S_\rho -\dfrac{4\beta}{3} \LCD_\mu T^\rho \LCD_\nu T_\rho + \ldots
\end{equation}
Expression \eqref{eq:decompDV} reveals that, for consistency, we have to impose two further conditions in order to satisfy $\Singpart{\sstar}{\disLCD_\mu\udis{T}^\rho\disLCD_\nu\udis{T}_\rho}=\Singpart{\sstar}{\disLCD_\mu\udis{S}^\rho\disLCD_\nu\udis{S}_\rho}=0$, namely:
\begin{align}
        \beta\jump{T_\mu}=0, \label{eq:consistency:continuity of beta T} \\
        \beta\jump{S_\mu}=0. \label{eq:consistency:continuity of beta S}
\end{align}
Notice that condition \eqref{eq:consistency:continuity of beta T} entails \eqref{eq:continuity of beta T perp}; on the contrary, condition \eqref{eq:consistency:continuity of beta S} is fully independent of \eqref{eq:continuity of alpha S}.

\subsection{Summary of consistency conditions.}
\label{cons:cond}

As per the results obtained thus far in the last three Subsections, the conditions required so as to avoid ill-defined terms in the distributional equations of motion are as follows:
\begin{align}
    0&=\Singpart{\sstar}{\udis{g}_{\mu\nu}}=\Singpart{\sstar}{\udis{T}_\mu}=\Singpart{\sstar}{\udis{S}_\mu},
        \vphantom{\left(\Leftrightarrow\myskip\Singpart{\szero}{\alpha\disLCD_\mu \udis{S}^\mu}=0 \right)} \label{eq:condNoSingTSg} \\
    0&=\jump{g_{\mu\nu}}
    \mybigskip\,\,\myskip\left(\Leftrightarrow\myskip\jump{h_{\mu\nu}}=0\myskip\Leftrightarrow\myskip\jump{h_{ab}}=0 \vphantom{\Singpart{\szero}{\udis{F}_A}}\right), \label{eq:condnojumpg}\\
    0&=\beta\jump{K}
    \mybigskip\,\,\,\myskip\left(\Leftrightarrow\myskip\Singpart{\szero}{\beta\udis{\LCR}}=0 \right), \label{eq:condJumpK0}\\
    0&=\beta\jump{T_\mu}
    \mybigskip\,\myskip\left(\Leftrightarrow\myskip\Singpart{\szero}{\beta\disLCD_\mu \udis{T}_\nu}=0 \right), \label{eq:condbjumpT0}\\
    0&=\beta\jump{S_\mu}
    \mybigskip\,\myskip\left(\Leftrightarrow\myskip\Singpart{\szero}{\beta\disLCD_\mu \udis{S}_\nu}=0 \right), \label{eq:condbjumpS0}\\
    0&= \alpha n^\mu\jump{S_\mu}
    \mybigskip\left(\Leftrightarrow\myskip\Singpart{\szero}{\alpha\disLCD_\mu \udis{S}^\mu}=0 \right). \label{eq:condJumpnS}
\end{align}

\section[Decomposition of the equations of motion. Singular spin and matter sources on \titlemath{$\Sigma$}.]{Decomposition of the equations of motion. Singular spin and matter sources on \titlebm{$\Sigma$}.}
\label{sec:decomp}

Having determined the conditions under which the equations of motion \eqref{eq:EoM1}--\eqref{eq:EoM3} of \BSPGT~are well-defined in the distributional sense, we now turn to explore the singular pieces making up their respective left-hand sides. In doing so, we intend to determine the allowed singular contributions to the distributional stress-energy tensor $\udis{\stress}_{\mu\nu}$ and spin currents $\udis{L}_\mu$ and $\udis{J}_\mu$ which source said equations. To accomplish the aforementioned goal, we must carefully examine all terms comprising Equations \eqref{eq:EoM1}--\eqref{eq:EoM3}. The results for each equation are encapsulated in Tables \ref{tab:termsEqT}--\ref{tab:termsEqMetric} below. Vector indices have been omitted for clarity almost everywhere, except for the combination $\alpha\LCD_\mu S^\mu$, in order to clearly signal where condition \eqref{eq:condJumpnS} has been used.

Our analysis consists of several successive steps. Each column in Tables \ref{tab:termsEqT}--\ref{tab:termsEqMetric} summarises one of these steps. In the first column, we show the different types of terms as they appear in the equations of motion. Those terms must be then expanded in terms of the basic fields of the theory ($g_{\mu\nu}$, $T_\mu$ and $S_\mu$), as stipulated in Section \ref{sec:procedure}; this is what we show in the second column of each table. Afterwards, in the third column, we identify the distributional structure of every term when promoted to distributions while subject to preliminary conditions \eqref{eq:condNoSingTSg}--\eqref{eq:condnojumpg}. To be precise, we enumerate the various multipolar pieces $\Delta^{\sell}$ each term contains besides its regular part $\varrho$. As explained before, some of the terms in Equation \eqref{eq:EoM3} are distributionally undefined (i.e.~they contain products of singular distributions) unless one further imposes consistency conditions \eqref{eq:condJumpK0}--\eqref{eq:condJumpnS}. Moreover, some multipolar pieces will vanish when considering this second set of consistency conditions. In the last column of each table we display the multipolar pieces surviving the enforcement of all consistency conditions \eqref{eq:condNoSingTSg}--\eqref{eq:condJumpnS}.

By inspection of the third columns in Tables \ref{tab:termsEqT}--\ref{tab:termsEqMetric}, we can deduce that the sources in Equations \eqref{eq:EoM1}--\eqref{eq:EoM3} are only allowed to have the following non-vanishing pieces:
\begin{align}
    \udis{\trcurrent}_\mu  &= \Regpart{\udis{\trcurrent}_\mu}+\Singpart{\szero}{\udis{\trcurrent}_\mu} , \\
    \udis{\axcurrent}_\mu  &= \Regpart{\udis{\axcurrent}_\mu}\, +\Singpart{\szero}{\udis{\axcurrent}_\mu}, \\
    \udis{\stress}_{\mu\nu} &= \Regpart{\udis{\stress}_{\mu\nu}}+\Singpart{\szero}{\udis{\stress}_{\mu\nu}} +\Singpart{\sone}{\udis{\stress}_{\mu\nu}}.
\end{align}
While the previous three expressions summarise the schematic distributional structure of the spin and matter sources in \BSPGT, our task shall be to find explicit expressions for the various singular pieces. For this reason, let us now study each of the singular contributions to $\udis{\trcurrent}_\mu$, $\udis{\axcurrent}_\mu$ and $\udis{\stress}_{\mu\nu}$ separately in Subsections \ref{sing:trace}, \ref{sing:axial} and \ref{sing:Einstein}, respectively.

~

\begin{table}[hbp!]
    \centering
    \renewcommand\arraystretch{1.2}
    \begin{tabular}{ m{2.5cm} m{3.5cm} m{3.5cm} m{3.5cm} }
    \hline
    \hline
    \multicolumn{1}{c}{\footnotesize Terms in \eqref{eq:EoM1}} & \multicolumn{1}{c}{\footnotesize In terms of basic fields} & \multicolumn{1}{c}{\footnotesize Pieces after \eqref{eq:condNoSingTSg}--\eqref{eq:condnojumpg}} & \multicolumn{1}{c}{\footnotesize Pieces after \eqref{eq:condNoSingTSg}--\eqref{eq:condJumpnS}}\\
    \hline
    \hline
    $M_T^2 T$&  $M_T^2 T$     & $\varrho$  & $\varrho$  \\ 
    \hline
    \multirow{3}{*}{$\beta T \MAR $} 
    &$\beta T \LCR$           & $\varrho + \Delta^{\szero}$  & $\varrho$ \\ 
    &$\beta T \LCD T$         & $\varrho + \Delta^{\szero}$  & $\varrho$ \\ 
    &$\beta SST, \beta TTT$   & $\varrho$                 & $\varrho$ (continuous) \\
    \hline
    \multirow{2}{*}{$\alpha S\holst $}
    &$\alpha S\LCD_\mu S^\mu$ & $\varrho + \Delta^{\szero}$  & $\varrho$ \\ 
    &$\alpha TSS$             & $\varrho $                & $\varrho$ \\ 
    \hline
    \multirow{3}{*}{$\beta \LCD R$} 
    &$\beta \LCD \LCR$        & $\varrho + \Delta^{\szero}+ \Delta^{\sone}$  & $\varrho+ \Delta^{\szero}$ \\ 
    &$\beta \LCD \LCD T$      & $\varrho + \Delta^{\szero}+ \Delta^{\sone}$  & $\varrho+ \Delta^{\szero}$ \\ 
    &$\beta S\LCD S$, $\beta T\LCD T$          & $\varrho + \Delta^{\szero}$  & $\varrho$ \\
    \hline
    \hline
\end{tabular}
    \renewcommand\arraystretch{1}
    \caption[Terms in the equation of $T_\mu$ in \BSPGT, and their respective regular and singular parts after distributional promotion.]{Terms in the equation of $T_\mu$, \eqref{eq:EoM1}, and their respective regular and singular parts after distributional promotion.}
    \label{tab:termsEqT}
\end{table}

~

\begin{table}[hbp!]
    \centering
    \renewcommand\arraystretch{1.2}
    \begin{tabular}{ m{2.5cm} m{3.5cm} m{3.5cm} m{3.5cm} } 
    \hline
    \hline
    \multicolumn{1}{c}{\footnotesize Terms in \eqref{eq:EoM2}} & \multicolumn{1}{c}{\footnotesize In terms of basic fields} & \multicolumn{1}{c}{\footnotesize Pieces after \eqref{eq:condNoSingTSg}--\eqref{eq:condnojumpg}} & \multicolumn{1}{c}{\footnotesize Pieces after \eqref{eq:condNoSingTSg}--\eqref{eq:condJumpnS}}\\
    \hline
    \hline
    $M_S^2 S$&  $M_S^2 S$       & $\varrho$  & $\varrho$  \\ 
    \hline
    \multirow{3}{*}{$\beta S \MAR $} 
    &$\beta S \LCR$             & $\varrho + \Delta^{\szero}$  & $\varrho$ \\ 
    &$\beta S \LCD T$           & $\varrho + \Delta^{\szero}$  & $\varrho$ \\ 
    &$\beta SSS, \beta STT$     & $\varrho$                 & $\varrho$ (continuous) \\
    \hline
    \multirow{2}{*}{$\alpha T\holst$}
    &$\alpha T\LCD_\mu S^\mu$   & $\varrho + \Delta^{\szero}$  & $\varrho$ \\ 
    &$\alpha STT$               & $\varrho $                & $\varrho$ \\ 
    \hline
    \multirow{2}{*}{$\alpha \LCD \holst$} 
    &$\alpha\LCD(\LCD_\mu S^\mu)$ & $\varrho + \Delta^{\szero}+ \Delta^{\sone}$  & $\varrho+ \Delta^{\szero}$ \\ 
    &$\alpha T\LCD S$, $\alpha S\LCD T$           & $\varrho + \Delta^{\szero}$  & $\varrho + \Delta^{\szero}$ \\
    \hline
    \hline
\end{tabular}
    \renewcommand\arraystretch{1}
    \caption[Terms in the equation of $S_\mu$ in \BSPGT, and their respective regular and singular parts after distributional promotion.]{Terms in the equation of $S_\mu$, \eqref{eq:EoM2}, and their respective regular and singular parts after distributional promotion.}
    \label{tab:termsEqS}
\end{table}

\begin{table}[htp!]
    \centering
    \renewcommand\arraystretch{1.2}
    \begin{tabular}{ m{2.5cm} m{3.5cm} m{3.5cm} m{3.5cm} } 
    \hline
    \hline
    \multicolumn{1}{c}{\footnotesize Terms in \eqref{eq:EoM3}} & \multicolumn{1}{c}{\footnotesize In terms of basic fields} & \multicolumn{1}{c}{\footnotesize Pieces after \eqref{eq:condNoSingTSg}--\eqref{eq:condnojumpg}} & \multicolumn{1}{c}{\footnotesize Pieces after \eqref{eq:condNoSingTSg}--\eqref{eq:condJumpnS}}\\
    \hline
    \hline
    $M_S^2 SS$, $M_T^2 TT$&  $M_S^2 SS$, $M_T^2 TT$& $\varrho$  & $\varrho$  \\ 
    \hline
    $\Mp^2\LCR_{\mu\nu}$&  $\Mp^2\LCR_{\mu\nu}$& $\varrho+ \Delta^{\szero}$  & $\varrho+ \Delta^{\szero}$  \\ 
    \hline
    \multirow{3}{*}{$\beta\LCR_{\mu\nu}\MAR$} 
    &$\beta\LCR_{\mu\nu}\LCR$     & distrib.~undefined  & $\varrho+ \Delta^{\szero}$ \\ 
    &$\beta\LCR_{\mu\nu} \LCD T$  & distrib.~undefined  & $\varrho$ \\ 
    &$\beta\LCR_{\mu\nu} TT$, $\beta\LCR_{\mu\nu} SS$   & $\varrho+ \Delta^{\szero}$                 & $\varrho+ \Delta^{\szero}$ \\
    \hline
    \multirow{3}{*}{$\beta TT \MAR $} 
    &$\beta TT \LCR$           & $\varrho + \Delta^{\szero}$  & $\varrho$ \\ 
    &$\beta TT \LCD T$         & $\varrho + \Delta^{\szero}$  & $\varrho$ \\ 
    &$\beta SSTT, \beta TTTT$   & $\varrho$                 & $\varrho$ (continuous) \\
    \hline
    \multirow{3}{*}{$\beta SS \MAR $} 
    &$\beta SS \LCR$           & $\varrho + \Delta^{\szero}$  & $\varrho$ \\ 
    &$\beta SS \LCD T$         & $\varrho + \Delta^{\szero}$  & $\varrho$ \\ 
    &$\beta SSSS, \beta SSTT$   & $\varrho$                 & $\varrho$ (continuous) \\ 
    \hline
    \multirow{2}{*}{$\alpha TS\holst $}
    &$\alpha TS\LCD_\mu S^\mu$ & $\varrho + \Delta^{\szero}$  & $\varrho$ \\ 
    &$\alpha TTSS$             & $\varrho $                & $\varrho$ \\  
    \hline
    \multirow{3}{*}{$\beta T\LCD \MAR$} 
    &$\beta T \LCD \LCR$        & $\varrho + \Delta^{\szero}+ \Delta^{\sone}$  & $\varrho+ \Delta^{\szero}$ \\ 
    &$\beta T \LCD \LCD T$      & $\varrho + \Delta^{\szero}+ \Delta^{\sone}$  & $\varrho+ \Delta^{\szero}$ \\ 
    &$\beta TS\LCD S$, $\beta TT\LCD T$          & $\varrho + \Delta^{\szero}$  & $\varrho$ \\
    \hline
    \multirow{2}{*}{$\alpha S \LCD \holst$} 
    &$\alpha S\LCD(\LCD_\mu S^\mu)$ & $\varrho + \Delta^{\szero}+ \Delta^{\sone}$  & $\varrho+ \Delta^{\szero}$ \\ 
    &$\alpha TS\LCD S$, $\alpha SS\LCD T$           & $\varrho + \Delta^{\szero}$  & $\varrho + \Delta^{\szero}$ \\ 
    \hline
    \multirow{3}{*}{$\alpha \holst^2$} 
    &$\alpha(\LCD_\mu S^\mu)^2$ & distrib.~undefined  & $\varrho$ \\ 
    &$\alpha ST\LCD_\mu S^\mu$  & $\varrho + \Delta^{\szero}$  & $\varrho$ \\
    &$\alpha SSTT$              & $\varrho$  & $\varrho$ \\
    \hline
    \multirow{6}{*}{$\beta \MAR^2$} 
    &$\beta \LCR^2$                   & distrib.~undefined  & $\varrho$ \\ 
    &$\beta \LCR\LCD T$               & distrib.~undefined  & $\varrho$ \\ 
    &$\beta SS\LCR$, $\beta TT\LCR$    & $\varrho + \Delta^{\szero}$  & $\varrho$ \\ 
    &$\beta \LCD T \LCD T$            & distrib.~undefined  & $\varrho$ \\ 
    &$\beta SS\LCD T$, $\beta TT\LCD T$& $\varrho + \Delta^{\szero}$  & $\varrho$ \\ 
    &$\beta \times \text{powers of}\,T\text{ or }S$ &$\varrho$  & $\varrho$ (continuous) \\
    \hline
    \multirow{4}{*}{$\beta \LCD\LCD \MAR$} 
    &$\beta \LCD\LCD \LCR$                   & $\varrho + \Delta^{\szero}+ \Delta^{\sone}+ \Delta^{\stwo}$ & $\varrho + \Delta^{\szero}+ \Delta^{\sone}$ \\ 
    &$\beta \LCD\LCD\LCD T$               & $\varrho + \Delta^{\szero}+ \Delta^{\sone}+ \Delta^{\stwo}$ & $\varrho + \Delta^{\szero}+ \Delta^{\sone}$ \\ 
    &$\beta S\LCD\LCD S$, $\beta T \LCD\LCD T$    & $\varrho + \Delta^{\szero}+ \Delta^{\sone}$  & $\varrho+ \Delta^{\szero}$ \\ 
    &$\beta \LCD T \LCD T$, $\beta \LCD S \LCD S$            & distrib.~undefined  & $\varrho$ \\ 
    \hline
    \hline
\end{tabular}
    \renewcommand\arraystretch{1}
    \caption[Terms in the equation of $g_{\mu\nu}$ in \BSPGT, and their respective regular and singular parts after distributional promotion.]{Terms in the equation of $g_{\mu\nu}$, \eqref{eq:EoM3}, and their respective regular and singular parts after distributional promotion.}
    \label{tab:termsEqMetric}
\end{table}

\newpage

\subsection[Singular part of the vector spin density \titlemath{$\udis{\trcurrent}_\mu$}.]{Singular part of the vector spin density \titlebm{$\udis{\trcurrent}_\mu$}.}
\label{sing:trace}

We start with the singular contributions to the equation of the torsion trace vector $T_\mu$, which will determine $\Singpart{\szero}{\udis{\trcurrent}_\mu}$. Keeping in mind the results in Table \ref{tab:termsEqT}, and rearranging its left-hand side, we find that equation \eqref{eq:EoM1} can be expressed as
\begin{align}    
    \trcurrent_\mu & = -4\beta\LCD_\mu \MAR+ \text{regular tensor terms} \nonumber\\
          & = -4\beta\LCD_\mu \LCR - 8\beta \LCD_\mu \LCD_\nu T^\nu + \text{regular tensor terms}
\end{align}
in terms of the basic fields and their derivatives. Distributional promotion leads to
\begin{equation}    
    \udis{\trcurrent}_\mu = -4\beta\disLCD_\mu \udis{\LCR} - 8\beta \disLCD_\mu \disLCD_\nu \udis{T}^\nu + \text{regular distributional terms};
\end{equation}
hence,
\begin{equation}    
    \Singpart{\szero}{\udis{\trcurrent}_\mu}  = -4\beta\Singpart{\szero}{\disLCD_\mu \udis{\LCR}} - 8\beta \Singpart{\szero}{\disLCD_\mu \disLCD_\nu \udis{T}^\nu}.
\end{equation}
Recalling consistency conditions \eqref{eq:condJumpK0} and \eqref{eq:condbjumpT0}, and using the traces of \eqref{eq:jumpDV} and \eqref{eq:Delta0DDV}, as well as \eqref{eq:Delta0DR} and
\begin{equation}
    \beta\jump{\LCR} +2\epsilon\beta n^\mu n^\nu \jump{\LCD_\mu T_\nu}=\beta\jump{\MAR}\label{eq:jumpRsimp}
\end{equation}
we finally obtain
\begin{equation} 
\label{D0Lmu}
    \Singpart{\szero}{\udis{\trcurrent}_\mu}  = -4\epsilon \beta \jump{\MAR} n_\mu \del .
\end{equation}
Thus, configurations with a non-vanishing monopolar vector spin density on $\Sigma$ are allowed for $\beta\neq 0$, if and only if $\jump{\MAR}\neq 0$.

\subsection[Singular part of the axial vector spin density \titlemath{$\udis{\axcurrent}_\mu$}.]{Singular part of the axial vector spin density \titlebm{$\udis{\axcurrent}_\mu$}.}
\label{sing:axial}

We now turn to study the singular contribution to the equation of the torsion axial vector $S_\mu$, which will determine $\Singpart{\szero}{\udis{\axcurrent}_\mu}$. Following the same approach as in the previous Subsection, the information provided in Table \ref{tab:termsEqS} allows us to expand $J_{\mu}$ as
\begin{align}    
    \axcurrent_\mu & = 2\alpha\LCD_\mu \holst + \text{regular tensor terms} \nonumber\\
          & = -2\alpha \LCD_\mu \LCD_\nu S^\nu +\dfrac{4\alpha}{3}(T^\nu \LCD_\mu S_\nu + S^\nu \LCD_\mu T_\nu) + \text{regular tensor terms}.
\end{align}
When promoted to distributions, the terms in the previous expression become
\begin{equation}    
    \udis{\axcurrent}_\mu = -2\alpha \disLCD_\mu \disLCD_\nu \udis{S}^\nu +\dfrac{4\alpha}{3}(\udis{T}^\nu \disLCD_\mu \udis{S}_\nu + \udis{S}^\nu \disLCD_\mu \udis{T}_\nu) + \text{regular distributional terms}\, ,
\end{equation}
where we have made use of the prescription \eqref{eq:prescTheDel}. Therefore,
\begin{equation}    
    \Singpart{\szero}{\udis{\axcurrent}_\mu}  = -2\alpha \Singpart{\szero}{\disLCD_\mu \disLCD_\nu \udis{S}^\nu} +\dfrac{4\alpha}{3}\left(T^{\Sigma\nu} \Singpart{\szero}{\disLCD_\mu \udis{S}_\nu} + S^{\Sigma\nu} \Singpart{\szero}{\disLCD_\mu \udis{T}_\nu}\right).
\end{equation}

Taking into account consistency condition \eqref{eq:condJumpnS}, and using the traces of \eqref{eq:Delta0DDV} and \eqref{eq:jumpDV}, as well as \eqref{eq:decompDV}, we finally get
\begin{equation}    
    \Singpart{\szero}{\udis{\axcurrent}_\mu}  =    2\epsilon \alpha \jump{\holst} n_\mu  \del.
\end{equation}
We thus conclude that configurations with a non-vanishing monopolar axial spin density on $\Sigma$ are possible if $\alpha\neq 0$, provided one additionally has $\jump{\holst}\neq 0$.

\subsection[Singular part of the energy-momentum tensor \titlemath{$\udis{\stress}_{\mu\nu}$}.]{Singular part of the energy-momentum tensor \titlebm{$\udis{\stress}_{\mu\nu}$}.}
\label{sing:Einstein}

Finally, we examine the singular pieces of the energy-momentum tensor. The information in Table \ref{tab:termsEqMetric} leads us to
\begin{align}    
    T_{\mu\nu} & = \Mp^2 \LCR_{\mu\nu} + 4 \beta \LCR_{\mu\nu} \LCR + \dfrac{\beta}{6}  \LCR_{\mu\nu} S_\rho S^\rho -  \dfrac{8 \beta}{3} \LCR_{\mu\nu} T_\rho T^\rho \nonumber\\
    &\quad
    + 8 \beta \LCR_{\mu\nu} \LCD_\rho T^\rho - 4 \beta \LCD_\mu\LCD_\nu \LCR - 8 \beta \LCD_\mu\LCD_\nu\LCD_\rho T^\rho \nonumber\\
    &\quad - \dfrac{\beta}{3} S^\rho \LCD_{(\mu}\LCD_{\nu)} S_\rho + \dfrac{16 \beta}{3} T^\rho \LCD_{(\mu}\LCD_{\nu)} T_\rho - 8 \beta T_{(\mu} \LCD_{\nu)} \LCR - 16 \beta T_{(\mu} \LCD_{\nu)}\LCD_\rho T^\rho \nonumber\\
    &\quad+ \dfrac{8 \alpha}{3} S^\rho S_{(\mu} \LCD_{\nu)} T_\rho+ \dfrac{8 \alpha}{3} T^\rho S_{(\mu} \LCD_{\nu)} S_\rho -  4\alpha S_{(\mu} \LCD_{\nu)}\LCD_\rho S^\rho  \nonumber\\   
    & \quad + 2 g_{\mu\nu} \Big(-\dfrac{\Mp^2}{4} \LCR + 2 \beta T^\rho \LCD_\rho\LCR  + 2 \beta  \LCbox \LCR + 4 \beta T^\rho \LCD_\rho\LCD_{\sigma}T^{\sigma}\nonumber\\
    &\qquad\qquad\hspace{8pt}+ \alpha S^\rho \LCD_\rho\LCD_{\sigma}S^{\sigma} -  \dfrac{2 \alpha}{3} S^\rho T^{\sigma} \LCD_\rho S_{\sigma} -  \dfrac{2 \alpha}{3} S^\rho S^{\sigma} \LCD_{\sigma}T_\rho \nonumber\\
    &\qquad\qquad\hspace{8pt}+ \dfrac{\beta}{6}  S^\rho \LCbox S_\rho -  \dfrac{8 \beta}{3} T^\rho \LCbox T_\rho + 4 \beta  \LCbox \LCD_\rho T^{\sigma}\Big)
+ \text{regular tensor terms}\, . \label{eq:Tmnsing}
\end{align}
As we shall see in the following, the singular part of $\udis{\stress}_{\mu\nu}$ consists of a dipolar (double-layer) component as well as of a monopolar (thin-shell) part.

\paragraph*{\textbf{Dipole.}}
When promoted to distributions, only the following terms in the stress-energy tensor provide double-layer contributions:
\begin{equation}    
    \Singpart{\sone}{\udis{T}_{\mu\nu}} = - 4 \beta (\delta^\rho_\mu \delta^\sigma_\nu - g_{\mu\nu} g^{\rho\sigma}) \left(
    \Singpart{\sone}{\disLCD_\rho \disLCD_\sigma \udis{\LCR}} + 2 \Singpart{\sone}{\disLCD_\rho\disLCD_\sigma \disLCD_\lambda \udis{T}^\lambda}\right),
\end{equation}
which, according to \eqref{eq:decomDDR2}, \eqref{eq:decomDDDT2} and \eqref{eq:jumpRsimp}, simplifies to
\begin{equation}    
    \Singpart{\sone}{\udis{T}_{\mu\nu}}  =  4\epsilon \beta\, \disLCD_\rho\left(
     \jump{\MAR} h_{\mu\nu} n^\rho \del \right).
\end{equation}

\paragraph*{\textbf{Monopole.}}
All the explicit terms in \eqref{eq:Tmnsing} will contribute to the monopolar part $\Singpart{\szero}{\udis{T}_{\mu\nu}}$. For convenience, we choose to split it into three parts,
\begin{equation}
    \Singpart{\szero}{\udis{T}_{\mu\nu}} \equiv \epsilon\left(\dfrac{H^{\sGR}_{\mu\nu}}{\kappa} + \beta\,  H^{\sbeta}_{\mu\nu} + \alpha\, H^{\salpha}_{\mu\nu}\right) \del ,\label{eq:sin0THs}
\end{equation}
with the different contributions being
\begin{align}
    H^{\sGR}_{\mu\nu} &=  - \jump{K_{\mu\nu}} + h_{\mu\nu}\jump{K},\label{eq:HGR}\\
    H^{\sbeta}_{\mu\nu} &=  -4 \MAR^\sSigma \jump{K_{\mu\nu}}-4 \Big(K^\sSigma_{\mu\nu} -\epsilon n_\mu n_\nu K^\sSigma+2 n_{(\mu}T^\sSigma_{\nu)}- g_{\mu\nu} T^\sSigma_{\perp} \Big) \jump{\MAR} \nonumber\\
    &\quad -4 \left(2 n_{(\mu}h^\rho{}_{\nu)}- h_{\mu\nu}n^\rho\right)\jump{\LCD_\rho \MAR} ,\label{eq:Hbeta}\\
    H^{\salpha}_{\mu\nu} &= 2\left(2 n_{(\mu} S^\sSigma_{\nu)} - g_{\mu\nu}  S^\sSigma_\perp\right) \jump{\holst}.\label{eq:Halpha}
\end{align}
Here, one should bear in mind that\footnote{
        Observe that, in the expression for $\MAR^\sSigma$, one may use \eqref{eq:ssigmaproduct} to simplify $\beta(S_\rho S^{\rho})^\sSigma=\beta S^\sSigma_\rho S^{\sSigma\rho}$ and $\beta(T_\rho T^{\rho})^\sSigma=\beta T^\sSigma_\rho T^{\sSigma\rho}$, because both  $S_\mu$ and $T_\mu$ are continuous when multiplied by $\beta$. }
\begin{equation}
    \MAR^\sSigma = \LCR^\sSigma + 2 (\LCD_\rho T^{\rho})^\sSigma + \dfrac{1}{24}(S_\rho S^{\rho})^\sSigma - \dfrac{2}{3}(T_\rho T^{ \rho})^\sSigma.
\end{equation}
Moreover, in deriving \eqref{eq:Hbeta}, we made use of the following useful simplification (where $V$ represents either $T$ or $S$):
\begin{equation}
    \beta V^\sSigma_\rho\jump{\LCD_\mu V^\rho}= \dfrac{\beta}{2}\jump{\LCD_\mu (V^\rho V_\rho)} + (\ldots)\underbrace{\beta\jump{V_\mu}}_0 = \dfrac{\beta}{2}\jump{\partial_\mu(V^\rho V_\rho)} = \partial_\mu(\underbrace{\beta\jump{V^\rho}}_0 V^\sSigma_\rho) = 0.
\end{equation}

\subsection{Summary of results.}

We now present a summary of the singular parts of the equations of motion. If we use the notation in \eqref{eq:defD0}--\eqref{eq:defD1} to write
\begin{align}
    \Singpart{\szero}{\udis{\trcurrent}_\mu} &=\trcurrent^\szero_\mu\del,
    & \Singpart{\szero}{\udis{T}_{\mu\nu}} &=T^\szero_{\mu\nu}\del,\nonumber\\   
    \Singpart{\szero}{\udis{\axcurrent}_\mu} &=\axcurrent^\szero_\mu\del,
    &\Singpart{\sone}{\udis{T}_{\mu\nu}}&=\disLCD_\rho (T^\sone_{\mu\nu} n^\rho \del),
\end{align}
then the various non-vanishing multipole densities in $\Sigma$ are
\begin{align}
    \trcurrent^\szero_\mu &= -4\epsilon \beta \jump{\MAR} n_\mu, \\
    \axcurrent^\szero_\mu &=    2\epsilon \alpha \jump{\holst} n_\mu ,\\
    T^\szero_{\mu\nu} & =\epsilon\left(\dfrac{H^{\sGR}_{\mu\nu}}{\kappa}  + \beta\,  H^{\sbeta}_{\mu\nu} + \alpha\, H^{\salpha}_{\mu\nu}\right),\\
    T^\sone_{\mu\nu} & =  4\epsilon \beta
     \jump{\MAR} h_{\mu\nu} , \label{eq:stress-energy double layer}
\end{align}
where $H^{\sGR}_{\mu\nu}$, $H^{\sbeta}_{\mu\nu}$ and $H^{\salpha}_{\mu\nu}$ are respectively given in \eqref{eq:HGR}, \eqref{eq:Hbeta} and \eqref{eq:Halpha}. Observe that the stress-energy dipolar density $\stress^{\sone}_{\mu\nu}$ vanishes identically when the spin monopolar density $\trcurrent_\mu^{\szero}$ is zero.

As done in Section \ref{sec:f(R) JCs} following References \cite{Senovilla:2013vra,Reina:2015gxa}, at this point it is also convenient to decompose $T^\szero_{\mu\nu}$ into its normal and tangential components,
\begin{equation} \label{eq:Tdecomp 0}
        \stress^\szero_{\mu\nu}=\tau_{\mu\nu}+2n_{(\mu}\tau_{\nu)}+\tau n_\mu n_\nu ,
    \end{equation}
where
\begin{equation}
    \tau_{\mu\nu}\equiv h^\rho{}_\mu h^\sigma{}_\nu T^\szero_{\rho\sigma},\mybigskip
    \tau_\mu\equiv\epsilon h^\rho{}_\mu n^\sigma T^\szero_{\rho\sigma},\mybigskip
     \tau\equiv n^\mu n^\nu T^\szero_{\mu\nu}.
\end{equation}
Recall that, by definition, the first two objects are purely tangential, i.e.~$n^\mu\tau_{\mu\nu}=0$ and $n^\mu\tau_\mu=0$. Each piece in the splitting of the energy-momentum monopole $\stress^\szero_{\mu\nu}$ reads:
\begin{align}
    \tau &= 4\beta (K^\sSigma-T_\perp^\sSigma)\jump{\MAR} + 2\alpha S_\perp^\sSigma \jump{\holst}, \label{eq:Tdecomp 1} \\
    \tau_\mu &= -4\epsilon\beta\left(h^\rho{}_\mu \jump{\LCD_\rho \MAR}+\tosigma{T}^\sSigma_\mu \jump{\MAR}\right) + 2\epsilon\alpha\tosigma{S}^\sSigma_\mu \jump{\holst}, \label{eq:Tdecomp 2} \\
    \tau_{\mu\nu} &=-\dfrac{\epsilon}{\kappa}\left(\jump{K_{\mu\nu}} - h_{\mu\nu}\jump{K}\right) -2\epsilon\alpha h_{\mu\nu} S^\sSigma_\perp \jump{\holst}  \nonumber\\
    & \quad -4\epsilon\beta \left[\MAR^\sSigma\jump{K_{\mu\nu}} +\left(K^\sSigma_{\mu\nu} - h_{\mu\nu} T^\sSigma_{\perp} \right)\jump{\MAR}  - h_{\mu\nu}n^\rho\jump{\LCD_\rho \MAR}\right].\label{eq:Tdecomp 3}
\end{align}
We remind the reader that the projection of any one-form $V_\mu$ in the direction tangential to $\Sigma$ is defined as $\tosigma{V}_\mu\equiv h^\rho{}_\mu V_\rho$, cf.~\eqref{eq:Vdecomp}. Hence, $\tosigma{T}_\mu^\sSigma=h^\rho{}_\mu T_\rho^\sSigma$ and, similarly, $\tosigma{S}_\mu^\sSigma=h^\rho{}_\mu S_\rho^\sSigma$.

\section{Junction conditions in generic \BSPGT~and some of its most relevant sub-cases.}
\label{sec:partcases}

We are finally in a position to establish the general junction conditions in \BSPGT, valid regardless of the values taken by parameters $\alpha$ and $\beta$ appearing in the theory's action. These junction conditions are summarised in the ensuing Result:

\begin{result}[name={[junction conditions in \BSPGT]}] \label{result:JCs in BSPGT}
    The junction conditions in bi-scalar Poincaré Gauge Gravity (\BSPGT) read as follows:
    \begin{coloritemize}
    \item The \BSPGT~field equations are only well-defined after distributional promotion provided that the following consistency requirements are all satisfied:
        \begin{itemize}
            \item The basic, dynamical fields of the theory---the metric $g_{\mu\nu}$ and the two torsion vectors $T_\mu$ and $S_\mu$---are all regular across $\Sigma$.
            \item $\jump{g_{\mu\nu}}=0$, i.e.~the metric is continuous across $\Sigma$; recall this condition is equivalent to $\jump{h_{\mu\nu}}=0$ $\Leftrightarrow$ $\jump{h_{ab}}=0$.
            \item $\beta\jump{K}=0$, $\beta\jump{T_\mu}=0$ and $\beta\jump{S_\mu}=0$, meaning that the extrinsic curvature and the two torsion vectors must be continuous unless $\beta=0$.
            \item $\alpha n^\mu\jump{S_\mu}=0$, meaning that the normal component of the axial part of torsion must be continuous across $\Sigma$ except in cases where $\alpha=0$.
        \end{itemize}
    \item A stress-energy monopole (thin shell) \smash{$\stress^\szero_{\mu\nu}$} given by \eqref{eq:sin0THs}--\eqref{eq:Halpha}---equivalently, by \eqref{eq:Tdecomp 0}--\eqref{eq:Tdecomp 3}---can be present on $\Sigma$ due to the matching. Since the expressions providing \smash{$\stress^\szero_{\mu\nu}$} are quite convoluted, there are a variety of circumstances under which the matter thin-shell might be vanishing. Ignoring exceptional cases whose specific values of $T_\mu^\sSigma$, $S_\mu^\sSigma$, $\MAR^\sSigma$ and $K_{\mu\nu}^\sSigma$ render \eqref{eq:sin0THs}--\eqref{eq:Halpha} or \eqref{eq:Tdecomp 0}--\eqref{eq:Tdecomp 3} equal to zero automatically, without imposing any further conditions, the absence of an energy-momentum thin shell on $\Sigma$ would require:
        \begin{itemize}
            \item $\jump{K_{\mu\nu}}=0$, i.e.~the continuity of the extrinsic curvature. Except in some very particular matchings where $\Mp^2+4\beta\MAR^\sSigma=0$ (see Subsection \ref{sec:betaneq0} below), this condition is otherwise unavoidable if one wishes to have a smooth matching at $\Sigma$ (regardless of $T_\mu^\sSigma$, $S_\mu^\sSigma$, $K_{\mu\nu}^\sSigma$ and $\alpha$), as in GR or metric \smash{$\LCf$} gravity.
            \item $\beta\jump{\MAR}=0$ and $\beta n^\mu\jump{\LCD_\mu\MAR}=0$, i.e.~that either $\beta=0$ or the torsionful Ricci scalar $\MAR$ and its normal derivative $n^\mu\LCD_\mu\MAR$ are continuous across $\Sigma$.
            \item $\alpha\jump{\holst}=0$, i.e.~that either $\alpha=0$ or the Holst pseudo-scalar $\holst$ is continuous across $\Sigma$.
        \end{itemize}
    \item A stress-energy dipole (double layer) \smash{$\stress^\sone_{\mu\nu}$} and a monopolar vector spin density \smash{$\trcurrent_\mu^\szero$} are induced on $\Sigma$ unless $\beta\jump{\MAR}=0$, i.e.~if either $\beta=0$ or $\MAR$ is continuous.
    \item Similarly, a monopolar axial-vector spin density \smash{$\axcurrent_\mu^\szero$} emerges on $\Sigma$ unless either $\alpha=0$ or $\holst$ is continuous across $\Sigma$, i.e.~if $\alpha\jump{\holst}=0$.
\end{coloritemize}
\end{result}

Having obtained the generic junction conditions for arbitrary $\alpha$ and $\beta$, we will now focus on some of the paradigmatic sub-cases mentioned in Result \ref{result:JCs in BSPGT}, in which these parameters take particular values. We shall then compare our results with those in the existing literature.

\subsection[Models lacking the Ricci-squared term (\titlemath{$\beta=0$}).]{Models lacking the Ricci-squared term (\titlebm{$\beta=0$}).}
\label{sec:betaeq0}

An immediate consequence of expression \eqref{eq:stress-energy double layer} is that all \BSPGT~models with $\beta=0$ are incompatible with double-layer configurations on the matching surface $\Sigma$, because the dipolar part of the energy-momentum tensor is always proportional to $\beta$.

There are two non-trivial cases with $\beta=0$. First, let us assume that, in addition to vanishing $\beta$, we also have $\alpha=0$. In such scenario, the theory ends up reducing to Einstein-Cartan theory with mass terms for the torsion vectors. As per Result \ref{result:JCs in BSPGT}, there are no extra consistency conditions besides the regularity of $T_\mu$ and $S_\mu$ as well as the regularity and continuity of the metric. Moreover, the two vector currents $\udis{\trcurrent}_\mu$ and $\udis{\axcurrent}_\mu$ become purely regular in this case (i.e.~there are no surface spin sources on $\Sigma$), and only the energy-momentum tensor $\udis{\stress}_{\mu\nu}$ may feature a monopolar singular part, which coincides with the one it has in GR, given back on Equation \eqref{eq:GR thin shell}.

If we now focus on the other possibility, namely $\beta=0$ but $\alpha\neq 0$, we find there is one additional consistency condition compared with the previous sub-case:
\begin{equation}
\label{jumpSperp}
    n^\mu\jump{S_\mu}=0.
\end{equation}
Unfortunately, condition \eqref{jumpSperp} does not allow one to further simplify any other expression. The only non-trivial singular sources present in this case are
\begin{align}
    \axcurrent^\szero_\mu &=   2\epsilon\alpha \jump{\holst} n_\mu, \label{eq:JCbeq0aneq0 1} \\
    \stress^\szero_{\mu\nu}  &= \epsilon\left(\Mp^2\, H^{\sGR}_{\mu\nu} + \alpha\,  H^{\salpha}_{\mu\nu}\right). \label{eq:JCbeq0aneq0 2}
\end{align}
Thus, the three pieces $\tau_{\mu\nu}$, $\tau_\mu$ and $\tau$ making up \smash{$\stress^\szero_{\mu\nu}$}---as defined in \eqref{eq:Tdecomp 1}--\eqref{eq:Tdecomp 3}---are, in general, non-trivial. Consequently, the absence of surface spin densities (\smash{$\axcurrent^\szero_\mu=0$}) and matter thin shells ($\tau_{\mu\nu}= \tau_\mu = \tau=0$) at $\Sigma$ leads to the following smooth-matching conditions:
\begin{equation} \label{eq:JCbeq0aneq0}
    \jump{K_{\mu\nu}}=0, \mybigskip
    \jump{\holst}=0.
\end{equation}

\subsection[Models with non-vanishing Ricci-squared term (\titlemath{$\beta\neq0$}).]{Models with non-vanishing Ricci-squared term (\titlebm{$\beta\neq0$}).}
\label{sec:betaneq0}

For \BSPGT~models having $\beta\neq 0$, besides the continuity of the metric and the regularity of the torsion vectors, we also find the following consistency conditions:
\begin{equation} \label{eq:condbneq0}
    \jump{K}=0,\mybigskip\jump{T_\mu}=0\mybigskip\jump{S_\mu}=0.
\end{equation}
This implies that the three basic fields of the theory, $g_{\mu\nu}$, $T_\mu$ and $S_\mu$ are all regular and continuous across the matching surface $\Sigma$. Furthermore, the continuity of the trace of the extrinsic curvature $K$ entails that the Levi-Civita Ricci scalar $\LCR$ is regular when $\beta\neq 0$, cf.~\eqref{eq:ricci scalar singular}. We also note that, for $\alpha\neq 0$, we get one additional, redundant condition, namely $n^\mu\jump{S_\mu}=0$. Therefore, requirements \eqref{eq:condbneq0} are certainly valid for arbitrary $\alpha$.

As a result of \eqref{eq:condbneq0}, for non-vanishing $\beta$ and arbitrary $\alpha$, the only non-trivial singular spin and matter sources on $\Sigma$ become:
\begin{align}
    \trcurrent^\szero_\mu&= -4\epsilon\beta \jump{\MAR} n_\mu , \label{eq:JCbneq0 1} \\
    \axcurrent^\szero_\mu &= 2\epsilon\alpha \jump{\holst} n_\mu , \label{eq:JCbneq0 2} \\
    T^\szero_{\mu\nu}  &= -\dfrac{\epsilon}{\kappa}\jump{K_{\mu\nu}} + \epsilon\beta\,  H^{\sbeta}_{\mu\nu} + 2\epsilon\alpha\left(2 n_{(\mu} S^\sSigma_{\nu)} - g_{\mu\nu}  S^\sSigma_\perp\right) \jump{\holst} , \label{eq:JCbneq0 3} \\
    T^\sone_{\mu\nu} & =  4\epsilon \beta
     \jump{\MAR} h_{\mu\nu} . \label{eq:JCbneq0 4}
\end{align}
Notice that, under the consistency conditions \eqref{eq:condbneq0}, we find that the discontinuity of the Holst pseudo-scalar across $\Sigma$ simplifies to
\begin{equation}
    \jump{\holst} = -\epsilon n^\mu n^\nu \jump{\LCD_\mu S_\nu}.
\end{equation}

From the expressions \eqref{eq:JCbneq0 1}--\eqref{eq:JCbneq0 4} for the singular sources, we observe that the existence of a non-vanishing monopolar contribution to $\udis{\trcurrent}_\mu$ is correlated with the presence of a non-trivial double layer in the energy-momentum tensor, since \smash{$T^\sone_{\mu\nu}=-n^\rho\trcurrent^\szero_\rho h_{\mu\nu}$}. Moreover, terms proportional to \smash{$n^\mu \udis{\trcurrent}_\mu^\szero$} and \smash{$n^\mu \udis{\axcurrent}_\mu^\szero$} contribute to the energy-momentum thin shell. In fact, even in cases where both spin monopoles are vanishing, i.e.~when $\jump{\MAR}=0$ and $\alpha\jump{\holst}=0$, the torsion vectors would continue to have an impact in the energy-momentum monopole; in particular, in the tangential energy-momentum:
\begin{equation}
    \tau_{\mu\nu} = -\epsilon (\Mp^2+4\beta \MAR^\sSigma)\jump{K_{\mu\nu}}, \mybigskip\text{but}\mybigskip
    \tau_\mu = 0 , \mybigskip
    \tau = 0.
\end{equation}
Therefore, for $\beta\neq 0$ and arbitrary $\alpha$, the absence of spin-density monopoles and stress-energy thin shells is guaranteed if and only if
\begin{equation}
    \jump{\MAR}=0, \mybigskip
    \alpha\jump{\holst}=0, \mybigskip
    (\Mp^2+4\beta \MAR^\sSigma)\jump{K_{\mu\nu}}=0.
\end{equation}
Full \BSPGT~corresponds to the case $\alpha\neq 0$; in that case, the previous smooth matching conditions lead to continuous total Ricci scalar $\MAR$ and Holst pseudo-scalar $\holst$, as well as the generic condition $\jump{K_{\mu\nu}}=0$, except for those very particular configurations where $\Mp^2+4\beta \MAR^\sSigma=0$, in which case the extrinsic curvature does not need to be continuous.

In principle, we could compare our results for the case $\alpha=0$---which corresponds to Palatini $f(\MAR)=\MAR+\beta\MAR^2$ gravity without non-metricity---with the ones obtained in \cite{Olmo:2020fri,Rosa:2021mln}. However, the formalism employed in said references requires some manipulations of the various quantities which are not compatible with our approach,\footnote{
    Throughout this Thesis, we have elaborated on the various ambiguities which appear when using distributions to study junction conditions. As we have already mentioned, it is most often necessary to resort to \emph{choices} and \emph{prescriptions} in order to deal with those subtleties. Thus, the formalisms employed by different authors might not be compatible with each other, causing direct comparisons to be impossible.}
as described in Section \ref{sec:procedure}. In particular, \cite{Olmo:2020fri,Rosa:2021mln} assume vanishing hypermomentum currents and integrate the equation of motion of the connection, resulting in denominators containing quantities that are to be promoted to distributions.

\subsection{Metric Ricci-squared models.}

For vanishing torsion vectors $T_\mu$ and $S_\mu$ and torsion masses $m_T$ and $m_S$, and in the absence of spin-density sources $\trcurrent_\mu$ and $\axcurrent_\mu$, \BSPGT~models with $\beta\neq 0$ reduce to metric $f(\LCR)=\LCR+\beta\LCR^2$ gravity. One might readily check that, in such case, the junction conditions presented in Result \ref{result:JCs in BSPGT} become those described in Section \ref{sec:f(R) JCs} (notice that we are dealing here with the exceptional case $\LCfppp=0$ for all $\LCR$). More precisely, we find that the only consistency condition surviving the torsionless limit is, certainly, $\jump{K}=0$, while Equations \eqref{eq:JCbneq0 3} and \eqref{eq:JCbneq0 4} respectively reduce to \eqref{eq:quadratic f(R) thin shell}--\eqref{eq:quadratic f(R) thin shell piece 3} and \eqref{eq:quadratic f(R) double layer}, as expected, after adapting the notational convention to $\beta \to \alpha/(2\kappa)$. We note that junction conditions such as \eqref{eq:JCbneq0 1} or \eqref{eq:JCbneq0 2} are meaningless in the absence of torsion, because they originate from an equation of motion which no longer exists when $T_\mu$ and $S_\mu$ are chosen to be vanishing from the start. Finally, it is worth mentioning that our previous results imply that the junction conditions in \BSPGT~reduce to those of GR when $\alpha=0$ and $\beta=0$.

\section{Comments on matter sectors depending on the tetrads.}
\label{sec:Dirac}

Interestingly, the paradigmatic matter Lagrangian giving rise to a non-trivial spin density, the Dirac Lagrangian, does not---automatically, at least---fit within the formalism developed in the previous Sections and Chapter \ref{chapter:Introduction: JCs}. The reason lies in the fact that such a matter Lagrangian requires introducing additional geometrical structures, forcing one to work in the tetrad formulation. Accordingly, the metric should no longer be a basic field of the theory, being replaced by the tetrad $e^I{}_\mu$, introduced back in Section \ref{sec:intro:Poincaré Gauge Gravity}, cf.~\eqref{eq:tetrad}. At the level of the action, would have
\begin{equation}
    S[e^I{}_\mu, T_\mu, S_\mu, t^\rho{}_{\mu\nu}, \Psi]=S_\mathrm{GFPG}[g^{\rho\sigma}(e^I{}_\mu), T_\mu, S_\mu, t^\rho{}_{\mu\nu}] + S_\matter[e^I{}_\mu, T_\mu, S_\mu, \Psi].
\end{equation}
Thus, instead of the equation of motion for the metric, one gets
\begin{equation}
    0=\dfrac{1}{|e|}\dfrac{\delta S}{\delta e^I{}_\mu} = -\dfrac{2}{|e|} g^{\mu (\rho} e_I{}^{\sigma)}\dfrac{\delta S_\mathrm{GFPG}}{\delta g^{\rho\sigma}} + \mathcal{T}^\mu{}_{I}
    = -g^{\mu\rho} e_I{}^{\sigma}\left[\mathcal{E}(g)_{\rho\sigma}-\mathcal{T}_{\rho\sigma}\right], \label{eq:EoMvielbein}
\end{equation}
where $e \equiv \det (e^I{}_\mu)$, $\mathcal{E}(g)_{\mu\nu}$ corresponds to the standard metric variation---cf.~\eqref{eq:var0}---and
\begin{equation}
    \mathcal{T}^\mu{}_{I}\equiv \dfrac{1}{|e|}\dfrac{\delta \tilde{S}_\matter}{\delta e^I{}_\mu},\myhugeskip \mathcal{T}_{\mu\nu} \equiv g_{\rho\mu}\, e^I{}_\nu\, \mathcal{T}^\rho{}_{I}.
\end{equation}
After contracting with the tetrad and the metric, we can split \eqref{eq:EoMvielbein} into symmetric and antisymmetric parts:
\begin{equation}
    \mathcal{E}(g)_{\mu\nu}=\mathcal{T}_{(\mu\nu)},\myhugeskip \mathcal{T}_{[\mu\nu]}=0.
\end{equation}
In the first equation, we can straightforwardly use all the previous results for symmetric energy-momentum tensors, since the gravitational part $\mathcal{E}(g)_{\mu\nu}$ is the same as in the metric formulation. The second equation, however, just tells us that all the singular parts coming from $\mathcal{T}_{[\mu\nu]}$ must vanish identically.

\section{Summary and conclusions.} \label{sec:JCsBSPGT conclusions}

In this Chapter, we have obtained the junction conditions for the ghost-free subclass of Poincaré Gauge Gravity which propagates one scalar and one pseudo-scalar field, in addition to the usual graviton of GR. To that end, we have particularised the formalism in Chapter \ref{chapter:Introduction: JCs} to the \BSPGT~field equations \eqref{eq:EoM1}--\eqref{eq:EoM3}. Since we have not assumed any specific value for the parameters of the theory, our results (synthesised in Result \ref{result:JCs in BSPGT}) are completely general. 

The junction conditions have been obtained in two steps. First, in order to avoid products of singular distributions, we determined the consistency conditions of the theory, which can be found in Section \ref{cons:cond}. Afterwards, we derived the relation between the matter content of the hypersurface and the allowed discontinuities in the basic fields of the theory under the aforementioned consistency conditions. To do so, we studied the singular contributions to the equations of motion, collected in Tables \ref{tab:termsEqT}-\ref{tab:termsEqMetric}. Then, in  Section \ref{sing:trace}, we analysed in detail the trace-vector equation, and showed that the only singular part allowed for the corresponding vector spin density is a monopole controlled by the combination $\beta\jump{\MAR}$. We also carried out an analogous analysis for the axial vector equation in Section \ref{sing:axial}, finding that only monopolar axial spin densities are allowed, with the corresponding singular part being characterised by the product $\alpha\jump{\holst}$. Next, concerning the equation of the metric, we showed in Section \ref{sing:Einstein} that both thin-shell (monopolar) and double-layer (dipolar) singular contributions are allowed for the stress-energy tensor. On the one hand, the dipolar part of the energy-momentum tensor vanishes identically for $\beta\jump{\MAR}=0$. On the other hand, the corresponding thin shell has non-vanishing irreducible parts, namely, tension scalar $\tau$, external momentum flux $\tau_\mu$ and surface energy-momentum tensor $\tau_{\mu\nu}$. The first two parts contain terms proportional to $\beta\jump{\MAR}$ and $\alpha\jump{\holst}$, while $\tau_{\mu\nu}$ acquires a contribution proportional to $\beta\jump{K_{\mu\nu}}$, in addition to terms proportional to $\beta\jump{\MAR}$ and $\alpha\jump{\holst}$ and a part which is identical to the one $\udis{\stress}_{\mu\nu}$ has in GR.

Finally, we have studied different sub-cases of the theory in Section \ref{sec:partcases}:
\begin{coloritemize}
    \item For $\alpha=\beta=0$, we have found the same junction conditions as in GR. For $\beta=0$ but $\alpha\neq0$, whose singular-layer conditions are collected in \eqref{eq:JCbeq0aneq0 1} and \eqref{eq:JCbeq0aneq0 2}, we have shown that the absence of surface sources of any kind requires the continuity of $n^\mu S_\mu$, $K_{\mu\nu}$ and $\holst$. Moreover, no double layers are possible in these models.

    \item For $\beta\neq 0$ and arbitrary $\alpha$, we have proven that the continuity of $T_\mu$, $S_\mu$ and $K$ is instrumental in order to avoid ill-defined terms involving products of singular distributions. The resulting singular-layer conditions are collected in \eqref{eq:JCbneq0 1}--\eqref{eq:JCbneq0 4}, and are valid for the bi-scalar case $\alpha\neq0$ and for the pure-Ricci theory $\alpha=0$. In the former case, a smooth matching at $\Sigma$ always entails the continuity of $\MAR$ and $\holst$, while a discontinuous $K_{\mu\nu}$ is only admissible in exceptional scenarios where $\Mp^2+4\beta\MAR^\sSigma=0$. In the latter case, the smooth-matching conditions are the same except for the continuity of $\holst$, which is no longer necessary.

    \item We have also provided details on the metric limit of the theory, namely, assuming that the torsion is vanishing from the beginning; the resulting theory is metric $\LCf$ with $\LCfppp=0$. We have compared our results with those in the existing literature \cite{Deruelle:2007pt,Senovilla:2013vra}, which we also summarised back in Section \ref{sec:f(R) JCs}. As anticipated, we have found they are in complete agreement with each other.
\end{coloritemize}

The junction conditions we have obtained could be used to explore physically relevant situations, such as spherical stellar collapse or brane-world scenarios. These studies can be used to understand better the dynamics of the bi-scalar theory. Even though we plan to undertake these issues in a future work, we can already make some educated guesses about the kind of results we might expect to find. For example, the wide range of possible singular sources in \BSPGT~opens up the possibility of finding a plethora of equally varied of thin-shell-wormhole solutions. Furthermore, given that the junction conditions in \BSPGT~are much more stringent than their GR counterparts, it is plausible that there exist constraining no-go results for the allowed exterior space-times outside collapsing stars in \BSPGT, akin to Results \ref{theoconstcurv}--\ref{theoanothernonstatic} found for metric $f(\LCR)$ gravity. Nonetheless, one could argue that, in generic cases in which $\beta\neq 0$, some of the tensions generated by junction condition $\jump{\LCR}=0$ in $f(\LCR)$ theories might be alleviated in \BSPGT, where the quantity which must be continuous across $\Sigma$ is the full Ricci scalar, $\MAR$. As such, the discontinuity of $\LCR$ could be compensated with the torsion terms in $\MAR$, cf.~\eqref{eq:jumpR}. In any case, it is clear that shedding light on these and many other questions concerning glued solutions in \BSPGT~will inevitably require conducting thorough investigations making use of the junction conditions explained herein.

\chapter{Non-viability of \titlebm{$\LCR_0$}-degenerate \titlebm{$\LCf$} models and their constant-curvature solutions.}
\label{chapter:constant curvature}

When modelling physical systems, it is often necessary to resort to simplifying assumptions, either to make the equations describing the problem more tractable or to gain further insight on the relevant physics. It seems obvious that such assumptions should be physically well-motivated and consistent with both experiments and the theoretical framework being employed. For example, within
    GR or metric $f(\LCR)$ gravity,
the explanation of a variety of cosmological observations relies on the crucial assumption that the Universe is approximately homogeneous and isotropic at sufficiently large scales.
    Except for some pathological examples, $f(\LCR)$ models are also mathematically consistent and can be reduced to standard GR in the appropriate limit, yielding predictions which are in accordance with observations. Indeed, appropriate choices of function $f$ lead to a correct description of both early- and late-universe physics, such as inflation and the dark-energy-dominated epoch \cite{Nojiri:2008nt}. As mentioned before, these phenomena cannot be explained within the framework of GR unless without introducing new, \emph{ad hoc} fields and dark-energy fluids in the theory, respectively.
$f(\LCR)$ models have also found applications in stellar physics, with some of them being compatible with neutron-star and gravitational-wave observations \cite{Olmo:2019flu,Feola:2019zqg,Astashenok:2017dpo,Astashenok:2020qds,Ananda:2007xh,Capozziello:2008fn,Bouhmadi-Lopez:2012piq}.

Nonetheless, outside of the highly-symmetric cosmological scenarios, it is often very complicated to solve the fourth-order equations of metric $f(\LCR)$ gravity without making further simplifying assumptions which, as mentioned before, should be well-motivated. Currently, one of the most popular choices is to find solutions with constant scalar curvature $\LCR$. Indeed, some of the most renowned solutions of GR have constant curvature; for instance, the Schwarzschild or Vaidya space-times, their generalisations including a cosmological constant, or the FLRW space-time sourced by radiation. It has been known for a long time that many $f(\LCR)$ gravity models host only the same vacuum constant-curvature solutions as GR \cite{delaCruz-Dombriz:2009pzc}. However, as will be further detailed in Section \ref{section:constant-curvature solutions}, there is a particular set of $f(\LCR)$ models---satisfying some additional assumptions \cite{Nzioki:2009av,Calza:2018ohl}---for which \emph{any} metric with a given constant Ricci scalar is a solution of said $f(\LCR)$ model. Because of this, in what follows we shall refer to these special $f(\LCR)$ models admitting all space-times with $\LCR=\LCR_0=\const$ as \emph{$\LCR_0$-degenerate $f(\LCR)$ models}. In order to find new constant-curvature solutions of $\LCR_0$-degenerate $f(\LCR)$ models which are not present in GR, one merely needs to solve the equation $\LCR=\LCR_0=\const$ for some particular metric \emph{Ansatz} and initial conditions.

For these reasons, it is almost immediate to obtain novel, $f(\LCR)$-exclusive constant-curvature solutions that provide an answer to virtually \emph{every} open problem in gravitational physics. For example, $f(\LCR)$-exclusive constant-curvature solutions describing wormholes made out of pure vacuum (and thus complying with the standard energy conditions), exotic black holes, or even space-times giving rise to the observed rotation curves of galaxies without introducing dark matter have been reported in the literature \cite{Duplessis:2015xva,Calza:2018ohl,Hendi:2020umk,Bertipagani:2020awe}. This approach has furthermore been generalised to other modified gravity theories, such as $f(Q)$, where a similar situation occurs \cite{Calza:2022mwt}.

In the present Chapter, we shall show that the vast majority of $\LCR_0$-degenerate $f(\LCR)$ models, as well as their constant-curvature solutions themselves, are most often pathological in nature, and thus not physically viable. In particular, we will prove that this special class of $f(\LCR)$ models only propagates one scalar degree of freedom at linear level (in contrast with generic $f(\LCR)$ models and GR), thus being incompatible with gravitational-wave observations. Furthermore, as we shall discuss, $\LCR_0$-degenerate $f(\LCR)$ models apparently lack predictability, because of the aforementioned infinite degeneracy of their constant-curvature solutions. Finally, to make matters worse, we will also show that most of the novel constant-curvature solutions are unstable and host a number of unphysical properties, such as regions in which the metric signature changes abruptly, naked curvature singularities, and more.
    This Chapter encapsulates the main results of the investigations carried out in Publication \cite{ThirdPaper}.

The contents of this Chapter will be organised as follows. First, in Section \ref{section:constant-curvature solutions}, we shall briefly review the conditions $f(\LCR)$ models must satisfy so as to harbour any metric with a given constant Ricci scalar. Second, the pathological character of the $f(\LCR)$ models within this special class will be discussed in Section \ref{Section: Pathologies of the models}. More precisely, the fact that the linearised spectrum of said models contains at most one massless scalar field only will be proven therein. Next, in Section \ref{Section: Stability of solutions}, we shall assess the stability of $f(\LCR)$-exclusive constant-curvature solutions under small perturbations of their Ricci scalar. Finally, Section \ref{introducing the solutions} will be devoted to a characterisation of several classes of novel constant-curvature solutions, some of which have not been yet reported in the literature, as far as we are concerned. We shall then conclude that most of the solutions analysed in the latter section display a variety of unphysical properties. Supplementary discussions and formulae are provided in Appendix \ref{appendix:constant-curvature}.

\section[Constant-curvature vacuum solutions of \titlemath{$f(\LCR)$} gravity.]{Constant-curvature vacuum solutions of \titlebm{$f(\LCR)$} gravity.} \label{section:constant-curvature solutions}

Throughout this Chapter, we shall define a \emph{constant-curvature space-time} as one represented by a metric whose Ricci scalar is constant, i.e.
\begin{equation}
    \LCR=\const\equiv\LCR_0.
\end{equation}
When the equations of motion of $f(\LCR)$ gravity \eqref{f(R) EOM} are evaluated in vacuum ($\stress_{\mu\nu}=0$) and solutions with constant scalar curvature $\LCR_0$ are sought, the last term on the left-hand side of \eqref{f(R) EOM} vanishes, as $f'(\LCR)=\const$ Thus, \eqref{f(R) EOM} reduce to
\begin{equation} \label{reduced f(R) EOM}
    f'(\LCR_0)\LCR_{\mu\nu}=\dfrac{f(\LCR_0)}{2}g_{\mu\nu}.
\end{equation}
Taking the trace of \eqref{reduced f(R) EOM}, one finds that vacuum constant-curvature solutions also satisfy
\begin{equation}
\label{reduced f(R) EOM trace}
    f'(\LCR_0)\LCR_0=2f(\LCR_0).
\end{equation}
Hence, in the event that vacuum solutions with constant-curvature $\LCR_0$ are present in a certain $f(\LCR)$ model, equations \eqref{reduced f(R) EOM} and \eqref{reduced f(R) EOM trace} hold simultaneously, giving rise to the following scenarios:
\begin{coloritemize}
    \item If $f'(\LCR_0)\neq 0$, we recover Einstein's equations \eqref{GR + Lambda EOM} with an effective cosmological constant $\Lambda_\eff$, which is related to $\LCR_0$:
    \begin{equation}
        \LCR_{\mu\nu}-\underbrace{\dfrac{f(\LCR_0)}{2 f'(\LCR_0)}}_{\equiv \Lambda_\eff}g_{\mu\nu}=0 \myskip \overset{\mathrm{trace}\,}{\Longrightarrow}\myskip \LCR_0=\dfrac{2f(\LCR_0)}{f'(\LCR_0)}=4\Lambda_\eff.
    \end{equation}
    Hence, in this case, the only possible constant-curvature solutions in the $f(\LCR)$ model are the same as in GR $+$ $\Lambda_\eff$. Notice that the existence of constant-curvature solutions with $\LCR_0=0$ (i.e.~$\Lambda_\eff=0$) inevitably requires the model to fulfil $f(0)=0$, as per all the previous equations.
    \item If $f'(\LCR_0)=0$, Equation \eqref{reduced f(R) EOM trace} entails that $f(\LCR_0)=0$, regardless of $\LCR_0$. Substituting these two conditions back on \eqref{reduced f(R) EOM}, one finds that the equations of motion are then automatically satisfied. As a result, \emph{any} metric with constant Ricci scalar $\LCR_0$ trivially solves the vacuum equations of motion of \emph{all} $f(\LCR)$ models such that
    \begin{equation} \label{constant-curvature conditions R0}
        f(\LCR_0)=0,\mybigskip\myskip f'(\LCR_0)=0.
    \end{equation}
    Analogously, \emph{every} metric with vanishing Ricci scalar is a vacuum solution of \emph{any} $f(\LCR)$ model satisfying
    \begin{equation} \label{constant-curvature conditions 0}
        f(0)=0,\mybigskip\myskip f'(0)=0,
    \end{equation}
    with the first condition---$f(0)=0$---being necessary for the model to harbour the vanishing-curvature solutions of GR.
\end{coloritemize}

$f(\LCR)$ models fulfilling conditions \eqref{constant-curvature conditions R0} or \eqref{constant-curvature conditions 0} shall thus be the object of study of the present Chapter. As explained in the introduction, we will generically refer to these special choices of function $f$ as \emph{$\LCR_0$-degenerate models}. In addition, we shall further distinguish between \emph{$(\LCR_0\neq 0)$-degenerate models}, which satisfy conditions \eqref{constant-curvature conditions R0}, and \emph{$(\LCR_0=0)$-degenerate models}, which comply with \eqref{constant-curvature conditions 0} instead. Finally, constant-curvature solutions exclusive to $\LCR_0$-degenerate $f(\LCR)$ models will be hereafter referred to as \emph{$\LCR_0$-degenerate} (in analogy with the models themselves), or as \emph{$f(\LCR)$-exclusive}, or even as \emph{novel} solutions.

The reader should note that it is straightforward to find non-trivial $\LCR_0$-degenerate $f(\LCR)$ models which appear to be, at least \emph{a priori}, physically well-motivated. The most paradigmatic examples would be the so-called `power-of-GR' models, $f(\LCR)\propto \LCR^{1+\delta}$, which satisfy conditions \eqref{constant-curvature conditions 0} provided that $\delta>0$. These models have interesting applications in cosmology and might be compatible with Solar-System experiments, depending on the value of $\delta$ \cite{Clifton:2005aj,Bajardi:2022ocw}. Another simple instance of $(\LCR_0\neq 0)$-degenerate model would be $f(\LCR)=\LCR-\LCR_0/2-\LCR^2/(2\LCR_0)$, which satisfies \eqref{constant-curvature conditions R0} while behaving as $\LCR-2\Lambda+\bigo(\LCR^2)$ for $\LCR\ll\LCR_0$, being $\LCR_0$ the only dimensional parameter. The fact that there exist $\LCR_0$-degenerate models which reduce to GR$+\Lambda$ in the appropriate limit is a remarkable result, given that GR with (or without) a cosmological constant is not $\LCR_0$-degenerate by itself (for any $\LCR_0$).

At this point, it is important to remark that $\LCR_0$-degenerate $f(\LCR)$ models may admit solutions which are not degenerate. For example, in Appendix \ref{Appendix: A simple model}, we present a simple $f(\LCR)$ model harbouring constant-curvature solutions for two different values of $\LCR_0$. However, the model is $\LCR_0$-degenerate for only one of these two $\LCR_0$s; for the other value, the $f(\LCR)$ model only hosts the (non-degenerate) constant-curvature solutions of $\text{GR}+\Lambda$.

Another important observation is that $f(\LCR)$ models do not need to be fine-tuned in order to become $\LCR_0$-degenerate. For example, the so-called `power-of-GR' models $f(\LCR)\propto \LCR^{1+\delta}$ with $\delta>0$ discussed above are always $(\LCR_0=0)$-degenerate. As previously mentioned these models have been well-studied in the literature in the context of cosmology and, certainly, were not purposefully designed to be $(\LCR_0=0)$-degenerate. Thus, as a matter of fact, whenever conditions \eqref{constant-curvature conditions R0} or \eqref{constant-curvature conditions 0} are satisfied---either accidentally or on purpose---the $f(\LCR)$ model in question is constant-curvature degenerate for the appropriate values of $\LCR_0$. It is true, however, that nothing prevents one from constructing a \emph{designer} $\LCR_0$-degenerate $f(\LCR)$ model hosting as many solutions as one wishes with a particular constant curvature $\LCR_0$.

Before closing this Section, one last brief comment concerning the predictability of $\LCR_0$-degenerate $f(\LCR)$ models is pertinent. As their name suggests, it is not clear whether $f(\LCR)$ models complying with either \eqref{constant-curvature conditions R0} or \eqref{constant-curvature conditions 0} have full predictive power. In any such $\LCR_0$-degenerate model, there is a set of initial conditions (namely, those requiring the Ricci scalar to be $\LCR_0$) whose evolution is not dictated by the vacuum equations of motion; recall equations \eqref{f(R) EOM} hold trivially for (the infinite number of) metrics with $\LCR=\LCR_0$. Moreover, there are indications that some metrics with the privileged Ricci scalar $\LCR_0$ can be smoothly glued to each other \cite{Casado-Turrion:2022xkl}, thus suggesting that $\LCR_0$-degenerate models might sometimes be unable to discern between its constant-curvature solutions.

\section[Strong coupling in \titlemath{$\LCR_0$}-degenerate \titlemath{$f(\LCR)$} models.]{Strong coupling in \titlebm{$\LCR_0$}-degenerate \titlebm{$f(\LCR)$} models.} \label{Section: Pathologies of the models}

Innocent as they might seem at first sight, the special class of $\LCR_0$-degenerate $f(\LCR)$ models---i.e.~those fulfilling either conditions \eqref{constant-curvature conditions R0} or \eqref{constant-curvature conditions 0}---can be shown to be inherently pathological, as stated in the introduction to this Chapter. In the following, we shall concentrate in a physically-relevant shortcoming of said models, namely, an apparent strong-coupling instability---i.e.~the non-propagation of all expected degrees of freedom at linear level around a flat background.

The linearised spectrum of a given gravity theory comprises all the independent fields which propagate on top of a suitable background when the equations of motion are expanded up to linear order in perturbations. The linearised spectrum around flat Minkowski space-time thus coincides with the possible gravitational-wave polarisation modes which can be observationally detected, since the weak-field approximation is appropriate at the location of current experimental settings.

It is well known that, generically, the gravitational wave spectrum of metric $f(\LCR)$ models consists of a massless and traceless graviton akin to that of GR (with two polarisation modes, the so-called `$+$' and `$\times$' polarisations), plus an additional longitudinal (i.e.~massive) scalar degree of freedom \cite{Capozziello:2008fn,Bouhmadi-Lopez:2012piq}, in consonance with the fact that $f(\LCR)$ theories of gravity are dynamically equivalent to a scalar-tensor theory \cite{Sotiriou:2008rp,DeFelice:2010aj}.\footnote{
    Some studies \cite{Alves:2009eg,Alves:2010ms,RizwanaKausar:2016zgi} claimed that the linearised spectrum of $f(\LCR)$ contained a second scalar polarisation mode, dubbed \emph{breathing mode}, in disagreement with previous results. The controversy was finally settled against the existence of such a breathing mode by resorting to the Hamiltonian formalism \cite{Liang:2017ahj} and gauge-invariant methods \cite{Moretti:2019yhs}.
}

The fact that \emph{most} $f(\LCR)$ models propagate a massless and traceless graviton renders them compatible with gravitational-wave observations \cite{Ezquiaga:2017ekz} (notice that no current gravitational-wave detectors are sensitive to non-tensorial modes, including scalar ones, which remain unobserved). However, it is important to remark that, in general, previous analyses of gravitational waves in $f(\LCR)$ gravity made no assumptions on function $f$ itself (apart from analyticity at $\LCR=0$, which is necessary to perform the linearisation of the field equations, as we shall see later). For these reasons, these investigations failed to conclude that \emph{not all} $f(\LCR)$ models propagate the expected linearised degrees of freedom (graviton $+$ scalar) around a Minkowski background. Indeed, we have found that $(\LCR_0=0)$-degenerate models feature such evanescence of the expected degrees of freedom, signalling the presence of a previously undiscovered strong-coupling instability.\footnote{
    A background is said to be \emph{strongly-coupled} whenever at least one of the expected perturbative degrees of freedom fails to propagate atop said background. In other words, the kinetic term(s) of the evanescent field(s) vanish when evaluated in the strongly-coupled background; equivalently, interaction terms blow up upon canonicalisation of the equations of motion, hence the name \emph{strong-coupling}. Comprehensive accounts of the generalities of strong-coupling phenomena may be found in works investigating the appearance of such instabilities in physical theories. We refer the interested reader to references such as \cite{BeltranJimenez:2020lee}, for instance.
} In particular, we have been able to establish the following two Results:

\begin{result}[name={[non-propagation of the graviton in $(\LCR_0=0)$-degenerate $f(\LCR)$ models]}] \label{Result: Non-propagation of gravity}
    At linear level in perturbations, $(\LCR_0=0)$-degenerate $f(\LCR)$ models---i.e.~those complying with conditions \eqref{constant-curvature conditions 0}---propagate, at most, one single massless scalar mode atop a flat background. In other words, these models do not contain the expected spin-2 graviton in their linearised spectrum, and thus Minkowski space-time is strongly-coupled in $(\LCR_0=0)$-degenerate $f(\LCR)$ models.
\end{result}

\begin{result}[name={[non-propagation of the scalaron in $(\LCR_0=0)$-degenerate $f(\LCR)$ models]}] \label{Result: Non-propagation of the scalaron}
    Around a Minkowski background, the linearised spectrum of $(\LCR_0=0)$-degenerate $f(\LCR)$ models satisfying $f''(0)=0$ does not contain any dynamical degrees of freedom whatsoever.
\end{result}

As mentioned before, Result \ref{Result: Non-propagation of gravity} puts into question the physical viability of $(\LCR_0=0)$-degenerate $f(\LCR)$ models, since the two polarisation modes corresponding to a massless and traceless spin-2 graviton have been detected in all gravitational-wave experiments carried out by the LIGO and VIRGO collaborations since 2015 \cite{LIGOScientific:2016aoc,LIGOScientific:2016lio}. We must also stress at this point that it is not very difficult to find $(\LCR_0=0)$-degenerate $f(\LCR)$ models that comply with the hypotheses of Result \ref{Result: Non-propagation of the scalaron}; for instance, all `power-of-GR' models $f(\LCR)\propto\LCR^{1+\delta}$ with $\delta>1$ (we shall also assume that $\delta$ is a natural number for the series expansion around $\LCR=0$ to exist).

In order to prove the assertions in Results \ref{Result: Non-propagation of gravity} and \ref{Result: Non-propagation of the scalaron}, we will proceed as follows. First, we will review the linearisation of the $f(\LCR)$ field equations \eqref{f(R) EOM} for any choice of function $f$. After that, we will particularise the aforementioned results to the special case of $(\LCR_0=0)$-degenerate models---i.e.~we will make use of conditions \eqref{constant-curvature conditions 0}---to demonstrate the existence aforementioned apparent strong-coupling instabilities.

\subsection[Linearisation of the \titlemath{$f(\LCR)$} equations of motion around a Minkowski background.]{Linearisation of the \titlebm{$f(\LCR)$} equations of motion around a Minkowski background.}

In order to perform the linear expansion of the $f(\LCR)$ equations of motion \eqref{f(R) EOM} around Minkowski space-time, one starts by choosing a suitable coordinate system in which the metric $g_{\mu\nu}$ can be decomposed as
\begin{equation}
    g_{\mu\nu}=\eta_{\mu\nu}+h_{\mu\nu},
\end{equation}
where $\eta_{\mu\nu}$ is the Minkowski background and $h_{\mu\nu}$ is the metric perturbation, i.e.~$|h_{\mu\nu}|\ll 1$ in the aforementioned coordinate system. As widely known, the following expressions are true at first order in $h_{\mu\nu}$: 
\begin{align}
    g^{\mu\nu} &= \eta^{\mu\nu}-h^{\mu\nu}+\bigo(h^2), \\
    \LCR_{\mu\nu} &= \LCR_{\mu\nu}^{\sone}+\bigo(h^2), \\
    \LCR &= \LCR^{\sone}+\bigo(h^2), \vphantom{\LCR_{\mu\nu}^{\sone}}
\end{align}
where $\bigo(h^2)$ represents terms which are quadratic in the metric perturbation $h_{\mu\nu}$, and we have introduced
\begin{align}
    \LCR_{\mu\nu}^{\sone} &\equiv \dfrac{1}{2}\left[2\partial_\rho\partial_{(\mu}h_{\hphantom{\rho}\nu)}^{\rho}-\LCbox h_{\mu\nu}-\partial_\mu\partial_\nu h\right], \label{linearised array ini} \\
    \LCR^{\sone} &\equiv \eta^{\mu\nu}R_{\mu\nu}^{\sone}=\partial_\mu\partial_\nu h^{\mu\nu}-\LCbox h, \label{R(1)} \\
    h^{\mu\nu} &\equiv \eta^{\mu\rho}\eta^{\nu\sigma}h_{\rho\sigma}\,,\myskip
    h^\mu_{\hphantom{\lambda}\nu}\equiv\eta^{\mu\rho}h_{\rho\nu}\,,\myskip
    h\equiv\eta^{\mu\nu}h_{\mu\nu}.\myskip\myskip \label{linearised array fin}
\end{align}
In expressions \eqref{linearised array ini}--\eqref{linearised array fin} and for the remainder of this Section, $\LCbox$ shall denote the Minkowski-space d'Alembertian, i.e.~$\LCbox=\eta^{\mu\nu}\partial_\mu\partial_\nu$.

The existence of a linearised regime in $f(\LCR)$ theories requires one additional assumption to be made, namely that $f$ must be analytic at $\LCR=0$, and therefore series-expandable up to to linear order in the Ricci-scalar perturbation $\LCR^{\sone}$, i.e.~up to linear order in metric perturbations. This is to guarantee that the resulting linearised equations of motion remain first-order in $h_{\mu\nu}$, something which is impossible whenever $f$ and its derivatives cannot be linearised in the first place.

For a generic $f(\LCR)$ model which is analytic around $\LCR=0$, taking into account all the previous considerations results in the following set of linearised vacuum equations of motion:
\begin{equation} \label{general f(R) linearised EOM}
    f'(0)\,\LCEin_{\mu\nu}^{\sone}-f''(0)\,(\partial_\mu\partial_\nu-\eta_{\mu\nu}\LCbox)\LCR^{\sone}+\bigo(h^2)=0,
\end{equation}
where we have defined the Einstein-like tensor
\begin{equation} \label{Einstein-like tensor}
    \LCEin_{\mu\nu}^{\sone}\equiv\LCR_{\mu\nu}^{\sone}-\dfrac{1}{2}\eta_{\mu\nu}\LCR^{\sone}.
\end{equation}
Taking the trace of \eqref{general f(R) linearised EOM}, one finds
\begin{equation} \label{R(1) Klein-Gordon}
    3f''(0)\,\LCbox\LCR^{\sone}-f'(0)\LCR^{\sone}+\bigo(h^2)=0.
\end{equation}
which is a non-canonical Klein-Gordon equation for $\LCR^{\sone}$ provided that $f''(0)\neq 0$. Direct inspection of this equation clearly reveals that the kinetic term for $\LCR^{\sone}$ vanishes if $f''(0)=0$. Therefore, expression \eqref{R(1) Klein-Gordon} alone suffices to conclude that only in cases where $f''(0)\neq 0$, the Ricci-scalar perturbation $R^{\sone}$ behaves as an independent, propagating scalar degree of freedom  at linearised level. One might then divide both sides of equation \eqref{R(1) Klein-Gordon} by $f''(0)$ so as to canonicalise the kinetic term, yielding
\begin{equation} \label{R(1) Klein-Gordon canonical}
    \LCbox\LCR^{\sone}-\dfrac{f'(0)}{3f''(0)}\LCR^{\sone}+\bigo(h^2)=0.
\end{equation}
Thus, for $f''(0)\neq 0$, the propagating scalar degree of freedom $\LCR^{\sone}$ has an effective mass $m_\mathrm{eff}$ given by
\begin{equation} \label{scalaron mass}
    m_\mathrm{eff}^2=\dfrac{f'(0)}{3f''(0)}.
\end{equation}

\subsection[Unravelling the strongly-coupled nature of Minkowski space-time in (\titlemath{$\LCR_0=0$})-degenerate \titlemath{$f(\LCR)$} models.]{Unravelling the strongly-coupled nature of Minkowski space-time in (\titlebm{$\LCR_0=0$})-degenerate \titlebm{$f(\LCR)$} models.}

Turning back to the full linearised equations of motion \eqref{general f(R) linearised EOM}, we are now in a position that will allow us to understand intuitively why the spin-2 sector of the theory does not propagate atop the Minkowski background in $(\LCR_0=0)$-degenerate $f(\LCR)$ models. Aside from the higher-order terms, equation \eqref{general f(R) linearised EOM} contains (i) the term proportional to $f'(0)$ and the Einstein-like tensor \smash{$\LCEin^{\sone}_{\mu\nu}$} given by \eqref{Einstein-like tensor}, which encapsulates all terms depending on $h_{\mu\nu}$ and its derivatives; and (ii) the term containing derivatives of the scalar mode $R^{\sone}$, which is proportional to $f''(0)$ and does \emph{not} depend on $h_{\mu\nu}$ nor its derivatives. From this, it is clear that:
\begin{coloritemize}
    \item The first term (i) vanishes whenever function $f$ is such $f'(0)=0$, which is precisely one of the defining conditions of $(\LCR_0=0)$-generate $f(\LCR)$ models, cf.~\eqref{constant-curvature conditions 0}. Given that \smash{$\LCEin^{\sone}_{\mu\nu}$} contains all the derivatives of $h_{\mu\nu}$ appearing in the equations of motion, the absence of this term entails that the spin-2 mode does not propagate, as stated in Result \ref{Result: Non-propagation of gravity}. The strong-coupling problem becomes evident once one notices that the interaction terms---i.e.~the $\bigo(h^2)$ terms---blow up when one divides equation \eqref{general f(R) linearised EOM} by $f'(0)$ (in order to canonicalise the graviton kinetic terms) and then takes the limit $f'(0)\rightarrow 0$. \hfill $\blacksquare$
    \item The second term (ii) will not be present either whenever $f''(0)=0$, as discussed above. Thus, for $(\LCR_0=0)$-generate $f(\LCR)$ models such that $f''(0)=0$, equations \eqref{general f(R) linearised EOM} and \eqref{R(1) Klein-Gordon} contain no kinetic terms at all, only the $\bigo(h^2)$ interaction terms survive, and thus those theories do not possess a linearised spectrum, as asserted in Result \ref{Result: Non-propagation of the scalaron}. \hfill $\blacksquare$
\end{coloritemize}

However, there is a more insightful (and more mathematically explicit) way of understanding why the spin-2 degree of freedom fully decouples when $f'(0)=0$, i.e.~for $(\LCR_0=0)$-degenerate $f(\LCR)$ models. The argument goes as follows. As in GR, in $f(\LCR)$ gravity it is possible \cite{Capozziello:2008fn} to define a new symmetric rank-two tensor field, $\bar{h}_{\mu\nu}$, such that equations \eqref{general f(R) linearised EOM} reduce to the wave equation
\begin{equation} \label{GW equation}
    \LCbox\bar{h}_{\mu\nu}+\bigo(h^2)=0
\end{equation}
in the de Donder gauge, i.e.~after setting
\begin{equation}
    \partial_\mu\bar{h}^{\mu\nu}=0
\end{equation}
using some of the available gauge freedom in the theory. The remaining gauge freedom is then employed to impose the transverse-traceless (TT) condition. In a generic $f(\LCR)$ model, after expanding \eqref{general f(R) linearised EOM} in the de Donder gauge and comparing the result with \eqref{GW equation}, one finds that $\bar{h}_{\mu\nu}$ is given by\footnote{
    Our expression \eqref{hbar} for $\bar{h}_{\mu\nu}$ is slightly different from that commonly found in the literature \cite{Capozziello:2008fn,RizwanaKausar:2016zgi},
    \begin{equation*}
        \bar{h}_{\mu\nu}=\bar{h}_{\mu\nu}^{\sGR}-\dfrac{f''(0)}{f'(0)}\LCR^{\sone}\eta_{\mu\nu},
    \end{equation*}
    which is not valid if $f'(0)=0$, i.e.~precisely in the case we are interested in.
}
\begin{equation} \label{hbar}
    \bar{h}_{\mu\nu}=f'(0)\,\bar{h}_{\mu\nu}^{\sGR}-f''(0)\,\LCR^{\sone}\,\eta_{\mu\nu},
\end{equation}
where
\begin{equation}
    \bar{h}_{\mu\nu}^{\sGR}\equiv h_{\mu\nu}-\dfrac{h}{2}\eta_{\mu\nu}
\end{equation}
is the usual spin-2 degree of freedom of GR. In consequence, Equations \eqref{GW equation} and \eqref{hbar} evince that what propagates at the speed of light (in vacuum) in a generic $f(\LCR)$ gravity model is a mixture of the GR spin-2 graviton and the extra scalar mode. As per \eqref{hbar}, such propagating mixture reduces to its scalar component provided that $f'(0)=0$. In such situation, $R^{\sone}$ becomes effectively massless, due to \eqref{scalaron mass}, and the Klein-Gordon equation \eqref{R(1) Klein-Gordon} becomes equivalent to wave equation \eqref{GW equation}. As a result, only the massless scalar degree of freedom propagates in $(\LCR_0=0)$-degenerate $f(\LCR)$ models such that $f''(0)\neq 0$, as previously stated in Result \ref{Result: Non-propagation of gravity}. Furthermore, one clearly sees that no propagating degree of freedom survives the limit $f''(0)\rightarrow 0$, in agreement with the strong-coupling instability described in Result \ref{Result: Non-propagation of the scalaron}.

\section{Stability of the degenerate constant-curvature solutions.} \label{Section: Stability of solutions}

Even though $\LCR_0$-degenerate $f(\LCR)$ models possess an infinite number of solutions having constant scalar curvature $\LCR=\LCR_0$, it is actually possible to study the stability of all such solutions at once within a given model, without needing to perform a case-by-case analysis. In order to do so, we will resort to the Einstein-frame (i.e.~scalar-tensor) representation of $f(\LCR)$ gravities, which is ideally suited to study stability against small perturbations about a given constant value of $\LCR$.

As previously stated, it is well-known that metric $f(\LCR)$ theories can be regarded as equivalent to a scalar-tensor gravitational theory. More precisely, in the so-called Einstein frame \eqref{Einstein-frame metric}, the action \eqref{f(R) action} of metric $f(\LCR)$ gravity transforms into that of GR plus a dynamical gravitational scalar field $\phi$, with the latter being given by \eqref{scalaron}. This scalar field, also known as the \emph{scalaron}, is subject to the $f(\LCR)$-model-dependent potential
\begin{equation} \label{scalaron potential}
    V(\LCR)=\dfrac{f'(\LCR) \LCR-f(\LCR)}{2\kappa f'^2(\LCR)}.
\end{equation}
Notice that the Ricci scalar $\LCR$ appearing in the previous expressions is that of the so-called Jordan-frame metric, i.e.~the original metric $g_{\mu\nu}$.

When working in the Einstein frame, the stability of Jordan-frame constant-curvature solutions will depend on whether $\LCR=\LCR_0$ is a minimum of the scalaron potential \eqref{scalaron potential}, and on whether such minimum is either global or local (in the latter case, the solution will only be metastable). Nonetheless, we must stress that some subtleties arise when using the Einstein frame in $\LCR_0$-degenerate models. The ones relevant to our work will be comprehensively discussed in Appendix \ref{Appendix: Caveats}. Also, it is straightforward to notice in \eqref{scalaron potential} that, if the $f(\LCR)$ model is $\LCR_0$-degenerate, a naive evaluation of $V(\LCR)$ at $\LCR=\LCR_0$ leads to a $0/0$ indetermination. The reason is that, in such case, both the numerator and the denominator in equation \eqref{scalaron potential} become zero when $R\rightarrow\LCR_0$, owing to conditions \eqref{constant-curvature conditions R0}. Therefore, the limit must be evaluated carefully.

There are two possible ways of computing limits which naively evaluate to indeterminations of the $0/0$ kind: (i) performing series expansions or (ii) applying L'H\^{o}pital's rule. The only difference between the aforementioned methods is their range of applicability: Taylor series require analyticity around the expansion point, while L'H\^{o}pital's rule only requires differentiability of the numerator and denominator. In our assessment of the stability of constant-curvature solutions in $\LCR_0$-degenerate models, we have made use of both methods, obtaining exactly the same outcomes, as shown in Results \ref{Result: R0 instability} and \ref{Result: R0=0 metastability} below (recall that analyticity requires differentiability, and thus the results obtained L'H\^{o}pital's rule imply those obtained using series expansions). However, for the sake of clarity, and to avoid cluttering up our explanation with long formulae, we shall only present the derivation using Taylor series, which produces shorter expressions at the expense of a more limited scope. However, we insist that the final results apply to non-analytic but differentiable $f$s as well.

It is worth noting that, regardless of method employed in the stability analysis, one is forced to assume that $f''(\LCR_0)\neq 0$ to prevent some denominators from blowing up at $\LCR=\LCR_0$.\footnote{
    Condition $f''(\LCR)\neq 0$ ensures that the correspondence between the Jordan and Einstein frames is well-posed \cite{Sotiriou:2008rp}. Moreover, the extrema of $V(\LCR)$ and $V(\phi)$ coincide if and only if $f''(\LCR)\neq 0$, as discussed in Appendix \ref{Appendix: Caveats}.
} In addition, it is well known that $f(\LCR)$ models having $f''(\LCR)<0$ develop the so-called Dolgov-Kawasaki matter instability \cite{Dolgov:2003px,Faraoni:2006sy,Seifert:2007fr}. As such, the additional constraint $f''(\LCR_0)>0$ is unavoidable on physical grounds.

\subsection[Instability of the (\titlemath{$\LCR_0\neq 0$})-degenerate constant-curvature solutions of \titlemath{$f(\LCR)$} gravity.]{Instability of the (\titlebm{$\LCR_0\neq 0$})-degenerate constant-curvature solutions of \titlebm{$f(\LCR)$} gravity.}

Having explained why we have chosen to present just the computations using the series-expansion method, let us proceed with the stability analysis. As stated above, apart from demanding conditions \eqref{constant-curvature conditions R0} to hold (so that the model is $\LCR_0$-degenerate and harbours any solution with constant scalar curvature $\LCR_0$), we shall only make two additional assumptions, in particular, that function $f$ is analytic around $\LCR=\LCR_0$ (so it can be Taylor-expanded around $\LCR_0$, as explained before), and also that $f''(R_0)>0$. Thus, it is possible to expand both the numerator and the denominator of the scalaron potential \eqref{scalaron potential} around $\LCR=\LCR_0$. Indeed, close to $\LCR_0$, the denominator of $V(\LCR)$ behaves as
\begin{equation}
    f'^{-2}(\LCR) \underset{\LCR=\LCR_0}{\sim}f''^{-2}(\LCR_0)(\LCR-\LCR_0)^{-2}+\dfrac{f'''(\LCR_0)}{f''^3(\LCR_0)}(\LCR-\LCR_0)^{-1}+\bigo[(\LCR-\LCR_0)^0].
\end{equation}
whereas the numerator can be expanded as
\begin{align}
    f'(\LCR) \LCR - f(\LCR) & \underset{\LCR=\LCR_0}{\sim} \LCR_0 f''(\LCR_0) (\LCR-\LCR_0)+\dfrac{f''(\LCR_0)+\LCR_0 f'''(\LCR_0)}{2}(\LCR-\LCR_0)^2 \nonumber \\
    &+\bigo[(\LCR-\LCR_0)^3]. \vphantom{\dfrac{1}{2}}
\end{align}
As a result, we have that
\begin{equation} \label{potential expansion}
    2\kappa V(\LCR)\underset
    {\LCR=\LCR_0}{\sim}\dfrac{\LCR_0}{f''(\LCR_0)}(\LCR-\LCR_0)^{-1}+\dfrac{f''(\LCR_0)-\LCR_0 f'''(\LCR_0)}{2f''^2(\LCR_0)}+\bigo(\LCR-\LCR_0).
\end{equation}
We can now clearly infer from this last expression that the limit of $V(\LCR)$ as $\LCR\rightarrow\LCR_0$ does not exist unless $\LCR_0=0$. Certainly, should $\LCR_0$ be different from zero, series expansion \eqref{potential expansion} would be dominated by the order $(\LCR-\LCR_0)^{-1}$ term, which is a hyperbola tending to either positive or negative infinity depending on whether $\LCR_0$ is approached from the left or the right. More precisely,
\begin{equation} \label{V limit at R=R0 no Dolgov}
    \lim_{\substack{\LCR\rightarrow\LCR_0^\pm \\ \LCR_0\neq 0}}V(\LCR)=\sign\left[\dfrac{\LCR_0}{f''(\LCR_0)}\right]\times(\pm\infty).
\end{equation}
Requiring the Dolgov-Kawasaki stability condition $f''(\LCR)>0$ to hold on physical grounds, this expression further simplifies to
\begin{equation} \label{V limit at R=R0}
    \lim_{\substack{\LCR\rightarrow\LCR_0^\pm \\ \LCR_0\neq 0}}V(\LCR)=\sign(\LCR_0)\times(\pm\infty),
\end{equation}
but, as we can see, the limit still does not exist under the new condition.

As mentioned earlier, even though the previous findings have been obtained using Taylor expansions (and thus under the assumption that $f$ is analytic at $\LCR=\LCR_0$), the equivalent computation using L'H\^{o}pital's rule yields exactly the same results \eqref{V limit at R=R0 no Dolgov}--\eqref{V limit at R=R0} for choices of $f$ which might not be analytic around $\LCR=\LCR_0$. We can therefore establish the following general result:

\begin{result}[name={[instability of $(\LCR_0\neq 0)$-degenerate solutions]}] \label{Result: R0 instability}
Consider an $(\LCR_0\neq 0)$-degenerate $f(\LCR)$ model---i.e.~one fulfilling conditions \eqref{constant-curvature conditions R0}---such that $f''(\LCR_0)\neq 0$. Then its infinitely many solutions with constant curvature $\LCR=\LCR_0$ are generically unstable.
\end{result}

\subsection[Metastability of the (\titlemath{$\LCR_0=0$})-degenerate constant-curvature solutions of \titlemath{$f(\LCR)$} gravity.]{Metastability of the (\titlebm{$\LCR_0=0$})-degenerate constant-curvature solutions of \titlebm{$f(\LCR)$} gravity.}

On the contrary, if $\LCR_0=0$, series expansion \eqref{potential expansion} yields
\begin{equation} \label{lim V R0 = 0}
    \lim_{\LCR\rightarrow 0}V(\LCR)=\dfrac{1}{4\kappa f''(0)},
\end{equation}
i.e.~the potential \eqref{scalaron potential} is analytic and perfectly well-defined at $\LCR=0$. Consequently, zero-scalar-curvature solutions will be stable whenever $\LCR=0$ is a global minimum of $V(\LCR)$. If $\LCR=0$ is a local but non-global minimum, the corresponding solutions will then be just metastable (i.e.~stable only under small-enough perturbations). In either case, the additional constraints one must impose on $V(\LCR)$ are the usual minimum conditions:
\begin{eqnarray}
    V'(0)&=&\lim_{\LCR\rightarrow 0} V'(R)=0, \label{Minimum Condition 1} \\
    V''(0)&=&\lim_{\LCR\rightarrow 0} V''(R)>0\,\text{ but finite}. \label{Minimum Condition 2}
\end{eqnarray}
Again, should the limits of $V'(\LCR)$ and $V''(\LCR)$ as $\LCR$ tends to zero be taken directly, indeterminations of the $0/0$-type would emerge in both cases. Consequently, we must proceed exactly as previously done when evaluating the limit of the potential itself. As before, we shall only assume that $f$ is analytic around $\LCR=\LCR_0=0$, and that $f''(0)\neq 0$ (and, eventually, that $f''(0)>0$).

Taking this into account, the limit of $V'(\LCR)$ as $\LCR\rightarrow 0$ may be again computed by expanding its numerator and denominator around $\LCR=0$, yielding
\begin{equation} \label{lim V' R0 = 0}
    \lim_{\LCR\rightarrow 0}V'(\LCR)=-\dfrac{f'''(0)}{12\kappa f''^2(0)}. \\
\end{equation}
As per local minimum condition \eqref{Minimum Condition 1}, metastability of the solutions with $\LCR=0$ implies
\begin{equation}
    f'''(0)=0.
\end{equation}
Similarly, we find that the limit of $V''(\LCR)$ as $\LCR\rightarrow 0$ is also well-defined, being
\begin{equation}
    \lim_{\LCR\rightarrow 0}V''(\LCR)=-\dfrac{f''''(0)}{24\kappa f''^2(0)}.
\end{equation}
As a result, local minimum condition \eqref{Minimum Condition 2} implies that an $(\LCR_0=0)$-degenerate $f(\LCR)$ model must be such that
\begin{equation} \label{last minimum condition on f}
    f''''(0)<0
\end{equation}
for its infinitely many vanishing-scalar-curvature solutions to be at least metastable. Once more, as in the $\LCR_0\neq 0$ case, results \eqref{V limit at R=R0 no Dolgov}, \eqref{V limit at R=R0} and \eqref{lim V' R0 = 0}--\eqref{last minimum condition on f} may be alternatively found using L'H\^{o}pital's rule, and therefore hold for functions $f$ which need not be analytic. In consequence, we can formulate the following generic result:

\begin{result}[name={[metastability criteria for $(\LCR_0\neq 0)$-degenerate solutions]}] \label{Result: R0=0 metastability}
    All vanishing-scalar-curvature space-times are at least metastable vacuum solutions of $(\LCR_0=0)$-degenerate $f(\LCR)$ models---i.e.~those satisfying conditions \eqref{constant-curvature conditions 0}---provided that function $f$ is such that (i) $f''(0)\neq 0$, (ii) $f'''(0)=0$, and (iii) $f''''(0)<0$. It should also be taken into account that compliance with the Dolgov-Kawasaki stability criterion forces $f''(\LCR)>0$ for all $\LCR$, and, \emph{a fortiori}, $f''(0)>0$, instead of condition (i) above.
\end{result}

Analytic $f(\LCR)$ models complying with all the assumptions of Result \ref{Result: R0=0 metastability} (including $f''(0)>0$) would admit a Taylor expansion around $\LCR=0$ of the form
\begin{equation}
    f(\LCR)=\alpha\LCR^2-\beta\LCR^4+\bigo(\LCR^5),
\end{equation}
where $\alpha>0$ and $\beta>0$ are real constants with the appropriate dimensions. The simplest such models are, evidently, the polynomial ones with
\begin{equation} \label{stable R = 0 polynomial f(R)}
    f(\LCR)=\alpha\LCR^2-\beta\LCR^4.
\end{equation}
It is not difficult to show that, for $\alpha>0$, $\LCR=0$ is a just a local minimum of the potential associated to model \eqref{stable R = 0 polynomial f(R)}. The only other extrema of $V(\LCR)$ are two maxima at $\LCR=\pm\sqrt{\alpha/6\beta}$. Thus, for $|\LCR|>\sqrt{\alpha/6\beta}$ the potential is monotonically decreasing, i.e.~$V(\LCR)\rightarrow-\infty$ as $\LCR\rightarrow\pm\infty$. In other words, the model has a potential not bounded below and hence lacks a ground state. As a result, the metastable solutions with $\LCR=0$ could be driven to infinite scalar curvature, should the perturbations applied on them be large enough. On the other hand, if $\alpha$ were negative, the extremum at $\LCR_0=0$ would be the true, global minimum of the potential; in such case, however, the Dolgov-Kawasaki stability condition $f''(\LCR)>0$ would be violated, and the model would be unstable.

\section[Some paradigmatic \titlemath{$f(\LCR)$}-exclusive constant-curvature solutions and their pathologies.]{Some paradigmatic \titlebm{$f(\LCR)$}-exclusive constant-curvature solutions and their pathologies.} \label{introducing the solutions}

As we saw before in Section \ref{section:constant-curvature solutions}, \emph{any} space-time with constant curvature $\LCR_0$ is a solution of \emph{any} $\LCR_0$-degenerate $f(\LCR)$ model. Therefore, the problem of obtaining new such solutions for such $\LCR_0$-degenerate models reduces to solving the differential equation $\LCR=\LCR_0$. This can be accomplished by postulating several simple \emph{Ans\"{a}tze} for the metric.

As a first approximation to the problem, one may require the constant-curvature solution to be static and spherically symmetric. In such case, it is always possible to choose `areal-radius' coordinates $(t,r,\theta,\varphi)$ such that the most generic static and spherically symmetric line element can be written as
\begin{equation} 
\label{static spherically symmetric}
    \dif s^2=-A(r)\,\dif t^2+B(r)\,\dif r^2+r^2\,\dif\Omega^2,
\end{equation}
where $A$ and $B$ are the only independent metric functions. The Ricci scalar of \eqref{static spherically symmetric} is
\begin{equation} \label{static spherically symmetric Ricci scalar}
    \LCR=-\dfrac{A''}{AB}+\dfrac{A'}{2AB}\left(\dfrac{A'}{A}+\dfrac{B'}{B}\right)-\dfrac{2}{r}\left(\dfrac{A'}{AB}-\dfrac{B'}{B^2}\right)+\dfrac{2}{r^2}\left(1-\dfrac{1}{B}\right).
\end{equation}
Notice that \eqref{static spherically symmetric} is directly expressed in the so-called Abreu-Nielsen-Visser gauge,
\begin{equation} \label{Abreu-Nielsen-Visser}
    \dif s^2 = -\e^{2\Phi}\left(1-\dfrac{2GM_\mathrm{MSH}}{r}\right)\dif t^2+\left(1-\dfrac{2GM_\mathrm{MSH}}{r}\right)^{-1}\dif r^2+r^2\,\dif\Omega^2,
\end{equation}
with Misner-Sharp-Hern\'{a}ndez (MSH) mass \cite{Misner:1964je,Hernandez:1966zia}
\begin{equation} \label{MSH mass}
    M_\mathrm{MSH}(r)=\dfrac{r}{2G}\left[1-\dfrac{1}{B(r)}\right]
\end{equation}
and anomalous redshift function
\begin{equation} \label{anomalous redshift}
    \Phi(r)=\dfrac{1}{2}\ln\left[A(r)B(r)\right].
\end{equation}
These two quantities will help us interpret the various solutions to be discussed in what follows.

Equation $\LCR=\LCR_0=\const$---with $\LCR$ given by expression \eqref{static spherically symmetric Ricci scalar}---has infinitely many solutions, i.e.~there is an infinite number of static, spherically symmetric space-times having constant Ricci scalar.\footnote{
    A remarkable observation is that, given any function $A(r)$, there is a closed-form expression for the precise function $B(r)$ which solves equation $\LCR=\LCR_0$, namely
    \begin{equation*}
        B(r)=\dfrac{I(r)}{B_0+B_1(r)},
    \end{equation*}
    where $B_0$ is an integration constant,
    \begin{equation*}
        B_1(r)\equiv\int_1^r\dfrac{\dif x}{x}\dfrac{4I(x)A(x)}{4A(x)+xA'(x)}\left(1-\dfrac{\LCR_0}{2}x^2\right),
    \end{equation*}
    and function $I(r)$ is defined as
    \begin{equation*}
        I(r)\equiv\exp\left[\int_1^r\dfrac{\dif x}{x}\dfrac{4A(x)+4xA'(x)-\frac{x^2A'^2(x)}{A(x)}+2x^2A''(x)}{4A(x)+xA'(x)}\right].
    \end{equation*}
} For this reason, we will focus on spherically symmetric space-times which have been previously reported in the literature \cite{Calza:2018ohl,Casado-Turrion:2022xkl}, or generalisations thereof. Even though some of these space-times might look like mere deviations from either Minkowski or Schwarzschild space-times, it is evident that hardly one family---namely, Class 4, to be discussed in Section \ref{Section: Class 4}---passes standard weak-field or Solar-System experiments. Thus, if they existed in Nature, the solutions we are considering in this work would describe other, more exotic compact objects, such as black holes, wormholes, etc. Moreover, it is important to remark that none of the infinitely many $\LCR_0$-degenerate solutions of $\LCR_0$-degenerate $f(\LCR)$ models is a priori privileged (on physical grounds) over the others---recall, for instance, that Birkhoff's theorem does not hold in $f(\LCR)$ theories of gravity in general, least of all in $\LCR_0$-degenerate models, which have an \emph{infinite} number of solutions with constant scalar curvature $\LCR_0$. As such, even the simplest such $\LCR_0$-degenerate solutions must be taken into full consideration within $\LCR_0$-degenerate $f(\LCR)$ models.\footnote{
    We would like to remind the reader that, in $f(\LCR)$ gravity, Birkhoff's theorem is absent and, in addition, Schwarzschild cannot describe the space-time outside \emph{any} matter source, and can only exist as a black hole \cite{Bueno:2017sui}. Therefore, novel exterior solutions must exist. In particular, nothing prevents, in principle, the exterior to a certain physical body to be described by any of the $\LCR_0$-degenerate solutions of degenerate $f(\LCR)$ models.
} Proving that these degenerate solutions display unphysical traits provides further evidence against the viability of their host $\LCR_0$-degenerate $f(\LCR)$ models (adding on to that already presented in Results \ref{Result: Non-propagation of gravity}--\ref{Result: R0=0 metastability} above). For these reasons, we will devote the remaining Sections of this Chapter to the study in detail the simple (yet archetypal) solutions we have chosen.

Once we have presented the paradigmatic solutions in Subsections \ref{Section: Class 1}--\ref{Section: Class 4}, we will proceed to characterise them. That is to say, we would like to know which kind of objects (wormholes, black holes, etc.) such solutions represent, and whether they have physically meaningful properties. With that purpose, we will analyse the following five aspects:
\begin{coloritemize}
    \item Apparent and Killing horizons.
    \item Coordinate singularities.
    \item Curvature singularities.
    \item Geodesic completeness, i.e.~whether the curvature singularities can be reached in finite time by freely falling observers.
    \item Regions in which the metric acquires an unphysical signature (e.g.~regions in which there are two time coordinates or the metric becomes Euclidean), and whether these pathological regions can be reached in finite time by freely-falling observers.
\end{coloritemize}
The precise definitions of the properties listed above may be found on Appendix \ref{Appendix: Characterisation}, along with useful formulae that will be employed during our characterisation of the solutions. For instance, Equation \eqref{Delta lambda} will allow us to determine whether a singularity or a region with unphysical metric signature is out of reach for causal observers.

The $\LCR_0$-degenerate solutions we are about to consider have been chosen because they allow for a fully analytic examination. Our findings concerning the points above are summarised in Table \ref{tab:solutions}. As may be deduced from these results, the paradigmatic solutions to be described in what follows exhibit a number of unusual characteristics which put their physical viability into question. As said before, given that the number of constant-curvature solutions a certain $\LCR_0$-degenerate model hosts is infinite, one expects the majority of those constant-curvature space-times to be pathological. This Section is thus intended to provide a limited but illustrative picture of the kind of issues one should expect to find when dealing with novel, $f(\LCR)$-exclusive constant-curvature solutions.

\begin{table}[p]
\scriptsize
\begin{minipage}{\textwidth}
\begin{tabularx}{\textwidth}{ccccL}
\hline\hline
\multirow{2}{*}{Type}&
Parameters &
\multirow{2}{*}{Subclass}&
\multirow{2}{*}{Metric}&
\multicolumn{1}{c}{\multirow{2}{*}{Issues and oddities}} \\
& (dimensions) & & & \\
\hline
\hline
\multirow{13}{*}{Class 1} & \multirow{13}{*}{$C\neq 0$ (L)} & \multirow{5}{*}{$C>0$} & \multirow{5}{*}{\eqref{Class 1.1}} & \tabitemize{\item Protected curvature singularity at the origin $r=0$ (i.e.~it cannot be reached by any causal observer in finite affine parameter). \item Can be smoothly matched with an interior Minkowski space-time at any given radius for every $C$ (see App.~\ref{staticsol}).} \\ \cline{3-5}
& & \multirow{8}{*}{$C<0$} & \multirow{8}{*}{\eqref{Class 1.2}} & \tabitemize{\item Protected curvature singularity at the origin $r=0$. \item Accessible curvature singularity at $r=|C|$ (i.e.~a causal observer can reach it in finite affine parameter). \item Surface $r=|C|$ also appears to be a Killing horizon, but it is not null. \item As their counterparts with $C<0$, they can be smoothly matched with an interior Minkowski space-time at any given radius for every $C$ (see App.~\ref{staticsol}).} \\ \hline
\multirow{9}{*}{Class 2} & \multirow{9}{*}{\shortstack[c]{$\LCR_0$ ($\mathrm{L}^{-2}$)
\\ $D$ ($\mathrm{L}^0$) \\ $C\neq 0$ ($\mathrm{L}$)
}} & \multirow{4}{*}{$\LCR_0>0$} & \multirow{4}{*}{\eqref{Class 2.1}} & \tabitemize{\item Infinite number of accessible curvature singularities. \item Infinite number of would-be (non-null) Killing horizons, all coincident with curvature singularities. \item Generically unstable, as per Result \ref{Result: R0 instability}.} \\ \cline{3-5}
& & \multirow{2}{*}{$\LCR_0<0$} & \multirow{2}{*}{\eqref{Class 2.2}} & \tabitemize{\item Protected curvature singularity at the origin. \item Generically unstable, as per Result \ref{Result: R0 instability}.} \\ \cline{3-5}
& & \multirow{3}{*}{$\LCR_0=0$
} & \multirow{3}{*}{\eqref{Class 2.0}} & \tabitemize{\item Protected curvature singularity at the origin. \item Describes a pair of disconnected parallel universes, one at each side of the central singularity.} \\ \hline
\multirow{37}{*}{Class 3} & \multirow{37}{*}{\shortstack[c]{$\LCR_0$ ($\mathrm{L}^{-2}$)
\\ $M$ ($\mathrm{L}^{-1}$)
}} & \multirow{8}{*}{$M=0$, $\LCR_0>0$} & \multirow{8}{*}{\eqref{Class 3 M = 0 R0 > 0}} & \tabitemize{\item The metric signature becomes unphysical, i.e.~$(+,+,-,-)$, for radii $r$ larger than the critical value $r=\sqrt{6/\LCR_0}$. \item Causal observers can reach the region with unphysical metric signature in finite proper time. \item $r=\sqrt{6/\LCR_0}$ is also an apparent horizon (a null surface in which $g^{rr}=0$). \item Generically unstable, as per Result \ref{Result: R0 instability}.} \\ \cline{3-5}
& & \multirow{2}{*}{$M=0$, $\LCR_0<0$} & \multirow{2}{*}{\eqref{Class 3 M = 0 R0 < 0}} & \tabitemize{\item Describes a traversable wormhole made out of vacuum. \item However, it is generically unstable, as per Result \ref{Result: R0 instability}.} \\ \cline{3-5}
& & \multirow{6}{*}{$M>0$, $\LCR_0=0$} & \multirow{6}{*}{\eqref{Class 3 M > 0 R0 = 0}} & \tabitemize{\item The metric signature becomes unphysical for $r<2GM$. \item Causal observers can reach the region with unphysical metric signature in finite proper time. \item $r=2GM$ is also the location of an apparent horizon. \item There is a curvature singularity located at $r=0$ (i.e.~within the unphysical region).} \\ \cline{3-5}
& & $M<0$, $\LCR_0=0$ & \eqref{Class 3 M < 0 R0 = 0} & \tabitemize{\item Accessible curvature singularity at the origin $r=0$.} \\ \cline{3-5}
& & \multirow{2}{*}{$M<0$, $\LCR_0<0$} & \multirow{2}{*}{\eqref{Class 3 M < 0 R0 < 0}} & \tabitemize{\item Accessible curvature singularity at the origin $r=0$. \item Generically unstable, as per Result \ref{Result: R0 instability}.} \\ \cline{3-5}
& & \multirow{5}{*}{$M>0$, $\LCR_0<0$} & \multirow{5}{*}{\eqref{Class 3 M > 0 R0 < 0}} & \tabitemize{\item The metric signature becomes unphysical for radii $r$ smaller than some critical value $r_\mathrm{ah}$ given by expression \eqref{Apparent horizon Class 3 M > 0 R0 < 0}. \item There is a curvature singularity located at $r=0$ (i.e.~within the unphysical region). \item Generically unstable, as per Result \ref{Result: R0 instability}.} \\ \cline{3-5}
& & \multirow{4}{*}{$M<0$, $\LCR_0>0$} & \multirow{4}{*}{\eqref{Class 3 M < 0 R0 > 0}} & \tabitemize{\item The metric signature becomes unphysical for radii $r$ larger than some critical value $r_\mathrm{ah}$ given by expression \eqref{Apparent horizon Class 3 M < 0 R0 > 0}. \item There is a curvature singularity located at $r=0$. \item Generically unstable, as per Result \ref{Result: R0 instability}.} \\ \cline{3-5}
& & \multirow{10}{*}{$M>0$, $\LCR_0>0$} & \multirow{10}{*}{\eqref{Class 3 M > 0 R0 > 0}} & \tabitemize{\item If $3GM\sqrt{\LCR_0/2}\geq 1$, the metric signature is unphysical for all $r$. If $3GM\sqrt{\LCR_0/2}=1$, then $r=3GM=\sqrt{2/\LCR_0}$ would be a wormhole throat (were the metric signature physical). \item If $0<3GM\sqrt{\LCR_0/2}<1$, the metric signature is unphysical for $r<r_0$ and for $r>r_1$, where $r_0$ and $r_1$ are two critical values given by expressions \eqref{Apparent horizon Class 3 M > 0 R0 > 0 r_0} and \eqref{Apparent horizon Class 3 M > 0 R0 > 0 r_1}, respectively. Surfaces $r=r_0$ and $r=r_1$ correspond to apparent horizons. \item There is a curvature singularity located at $r=0$ (i.e.~within the region with unphysical metric signature). \item Generically unstable, as per Result \ref{Result: R0 instability}.} \\ \hline
\multirow{6}{*}{Class 4} & \multirow{6}{*}{\shortstack[c]{$z\neq 0$ (L$^0$)
\\ $c>0$ (L$^b$)
\\ $r_0>0$ (L)
}} & \multirow{6}{*}{\shortstack[c]{$b=b_\pm$ \\ (both)}} & \multirow{6}{*}{\eqref{Class 4}} & \tabitemize{\item They are not asymptotically flat (they do not describe fully isolated objects), but they comply with Solar System tests for $b=b_-$ and $|z|\ll 1$. \item Central curvature singularity at $r=0$, which is unprotected for $z>-10+4\sqrt{6}=-0.202...$, but protected if, instead, $z<-10-4\sqrt{6}=-19.798...$} \\
\hline
\hline
\end{tabularx}
\caption[Overview of the $f(\LCR)$-exclusive constant-curvature solutions considered in Chapter \ref{chapter:constant curvature} and their pathological traits.]{Overview of the $f(\LCR)$-exclusive constant-curvature solutions considered in this Chapter and their pathological characteristics.}
\label{tab:solutions}
\end{minipage}
\end{table}

\subsection{Novel solutions of Class 1.} \label{Section: Class 1}

The first set of $f(\LCR)$-exclusive constant-curvature solutions we would like to discuss (hereafter to be known as Class 1 solutions) shall be those metrics whose line element can be expressed as
\begin{equation} \label{Class 1.1}
    \dif s^2=-\left(1+\dfrac{C}{r}\right)^2\dif t^2+\dif r^2+r^2\dif\Omega^2,
\end{equation}
where $C$ is a free parameter with dimensions of length, which might be either positive or negative. These space-times have vanishing Ricci scalar for any value of $C$, and thus solve the equations of motion of all $f(\LCR)$ models complying with conditions \eqref{constant-curvature conditions 0}. An interesting property of Class 1 space-times is that they can be smoothly matched to a Minkowski interior at any given spherical surface $r=r_*=\const$, as we showed in Publication \cite{Casado-Turrion:2022xkl} (see also Appendix \ref{staticsol}). Metric \eqref{Class 1.1} thus represents space-time outside of such a static vacuole solution.

The most straightforward way of obtaining line element \eqref{Class 1.1} consists in setting $\LCR=0$ in equation \eqref{static spherically symmetric Ricci scalar} and solving for $A(r)$ using the simple \emph{Ansatz} $B(r)=1$. In doing so, one finds that the general solution is given by the (a priori) two-parameter family of metrics
\begin{equation} \label{Class 1 general solution}
    \dif s^2=-\left(D+\dfrac{C}{r}\right)^2\dif t^2+\dif r^2+r^2\dif\Omega^2,
\end{equation}
where $C$ and $D$ are, again, real constants, with $D$ being dimensionless. However, it is not difficult to realise that parameter $D$ can always be removed from the metric. If $D\neq 0$, it is always possible to transform line element \eqref{Class 1 general solution} into the Class 1 form \eqref{Class 1.1} presented above by redefining $C/D\rightarrow C$ and $D^2\,\dif t^2\rightarrow\dif t^2$. Nonetheless, if $D=0$, metric \eqref{Class 1 general solution} becomes
\begin{equation} \label{Class 2.0}
    \dif s^2=-\left(\dfrac{C}{r}\right)^2\dif t^2+\dif r^2+r^2\dif\Omega^2,
\end{equation}
space-times of this form \eqref{Class 2.0} are a particular instance of one of the families of $f(\LCR)$-exclusive constant-curvature solutions originally discovered by Calz\`{a}, Rinaldi and Sebastiani in Reference \cite{Calza:2018ohl}.\footnote{More precisely, line element \eqref{Class 2.0} can be obtained by setting $b=2$, $z=-2$ and $c_0=0$ in equations (33)--(35) of \cite{Calza:2018ohl}.} We shall consider metrics of the form \eqref{Class 2.0} to be solutions of Class 2 below, given that they cannot be recovered from the standard Class 1 form \eqref{Class 1.1}, and for further reasons that will become apparent in subsection \ref{Section: Class 2}. Consequently, in the present subsection we shall focus on analysing only the properties of metrics of the form \eqref{Class 1.1}.

\subsubsection{General properties of Class 1 solutions.}

It is clear from line element \eqref{Class 1.1} that Class 1 solutions are asymptotically flat, and that they also reduce to Minkowski space-time in the limit $C\rightarrow 0$. Using \eqref{MSH mass} and \eqref{anomalous redshift}, one might immediately deduce that space-times of the form \eqref{Class 1.1} have vanishing MSH mass and anomalous redshift function
\begin{equation} \label{vacuole exterior redshift}
    \Phi(r)=\ln\left|1+\dfrac{C}{r}\right|.
\end{equation}
In consequence, Class 1 solutions may be regarded as metrics deviating from Minkowski space-time only in a gravitationally-induced anomalous redshift. The interpretation of length scale $C$ appearing in \eqref{Class 1.1} is far less clear. As shown in Appendix \ref{staticsol}, the value of this parameter cannot be fixed by gluing \eqref{Class 1.1} to a Minkowski space-time at a given spherical surface $r=r_*=\const$ Class 1 solutions are also strikingly similar to the extremal Reissner-Nordstr\"{o}m space-time, sharing a common Newtonian limit upon identification of $C$ with (Newton's constant times) the mass or the charge of the black hole. In spite of this, Class 1 metrics are not sourced by any electromagnetic field. Instead, as mentioned before, the correction to the point-like-mass potential in the weak field limit of \eqref{Class 1.1} is entirely due to anomalous gravitational effects (recall this is a vacuum solution). Therefore, the interpretation of $C$ as an `extremal charge' seems not to be appropriate.

The remaining properties of Class 1 solutions depend crucially on the sign of $C$. As such, we will study the subclass of \eqref{Class 1.1} having $C>0$  separately from the subclass with $C<0$.

\subsubsection{Class 1 solutions with \titlebm{$C>0$}.}

Direct inspection of line element \eqref{Class 1.1} reveals that Class 1 solutions with $C>0$ only harbour one coordinate singularity, which is located at $r=0$; the metric remains regular for any other $r>0$. Moreover there are neither apparent or Killing horizons when $C>0$, nor any regions in which the signature of line element \eqref{Class 1.1} becomes unphysical.

Given that the Kretschmann scalar corresponding to \eqref{Class 1.1} is
\begin{equation}
    \LCKre=\dfrac{24 C^2}{r^6}\left(1+\dfrac{C}{r}\right)^{-2},
\end{equation}
one immediately realises that the coordinate singularity at the origin is an actual curvature singularity. Despite this, the singularity is inaccessible for any causal observer. Integral \eqref{Delta lambda} may be computed for Class 1 solutions with $C>0$ directly, yielding
\begin{equation}
    \Delta\lambda(r_\mathrm{ini}\rightarrow r_\mathrm{fin})=\left|r_\mathrm{fin}-r_\mathrm{ini}+C\ln\left(\dfrac{r_\mathrm{fin}}{r_\mathrm{ini}}\right)\right|.
\end{equation}
It is thus apparent that, along a radial null geodesic, the affine parameter separating any $r>0$ from the central singularity is infinite. Because photons take an infinite amount of affine parameter to reach the central singularity, no other particle can reach $r=0$ in finite proper time either. In consequence, for all practical purposes, the singularity at $r=0$ is inaccessible, and space-time \eqref{Class 1.1} is geodesically complete for $C>0$.

The previous result can be easily understood in view of the form of the anomalous redshift function \eqref{vacuole exterior redshift} for Class 1 space-times with $C>0$. As we can clearly deduce from this expression, the anomalous redshift becomes infinite at $r=0$. In other words, modified-gravity effects induce an ever-increasing redshift function which \emph{protects} observers from the curvature singularity.

\subsubsection{Class 1 solutions with \titlebm{$C<0$}.}

The metric of Class 1 space-times with $C<0$, viz.
\begin{equation} \label{Class 1.2}
    \dif s^2=-\left(1-\dfrac{|C|}{r}\right)^2\dif t^2+\dif r^2+r^2\dif\Omega^2,
\end{equation}
exhibits two coordinate singularities: the one at $r=0$ (which was already present when $C>0$), and a second one at $r=|C|$. Both prove to be curvature singularities, since the Kretschmann scalar becomes
\begin{equation}
    \LCKre=\dfrac{24C^2}{r^6}\left(1-\dfrac{|C|}{r}\right)^{-2}
\end{equation}
for $C<0$. As we shall see now, the singularity at $r=0$ remains inaccessible for radially infalling photons. However, the singularity at $r=|C|$ is causally connected to the rest of the space-time.

On the one hand, for $r_\mathrm{ini}$, $r_\mathrm{fin}<|C|$, integral \eqref{Delta lambda} yields
\begin{equation}
    \Delta\lambda(r_\mathrm{ini}<|C|\rightarrow r_\mathrm{fin}<|C|)=\left|r_\mathrm{fin}-r_\mathrm{ini}-|C|\ln\left(\dfrac{r_\mathrm{fin}}{r_\mathrm{ini}}\right)\right|.
\end{equation}
We may readily substitute $r_\mathrm{fin}=0$ in this expression to find that the curvature singularity at $r=0$ cannot be reached in finite time by causal observers. In parallel with the $C>0$ case, the curvature singularity is protected by an $f(\LCR)$-induced infinite redshift.

On the other hand, if $r_\mathrm{ini}\neq |C|$, but $r_\mathrm{fin}=|C|$, then integral \eqref{Delta lambda} becomes
\begin{equation}
    \Delta\lambda(r_\mathrm{ini}\rightarrow |C|)=\left||C|-r_\mathrm{ini}-|C|\ln\left|\dfrac{C}{r_\mathrm{ini}}\right|\right|,
\end{equation}
which is finite for every $r_\mathrm{ini}\neq 0$. Therefore, a photon travelling along a null radial geodesic will reach the singularity at $r=|C|$ in finite time, should it not be `emitted' at the other singularity at the origin.

For completeness, we shall also compute $\Delta\lambda(r_\mathrm{ini}\rightarrow r_\mathrm{fin})$ for $r_\mathrm{ini}>|C|$ and $r_\mathrm{fin}\leq|C|$ (without loss of generality\footnote{
    As mentioned in Appendix \ref{Appendix: Characterisation}, the absolute value in \eqref{Delta lambda} entails that it is always possible to exchange $r_\mathrm{ini}\leftrightarrow r_\mathrm{fin}$, because $\Delta\lambda(r_\mathrm{ini}\rightarrow r_\mathrm{fin})=\Delta\lambda(r_\mathrm{fin}\rightarrow r_\mathrm{ini})$.
}). To do so, we must first remark that the integral in \eqref{Delta lambda} breaks into two pieces in this case:
\begin{equation}
    \Delta\lambda(r_\mathrm{ini}\rightarrow r_\mathrm{fin})=\left|\int_{r_\mathrm{ini}}^{|C|}\dif r\,\left(1-\dfrac{|C|}{r}\right)-\int_{|C|}^{r_\mathrm{fin}}\dif r\,\left(1-\dfrac{|C|}{r}\right)\right|.
\end{equation}
As a result,
\begin{equation}
    \Delta\lambda(r_\mathrm{ini}\rightarrow r_\mathrm{fin})=\left|r_\mathrm{ini}+r_\mathrm{fin}-2|C|\left[1+\ln\left(\dfrac{\sqrt{r_\mathrm{ini}\,r_\mathrm{fin}}}{|C|}\right)\right]\right|.
\end{equation}
This expression attests once more that, whatsoever value $r_\mathrm{ini}$ takes, $\Delta\lambda(r_\mathrm{ini}\rightarrow r_\mathrm{fin})$ remains finite unless $r_\mathrm{fin}=0$. Thus, the central singularity is protected, with the one at $r=|C|$ being truly naked.

Another surprising feature of surface $r=|C|$ is that it appears to be a Killing horizon for $\xi=\partial/\partial t$, since
\begin{equation}
    g_{\mu\nu}\xi^\mu\xi^\nu=-\left(1-\dfrac{|C|}{r}\right)^2
\end{equation}
vanishes at $r=|C|$. However, $r=|C|$ is not a null surface, because its normal vector $n_\mu=\partial_\mu r=\delta^r_{\hphantom{r}\mu}$ is everywhere time-like. Therefore, strictly speaking, it cannot be a Killing horizon.\footnote{
    Since $r=|C|$ is also a curvature singularity, as seen before, it is technically not part of the space-time.
} Given that $g_{\mu\nu}\xi^\mu\xi^\nu>0$ for any $r\neq|C|$, the would-be Killing horizon at $r=|C|$ would be degenerate, in analogy with the true Killing horizon of extremal Reissner-Nordstr\"{o}m black holes. A remarkable difference between Class 1 solutions with $C<0$ and extremal Reissner-Nordstr\"{o}m black holes is that the former do not possess any apparent horizons, while the Killing horizon of extremal Reissner-Nordstr\"{o}m black holes is also an apparent horizon.\footnote{Furthermore, the horizon of extremal Reissner-Nordstr\"{o}m black holes is regular, i.e.~free of curvature singularities.}

\subsection{Novel solutions of Class 2.} \label{Section: Class 2}

Class 2 shall encompass three different kinds of constant-curvature solutions, all of which are characterised by their compliance with the simple \emph{Ansatz} $B(r)=1$, as those of Class 1. As a result, all Class 2 space-times will have vanishing MSH mass, and thus only differ from Minkowski space-time in an $f(\LCR)$-induced anomalous redshift factor.

The first two subclasses within Class 2 will contain, respectively, the solutions of equation $\LCR=\LCR_0$ with $B(r)=1$ and either positive (first subclass) or negative (second subclass) constant curvature $R_0$---recall that the Ricci scalar of static space-times is given by expression \eqref{static spherically symmetric Ricci scalar}. The third subclass will consist of those space-times resulting from setting $\LCR_0\rightarrow 0$ in the aforementioned Class 2 solutions with $\LCR_0\neq 0$.

As done in the previous Section, we will analyse each subclass within Class 2 separately.

\subsubsection{Class 2 solutions with \titlebm{$\LCR_0>0$}.}

Substituting the simple \emph{Ansatz} $B(r)=1$ in \eqref{static spherically symmetric Ricci scalar} and solving the equation $\LCR=\LCR_0>0$ for $A(r)$, one finds the following family of constant-curvature solutions of $(\LCR_0\neq 0)$-degenerate $f(\LCR)$ gravity models:
\begin{equation} \label{Class 2.1}
    \dif s^2=-\left(\dfrac{C}{r}\right)^2\cos^2\left(\sqrt{\dfrac{\LCR_0}{2}}\,r+D\right)\dif t^2+\dif r^2+r^2\dif\Omega^2,
\end{equation}
where $C\neq 0$ and $D$ are integration constants, $C$ having dimensions of length, and $D$ being dimensionless. Notice that, even though $C$ can always be removed from the line element through the coordinate redefinition $t\rightarrow t/C$, we have kept it in \eqref{Class 2.1} for dimensional reasons (i.e.~the re-scaled time coordinate would be dimensionless, while the re-scaled $g_{tt}$ would have dimensions of length squared).

Class 2 space-times with $\LCR_0>0$ are highly pathological; despite their simple appearance, they are, perhaps, the least physically well-founded solutions we will consider in this Chapter. For instance, it is immediate to infer from expression \eqref{Class 2.1} that these solutions have an infinite number of coordinate singularities. One of them is located at $r=0$, where $g_{tt}\rightarrow\infty$, while the remaining ones are located at the points in which the cosine vanishes, that is to say, at
\begin{equation} \label{r_n}
    r_n=\sqrt{\dfrac{2}{\LCR_0}}\left[\dfrac{\pi}{2}(2n+1)-D\right],
\end{equation}
with $n$ an integer.\footnote{We remark that the allowed values of $n$ depend on the value of $D$. In particular, the minimum possible $n$ should be such that $r_n>0$ for all permitted $n$s.} Direct computation of the Kretschmann scalar provides a long expression (which we shall not include here) that diverges at $r=0$ and all the $r_n$s above. As a result, the infinite coordinate singularities represent actual curvature singulatities.

What is worse, only the curvature singularity at $r=0$ can be out of reach for any causal observer, and only in very particular circumstances, as we shall see. To obtain this result, we first recall that $\Delta\lambda(r_\mathrm{ini}\rightarrow r_\mathrm{fin})$ can always be expressed in terms of the primitive \eqref{lambda primitive}, evaluated at some particular values of $r$.\footnote{
    Note that expression \eqref{lambda difference} does not hold for arbitrary pairs $\{r_\mathrm{ini}$, $r_\mathrm{fin}\}$ in this case. The reason is that, for Class 2 solutions with $\LCR_0>0$, the integrand in \eqref{lambda primitive} is a cosine, which changes sign periodically.
} For solutions of the form \eqref{Class 2.1}, it turns out that
\begin{equation} \label{lambda Class 2.1}
    \lambda(r)=|C|\left[\sin D\,\Si\left(\sqrt{\dfrac{\LCR_0}{2}}r\right)+\cos D\, \Ci\left(\sqrt{\dfrac{\LCR_0}{2}}r\right)\right],
\end{equation}
where $\Si$ and $\Ci$ are the so-called sine integral and cosine integral functions, respectively:
\begin{gather}
    \Si(z)=\int_0^z \dfrac{\dif x}{x}\,\sin x, \label{Si} \\
    \Ci(z)=\gamma+\ln z+\int_0^z \dfrac{\dif x}{x}\,(\cos x-1). \label{Ci}
\end{gather}
Here, $\gamma$ is the Euler-Mascheroni constant. The sine integral function is regular for all $r$, while the cosine integral only blows up (to negative infinity) when $r=0$. As a result, $\lambda(r)$ solely becomes infinite at $r=0$ provided that $\cos D\neq 0$. There are, as a result, two different scenarios:
\begin{coloritemize}
    \item If $\cos D\neq 0$, then $\Delta\lambda(r_\mathrm{ini}\rightarrow r_\mathrm{fin})$ blows up if and only if either $r_\mathrm{ini}$ or $r_\mathrm{fin}$ is equal to zero, and thus `only' the infinite number of curvature singularities located at $r_n$---with $r_n$ given by \eqref{r_n}---are accessible for radially falling causal observers.
    \item If $\cos D=0$, then $\Delta\lambda(r_\mathrm{ini}\rightarrow r_\mathrm{fin})$ remains finite regardless of the values of $r_\mathrm{ini}$ and $r_\mathrm{fin}$. As a result, all curvature singularities can be reached in finite time by causal observers.
\end{coloritemize}
In either case, causally-propagating particles in Class 2 space-times with $\LCR_0>0$ may encounter an infinite number of curvature singularities in finite time, and thus we can immediately assert that these solutions are physically unjustifiable just for this reason. Other pathological properties of Class 2 solutions with $\LCR_0>0$ include the existence of infinite would-be Killing horizons located at each of the $r_n$s given by \eqref{r_n} (which cannot be true Killing horizons since their normal remains time-like for all $r$, as in Class 1 solutions with $C<0$), as well as their inherent instability, due to them being $f(\LCR)$-exclusive constant-curvature solutions with $R_0\neq 0$ (cf.~Result \ref{Result: R0 instability}).

\subsubsection{Class 2 solutions with \titlebm{$\LCR_0<0$}.}

Setting $B(r)=1$ and solving equation $\LCR=\LCR_0<0$ for $A(r)$, one obtains instead another set of novel solutions, characterised by
\begin{equation} \label{Class 2.2}
    \dif s^2=-\left(\dfrac{C}{r}\right)^2\cosh^2\left(\sqrt{\dfrac{|\LCR_0|}{2}}\,r+D\right)\dif t^2+\dif r^2+r^2\dif\Omega^2,
\end{equation}
where $C\neq 0$ and $D$ are once again integration constants with the appropriate dimensions. Since the hyperbolic cosine never vanishes, \eqref{Class 2.2} only possesses one coordinate singularity, located at the origin $r=0$. As in the $\LCR_0>0$ case, this coordinate singularity turns out to be a curvature singularity. The Kretschmann scalar for Class 2 solutions with $\LCR_0<0$ is given again by a convoluted expression (which we shall not include here); careful examination of such expression reveals that there are no more curvature singularities apart from the one at $r=0$.

Once more, we turn to determine whether the central curvature singularity is accessible to causal observers in space-times of the form \eqref{Class 2.2}. In this particular case, primitive \eqref{lambda primitive} might be readily expressed in terms of special functions, namely the hyperbolic sine integral (Shi) and the hyperbolic cosine integral (Chi), as\footnote{Notice that the integrand leading to \eqref{lambda Class 2.2} is a hyperbolic cosine, which is always positive. Thus, for Class 2 solutions with $\LCR_0<0$, identity \eqref{lambda difference} holds for whatsoever $r_\mathrm{ini}$, $r_\mathrm{fin}$.}
\begin{equation} \label{lambda Class 2.2}
    \lambda(r)=|C|\left[\sinh D\,\Shi\left(\sqrt{\dfrac{|\LCR_0|}{2}}r\right)+\cosh D\, \Chi\left(\sqrt{\dfrac{|\LCR_0|}{2}}r\right)\right].
\end{equation}
Said special functions are defined as
\begin{gather}
    \Shi(z)=\int_0^z \dfrac{\dif x}{x}\,\sinh x, \\
    \Chi(z)=\gamma+\ln z-\int_0^z \dfrac{\dif x}{x}\,(\cosh x-1),
\end{gather}
in analogy with their trigonometric counterparts \eqref{Si} and \eqref{Ci}. Moreover, Shi, as Si, is regular for all $r$, while Chi, as Ci, diverges at $r=0$. Hence, $\lambda(r)$---and thus $\Delta\lambda(r_\mathrm{ini}\rightarrow r_\mathrm{fin})$---only diverges if either $r_\mathrm{ini}$ or $r_\mathrm{fin}$ is zero (notice that, this time, the pre-factor accompanying Chi in \eqref{lambda Class 2.2}, which is $\cosh D$, does not vanish for any $D$). As a result, the only singularity in \eqref{Class 2.2} is out of reach of radially infalling photons, and for this reason causal observers cannot encounter it in finite time.

The metric components of Class 2 solutions with $\LCR_0>0$ are always positive, so these metrics contain neither apparent nor Killing horizons. As such, they describe the space-time outside the singularity at $r=0$. Since this singularity cannot interfere in finite time with causal observers, space-times of the form \eqref{Class 2.2} are much more physically well-founded than their counterparts \eqref{Class 2.1} with $\LCR_0<0$. Nonetheless, we insist that Class 2 space-times with $\LCR_0\neq 0$ are automatically unstable, as per Result \ref{Result: R0 instability}.

\subsubsection{Class 2 solutions with \titlebm{$\LCR_0=0$}.}

The only Class 2 solutions which can be stable (in accordance with Result \ref{Result: R0=0 metastability}) are the ones which result from taking the limit of \eqref{Class 2.1} or \eqref{Class 2.2} as $\LCR_0\rightarrow 0$. It turns out that those solutions have a line element given by expression \eqref{Class 2.0}. This is the reason why we have grouped these solutions within Class 2 and not Class 1.

Line element \eqref{Class 2.0} appears to describe a wormhole with throat at $r=0$ upon extension of coordinate $r$ to negative values \cite{Calza:2018ohl}. This would be a remarkable result, since such a wormhole would not require negative energies to form; in fact, it would be a \emph{vacuum} solution of the equations of motion.

However, Class 2 solutions with $\LCR_0=0$ cannot describe traversable wormholes since (i) they contain a naked singularity at $r=0$, and (ii) light rays cannot arrive at the locus of such singularity in finite affine parameter by following radial null geodesics. On the one hand, feature (ii) implies that the singularity is \emph{protected}; however, it also entails that, even if there were no curvature singularity, light rays would never be able cross the wormhole throat in finite time.

The previous assertions (i) and (ii) are easy to verify. Indeed, the Kretschmann scalar corresponding to line element \eqref{Class 2.0} evaluates to
\begin{equation}
    \LCKre=\dfrac{24}{r^4},
\end{equation}
which diverges at $r=0$ independently of the value of $C$, as expected.\footnote{
    Notice that constant $C$ in line element \eqref{Class 2.0} can always be removed by redefining $C^2\,\dif t^2\rightarrow\dif t^2$; however, as we have done before for the other Class 2 solutions, we have explicitly included it in the metric for dimensional reasons.
} Using \eqref{Delta lambda}, we find that
\begin{equation}
    \Delta\lambda(r_\mathrm{ini}\rightarrow r_\mathrm{fin})=|C|\,\ln\left|\dfrac{r_\mathrm{fin}}{r_\mathrm{ini}}\right|,
\end{equation}
which only diverges when either $r_\mathrm{ini}$ or $r_\mathrm{fin}$ equals zero. Therefore, we conclude that these solutions cannot describe a traversable wormhole. In reality, what they model is two parallel and causally disconnected universes which are separated by an unreachable curvature singularity at the origin.

Apart from the central singularity, Class 2 solutions with $R_0=0$ do not have any other remarkable features; in particular, they lack either apparent or Killing horizons.

\subsection{Solutions of Class 3.} \label{Section: Class 3}

Class 3 solutions, discovered in Reference \cite{Calza:2018ohl}, can be obtained by solving equation $\LCR=\LCR_0$---with $\LCR_0$ given, as always, by expression \eqref{static spherically symmetric Ricci scalar}---for $B(r)$, under the simple assumption that $A(r)=1$. Class 3 comprises a family of Kottler\footnote{The Kottler space-time is also known as Schwarzschild-de Sitter or Schwarzschild-Anti de Sitter, depending on whether $\LCR_0$ is positive or negative, respectively.} lookalikes,
\begin{equation} \label{Class 3}
    \dif s^2=-\dif t^2+\left(1-\dfrac{2GM}{r}-\dfrac{\LCR_0}{6}r^2\right)^{-1}\dif r^2+r^2\dif\Omega^2,
\end{equation}
where $M$ is a free parameter with units of mass, which can be (in principle) either positive or negative. Class 3 solutions with constant curvature $\LCR_0\neq 0$ are generically unstable as per Result \ref{Result: R0 instability}; Class 3 solutions with $\LCR_0=0$ could be at least metastable should the hypotheses of Result \ref{Result: R0 instability} hold.

Even though line element \eqref{Class 3} is reminiscent of the Kottler metric, there are two notable differences between them. First, unlike the latter, the former have an anomalous redshift function distinct from unity; specifically,
\begin{equation} \label{Class 3 redshift}
    \Phi(r)=-\dfrac{1}{2}\ln\left(1-\dfrac{2GM}{r}-\dfrac{\LCR_0}{6}r^2\right).
\end{equation}
This modified-gravity-induced redshift function \eqref{Class 3 redshift} allows \eqref{Class 3} to have $g_{tt}=1$ instead of $g_{tt}g_{rr}=-1$. Second, the `cosmological-constant term' of Class 3 solutions, $\LCR_0 r^2/6$, is exactly half of the one present in Kottler space-time, namely $\LCR_0 r^2/12$. As a result, Class 3 solutions have MSH mass
\begin{equation}
    M_\mathrm{MSH}(r)=M+\dfrac{\LCR_0}{12G}r^3.
\end{equation}
Notice that, while Class 3 solutions might harbour apparent horizons, they do not exhibit any Killing horizon corresponding to the generator of time translations $\xi=\partial/\partial t$, since its norm remains equal to $-1$ throughout all space-time, as per \eqref{Killing norm} and \eqref{Class 3}.

While all members of this family were originally thought to describe traversable wormholes \cite{Calza:2018ohl}, we shall see that they might exhibit several unphysical properties depending on the values of $M$ and $\LCR_0$. We shall sort Class 3 solutions into subclasses according to the signs of their free parameters $M$ and $\LCR_0$. We shall also consider the simple cases $M=0$, $\LCR_0\neq 0$ and $M\neq 0$ and $\LCR_0=0$ separately.

\subsubsection{Class 3 solutions with \titlebm{$M=0$} and \titlebm{$\LCR_0>0$}.}

In the simple case in which parameter $M$ vanishes and $R_0$ is positive, the metric reduces to a de Sitter lookalike:
\begin{equation} \label{Class 3 M = 0 R0 > 0}
    \dif s^2=-\dif t^2+\left(1-\dfrac{\LCR_0}{6}r^2\right)^{-1}\dif r^2+r^2\dif\Omega^2,
\end{equation}
Class 3 solutions with $M=0$ have no curvature singularities, since their Kretschmann scalar, $\LCKre=\LCR_0^2/3$, remains constant for all $r$. Nonetheless, direct inspection of line element \eqref{Class 3 M = 0 R0 > 0} reveals that these space-times suffer from an evident physical pathology: if $r>r_\mathrm{ah}$, where
\begin{equation}
    r_\mathrm{ah}=\sqrt{\dfrac{6}{\LCR_0}},
\end{equation}
then the metric has two time coordinates, $t$ and $r$, since $g_{rr}$ becomes negative---notice that $r_\mathrm{ah}$ corresponds to an apparent horizon of \eqref{Class 3 M = 0 R0 > 0}, being a zero of $g^{rr}$. What is more, the region with unphysical metric signature can be reached by causal observers in finite time. When evaluated for space-time \eqref{Class 3 M = 0 R0 > 0} and $r_\mathrm{ini}$, $r_\mathrm{fin}\leq r_\mathrm{ah}$, integral \eqref{Delta lambda} can be computed using \eqref{lambda difference}, with primitive \eqref{lambda primitive} being given by
\begin{equation}
    \lambda(r)=\arcsin\left(\sqrt{\dfrac{\LCR_0}{6}}r\right).
\end{equation}
It is evident from this expression that $\Delta\lambda(r_\mathrm{ini}\rightarrow r_\mathrm{fin})$ remains finite when either $r_\mathrm{ini}$ or $r_\mathrm{fin}$ is equal to $r_\mathrm{ah}$. Space-times of the form \eqref{Class 3 M = 0 R0 > 0} are also unstable, due to Result \ref{Result: R0 instability}. Hence, they cannot be considered to be physically viable.

\subsubsection{Class 3 solutions with \titlebm{$M=0$} and \titlebm{$\LCR_0<0$}.}

One way to cure the pathology in the metric signature exhibited by \eqref{Class 3} is to change the sign of the constant curvature $\LCR_0$. In such case, one obtains a line element given by
\begin{equation} \label{Class 3 M = 0 R0 < 0}
    \dif s^2=-\dif t^2+\left(1+\dfrac{\LCR_0}{6}r^2\right)^{-1}\dif r^2+r^2\dif\Omega^2,
\end{equation}
whose components remain non-negative for all $r$, and has neither horizons nor curvature singularities.\footnote{Notice that, as in the previous case, the Kretschmann scalar is constant and equal to $\LCKre=R_0^2/3$. As a result, the usual coordinate singularity at $r=0$ is not an actual curvature singularity, but an artefact caused by the choice of `areal-radius' coordinates.} In fact, \eqref{Class 3 M = 0 R0 < 0} is the only Class 3 solution which truly describes a traversable wormhole centred at the origin, since the metric can be effortlessly extended to negative values of $r$. The remarkable fact that a traversable wormhole can exist as a vacuum solution of an infinite number of $f(\LCR)$ models is downplayed by the fact that such wormhole is generically unstable, as per Result \ref{Result: R0 instability}.

\subsubsection{Class 3 solutions with \titlebm{$M>0$} and \titlebm{$\LCR_0=0$}.}

In this case, the metric is a Schwarzschild lookalike:
\begin{equation} \label{Class 3 M > 0 R0 = 0}
    \dif s^2=-\dif t^2+\left(1-\dfrac{2GM}{r}\right)^{-1}\dif r^2+r^2\dif\Omega^2.
\end{equation}
The metric signature changes to $(-,-,+,+)$ if $r<2GM$. Within this region also lies a curvature singularity, located at the origin $r=0$, as revealed by the form of the Kretschmann scalar:
\begin{equation} \label{Kretschmann Class 3 R0 = 0}
    \LCKre=\dfrac{24G^2M^2}{r^6}.
\end{equation}
Exactly as in the $M=0$, $R_0>0$ case, the unphysical region can be reached by radially infalling photons in a finite amount of affine parameter. This is because, for space-time \eqref{Class 3 M > 0 R0 = 0}, integral \eqref{Delta lambda} is of the form \eqref{lambda difference}, where primitive \eqref{lambda primitive} is now given by
\begin{equation}
    \lambda(r)=\sqrt{r(r-2GM)}+GM\ln\left[r-GM+\sqrt{r(r-2GM)}\right].
\end{equation}
As we one may readily infer from this expression, $\lambda(r)$ remains finite and positive for all $r_\mathrm{ini}$ and $r_\mathrm{fin}$ equal or larger than $2GM$.\footnote{Notice that $\lambda(r)$ becomes complex for $r<2GM$, in yet another example of the pathological character of this subclass of space-times.} In particular,
\begin{equation}
    \lambda(2GM)=GM\ln(GM).
\end{equation}
Hence, we conclude that line element \eqref{Class 3 M > 0 R0 = 0} is physically unsatisfactory, as causal observers can access the region with unphysical metric signature in finite time.

\subsubsection{Class 3 solutions with \titlebm{$M<0$} and \titlebm{$\LCR_0=0$}.}

A change in the sign of the mass parameter turns line element \eqref{Class 3 M = 0 R0 > 0} into
\begin{equation} \label{Class 3 M < 0 R0 = 0}
    \dif s^2=-\dif t^2+\left(1+\dfrac{2G|M|}{r}\right)^{-1}\dif r^2+r^2\dif\Omega^2.
\end{equation}
As a result, $g_{rr}$ is positive for all $r>0$, and thus the metric signature remains physical throughout all space-time. Space-time \eqref{Class 3 M < 0 R0 = 0} is nonetheless pathological, possessing a curvature singularity at the origin---the Kretschmann scalar is still of the form \eqref{Kretschmann Class 3 R0 = 0}---which is naked and reachable in finite proper time by causal observers. The affine-parameter interval between any two radii is given again by \eqref{lambda difference}, but with
\begin{equation}
    \lambda(r)=\sqrt{r(r+2G|M|)}-G|M|\ln\left[r+G|M|+\sqrt{r(r+2G|M|)}\right].
\end{equation}
This quantity remains finite for all $r$ and, in particular,
\begin{equation}
    \lambda(0)=-G|M|\ln(G|M|);
\end{equation}
hence, the central singularity is causally connected to the rest of the space-time. From this it is clear that, just as its positive-mass counterpart \eqref{Class 3 M > 0 R0 = 0}, metric \eqref{Class 3 M < 0 R0 = 0} is not well-founded from a physical point of view.

\subsubsection{Class 3 solutions with \titlebm{$M<0$} and \titlebm{$\LCR_0<0$}.}

When both the mass scale $M$ and the constant curvature $\LCR_0$ in \eqref{Class 3} are negative, the line element becomes 
\begin{equation} \label{Class 3 M < 0 R0 < 0}
    \dif s^2=-\dif t^2+\left(1+\dfrac{2G|M|}{r}+\dfrac{|\LCR_0|}{6}r^2\right)^{-1}\dif r^2+r^2\dif\Omega^2.
\end{equation}
Thus, in this case, $g_{rr}$ is positive for all positive radii, and the metric signature remains physical through all space-time. However, the metric harbours a curvature singularity at $r=0$, where the Kretschmann scalar,
\begin{equation} \label{Class 3 same sign Kretschmann}
    \LCKre=\dfrac{24 G^2 M^2}{r^6}+\dfrac{\LCR_0^2}{3},
\end{equation}
diverges. Radially infalling photons can have access this singularity in finite affine parameter. This is because
\begin{eqnarray}
    \Delta\lambda(r_\mathrm{ini}\rightarrow 0)&=&\left|\int_{r_\mathrm{ini}}^0 \dif r\,\left(1+\dfrac{2G|M|}{r}+\dfrac{|\LCR_0|}{6}r^2\right)^{-1/2}\right| \nonumber \\
    &<&\left|\int_{r_\mathrm{ini}}^0 \dif r\,\right|=|r_\mathrm{ini}|<\infty
\end{eqnarray}
for any $r_\mathrm{ini}<\infty$. Therefore, this subclass is also pathological.

\subsubsection{Class 3 space-times with \titlebm{$M>0$} and \titlebm{$\LCR_0<0$}.}

In this case, the metric is
\begin{equation} \label{Class 3 M > 0 R0 < 0}
    \dif s^2=-\dif t^2+\left(1-\dfrac{2GM}{r}+\dfrac{|\LCR_0|}{6}r^2\right)^{-1}\dif r^2+r^2\dif\Omega^2.
\end{equation}
Apart from the coordinate singularity at the origin, \eqref{Class 3 M > 0 R0 < 0} also has a coordinate singularity whenever there is an apparent horizon, i.e.~when $g^{rr}=0$. This occurs only at
\begin{equation} \label{Apparent horizon Class 3 M > 0 R0 < 0}
    r_\mathrm{ah}=\sqrt{\dfrac{8}{|\LCR_0|}}\sinh\left[\dfrac{1}{3}\arcsinh\left(3GM\sqrt{\dfrac{|\LCR_0|}{2}}\right)\right],
\end{equation}
$g^{rr}$ may be readily shown to be a strictly increasing function of $r$ for $r>0$. The monotonocity of $g^{rr}$, together with the fact that \eqref{Apparent horizon Class 3 M > 0 R0 < 0} is its only root, guarantees that $g^{rr}<0$ for $r<r_\mathrm{ah}$, and thus the metric signature changes inside the apparent horizon.

\subsubsection{Class 3 space-times with \titlebm{$M<0$} and \titlebm{$\LCR_0>0$}.}

This case is completely analogous to the previous one. The metric now reads
\begin{equation} \label{Class 3 M < 0 R0 > 0}
    \dif s^2=-\dif t^2+\left(1+\dfrac{2G|M|}{r}-\dfrac{\LCR_0}{6}r^2\right)^{-1}\dif r^2+r^2\dif\Omega^2,
\end{equation}
and has a curvature singularity at $r=0$ and a coordinate singularity at
\begin{equation} \label{Apparent horizon Class 3 M < 0 R0 > 0}
    r_\mathrm{ah}=\sqrt{\dfrac{8}{\LCR_0}}\cosh\left[\dfrac{1}{3}\arccosh\left(3GM\sqrt{\dfrac{\LCR_0}{2}}\right)\right],
\end{equation}
which is also the only zero of $g^{rr}$, i.e.~an apparent horizon. Because $g^{rr}$ is monotonically decreasing for all $r$, we find that the metric signature changes for $r>r_\mathrm{ah}$.

\subsubsection{Class 3 space-times with \titlebm{$M>0$} and \titlebm{$\LCR_0>0$}.}

To conclude the investigation of Class 3 space-times, we study the case in which the mass and the scalar curvature of the Kottler lookalike are both positive. The line element becomes
\begin{equation} \label{Class 3 M > 0 R0 > 0}
    \dif s^2=-\dif t^2+\left(1-\dfrac{2GM}{r}-\dfrac{\LCR_0}{6}r^2\right)^{-1}\dif r^2+r^2\dif\Omega^2.
\end{equation}
After some computations, one might realise that:
\begin{coloritemize}
    \item If $3GM\sqrt{\LCR_0/2}>1$, then $g^{rr}$ does not vanish for any $r$.
    \item If $3GM\sqrt{\LCR_0/2}=1$, then $g^{rr}$ has a double zero (i.e.~a wormhole throat) located at $r=3GM=\sqrt{2/\LCR_0}$, which is also a coordinate singularity.
    \item If $0<3GM\sqrt{\LCR_0/2}<1$, then $g^{rr}$ vanishes at
    \begin{gather}
        r_0=\sqrt{\dfrac{8}{\LCR_0}}\sin\psi, \label{Apparent horizon Class 3 M > 0 R0 > 0 r_0} \\
        r_1=\sqrt{\dfrac{2}{\LCR_0}}\left(\sqrt{3}\cos\psi-\sin\psi\right), \label{Apparent horizon Class 3 M > 0 R0 > 0 r_1}
    \end{gather}
    where
    \begin{equation}
        \psi\equiv\dfrac{1}{3}\arcsinh\left(3GM\sqrt{\dfrac{\LCR_0}{2}}\right).
    \end{equation}
    Both $r_0$ and $r_1$ are apparent horizons and coordinate singularities.
\end{coloritemize}
Furthermore, it is straightforward to show that $g^{rr}$ always has a maximum located at $r_*=(6GM/\LCR_0)^{1/3}$, with $g^{rr}(r_*)=1-3GM\sqrt{\LCR_0/2}$. In consequence,
\begin{coloritemize}
    \item If $3GM\sqrt{\LCR_0/2}>1$, then $g^{rr}(r_*)<0$, and thus $g^{rr}<0$ for all $r$ (i.e.~the metric signature is everywhere unphysical), since $r_*$ is a global maximum. Thus, this subclass of solutions is not acceptable from a physical point of view.
    \item If $3GM\sqrt{\LCR_0/2}=1$, then $g^{rr}(r_*)=0$, and again the metric signature is unphysical for all $r$, due to the fact that $r_*$ is a maximum. For this reason, this subclass of solutions is also not acceptable from a physical point of view.
    \item If $0<3GM\sqrt{\LCR_0/2}<1$, then $g^{rr}(r_*)>0$. Thus, because $r_0$ and $r_1$---as given by \eqref{Apparent horizon Class 3 M > 0 R0 > 0 r_0} and \eqref{Apparent horizon Class 3 M > 0 R0 > 0 r_0}, respectively---are zeroes of $g^{rr}$ (with $r_0<r_1$), we have that $g^{rr}$ must become negative for $r<r_0$ and $r>r_1$. Therefore, the metric signature is physical only for $r_0<r<r_1$.
\end{coloritemize}
Notice that all Class 3 solutions with $M>0$ and $\LCR_0>0$ also harbour a curvature singularity at the origin, as their Kretschmann scalar is given by \eqref{Class 3 same sign Kretschmann}. However, this curvature singularity lies within the region with unphysical metric signature.

\subsection{Solutions of Class 4.}  \label{Section: Class 4}

Last but not least, we proceed to analyse the $f(\LCR)$-exclusive, vanishing-scalar-curvature solutions
\begin{equation} \label{Class 4}
    \dif s^2=-\left(\dfrac{r}{r_0}\right)^z\left(k-\dfrac{c}{r^b} \vphantom{\dfrac{r}{r_0}}\right)\dif t^2+\left(k-\dfrac{c}{r^b} \vphantom{\dfrac{r}{r_0}}\right)^{-1}\dif r^2+r^2\,\dif\Omega^2,
\end{equation}
with $z\neq 0$ a dimensionless real parameter, $k$ given by
\begin{equation}
    k=\dfrac{4}{z^2+2z+4},
\end{equation}
exponent $b$ being allowed to take the $z$-dependent values
\begin{equation}
    b_\pm=\dfrac{1}{4}\left(3z+6\pm\sqrt{z^2+20z+4}\right), \label{Class 4 b+-}
\end{equation}
$c>0$ being an integration constant (with units of length to the power of $b$) and $r_0$ a free parameter which is left in line element \eqref{Class 4} for dimensional purposes only. Requiring exponents $b=b_\pm$ to be real imposes two additional constraints: $z<-10-4\sqrt{6}$ and $z>-10+4\sqrt{6}$ (numerically, $z<-19.798...$ and $z>-0.202...$).

Metrics of the form \eqref{Class 4} are a particular instance of some novel black-hole solutions with spherical topology discovered in \cite{Calza:2018ohl}. One could in principle distinguish between two subclasses within Class 4, depending on whether $b=b_+$ or $b=b_-$. In spite of this, this distinction will be of little help in practice, as both subtypes exhibit the same key features. The main advantage of Class 4 solutions as compared with those of Classes 1--3 is that they are able to pass weak-field and Solar System tests for $b=b_-$ and $|z|\ll 1$. Moreover, their vanishing Ricci scalar allows them to be at least metastable solutions of $(\LCR_0=0)$-degenerate $f(\LCR)$ models. However, no Class 4 space-time is asymptotically flat, as may be immediately deduced from line element \eqref{Class 4}. This entails that they do not truly describe fully isolated black holes (except in the GR limit $z=0$). In any case, Class 4 solutions are certainly among the less pathological space-times we are considering in this Chapter.

As mentioned before, Class 4 space-times describe spherically-symmetric black holes. This is because
\begin{equation}
    r_\mathrm{hor}=\left(\dfrac{c}{k}\right)^{1/b},
\end{equation}
the only positive root of equation $g^{rr}=0$, turns out to be both an apparent and a Killing horizon for Killing vector $\xi=\partial/\partial t$.\footnote{
    Note that $g^{rr}$ can be shown to be a monotonic function of $r$, and one can thus guarantee that $r_\mathrm{hor}$ is the only horizon of \eqref{Class 4}. This is a null surface; as in Schwarzschild space-time, coordinates $t$ and $r$ exchange roles at $r=r_\mathrm{hor}$, so there is no region with unphysical metric signature.
} This horizon is regular; the Kretschmann scalar associated to \eqref{Class 4} can be shown to be of the form
\begin{equation}
    \LCKre(r)=\dfrac{1}{r^4}\left[\LCKre_0(z)+\left(\dfrac{c}{r^b}\right)\LCKre_1(z)+\left(\dfrac{c}{r^b}\right)^2\LCKre_2(z)\right],
\end{equation}
where functions $\LCKre_i=\LCKre_i(z)$ do not depend on $r$. Thus, we clearly see that the only curvature singularity is the one at the origin $r=0$, and is hidden inside the horizon. As per \eqref{Delta lambda}, the difference in affine parameter between any $r_\mathrm{ini}>0$ and $r_\mathrm{fin}=0$ is given by \eqref{lambda difference}, with primitive \eqref{lambda primitive} being\footnote{We note that the exceptional value $z=-2$, for which $\lambda(r)=r_0\ln(r/r_0)$, lies within the forbidden parameter band $-10-4\sqrt{6}<z<-10+4\sqrt{6}$.}
\begin{equation}
    \lambda(r)=\dfrac{2r_0}{z+2}\left(\dfrac{r}{r_0}\right)^{(z+2)/2},
\end{equation}
which is finite for every $r>0$, regardless of $z$. Considering only the allowed values of $z$, we deduce that $\lambda(0)$---and thus $\Delta\lambda(r_\mathrm{ini}\rightarrow 0)$---remains finite for $z>-10+4\sqrt{6}$, but diverges for $z<-10-4\sqrt{6}$, in which case a photon falling radially into the black hole would never encounter the central singularity (i.e.~it would be protected).

\section{Conclusions and outlook.} \label{Section: Conclusions}

In this Chapter we have assessed the physical viability of $\LCR_0$-degenerate $f(\LCR)$ models, as well as of their---infinitely many---constant-curvature solutions. As our results demonstrate, there are reasons to believe these models cannot be successful in describing Nature, even when experimental uncertainties or other theoretical viability constraints do not rule them out directly. It is important to emphasise that $f(\LCR)$ models which are $\LCR_0$-degenerate for some values of $\LCR_0$ are not difficult to find even without fine-tuning or \emph{model-engineering}. In fact, some of these $f(\LCR)$ models are compatible with both Solar-System experiments and cosmological observations, and may host non-degenerate and pathology-exempt solutions for Ricci scalars different from the degenerate value $\LCR_0$.

One of the first anomalies we have detected is that $(\LCR_0=0)$-degenerate $f(\LCR)$ models feature an strongly-coupled Minkowski background (Result \ref{Result: Non-propagation of gravity}), meaning that the expected massless and traceless graviton does not propagate on top of such background. This result renders $(\LCR_0=0)$-dependent $f(\LCR)$ models incompatible with gravitational-wave observations. Result \ref{Result: Non-propagation of gravity} applies, for instance, to the so-called `power-of-GR' models $f(\LCR)\propto R^{1+\delta}$ with integer $\delta>0$, which have been widely employed in cosmology and other scenarios. We have also found that some $(\LCR_0=0)$-degenerate $f(\LCR)$ models do not even propagate the scalar degree of freedom atop a Minkowski background (Result \ref{Result: Non-propagation of the scalaron}), and thus their linearised spectrum does not contain any polarisation modes whatsoever.

Another important weakness of $\LCR_0$-degenerate $f(\LCR)$ models (either with $\LCR_0=0$ or $\LCR_0\neq 0$) is their apparent lack of predictive power, given the infinite degeneracy of their constant-curvature solutions and the triviality of the equations of motion \eqref{f(R) EOM} when evaluated for $\LCR_0$-degenerate constant-curvature solutions. More precisely, there is a subset of all possible initial conditions for the metric (namely, requiring $\LCR=\LCR_0$) whose evolution is not determined by the equations of motion of $\LCR_0$-degenerate models, which hold automatically for those initial conditions.

In relation to the previous point, there are reasons to believe that some $f(\LCR)$-exclusive constant-curvature solutions can be easily matched to each other in $\LCR_0$-degenerate models \cite{Casado-Turrion:2022xkl}. This result is actually more disturbing than it seems, given that one expects most of the (infinitely many) constant-curvature solutions to exhibit all sorts of physically undesirable properties. In particular, the Class 1 models presented in Section \ref{Section: Class 1}, which are not exempt from pathologies, are known to smoothly match with Minkowski space-time, forming a vacuole-like solution. The fact that one may freely choose the radius where solutions \eqref{Class 1.1} matches the interior Minkowski space-time raises the question of whether a degenerate $f(\LCR)$ model can actually predict the boundaries at which space-time ceases to be described by one of its constant-curvature solution and makes way for another constant-curvature solution.

Regarding the $\LCR_0$-degenerate constant-curvature solutions themselves, we have analysed their stability, finding that, in general terms, the constant-curvature solutions of $(\LCR_0\neq 0)$-degenerate models are all unstable (Result \ref{Result: R0 instability}). The constant-curvature solutions of $(\LCR_0=0)$-degenerate $f(\LCR)$ models can be metastable provided that function $f$ satisfies some constraints (Result \ref{Result: R0=0 metastability}). This can be a problem, however, since the conditions required for the stability of constant-curvature solutions may be incompatible with those guaranteeing the absence of other instabilities (such as the Dolgov-Kawasaki instability). Moreover, the latter result also implies that constant-curvature solutions exhibiting pathological features can be stable.

In order to exemplify the kind of pathologies one may find in $f(\LCR)$-exclusive constant-curvature solutions, we have chosen four representative classes of such space-times, so as to thoroughly investigate four key aspects determining their physical viability: coordinate and curvature singularities, regions in which the metric acquires an unphysical signature (i.e.~the metric determinant changes sign due to one spatial coordinate abruptly becoming time-like), and geodesic completeness (i.e.~whether causal observers can or cannot encounter any of the previous pathologies in finite proper time). As the results in Table \ref{tab:solutions} reveal, all the solutions considered in this work exhibit unphysical properties. These unsubstantiated traits suffice to conclude that these solutions are unlikely to exist in Nature.

Finally, we must stress that there are some issues we would like to study in more detail in the future. For example, we have not investigated the linearised spectrum of $(\LCR_0\neq 0)$-degenerate $f(\LCR)$ theories, in which the natural background is no longer Minkowski space-time, but (Anti-)de Sitter (depending on the sign of $\LCR_0$). Similarly, we have not performed a perturbative expansion around any of the novel, $f(\LCR)$-exclusive constant-curvature solutions, since we have only been concerned with obtaining the linearised spectrum of the theory far from any gravitational-wave source. To shed more light on these issues, we intend to carry out a general analysis of strong-coupling instabilities in $f(\LCR)$ gravity theories.

\chapter*{Conclusions.}
\addcontentsline{toc}{chapter}{Conclusions.}

With the investigations \cite{Casado-Turrion:2022xkl,SecondPaper,ThirdPaper} leading to the elaboration of this Thesis, we have attempted to provide novel theoretical insights into the study of compact objects in modified gravity theories. Owing to the unrelenting advances in gravitational-wave experiments, the study of compact objects is bound to play a central role in testing the validity of both GR and its competitor theories in the forthcoming decades. Additionally, the vast amount of alternative gravity theories which have been considered in the literature so far demands the consideration of additional criteria to constrain their respective parameter spaces, if not to put into question their physical viability altogether. For these reasons, it becomes indispensable to ensure that the new models for gravity are mathematically- and observationally-consistent, as well as capable of producing precise, well-founded predictions, to be eventually confronted to experiment.

Accordingly, most of the Thesis has been devoted to the study of junction conditions in modified gravity theories, more specifically, in $f(R)$ gravity and \BSPGT~theories. As explained in Chapter \ref{chapter:Introduction: Modified Gravity}, these alternatives to GR are not only grounded on solid theoretical arguments, but also on their ability to account for physical phenomena (such as dark energy and inflation in the case of $f(R)$, or the fact that spinor fields exist in Nature and necessarily induce space-time torsion, in the case of \BSPGT). Junction conditions are of paramount importance when studying stars, thin-shell wormholes and several black-hole mimickers, as well as gravitational collapse. However, progress on these areas has been historically hampered by a lack of comprehension of the junction conditions, either for their increased complexity in comparison with GR, as in $f(R)$ gravity, or because they had never been rigorously derived anywhere else in the literature, as was the case in \BSPGT. Furthermore, the computation and handling of junction conditions requires resolving several mathematical conundrums and subtleties.

As such, on Chapter \ref{chapter:Introduction: JCs} and related Appendices \ref{app:subtleties} and \ref{app:regsing}, we have offered a clear, mathematically-consistent formulation of the theory of junction conditions in any field theory, including geometrical depictions of gravity. In particular, we have addressed the issue of distributional multiplication of terms of the form $\udis{\Theta}\,\del$, ubiquitous in most relevant cases, and prescribed a deterministic algorithm which allows for the correct determination of the junction conditions of a given theory, summarised in Section \ref{sec:procedure}.

As our first application of the junction-condition formalism to physical scenarios, we have devoted Chapter \ref{chapter:f(R) collapse} and Appendix \ref{Appendix:stellar collapse formulae} to the study of gravitational collapse in $f(R)$ gravity theories. To that end, we have focused on the simple---yet illustrative---case of a collapsing, uniform-density dust star, i.e.~the counterpart of the renowned Oppenheimer-Snyder model of GR. On the one hand, in the metric formulation of the theory, we have discovered that most difficulties arise from the lack of a Schwarzschild exterior in metric $f(\LCR)$ gravities. What is more, we have found that the space-time outside the collapsing dust star must be diametrically dissimilar to the Schwarzschild solution one has in GR, being necessarily time-dependent and possessing a non-constant Ricci scalar and non-trivial metric components $g_{tt} g_{rr}\neq -1$ in so-called `areal-radius' coordinates (Results \ref{theoconstcurv}--\ref{theoanothernonstatic}). On the other hand, in Palatini $f(\MAR)$ gravity, we have observed that dust stars are incompatible with the junction conditions (Result \ref{result:OS Palatini}), and hence the Oppenheimer-Snyder construction is impossible in this case, even though Schwarzschild is the appropriate exterior, as in GR. While it is true that requiring the star to be made of dust might constitute an oversimplification of the problem (specially in the Palatini case), this model of collapse is arguably the simplest possible. Given the inherent difficulty of treating junction conditions numerically, having an exact, analytical model encapsulating the main features of gravitational collapse is important in order to understand this phenomena on general terms. Furthermore, in the case of metric $f(\LCR)$, we have argued that the problem of collapse is tied to the no-hair theorems, so a better understanding of the former could entail a better understanding of the latter. Because of this, further research is necessary to shed light on the issue.

One of the possible solutions to the problems described in the previous paragraph could be by changing the $f(R)$ action to include more terms and, in consequence, modifying the junction conditions. One among the simplest, metric-affine gravities reducing to quadratic $f(R)$ in the appropriate limits are the aforementioned \BSPGT~models. Despite the complexity of the field equations, the power of our formalism---as described in Section \ref{sec:procedure}---has allowed us to derive the junction conditions in \BSPGT~theories (Result \ref{result:JCs in BSPGT}) in Chapter \ref{chapter:BSPGT JCs}. As expected, we have found that the junction conditions of \BSPGT~allow for a richer variety of singular structures which are allowed on the matching surface, in comparison with GR and $f(R)$ gravity. In particular, we have found that it is possible to have monopolar contributions to the spin currents, as well as monopolar (thin-shell) and dipolar (double-layer) terms in the stress-energy tensor. We have also verified that the junction conditions in \BSPGT~reduce to the previously-known results in the appropriate cases. Applications of our results are to be expected in the future, for example, in stellar dynamics, gravitational collapse and black-hole formation, or thin-shell wormholes. 

Finally, in Chapter \ref{chapter:constant curvature} and related Appendix \ref{appendix:constant-curvature}, we have resorted to other viability criteria---such as gravitational-wave spectra or absence of instabilities---to conclude that $\LCR_0$-degenerate $f(\LCR)$ models and their constant-curvature solutions---as defined in Section \ref{section:constant-curvature solutions}---are not physically well-founded. Thus, they should not be considered as an appropriate description of Nature, in spite of the fact that some of them had found great success in describing cosmological observations, the rotation curves of galaxies, or even giving rise to interesting exotic compact objects, such as vacuum wormholes and novel black-hole solutions. In particular, we have shown that said models feature a strongly-coupled Minkowski background (Results \ref{Result: Non-propagation of gravity} and \ref{Result: Non-propagation of the scalaron}), rendering them incompatible with gravitational-wave observations. Additionally, we have found that all $f(\LCR)$-exclusive, $(\LCR_0\neq 0)$-degenerate solutions are generically unstable (Result \ref{Result: R0 instability}), while $(\LCR_0=0)$-degenerate solutions can be metastable (Result \ref{Result: R0=0 metastability}), but sometimes at the expense of violating the all-important Dolgov-Kawasaki stability condition. In any case, as evinced by the results in Table \ref{tab:solutions} and the discussion in Section \ref{section:constant-curvature solutions}, it is expected that most degenerate, constant-curvature solutions feature unphysical traits, such as accessible naked singularities, regions where the space-time signature ceases to be Lorentzian, etc.

To sum up, the results herein evince that very basic theoretical considerations can have a profound impact on our comprehension of physical theories and the compact-object dynamics they can give rise to. Our contributions to the literature intend to improve our understanding of gravitational collapse and stellar configurations in modified gravity, as well as more exotic compact objects that could exist in these theories.

\begin{appendices}

\renewcommand\thesubsection{\Alph{chapter}.\arabic{section}.\arabic{subsection}}
\renewcommand\thesubsubsection{\Alph{chapter}.\arabic{section}.\arabic{subsection}.\arabic{subsubsection}}

\pagestyle{MyPagestyleAppendix}

\renewcommand*{\printparttitle}{\partnamefont\textcolor{complutense}}

\renewcommand{\afterpartskip}{\par\noindent\parbox{\linewidth}{\hrulefill}\vfill}

\cleardoublepage 
\phantomsection 
\addcontentsline{toc}{part}{Appendices.}
\part*{Appendices.}
\label{Part:Appendices}

\chapter{On the derivation of the equations of motion of \BSPGT.}
\label{app:EoM}

\section{Variations of the gravitational sector.} \label{app:BSPGT variations}

In this Appendix we shall obtain the equations of motion of \BSPGT~by varying the theory's action, \eqref{eq:fullaction}. Let us first introduce the following notation for the variations of the pruely-gravitational part of the action, $S_\mathrm{GFPG}$, as given by \eqref{eq:gravaction}:
\begin{align} 
    \mathcal{E}(T)_\mu &\equiv \dfrac{1}{\sqrt{-g}}\dfrac{\delta S_\mathrm{GFPG}}{\delta T^\mu},
    &
    \mathcal{E}(S)_\mu &\equiv \dfrac{1}{\sqrt{-g}}\dfrac{\delta S^\mathrm{GFPG}}{\delta S_\mu}, \nonumber\\
    \mathcal{E}(t)^\rho{}_{\mu\nu} &\equiv \dfrac{1}{\sqrt{-g}} \dfrac{\delta S_\mathrm{GFPG}}{\delta t_\rho{}^{\mu\nu}},
    &
    \mathcal{E}(g)_{\mu\nu} &\equiv \dfrac{1}{\sqrt{-g}}\dfrac{\delta S_\mathrm{GFPG}}{\delta g^{\mu\nu}}. \label{eq:var0}
\end{align}
The aforementioned variations yield
\begin{align} 
    \mathcal{E}(T)_\mu&=\left(M_T^2-\dfrac{8\beta}{3}\MAR\right)T_\mu +\dfrac{4\alpha}{3}\holst S_\mu -4\beta \LCD_\mu\MAR, \label{eq:varT0}\\
    \mathcal{E}(S)_\mu&= \left(M_S^2+\dfrac{\beta}{6} \MAR\right)S_\mu +\dfrac{4\alpha}{3}\holst T_\mu +2\alpha \LCD_\mu \holst    , \label{eq:varS0}\\
    \mathcal{E}(t)^\lambda{}_{\mu\nu}&= \left(M_t^2+2\beta \MAR\right)t^\lambda{}_{\mu\nu} + 2\alpha\holst  \LCten_{\mu\nu\rho\sigma}t^{\lambda\rho\sigma}, \label{eq:vart0}\\
    \mathcal{E}(g)_{\mu\nu}&=
    \dfrac{\Mp^2}{2} \left(E_{\mu\nu}-\dfrac{1}{2}g_{\mu\nu}E\right)-2\beta (\LCD_{\mu}\LCD_{\nu}- g_{\mu\nu}\LCbox)\MAR  + P_{\mu\nu}, \label{eq:varg0}
\end{align}
where $M_T$ and $M_S$ are defined in \eqref{eq:defMSMT}, 
\begin{equation}
    M_t^2 \equiv m_t^2 + \dfrac{\Mp^2}{2},
\end{equation}
and
\begin{align}
   \Mp^2 E_{\mu\nu}&\equiv \Mp^2\mathring{R}_{\mu\nu} + 4\beta \LCR_{\mu\nu}\MAR + 4 \alpha S_{(\mu}\LCD_{\nu)}\holst - 8\beta T_{(\mu}\LCD_{\nu)}\MAR \nonumber \\
     &
    \quad + \left(M_S^2+ \dfrac{\beta}{6} \MAR\right)S_\mu S_\nu + \dfrac{8}{3}\alpha \holst  S_{(\mu} T_{\nu)} + \left(M_T^2- \dfrac{8\beta}{3} \MAR\right)T_\mu T_\nu\nonumber \\
     &
    \quad -g_{\mu\nu}(\alpha \holst ^2 + \beta \MAR^2),\label{eq:deftensorE}   \\ 
   P_{\mu\nu}&\equiv \left(\dfrac{M_t^2}{2} + \beta \MAR\right)\left(2t^{\rho\sigma}{}_\mu t_{\rho\sigma\nu}-t_\mu{}^{\rho\sigma}t_{\nu\rho\sigma} -\dfrac{1}{2}g_{\mu\nu} t^{\rho\sigma\lambda}t_{\rho\sigma\lambda}   
   \right)\nonumber\\
   &\quad - \alpha\holst   \left[
   \LCten_{\rho\sigma\lambda\tau}(t_{\mu}{}^{\rho\sigma}t_{\nu}{}^{\lambda\tau}+g_{\mu\nu}t_{\gamma}{}^{\rho\sigma}t^{\gamma\lambda\tau})+ 4 t_{\gamma}{}^{\rho\sigma}t^{\gamma\lambda}{}_{(\mu} \LCten_{\nu)\lambda\rho\sigma}
   \right].\label{eq:deftensorP}
\end{align}
Note that $P_{\mu\nu}$ vanishes when $t^\rho{}_{\mu\nu}=0$.\footnote{To find the expression \eqref{eq:deftensorP} with xAct, we had to perform some simplifications based on the following fact. For any tensor $Z_{\mu\nu\rho\sigma}$, with the same symmetries as the Riemann tensor (antisymmetry in both pairs and symmetry under pair exchange), it can be shown that, in four dimensions, $Z_{\mu[\nu\rho\sigma]}= Z_{[\mu\nu\rho\sigma]}$, which leads to the useful identity:
\begin{equation*}
    Z^{\mu[\rho\sigma\lambda]}Z_{\nu[\rho\sigma\lambda]}=\dfrac{1}{4}\delta^\mu_{\nu}Z^{\tau\rho\sigma\lambda}Z_{\tau[\rho\sigma\lambda]}.
\end{equation*}
We rearranged it as 
\begin{equation*}
    Z_\mu{}^{\lambda\rho\sigma}Z_{\nu\rho\sigma\lambda} = -\dfrac{1}{2}Z_\mu{}^{\rho\sigma\lambda}Z_{\nu\rho\sigma\lambda}+\dfrac{3}{8}g_{\mu\nu}Z^{\tau\rho\sigma\lambda}Z_{\tau[\rho\sigma\lambda]}
\end{equation*}
and use it for $Z_{\mu\nu\rho\sigma} = t_{\lambda\mu\nu}t^{\lambda}{}_{\rho\sigma}$.}

\section{On the equation of motion of the tensor part of the torsion.}
\label{app:EoMt}

We now turn to analyse the equation of $t^\rho{}_{\mu\nu}$. From \eqref{eq:vart0}, ignoring the matter contribution, we simply get
\begin{equation}
    0=\mathcal{E}(t)_{\lambda\mu\nu}= C_1\,t_{\lambda\mu\nu} + C_2\,\LCten_{\mu\nu}{}^{\rho\sigma}t_{\lambda\rho\sigma},
\end{equation}
with $C_1 \equiv M_t^2+2\beta R$ and $C_2\equiv2\alpha\holst $. First, we notice that, if $C_1=0$ and $C_2\neq 0$, or if $C_1\neq0$ and $C_2= 0$, we automatically have $t_{\lambda\mu\nu}=0$. If both are non-vanishing, then one can construct the combination
\begin{equation}
    0=C_1\,\mathcal{E}(t)_{\lambda\mu\nu}- C_2\,\LCten_{\mu\nu}{}^{\rho\sigma}\mathcal{E}(t)_{\lambda\rho\sigma} = (C_1^2+ 4 C_2^2)\,t_{\lambda\mu\nu},
\end{equation}
which leads to $t_{\lambda \mu\nu}=0$. Finally, in the case in which both $C_1$ and $C_2$ vanish,
\begin{equation}
    \beta \MAR= - \dfrac{M_t^2}{2}, \mybigskip \alpha\holst =0. \label{eq:condC1C2}
\end{equation}
Under conditions \eqref{eq:condC1C2}, tensor $t^\rho{}_{\mu\nu}$ remains (dynamically) undetermined and only subjected to the previous restrictions. Indeed, $t^\rho{}_{\mu\nu}$ disappears from the rest of the equations; for example, notice that $P_{\mu\nu}$ in \eqref{eq:deftensorP} vanishes identically under \eqref{eq:condC1C2}. After introducing the matter currents \eqref{eq:defLJT} and making use of \eqref{eq:condC1C2}, the remaining equations of motion become
\begin{align} 
    L_\mu&= \left(M_T^2+\dfrac{4M_t^2}{3}\right) T_\mu  = \left(m_T^2+\dfrac{4m_t^2}{3}\right) T_\mu ,\\
   J_\mu&= \left(M_S^2-\dfrac{M_t^2}{12} \right)S_\mu = \left(m_S^2-\dfrac{m_t^2}{12} \right)S_\mu, \\
    T_{\mu\nu}&=
    \left(\Mp^2-2M_t^2\right)\mathring{G}_{\mu\nu}  -\dfrac{M_t^2}{2}g_{\mu\nu}R  \nonumber\\
    &\quad +  \dfrac{1}{2}\left[ 2J_{(\mu} S_{\nu)}+ 2L_{(\mu} T_{\nu)} -g_{\mu\nu}(J_\lambda S^\lambda+L_\lambda T^\lambda)\right].
\end{align}
We see that the trace and axial vectors become non-dynamical and that, in vacuum, they vanish for generic values of the parameters (except in cases where the combinations in brackets vanish). For $\beta=0$, we get $M_t=0$, corresponding to GR sourced by an effective energy-momentum tensor, i.e.~to $\mathring{G}_{\mu\nu} = \Mp^{-2} T_{\mu\nu}^\text{eff}$, with
\begin{equation}
    T_{\mu\nu}^\text{eff}\equiv T_{\mu\nu} -  \dfrac{1}{2}\left[ 2J_{(\mu} S_{\nu)}+ 2L_{(\mu} T_{\nu)} -g_{\mu\nu}(J_\lambda S^\lambda+L_\lambda T^\lambda)\right].
\end{equation}
If $\beta\neq 0$, we also get an effective cosmological constant and a correction to Newton's constant:
\begin{align}
    \left(\Mp^2-2M_t^2\right)\mathring{G}_{\mu\nu}  +\dfrac{M_t^4}{4\beta}g_{\mu\nu}  = T_{\mu\nu}^\text{eff}.
\end{align}
In both cases, the tensor part of the torsion drops from the whole set of equations of motion.

Judging by the previous calculations, we can state that the only way for the tensor part of torsion to contribute dynamically to the field equations is by having a non-minimal derivative coupling with matter. Otherwise, it can only contribute as a term that can be absorbed in the energy-momentum tensor. Therefore, we could assume that all the dynamics would be encoded in the axial and trace vectors and omit the tensorial part from the beginning. This agrees with the considerations in \cite{BeltranJimenez:2019hrm}, where the authors justify this choice by performing a Legendre transformation on the action of the theory.

\chapter{Some subtle aspects regarding our prescription for distributional promotion.}
\label{app:subtleties}

For equations of motion leading---upon distributional promotion---to products of the form $\udis{X}\,\udis{Y}$, where $\udis{X}$ is regular and $\udis{Y}$ possesses a non-trivial monopolar part (but no other singular contribution), our prescription \eqref{eq:prescTheDel} would entail
\begin{equation}
    X Y\myskip\overset{\text{def.}}{\longrightarrow}\myskip\udis{X}\,\udis{Y} \equiv X^+ Y^+ \,\heavidis{+} + X^-Y^-\,\heavidis{-}+X^\sSigma \singpart{\szero}{\udis{Y}}. 
\end{equation}
This is the same result we would have obtained through the naive promotion
\begin{equation}
     XY\myskip\overset{\text{naively}}{\longrightarrow}\myskip\regpart{X}(\regpart{Y}+\singpart{\szero}{Y}),
\end{equation}
as well as the identifications    
\begin{equation} \label{eq:identif}
    \heavidis{\pm}\heavidis{\pm}=\heavidis{\pm},\mybigskip \heavidis{\pm}\heavidis{\mp}=0,\mybigskip\heavidis{\pm}\del=\frac{1}{2}\del.
\end{equation}
However, when the equations of motion involve products of three objects, two of which are regular and the remaining one monopolar (as is the case in \BSPGT, whose equations of motion contain products such as $T^\mu S^\nu\LCD_\nu S_\mu$), one might be tempted to think that the naive identifications in \eqref{eq:identif} would still work. However, there is a problem: the product therein is not associative. Compare, for example,
\begin{align}
    (\heavidis{\pm}\heavidis{\pm})\del &= \heavidis{\pm}\del = \frac{1}{2}\del, \nonumber\\
     \heavidis{\pm}(\heavidis{\pm}\del) &= \frac{1}{2}\heavidis{\pm}\del = \frac{1}{4}\del.        
\end{align}
Thus, one is forced to make a choice. Our prescription \eqref{eq:prescTheDel} corresponds to `multiplying the $\udis{\Theta}$s by the $\del$ one by one and not among themselves.' In the case of two regular tensor fields $X_1$ and $X_2$, i.e.~$X=X_1 X_2$, the alternative choice would have led to    
\begin{equation}
        X_1 X_2 Y \myskip\overset{\text{alt.}}{\longrightarrow}\myskip X^+_1 X^+_2 Y^+ \heavidis{+}\ +\ X^-_1 X^-_2 Y^-\heavidis{-}\ +\ (X_1 X_2)^\sSigma \singpart{0}{Y}. 
\end{equation}
Notice that the last term is, in general, different from the one obtained using our prescription \eqref{eq:prescTheDel}, which is $X^\sSigma_1 X^\sSigma_2\singpart{0}{Y}$, cf.~\eqref{eq:ssigmaproduct}. Observe that the ambiguity is even worse when there are more than two $X_i$. For example, for three of them, there are five possibilities: $(X_1 X_2)^\sSigma X^\sSigma_3$,  $(X_3 X_2)^\sSigma X^\sSigma_1$, $(X_1 X_3)^\sSigma X^\sSigma_2$, $(X_1 X_2 X_3)^\sSigma$ and $X^\sSigma_1X^\sSigma_2X^\sSigma_3$. In spite of this, only the case of two $X_i$s are relevant for the purposes of this Thesis.
    
Finally, let us justify why our choice is \emph{sensible}, by defining it in a different way. The underlying idea is that we can treat the Heaviside distributions $\heavidis{\pm}$ inside the $\udis{X}_i$ simply as if they were the Heaviside \emph{functions} $\Theta^\pm$. In other words, we are identifying the regular distributions $\udis{X}_i$ with their associated tensors $X_i$. Intuitively, this makes sense, since regular tensor distributions do not present singular contributions anywhere (of course, this is not mathematically rigorous, as $\udis{X}_i$ and $X_i$ belong to different spaces). This leads us to define the product $\udis{X}_1 \udis{X}_2 \udis{Y}$ as the distribution acting on test functions $\varphi$ as follows:
 \begin{align}
    \langle\udis{X}_1\udis{X}_2\udis{Y},\varphi\rangle &\equiv \langle X_1 X_2 \udis{Y},\, \varphi \rangle\nonumber\\
        &= \langle  \udis{Y},\, X_1 X_2 \varphi \rangle\nonumber\\
        &= \langle  \regpart{\udis{Y}}, X_1 X_2 \varphi \rangle + \langle  \singpart{0}{\udis{Y}}, X_1 X_2 \varphi \rangle\nonumber\\
        &= \langle  \heavidis{+}, X_1 X_2 Y^+ \varphi \rangle + \langle  \heavidis{-}, X_1 X_2 Y^- \varphi \rangle+ \langle  \singpart{0}{\udis{Y}}, (X_1 X_2 \varphi)|_\Sigma \rangle\nonumber\\
        &= \Big\langle  (X^+_1 X^+_2 Y^+)\,\heavidis{+}, \varphi \Big\rangle + \Big\langle  (X^-_1 X^-_2 Y^-)\,\heavidis{-}, \varphi \Big\rangle + \Big\langle  \singpart{0}{\udis{Y}}, X_1|_\Sigma X_2|_\Sigma \varphi \Big\rangle\nonumber\\
        &= \Big\langle  \Big[(X^+_1 X^+_2 Y^+)\,\heavidis{+} + (X^-_1 X^-_2 Y^-)\,\heavidis{-} + X^\sSigma_1 X^\sSigma_2\singpart{0}{\udis{Y}}\Big], \varphi \Big\rangle.\label{eq:XXYint}
\end{align}
This is compatible with our prescription \eqref{eq:prescTheDel}. In the last step above in \eqref{eq:XXYint}, we have used $X_i|_\Sigma = X^\sSigma_i$, which holds for cases relevant to the computations in Chapter \ref{chapter:BSPGT JCs}: $X_i = T_\mu$, $S_\mu$. Note that, except for the definition in the first line, all the steps in \eqref{eq:XXYint} are mathematically rigorous.

\chapter{Regular and singular parts. Useful formulae.} \label{app:regsing}

\section{Formulae involving discontinuities.}

The discontinuity across $\Sigma$ of the product of a pair of regular tensor fields, $A$ and $B$ (we supress their arbitrary indices for simplicity), is given by
\begin{equation}
    \jump{AB}=A^\sSigma\jump{B}+\jump{A} B^\sSigma.
\end{equation}

For the derivative of a regular scalar $f$ and a regular one-form $V_\mu$ we find:\footnote{See \cite[App. D.2]{Reina:2015gxa} for the derivation of \eqref{eq:jumpDV} for $\epsilon=+1$ (i.e.~for time-like $\Sigma$).}
\begin{align}
   \jump{\LCD_\mu f} &= \jump{\partial_\mu f} = \partial_\mu\jump{f}, \label{eq:jumpDS} \\
    \jump{\LCD_\mu V_\nu} &= \epsilon n_\mu n^\rho \jump{\LCD_\rho V_\nu} + h^\rho{}_\mu \LCD^\sSigma_\rho \jump{V_\nu} +\epsilon V^\sSigma_\rho \left(n^\rho\jump{K_{\mu\nu}}-h^{\rho\sigma}\jump{K_{\mu\sigma}}n_\nu\right), \label{eq:jumpDV}
\end{align}
where $\LCD_\mu^\sSigma$ is the covariant derivative taken with respect to $\LCG^{\sSigma\rho}{}_{\mu\nu}$.

As an application of \eqref{eq:jumpDV}, it can be proven that, for vanishing $Q_{\rho\mu\nu}$ and $t^\rho{}_{\mu\nu}$,
\begin{align}
    \jump{\MAR}&= \jump{\LCR}-\dfrac{2}{3}\jump{T_\mu T^\mu}+\dfrac{1}{24}\jump{S_\mu S^\mu} \nonumber \\ 
        &\quad+2\epsilon n^\mu n^\nu \jump{\LCD_\mu T_\nu} +2 h^{\mu\nu}\LCD^\sSigma_\mu \jump{T_\nu}+2\epsilon T^\sSigma_\perp \jump{K}, \vphantom{\dfrac{2}{3}} \label{eq:jumpR} \\
    \jump{\holst}&= \dfrac{2}{3}\jump{S_\mu T^\mu} - \epsilon n^\mu n^\nu \jump{\LCD_\mu S_\nu} -h^{\mu\nu}\LCD^\sSigma_\mu \jump{S_\nu}-\epsilon S^\sSigma_\perp \jump{K}. \label{eq:jumpH}
\end{align}
These results are fundamental for the simplification of the junction conditions in \BSPGT.

\section{First derivative of a monopolar distribution.}

Let us focus on a tensor distribution whose singular part is purely monopolar, i.e.
\begin{equation} \label{eq:decompS}
    \udis{F}_A  = \Regpart{\udis{F}_A} + \Singpart{\szero}{\udis{F}_A},
\end{equation}
where each of the pieces are, by definition, of the form
\begin{align}
    \Regpart{\udis{F}_A} &= F^{+}_A\,\heavidis{+} + F^{-}_A\,\heavidis{-},\\
   \Singpart{\szero}{\udis{F}_A}&= F^\szero_A \del.
\end{align}
The first derivative of \eqref{eq:decompS} decomposes---in complete agreement with \eqref{eq:schemeD}---as
\begin{align}
    \disLCD_\mu \udis{F}_A  &= \Regpart{\disLCD_\mu \udis{F}_A} + \Singpart{\szero}{\disLCD_\mu \udis{F}_A}+ \Singpart{\sone}{\disLCD_\mu \udis{F}_A},
\end{align}
with each part being given by
\begin{align}
    \Regpart{\disLCD_\mu \udis{F}_A} &= (\LCD_\mu F_A)^{+} \, \heavidis{+} + (\LCD_\mu F_A)^{-} \,\heavidis{-}, \label{eq:decomDS1} \\
   \Singpart{\szero}{\disLCD_\mu \udis{F}_A}&= \left( \epsilon n_\mu \big(\jump{F_A }- K^\sSigma F^\szero_A\big)+h^\rho{}_\mu \LCD_\rho F^\szero_A\right) \del ,\label{eq:decomDS2}\\
   \Singpart{\sone}{\disLCD_\mu \udis{F}_A}&= \disLCD_\rho \left(\epsilon F^\szero_A n_\mu n^\rho \del\right).\label{eq:decomDS3}
\end{align}
We shall collect some special cases of the above formulae in the remainder of this Appendix. The expressions below are particularly useful in the derivation of the junction conditions in \BSPGT~(Chapter \ref{chapter:BSPGT JCs}).

\subsection{Derivative of the Levi-Civita Ricci scalar.}

For the Ricci scalar of the Levi-Civita connection, i.e.~$\udis{F}_A\rightarrow\udis{\LCR}$ and $F^\szero_A\rightarrow-2\epsilon\jump{K}$, cf.~\eqref{eq:ricci scalar singular}, its corresponding first distributional derivative,
\begin{equation} \label{eq:LCR derivative}
    \disLCD_\mu \LCR = \Regpart{\disLCD_\mu \LCR} + \Singpart{\szero}{\disLCD_\mu \LCR}+ \Singpart{\sone}{\disLCD_\mu \LCR}, 
\end{equation}
has the following decomposition:
 \begin{align}
    \Regpart{\disLCD_\mu \LCR} &= (\LCD_\mu\LCR)^{+} \, \heavidis{+} + (\LCD_\mu \LCR)^{-} \,\heavidis{-} ,\\
   \Singpart{\szero}{\disLCD_\mu \LCR}&= \epsilon\left( n_\mu \big(\jump{\LCR} + 2\epsilon K^\sSigma \jump{K}\big)- 2h^\rho{}_\mu \LCD_\rho \jump{K}\right) \del , \label{eq:Delta0DR} \\
   \Singpart{\sone}{\disLCD_\mu\LCR}&=-2 \disLCD_\rho \left(\jump{K} n_\mu n^\rho\del\right). \label{eq:Delta1DR}
\end{align}

\subsection{Second derivative of a regular one-form.}

Consider a regular one-form distribution $\udis{V}_\mu$, i.e.~one which can be written as
\begin{equation}
    \udis{V}_\mu =  \Regpart{\udis{V}_\mu} =  V^{+}_\mu \, \heavidis{+} + V^{-}_\mu \,\heavidis{-}.
\end{equation}
For instance, the torsion trace and axial vectors $\udis{T}_\mu$ and $\udis{S}_\mu$ of \BSPGT~are tensor distributions of this type (as discussed in Section \ref{sec:preconsistency}), so the following four equations---namely, \eqref{eq:decompDV}--\eqref{eq:Delta1DDV}---are valid for them. The first distributional derivative of $\udis{V}_\rho$,
\begin{equation}
    \disLCD_\mu\udis{V}_\rho  =  (\LCD_\mu V_\rho)^{+} \, \heavidis{+} + (\LCD_\mu V_\rho)^{-} \,\heavidis{-} + \epsilon n_\mu \jump{V_\rho}\del, \label{eq:decompDV}
\end{equation}
is a distribution of the form \eqref{eq:decompS}; in other words, its singular part only contains a contribution of the type $\Delta^{\szero}$. Therefore, to compute the second distributional derivative, one may use the general formulae \eqref{eq:decomDS1}--\eqref{eq:decomDS3} with the substitutions \smash{$\udis{F}_A  \to \disLCD_\nu \udis{V}_\rho$}, \smash{$F^{\pm}_A \to \LCD_\nu^\pm V^{\pm}_\rho$} and \smash{$F^\szero_A \to \epsilon n_\nu \jump{V_\rho}$}. The result reads
 \begin{align}
    \Regpart{\disLCD_\mu  \disLCD_\nu \udis{V}_\rho} &= (\LCD_\mu\LCD_\nu V_\rho)^{+} \, \heavidis{+} + (\LCD_\mu\LCD_\nu V_\rho)^{-} \,\heavidis{-} ,\label{eq:regDDV}\\
    \Singpart{\szero}{\disLCD_\mu \disLCD_\nu \udis{V}_\rho}&= 
    \epsilon\left[n_\mu\jump{\LCD_\nu V_\rho}+(K^\sSigma_{\mu\nu}-\epsilon n_\mu n_\nu K^\sSigma)\jump{V_\rho}+n_\nu h^\sigma{}_\mu \LCD_\sigma   \jump{V_\rho}\right] \del
   ,\label{eq:Delta0DDV}\\
   \Singpart{\sone}{\disLCD_\mu  \disLCD_\nu \udis{V}_\rho}&= \disLCD_\sigma \left(\jump{V_\rho} n_\mu n_\nu  n^\sigma\del\right).\label{eq:Delta1DDV}
\end{align}
The second one can be further expanded by using \eqref{eq:jumpDV}.

\section{First and second derivative of a regular distribution.}

In \BSPGT's equation of the metric, \eqref{eq:EoM3}, terms of the form $\beta\LCD_\mu \LCD_\nu \LCR$ and $\beta\LCD_\mu \LCD_\nu \LCD_\rho T^\rho$ are present. According to the results in the previous Section, these terms would---in principle---require one to consider a singular structure that goes up terms of the type $\Delta^{\stwo}$, since both $\LCR$ and $\LCD_\rho T^\rho$ could potentially contain terms up to $\Delta^{\szero}$. However, owing to the consistency conditions in Section \ref{cons:cond}, $\beta\LCR$ and $\beta\LCD_\rho T^\rho$ are purely regular. For this reason, only the decompositions of the first and second derivatives of regular distributions are relevant for the analysis of those terms.

We start by considering a regular (but not necessarily continuous) distribution:
\begin{equation}
    \udis{F}_A  = \Regpart{\udis{F}_A} = F^{+}_A \,\udis{\Theta}^{+} + F^{-}_A \, \udis{\Theta}^{-} . \label{eq:decompR}
\end{equation}
This is a particular case of \eqref{eq:decompS} when $F^\szero_A=0$. In practice, we are interested in scalar distributions $\udis{F}_A \to \udis{F}$, since $\LCR$ and  $\LCD_\mu T^\mu$ are scalars, but we provide the expressions for objects with arbitrary tensor indices for the sake of completeness. 

The first and second derivatives of \eqref{eq:decompR} will have the following singular structure, cf.~\eqref{eq:schemeD}:
\begin{align}
    \disLCD_\mu \udis{F}_A  &= \Regpart{\disLCD_\mu \udis{F}_A} + \Singpart{\szero}{\disLCD_\mu \udis{F}_A}, \\   
    \disLCD_\mu \disLCD_\nu \udis{F}_A  &= \Regpart{\disLCD_\mu\disLCD_\nu \udis{F}_A} + \Singpart{\szero}{\disLCD_\mu\disLCD_\nu \udis{F}_A}+ \Singpart{\sone}{\disLCD_\mu \disLCD_\nu \udis{F}_A}.
\end{align}
For the first distributional derivative, the decomposition is simply \eqref{eq:decomDS1}--\eqref{eq:decomDS2} for $F^\szero_A=0$, whereas for the second derivative we get
\begin{align}
    \Regpart{\disLCD_\mu\disLCD_\nu \udis{F}_A} &= (\LCD_\mu\LCD_\nu  F_A)^{+} \, \udis{\Theta}^{+} + (\LCD_\mu \LCD_\nu F_A)^{-} \,\udis{\Theta}^{-} , \label{eq:decomDDReg1}\\
    \Singpart{\szero}{\disLCD_\mu\disLCD_\nu \udis{F}_A}&=  \epsilon \left[n_\mu \jump{\LCD_\nu F_A}+n_\nu h^\rho{}_\mu \LCD_\rho\jump{F_A}+(K^\sSigma_{\mu\nu}-\epsilon n_\mu n_\nu K^\sSigma)\jump{F_A}\right] \del , \label{eq:decomDDReg2} \\
    \Singpart{\sone}{\disLCD_\mu\disLCD_\nu \udis{F}_A}&=   \disLCD_\rho \left( \jump{F_A} n_\mu n_\nu n^\rho \del\right) .\label{eq:decomDDReg3}
\end{align}
Observe that \eqref{eq:regDDV}--\eqref{eq:Delta1DDV} are particular cases of these expressions. Another interesting property that we can read from \eqref{eq:decomDDReg3} is that, for regular $\udis{F}_A$,
\begin{equation}
\Singpart{\sone}{\disLCD_\mu\disLCD_\nu \udis{F}_A} =   \epsilon\disLCD_\rho \left( \Singpart{\szero}{\disLCD_{(\mu|} \udis{F}_{A}} n_{|\nu)} n^\rho \right).
\end{equation}

Below we provide the decomposition of the singular parts of the highest-order derivative terms in the equation of the metric \eqref{eq:EoM3}.

\subsection{Singular parts of the second derivative of the (regularised) Ricci scalar in \BSPGT.}

If, in \BSPGT, we assume $\beta \jump{K}=0$, then the singular part of the combination $\beta\disLCD_\mu\disLCD_\nu \udis{\LCR}$ can be decomposed as follows into two parts:
\begin{align}
   \Singpart{\szero}{\beta\disLCD_\mu\disLCD_\nu \udis{\LCR}}&=  \epsilon \beta \left[n_\mu\partial_\nu\jump{\LCR}+ n_\nu h^\rho{}_\mu \partial_\rho\jump{\LCR}+(K^\sSigma_{\mu\nu}-\epsilon n_\mu n_\nu K^\sSigma)\jump{\LCR}\right] \del ,\label{eq:decomDDR1}\\
   \Singpart{\sone}{\beta\disLCD_\mu\disLCD_\nu \udis{\LCR}}&=  \beta\disLCD_\rho \left( \jump{\LCR} n_\mu n_\nu n^\rho \del\right) ,\label{eq:decomDDR2}
\end{align}
where we have used \eqref{eq:jumpDS}. In particular, if we contract with the metric, we get:
\begin{align}
   \Singpart{\szero}{\beta\disLCD_\mu\disLCD^\mu \udis{\LCR}}&=  \epsilon \beta n^\mu \partial_\mu \jump{\LCR}  \del ,\label{eq:decomtrDDR1}\\
   \Singpart{\sone}{\beta\disLCD_\mu\disLCD^\mu \udis{\LCR}}&=  \epsilon\beta\disLCD_\mu \left( \jump{\LCR} n^\mu \del\right) .\label{eq:decomtrDDR2}
\end{align}

\subsection{Singular parts of the second derivative of the (regularised) divergence of the torsion trace in \BSPGT.}

For the singular part of $\beta\disLCD_\mu\disLCD_\nu \disLCD_\rho\udis{T}^\rho$ we get
\begin{align}
   \Singpart{\szero}{\beta\disLCD_\mu\disLCD_\nu \disLCD_\rho\udis{T}^\rho}&=  \epsilon \beta\Big[n_\mu \left(\partial_\nu\jump{ \LCD_\rho T^\rho}-\epsilon  K^\sSigma n_\nu\jump{\LCD_\rho T^\rho}\right) \nonumber\\ 
   &\quad\qquad +K^\sSigma_{\mu\nu}\jump{\LCD_\rho T^\rho} + n_\nu h^\sigma{}_\mu \LCD_\sigma\jump{\LCD_\rho T^\rho}\Big] \del ,\label{eq:decomDDDT1}\\
   \Singpart{\sone}{\beta\disLCD_\mu\disLCD_\nu \disLCD_\rho\udis{T}^\rho}&=  \beta \disLCD_\sigma \left( \jump{\LCD_\rho T^\rho} n_\mu n_\nu n^\sigma \del\right) ,\label{eq:decomDDDT2}
\end{align}
whose traces are:
\begin{align}
   \Singpart{\szero}{\beta\disLCD_\mu\disLCD^\mu \disLCD_\nu\udis{T}^\nu}&=  \epsilon \beta n^\mu \partial_\mu\jump{ \LCD_\nu T^\nu}  \del ,\label{eq:decomtrDDDT1}\\
   \Singpart{\sone}{\beta\disLCD_\mu\disLCD^\mu \disLCD_\nu\udis{T}^\nu}&=  \epsilon\beta \disLCD_\mu \left( \jump{\LCD_\nu T^\nu} n^\mu \del\right) .\label{eq:decomtrDDDT2}
\end{align}
When using these expressions, it is useful to keep in mind that, from \eqref{eq:jumpDV}, and under $\beta \jump{K}=0$ and $\beta \jump{T_\mu}=0$, the following expression holds
\begin{equation}
    \beta\jump{\LCD_\mu T^\mu}= \epsilon \beta n^\mu n^\nu \jump{\LCD_\mu T_\nu} . \label{eq:jumpBetaDivT}
\end{equation}

\chapter{Junction conditions for collapsing dust stars.}
\label{Appendix:stellar collapse formulae}

In this Appendix, we shall obtain the junction conditions resulting from smoothly matching an interior FLRW space-time \eqref{FLRW} with the most general spherically-symmetric line element, across the time-like boundary $\Sigma_*$ given by \eqref{intstellarsurface} in interior coordinates $x_-^\mu=(\tau,\chi,\theta,\varphi)$. The formulae herein shall be extensively used in Chapter \ref{chapter:f(R) collapse}.

\section{Using the areal radius as a coordinate.} \label{ArealRadiusJCAppendix}

We first note that one can always choose `areal-radius' coordinates $x_+^\mu=(t,r,\theta,\varphi)$ such that the exterior metric takes the form
\begin{equation} \label{sphericallysymmetric appendix}
    \dif s^2_+=-A(t,r)\,\dif t^2+B(t,r)\,\dif r^2+r^2\,\dif \Omega^2,
\end{equation}
where $A$ and $B$ are two functions which completely characterise the exterior space-time. In these `areal-radius' coordinates, the matching surface is given by expressions \eqref{stellarsurface out r}. We shall closely follow the treatment of the problem given in Reference \cite{Poisson:2009pwt}.

\subsection{First junction condition.}

On the one hand, as seen from the interior of the star,
\begin{equation} \label{dx-/dy}
    \dfrac{\partial x_-^\mu}{\partial y^a}=\delta^\mu_a,
\end{equation}
and thus the induced metric on the inner side of $\Sigma_*$ is
\begin{equation} \label{h-}
    \dif s_{\Sigma^-_*}^2=-\dif\tau^2+a^2\chi_*^2\,\dif\Omega^2.
\end{equation}
On the other hand, from the exterior,
\begin{equation} \label{dx+/dy}
    \dfrac{\partial x_+^\mu}{\partial y^a}=(\dot{t}_*\,\delta^\mu_t+\dot{r}_*\,\delta^\mu_r)\,\delta^\tau_a+\delta^\mu_\theta\,\delta^\theta_a+\delta^\mu_\varphi\,\delta^\varphi_a.
\end{equation}
Consequently, the induced metric on the outer side of $\Sigma_*$ is given by
\begin{equation} \label{h+}
    \dif s_{\Sigma^+_*}^2=-(A_*\dot{t}_*^2-B_*\dot{r}_*^2)\,\dif\tau^2+r_*^2\,\dif\Omega^2,
\end{equation}
where $A_*(\tau)\equiv A(t_*(\tau),r_*(\tau))$ and $B_*(\tau)\equiv B(t_*(\tau),r_*(\tau))$. The equality of the two induced metrics \eqref{h-} and \eqref{h+} at both sides of $\Sigma_*$ imposes two conditions on the metric functions, namely \eqref{1.1 f(R)} and \eqref{1.2 f(R)}. Equation \eqref{1.2 f(R)} can be conveniently rearranged to produce expression \eqref{betadeff(R)}, which serves as the definition of function $\beta=\beta(\tau)$.

\subsection{Second junction condition.} \label{JC 2 r Appendix}

In order to compute the extrinsic curvature at the junction surface, we first need to determine the unit normal to $\Sigma_*$, $n_\mu$. If $u^\mu$ denotes the four-velocity any fluid element, then $n_\mu$ is completely characterised by spherical symmetry together with the normalisation and orthogonality conditions $g^{\mu\nu} n_\mu n_\nu=1$ and $u^\mu n_\mu=0$ (respectively). In interior (comoving) coordinates,
\begin{equation}
    u_-^\mu=\dfrac{\partial x_-^\mu}{\partial\tau}=\delta^\mu_\tau,
\end{equation}
and $n_\mu^-$ is thus fixed to be
\begin{equation} \label{n-}
    n^-_\mu=\dfrac{a}{\sqrt{1-k\chi_*^2}}\delta^\chi_\mu.
\end{equation}

Combining \eqref{eq:K_ab+-}, \eqref{dx-/dy} and \eqref{n-} one finds that the extrinsic curvature of $\Sigma_*$, as seen from the inside, is
\begin{equation}
    K^-_{ab}=-\dfrac{a}{\sqrt{1-k\chi_*^2}}\Gamma^\chi_{-*ab}.
\end{equation}
Because the only relevant and non-vanishing Christoffel symbol of the interior FLRW space-time is
\begin{equation}
    \Gamma^\chi_{-\theta\theta}=\dfrac{\Gamma^\chi_{-\varphi\varphi}}{\sin^2\theta}=-\chi\,(1-k\chi^2),
\end{equation}
the only non-zero components of $K^-_{ab}$ are
\begin{equation} \label{K-angularOS}
    K^-_{\theta\theta}=\dfrac{K^-_{\varphi\varphi}}{\sin^2\theta}=a\chi_*\sqrt{1-k\chi_*^2}.
\end{equation}

In exterior coordinates, the four-velocity of the fluid is
\begin{equation}
    u_+^\mu=\dfrac{\partial x_+^\mu}{\partial\tau}=\dot{t}_*\delta^\mu_t+\dot{r}_*\delta^\mu_r,
\end{equation}
so the (properly normalised) normal vector is
\begin{equation} \label{unormal+}
    n^+_\mu=\sqrt{A_* B_*}\,(-\dot{r}_*\delta^\mu_t+\dot{t}_*\delta^\mu_r),
\end{equation}
where we have also made a consistent choice of the overall sign. Hence, as seen from outside, the extrinsic curvature of $\Sigma_*$ becomes
\begin{equation}
    \dfrac{K^+_{ab}}{\sqrt{A_* B_*}}=(\ddot{t}_*\dot{r}_*-\dot{t}_*\ddot{r}_*)\delta^\tau_a\delta^\tau_b+(\dot{r}_*\Gamma^t_{+*\rho\sigma}-\dot{t}_*\Gamma^r_{+*\rho\sigma})\dfrac{\partial x_+^\rho}{\partial y^a}\dfrac{\partial x_+^\sigma}{\partial y^b}.
\end{equation}
In order to compute the components of $K_{ab}^+$, we make use of \eqref{dx+/dy} and take into account that the only relevant, non-vanishing Chistoffel symbols are
\begin{equation}
    \begin{gathered}
    \Gamma^t_{+*tt}=\dfrac{A_{t*}}{2A_*},\myhugeskip\Gamma^r_{+*rr}=\dfrac{B_{r*}}{2B_*},\myhugeskip\Gamma^r_{+*\theta\theta}=-\dfrac{r_*}{B_*}, \\
    \Gamma^t_{+*tr}=\Gamma^t_{+*rt}=\dfrac{A_{r*}}{2A_*}, \myhugeskip\myhugeskip\Gamma^t_{+*rr}=\dfrac{B_{t*}}{2A_*}, \\
    \Gamma^r_{+*tr}=\Gamma^r_{+*rt}=\dfrac{B_{t*}}{2B_*},\myhugeskip\myhugeskip\Gamma^r_{+*tt}=\dfrac{A_{r*}}{2B_*},
\end{gathered}
\end{equation}
where subindices $t$ and $r$ respectively denote partial differentiation with respect to $t$ and $r$. Hence, for instance, $B_t(t,r)\equiv\partial B(t,r)/\partial t$ while $B_{t*}(\tau)\equiv B_t(t_*(\tau),r_*(\tau))$, and so forth. (For the remainder of this Appendix, and also in Chapter \ref{chapter:f(R) collapse}, subindices shall denote partial differentiation). After some calculations, we obtain equations
\eqref{2.1 f(R)} and \eqref{2.2 f(R)} corresponding to the second junction condition, with function $\beta$ and parameter $\beta_0$ being respectively given by \eqref{betadeff(R)} and \eqref{beta0 f(R)}.

\subsection{Interlude: Oppenheimer-Snyder collapse in GR.} \label{Oppenheimer-Snyder Appendix}

As explained before, GR has only two junction conditions: the first one, $\jump{h_{ab}}=0$, and the second one, $\jump{K_{ab}}=0$. Therefore, the relevant junction conditions in this case are equations \eqref{1.1 f(R)} and \eqref{1.2 f(R)}, \eqref{2.1 f(R)} and \eqref{2.2 f(R)}. Moreover, in GR, the exterior can only be Schwarzschild, which satisfies
\begin{equation}
    A(t,r)=\dfrac{1}{B(t,r)}=1-\dfrac{2GM}{r},
\end{equation}
and thus function $\beta$, defined in \eqref{betadeff(R)}, reduces to
\begin{equation} \label{beta OS appendix 1}
    \beta(\tau)=\sqrt{\dot{r}_*^2(\tau)+A_*(\tau)}.
\end{equation}
The Schwarzschild metric is independent of $t$. As a result, \eqref{2.1 f(R)} yields $\dot{\beta}=0$. This is compatible with the other constraint coming from the second junction condition, \eqref{2.2 f(R)}, which reduces to
\begin{equation} \label{beta OS appendix 2}
    \beta(\tau)=\beta_0=\sqrt{1-k\chi_*^2}=\const
\end{equation}
Combining \eqref{beta OS appendix 1} with \eqref{beta OS appendix 2}, one finds that, in Oppenheimer-Snyder collapse,
\begin{equation} \label{JC 2 OS}
    \dot{r}_*^2(\tau)=\beta_0^2-A_*(\tau)=-k\chi_*^2+\dfrac{2GM}{r_*(\tau)}.
\end{equation}
Substituting \eqref{1.1 f(R)} and the equation for $a(\tau)$, which in GR is \eqref{cycloid}, one finally finds that junction conditions require the following algebraic constraint to hold:
\begin{equation}
    M=\dfrac{k}{2G}\chi_*^3=\dfrac{4\pi}{3}\rho_0\chi_*^3.
\end{equation}

\subsection{Third and fourth junction conditions.}

The first new junction condition arising from $f(\LCR)$ gravity is the continuity of the Ricci scalar at $\Sigma_*$, i.e.~$\jump{\LCR}=0$.\footnote{It is worth mentioning that the Ricci scalar of the exterior solution is given in terms of functions $A$ and $B$ by
\begin{equation*}
    \LCR^{+}=-\dfrac{A_{rr}}{AB}+\dfrac{A_r}{2AB}\left(\dfrac{A_r}{A}+\dfrac{B_r}{B}\right)-\dfrac{2}{r}\left(\dfrac{A_r}{AB}-\dfrac{B_r}{B^2}\right)+\dfrac{2}{r^2}\left(1-\dfrac{1}{B}\right)+\dfrac{B_{tt}}{AB}-\dfrac{B_t}{2AB}\left(\dfrac{A_t}{A}+\dfrac{B_t}{B}\right).
\end{equation*}
} In terms of the scale factor of the interior FLRW space-time, this condition reads
\begin{equation} \label{3 f(R) appendix partial}
    \LCR^+_*=\LCR^-_*=6\left(\dfrac{\dot{a}^2+k}{a^2}+\dfrac{\ddot{a}}{a}\right),
\end{equation}
Equation \eqref{1.1 f(R)} can be employed to reexpress \eqref{3 f(R) appendix partial} in terms of $r_*$ and its derivatives, yielding \eqref{3 f(R)}.

The other novel junction condition coming from $f(\LCR)$ gravity is the continuity of the normal derivative of the Ricci scalar at the stellar surface, i.e.~$n^\mu \jump{\LCD_\mu\LCR}=0$. As seen from inside the star, this normal derivative is simply
\begin{equation}
    g_{-*}^{\mu\nu}n_\mu^-\partial_\nu\LCR^-_*=\LCR^-_{\chi*}=0.
\end{equation}
As seen from the exterior, the normal derivative of $\LCR_+$ does not vanish in principle:
\begin{equation}
    g_{+*}^{\mu\nu}n_\mu^+\partial_\nu \LCR^+_*=-\sqrt{A_* B_*}\left(\dfrac{\dot{r}_*}{A_*}\LCR_{r*}^++\dfrac{\dot{t}_*}{B_*}\LCR_{t*}^+\right).
\end{equation}
This forces us to require \eqref{4 f(R)} for the fourth junction condition to be accomplished.

\section{Without using the areal radius as a coordinate.}
\label{NOTArealRadiusJCAppendix}

There most general, spherically-symmetric exterior line element can always be cast as
\begin{equation} \label{exteriorx}
    \dif s^2_+=-C(\eta,\xi)\,\dif \eta^2+D(\eta,\xi)\,\dif\xi^2+r^2(\eta,\xi)\,\dif \Omega^2.
\end{equation}
As we can see, the difference between \eqref{exteriorx} and \eqref{sphericallysymmetric appendix} is that the areal radius $r$ is not a coordinate, but a function of the new time variable $\eta$ and the new spatial coordinate $\xi$ instead. We intend to develop the junction conditions resulting from gluing \eqref{FLRW} and \eqref{exteriorx} across the stellar surface $\Sigma_*$ using coordinates $(\eta,\xi,\theta,\varphi)$, and then compare the results with those of Appendix \ref{ArealRadiusJCAppendix}.

As our starting point, we must note that the stellar surface is now given by
\begin{equation} \label{sigmax}
    \eta=\eta_*(\tau),\mybigskip \xi=\xi_*(\tau)
\end{equation}
in these coordinates. Additionally, we now have that
\begin{equation} \label{r(t,r)}
    r=r(\eta,\xi)\myskip\Rightarrow\myskip r_*(\tau)=r_*(\eta_*(\tau),\xi_*(\tau)).
\end{equation}

\subsection{First junction condition.}

Since, as seen from outside the star,
\begin{equation} \label{dx+/dy x}
    \dfrac{\partial x_+^\mu}{\partial y^a}=(\dot{\eta}_*\,\delta^\mu_\eta+\dot{\xi}_*\,\delta^\mu_\xi)\,\delta^\tau_a+\delta^\mu_\theta\,\delta^\theta_a+\delta^\mu_\varphi\,\delta^\varphi_a,
\end{equation}
the induced metric on the exterior side of $\Sigma_*$ will be
\begin{equation} \label{h+ x}
    \dif s_{\Sigma^+_*}^2=-(C_*\dot{\eta}_*^2-D_*\dot{\xi}_*^2)\,\dif\tau^2+r_*^2\,\dif\Omega^2,
\end{equation}
where $C_*\equiv C(\eta_*(\tau),\xi_*(\tau))$ and $D_*\equiv D(\eta_*(\tau),\xi_*(\tau))$. Equalling the induced metrics \eqref{h-} and \eqref{h+ x}, one obtains two conditions on the metric functions, namely equation \eqref{1.1 f(R)}---which remains unchanged---and
\begin{equation} \label{1.2 f(R) x}
    C_*\dot{\eta}_*^2-D_*\dot{\xi}_*^2=1.
\end{equation}

Equation \eqref{1.2 f(R) x} can be conveniently rearranged as
\begin{equation} \label{betadeff(R) x}
    C_*\,\dot{\eta}_*=\sqrt{C_*+C_*D_*\dot{\xi}_*^2}\equiv\tilde{\beta},
\end{equation}
which serves as the definition of function $\tilde{\beta}=\tilde{\beta}(\tau)$.

\subsection{Second junction condition.}

The four-velocity of the fluid is now
\begin{equation}
    u_+^\mu=\dfrac{\partial x_+^\mu}{\partial\tau}=\dot{\eta}_*\delta^\mu_\eta+\dot{\xi}_*\delta^\mu_\xi,
\end{equation}
so the (properly normalised) normal vector is
\begin{equation} \label{unormal+ x}
    n^+_\mu=\sqrt{C_* D_*}\,(-\dot{\xi}_*\delta^\mu_\eta+\dot{\eta}_*\delta^\mu_\xi).
\end{equation}
Therefore, the extrinsic curvature of $\Sigma_*$ expressed in exterior coordinates is
\begin{equation} \label{K+tr x}
    \dfrac{K^+_{ab}}{\sqrt{C_* D_*}}=(\ddot{\eta}_*\dot{\xi}_*-\dot{\eta}_*\ddot{\xi}_*)\delta^\tau_a\delta^\tau_b+(\dot{\xi}_*\Gamma^\eta_{+*\alpha\beta}-\dot{\eta}_*\Gamma^\xi_{+*\alpha\beta})\dfrac{\partial x_+^\alpha}{\partial y^a}\dfrac{\partial x_+^\beta}{\partial y^b}.
\end{equation}
Proceeding exactly as we did back in Appendix \ref{JC 2 r Appendix}, we obtain, after a rather long computation, the following two independent equations for the second junction condition:
\begin{gather}
    \dot{\tilde{\beta}}=\dfrac{C_{\eta*}\dot{\eta}_*^2-D_{\eta*}\dot{\xi}_*^2}{2}, \label{2.1 f(R) x} \\
    \tilde{\beta}=\dfrac{\beta_0\sqrt{C_* D_*}-r_{\eta*}D_*\dot{\xi}_*}{r_{\xi*}}, \label{2.2 f(R) x}
\end{gather}
where $\beta_0$ is given again by expression \eqref{beta0 f(R)}, $C_\eta\equiv\partial C/\partial\eta$, $D_\eta\equiv\partial D/\partial\eta$, $r_\eta\equiv\partial r/\partial\eta$ and $r_\xi\equiv\partial r/\partial\xi$.

\subsection{Third and fourth junction conditions.}

It is almost immediate to check that the third junction condition still leads to equation \eqref{3 f(R)}.\footnote{Obviously, \eqref{r(t,r)} must be taken into account in this case. \eqref{r(t,r)} entails that $\dot{r}_*=r_{\eta*}\dot{\eta}_*+r_{\xi*}\dot{\xi}_*$, so the third junction condition is much more convoluted in these coordinates.} Finally, the fourth junction condition now yields\footnote{We would like to remark that, in these coordinates, the Ricci scalar of the exterior solution is given by
\begin{align*}
    \LCR^{+}&=-\dfrac{C_{\xi\xi}}{CD}+\dfrac{C_\xi}{2CD}\left(\dfrac{C_\xi}{C}+\dfrac{D_\xi}{D}\right)-\dfrac{2r_\xi}{r}\left(\dfrac{C_\xi}{CD}-\dfrac{D_\xi}{D^2}\right)+\dfrac{2}{r^2}\left(1+\dfrac{r_\eta^2}{C}-\dfrac{r_\xi^2}{D}\right)+\dfrac{D_{\eta\eta}}{CD}-\dfrac{D_\eta}{2CD}\left(\dfrac{C_\eta}{C}+\dfrac{D_\eta}{D}\right)\nonumber\\ &\quad+\dfrac{4}{r}\left(\dfrac{r_{\eta\eta}}{C}-\dfrac{r_{\xi\xi}}{D}\right)-\dfrac{2}{r^2}\left(\dfrac{C_\eta}{C^2}-\dfrac{D_\eta}{CD}\right).
\end{align*}}
\begin{equation} \label{4 f(R) x}
    \dfrac{\dot{\xi}_*}{C_*}\LCR_{\xi*}^++\dfrac{\dot{\eta}_*}{D_*}\LCR_{\eta*}^+.
\end{equation}
As we can clearly see, the junction conditions for \eqref{exteriorx} reduce to \eqref{1.1 f(R)}--\eqref{2.2 f(R)}, \eqref{3 f(R)} and \eqref{4 f(R)} if one sets $r=\xi$ and performs the substitutions $\eta\rightarrow t$, $C\rightarrow A$ and $D\rightarrow B$, since $\tilde{\beta}$ also reduces to $\beta$---i.e.~to expression \eqref{betadeff(R)}---under these circumstances.

\subsection{Summary and comparison with the results of Appendix \ref{ArealRadiusJCAppendix}.}

To sum up, the relevant junction conditions arising from the smooth matching of a FLRW dust star interior \eqref{FLRW} and \eqref{exteriorx} at \eqref{sigmax} are equations \eqref{1.1 f(R)}, \eqref{3 f(R)}, \eqref{1.2 f(R) x} and \eqref{2.1 f(R) x}--\eqref{4 f(R) x}. These constraints are to be complemented with the definition of $\tilde{\beta}$, equation \eqref{betadeff(R) x}.

Junction conditions \eqref{1.2 f(R) x} and \eqref{2.1 f(R) x}--\eqref{4 f(R) x} are more complex than their counterparts \eqref{1.2 f(R)}--\eqref{2.2 f(R)} and \eqref{4 f(R)}. This was to be expected, since coordinates $(\eta,\xi,\theta,\varphi)$ are more general than $(t,r,\theta,\varphi)$. As a result, computations should be---in principle---much more difficult to perform when the exterior space-time is expressed as in \eqref{exteriorx}.

For example, the expressions for $\dot{\eta}_*$ and $\dot{\xi}_*$ in terms of $C_*$ and $D_*$ are not as simple as the expressions \eqref{tp} and \eqref{rp} for $\dot{t}_*$ and $\dot{r}_*$ in terms of $A_*$ and $B_*$. This is because equation \eqref{2.2 f(R) x} includes a term proportional to $\dot{\xi}_*$ in its right-hand side; in contrast, \eqref{2.2 f(R)} does not contain any term proportional to $\dot{r}_*$. We thus have a quadratic equation for $\dot{\xi}_*$ after substituting \eqref{2.2 f(R) x} in \eqref{betadeff(R) x}. From this quadratic equation one obtains the following two solutions for $\dot{\xi}_*$:
\begin{equation} \label{xistar}
    \dot{\xi}_*=\dfrac{\sqrt{C_*D_*}\left(r_{\eta*} \beta_0\pm r_{\xi*}\sqrt{\dfrac{r_{\eta*}^2+C_*\beta_0^2}{D_*}-\dfrac{C_*}{D_*^2}r_{\xi*}^2}\right)}{D_* r_{\eta *}^2-C_* r_{\xi *}^2}.
\end{equation}
Substituting \eqref{xistar} in \eqref{betadeff(R) x} yields the two corresponding solutions for $\dot{\eta}_*$. Again, it is straightforward to check that both solutions for $\dot{\xi}_*$ and $\dot{\eta}_*$ respectively reduce to \eqref{rp} and \eqref{tp} if one sets $r=\xi$ and performs the substitutions $\eta\rightarrow t$, $C\rightarrow A$ and $D\rightarrow B$. Nonetheless, the highly convoluted appearance of $\dot{\xi}_*$ and $\dot{\eta}_*$ implies that using them to build a system of equations for metric functions $C$, $D$ and $r$ is in principle much more complicated than obtaining a system of equations for functions $A$ and $B$ using \eqref{rp} and \eqref{tp}, as we did back in Section \ref{ruling out}.

Accordingly, we clearly see that `areal-radius' coordinates $(t,r,\theta,\varphi)$ are more natural, in the sense that $r_*$ is always the quantity which becomes proportional to the interior scale factor $a$ as per the first junction condition \eqref{1.1 f(R)}, which remains unmodified when one abandons `areal-radius' coordinates and switches to $(\eta,\xi,\theta,\varphi)$. Junction conditions are also more difficult to handle when one uses the alternative coordinate system $(\eta,\xi,\theta,\varphi)$. This could potentially cause problems when the exterior space-time cannot be analytically cast in the form \eqref{sphericallysymmetric appendix} using a coordinate transformation.

\chapter{\titlebm{$\LCR_0$}-degenerate \titlebm{$f(\LCR)$} models. Discussions and useful formulae.}
\label{appendix:constant-curvature}

\section[Some caveats concerning the use of the Einstein frame in \titlemath{$\LCR_0$}-degenerate \titlemath{$f(\LCR)$} models.]{Some caveats concerning the use of the Einstein frame in \titlebm{$\LCR_0$}-degenerate \titlebm{$f(\LCR)$} models.} \label{Appendix: Caveats}

In this Appendix, we shall present some of the subtleties arising when using the Einstein-frame representation \eqref{Einstein-frame metric}--\eqref{scalaron potential} of $\LCR_0$-degenerate $f(\LCR)$ models, and in particular when dealing with their constant-curvature solutions and their stability.

The first caveat is that conformal transformation \eqref{Einstein-frame metric} leading to the Einstein frame becomes singular when evaluated on $f(\LCR)$-exclusive constant-curvature solutions, since the existence of those solutions requires $f'(\LCR_0)=0$. In consequence, the dynamical equivalence between the Einstein and Jordan frames appears to break down in this scenario. However, given that the transformation between frames becomes singular only if $\LCR=\LCR_0$ (i.e.~in a null-measure set of values of $\LCR$), one may argue that the transformation does not indeed fail provided that all the relevant physical quantities remain well-defined after taking the limit $\LCR\rightarrow\LCR_0$.\footnote{Under this scope, the fact that the conformal transformation \eqref{Einstein-frame metric} becomes singular at $\LCR=\LCR_0$ is simply a reflection of the fact that the scalaron \eqref{scalaron} tends to negative infinity as $\LCR\rightarrow\LCR_0$.} In particular, the scalaron potential naively evaluates to a $0/0$ indetermination for $\LCR_0$-degenerate constant-curvature solutions (as mentioned in Section \ref{Section: Stability of solutions}). Only in cases where said indetermination can be resolved we shall consider the Einstein-frame to be a valid representation of the $f(\LCR)$ dynamics.

A second crucial observation is that, in our stability analysis, we have treated the scalaron potential $V$ as a function of $\LCR$ instead of as a function of the scalaron $\phi$, the reason being that it is more enlightening to perturb the scalar curvature instead of the abstract scalaron. However, in doing so, one must check that the extrema of $V$ as a function of $\LCR$ coincide with those of $V$ seen as a function of $\phi$. This is indeed the case provided that $f''(\LCR)\neq 0$. More precisely, given that
\begin{equation}
    \dfrac{\dif V}{\dif\LCR}=\dfrac{\dif\phi}{\dif\LCR}\dfrac{\dif V}{\dif\phi}=\sqrt{\dfrac{3}{2\kappa}}\dfrac{f''(\LCR)}{f'(\LCR)}\dfrac{\dif V}{\dif\phi},
\end{equation}
the zeroes of $\dif V/\dif R$ coincide with those of $\dif V/\dif\phi$ if and only if $f''(\LCR)\neq 0$. Moreover, one also has that
\begin{equation}
    \dfrac{\dif^2 V}{\dif\LCR^2}=\dfrac{\dif^2\phi}{\dif\LCR^2}\dfrac{\dif V}{\dif\phi}+\left(\dfrac{\dif\phi}{\dif\LCR}\right)^2\dfrac{\dif^2 V}{\dif\phi^2},
\end{equation}
by virtue of which
\begin{equation}
    \sign\left[\left.\dfrac{\dif^2 V}{\dif\LCR^2}\right|_{\frac{\dif V}{\dif\phi}=0}\right]=\sign\left[\dfrac{\dif^2 V}{\dif\phi^2}\right],
\end{equation}
i.e.~the character of the extremum (maximum, minimum or saddle point) does not change when one regards the scalaron potential to be a function of $\LCR$ instead of $\phi$.

Finally, it is worth mentioning that
\begin{equation}
    \dfrac{\dif V}{\dif\LCR}=\dfrac{f''(\LCR)}{2\kappa}\dfrac{2f(\LCR)-f'(\LCR) \LCR}{f'^3(\LCR)}.
\end{equation}
Thus, a constant-curvature solution with $\LCR=\LCR_0$ will extremise the scalaron potential---with respect to both $\LCR$ and $\phi$---provided that (i) $f''(\LCR_0)\neq 0$, (ii) $f'(\LCR_0)\neq 0$ and (iii) the trace \eqref{reduced f(R) EOM trace} of the equations of motion holds. If $f'(\LCR_0)=0$, i.e.~in $\LCR_0$-degenerate models, the stability analysis is more complicated and must be performed as done in Section \ref{Section: Stability of solutions}, as the potential and its derivatives might not be well defined in the limit $\LCR\rightarrow\LCR_0$. Moreover, when $f'(\LCR_0)=0$, mere compliance with equation \eqref{reduced f(R) EOM trace}---which now holds automatically---does not guarantee that $\LCR_0$-degenerate solutions extremise the potential. Results \ref{Result: R0 instability} and \ref{Result: R0=0 metastability} provide the conditions under which $\LCR_0$-degenerate constant-curvature solutions are not stable.

In order to shed more light on these issues, in Appendix \ref{Appendix: A simple model} below we shall consider a particular instance of $f(\LCR)$ model hosting both unstable, degenerate constant-curvature solutions and stable, non-degenerate constant-curvature solutions.

\section{A simple illustrative model.} \label{Appendix: A simple model}

For illustrative purposes, let us consider the simple, one-parameter model
\begin{equation} \label{simple model}
    f(\LCR)=\LCR_*-\LCR+\LCR\ln\left(\dfrac{\LCR}{\LCR_*}\right),
\end{equation}
where $\LCR_*>0$ is a constant with units of inverse length squared. This model harbours constant-curvature solutions for two different values of $\LCR$, namely:
\begin{coloritemize}
    \item $\LCR_0=\LCR_*$. Since $f(\LCR_*)=0$ and $f'(\LCR_*)=0$, all space-times with $\LCR_0=\LCR_*$ trivially solve the equations of motion associated to \eqref{simple model}, i.e.~the model is $\LCR_*$-degenerate.
    \item $\LCR_0=\eta\LCR_*$, with $\eta=4.92155...$ being the non-trivial solution of $2-2\eta+\eta\ln\eta=0$ ($\eta=1$ is the trivial solution of this transcendental equation, leading to the case $\LCR_0=\LCR_*$ considered above). In this scenario, $f(\eta\LCR_*)\neq 0$ and $f'(\eta\LCR_*)\neq 0$, but the trace of the equations of motion \eqref{reduced f(R) EOM trace} holds. As a result, the $f(\LCR)$ equations of motion \eqref{f(R) EOM} reduce to the Einstein equations with cosmological constant \eqref{reduced f(R) EOM}, and model \eqref{simple model} admits the same constant-curvature solutions as $\text{GR}+\Lambda$, provided that $\Lambda=\eta\LCR_*/4$, without being $(\eta\LCR_*)$-degenerate.
\end{coloritemize}
Due to Result \ref{Result: R0 instability}, the constant-curvature solutions of \eqref{simple model} having $\LCR_0=\LCR_*$ are all unstable. However, the non-degenerate constant-curvature solutions with $\LCR_0=\eta\LCR_*$ are all stable. Indeed, it is straightforward to check that (i) $f''(\eta\LCR_*)\neq 0$, (ii) $f'(\eta\LCR_*)\neq 0$ and (iii) equation \eqref{reduced f(R) EOM trace} is satisfied for constant-curvature solutions with $\LCR_0=\eta\LCR_*$. Thus, as per the discussion in Appendix \ref{Appendix: Caveats}, non-degenerate solutions having $\LCR_0=\eta\LCR_*$ extremise the scalaron potential, regardless of whether it is understood to be a function of $\LCR$ or of $\phi$. Moreover,
\begin{equation}
    \left.\dfrac{\dif V}{\dif\LCR}\right|_{\LCR=\eta\LCR_*}=\dfrac{\eta}{8\LCR_*}\dfrac{\eta-2}{(\eta-1)^3}>0,
\end{equation}
given that $\LCR_*>0$ and $\eta>2$. Thus, $\LCR_0=\eta\LCR_*$ is a minimum of the scalaron potential. It can be shown that there are no other extrema in the region $\LCR>\LCR_0$, and given that $V(\LCR)\rightarrow+\infty$ as $\LCR\rightarrow\LCR_0^+$, we conclude that non-degenerate constant-curvature solutions of model \eqref{simple model} having $\LCR_0=\eta\LCR_*$ are stable.

\section{Quantities and formulae of interest for the physical characterisation of solutions.} \label{Appendix: Characterisation}

Given the extraordinary amount of symmetry exhibited by static, spherically symmetric space-times of the form \eqref{static spherically symmetric}, the study of all the points listed at the beginning of Section \ref{section:constant-curvature solutions} simplifies considerably. For any given solution, one essentially needs to determine the values of $r$ where functions $A(r)$ and $B(r)$ either vanish or become infinite, and which of these points or regions are pathological, as explained in what follows. As stated in the bulk of the text, this Appendix is also meant to further clarify our nomenclature and symbol conventions.

\subsection{Apparent and Killing horizons.}

It can be shown that, when a static, spherically symmetric space-time is expressed in the Abreu-Nielsen-Visser gauge \eqref{Abreu-Nielsen-Visser} or, equivalently, using standard areal-radius coordinates \eqref{static spherically symmetric}, its apparent horizons are located at the \textit{simple} roots $r_\mathrm{ah}$ of the algebraic equation $g^{rr}(r_\mathrm{ah})=0$.\footnote{In addition, space-times \eqref{Abreu-Nielsen-Visser} possess a wormhole throat wherever $g^{rr}$ has a \textit{double} root.} In other words, \eqref{static spherically symmetric} has an apparent horizon at $r=r_\mathrm{ah}$ provided that $B(r_\mathrm{ah})\rightarrow\infty$. This occurs whenever
\begin{equation} 
\label{apparent horizon ANV gauge}
    r_\mathrm{ah}=2G M_\mathrm{MSH}(r_\mathrm{ah})
\end{equation}
has a single root.

On the other hand, $\xi=\partial/\partial t$ is a Killing vector of every line element of the form \eqref{static spherically symmetric}, since they are all static. As a result, these space-times will harbour a Killing horizon provided that Killing vector $\xi$ becomes null in some region of space-time. In areal-radius coordinates $(t,r,\theta,\varphi)$, $\xi^\mu=\delta^\mu_{\hphantom{\mu}t}$, so the norm of $\xi$ is given by
\begin{equation} \label{Killing norm}
    g_{\mu\nu}\xi^\mu\xi^\nu=g_{tt}=-A(r).
\end{equation}
We then conclude that space-times of the form \eqref{static spherically symmetric} will host a Killing horizon at $r=r_\mathrm{Kh}$ provided that $r_\mathrm{Kh}$ is a root of
\begin{equation} 
\label{Killing horizon static spherically symmetric areal}
    A(r_\mathrm{Kh})=0.
\end{equation}
Expressions \eqref{apparent horizon ANV gauge} and \eqref{Killing horizon static spherically symmetric areal} imply that line element \eqref{static spherically symmetric} exhibits coordinate singularities at the locations of the apparent and Killing horizons, respectively. However, we must stress that the aforementioned conditions \eqref{apparent horizon ANV gauge} and \eqref{Killing horizon static spherically symmetric areal} are obtained by computing scalar quantities, which should remain invariant even when they are expressed in singular coordinates, such as the areal radius coordinates. For the purpose of our analysis, this level of rigour shall be sufficient; we are nonetheless aware that a more detailed and mathematically precise computation can be performed.

\subsection{Singularities and geodesic completeness.}

The existence of coordinate singularities---points where $A(r)$ and/or $B(r)$ become zero or infinite---may also point out the presence of curvature singularities. The existence of such curvature singularities, however, must be determined in a coordinate-invariant way; for example, we will compute the Kretschmann scalar
\begin{equation}
    \LCKre=\LCR_{\mu\nu\rho\sigma}\LCR^{\mu\nu\rho\sigma}
\end{equation}
for each solution, and then determine the values of $r$ where this quantity diverges.

The mere existence of curvature singularities is not a sign of unphysical dynamics \textit{per se}, even though they represent points or regions in which tidal forces become infinite. Should these singularities be unreachable in finite affine parameter for causal observers, then none of such observers would experience infinite tidal forces at any point along their world-lines.

In order to compute whether a photon can reach a singularity for a finite value of the affine parameter, we will analyse the corresponding geodesic equation. Given that all the space-times we are considering are spherically symmetric, we may always choose, without loss of generality, to perform all computations on the equatorial plane ($\theta=\pi/2$). As a result, the equation for the null geodesics of \eqref{static spherically symmetric} reduces to
\begin{equation} \label{static spherically symmetric null geodesic eq equatorial}
    \left(\dfrac{\dif r}{\dif\lambda}\right)^2=\dfrac{E^2}{A(r)B(r)}-\dfrac{L^2}{r^2B(r)}\,,
\end{equation}
where $\lambda$ is the affine parameter of the trajectory and $E$ and $L$ are the observer's conserved quantities, namely
\begin{gather}
    E=A(r)\,\dfrac{\dif t}{\dif\lambda}=\const\,, \label{geodesic energy} \\
    L=r^2\,\dfrac{\dif\varphi}{\dif\lambda}=\const
\end{gather}
From expression \eqref{geodesic energy}, it is evident that one can always rescale $\lambda$ in such a way that $E=1$. Accordingly, a photon with unit energy travelling from a given $r_\mathrm{ini}$ following a radial geodesic ($L=0$) takes the following variation of affine parameter $\lambda$ to reach any other value of the areal radius $r_\mathrm{fin}$:
\begin{equation} \label{Delta lambda}
    \Delta\lambda(r_\mathrm{ini}\rightarrow r_\mathrm{fin})=\left|\int_{r_\mathrm{ini}}^{r_\mathrm{fin}}\dif r\,\sqrt{A(r)B(r)}\right|\,.
\end{equation}
(Notice that an absolute value has been intentionally included in the previous expression in order to produce the same value of $\Delta\lambda(r_\mathrm{ini}\rightarrow r_\mathrm{fin})$ regardless of whether $r_\mathrm{ini}<r_\mathrm{fin}$ or $r_\mathrm{ini}>r_\mathrm{fin}$.) In most practical cases, the evaluation of \eqref{Delta lambda} will reduce to
\begin{equation} \label{lambda difference}
    \Delta\lambda(r_\mathrm{ini}\rightarrow r_\mathrm{fin})=|\lambda(r_\mathrm{fin})-\lambda(r_\mathrm{ini})|\,,
\end{equation}
where we have introduced the primitive
\begin{equation} \label{lambda primitive}
    \lambda(r)\equiv\int\dif r\,\sqrt{A(r)B(r)},
\end{equation}
defined up to an arbitrary constant, which is relevant for dimensional purposes, but disappears when computing $\Delta\lambda(r_\mathrm{ini}\rightarrow r_\mathrm{fin})$. It is evident that \eqref{lambda difference} does not hold when the integrand $\sqrt{A(r)B(r)}$ changes sign within the integration interval, as is the case with some of the space-times and intervals considered in Section \ref{section:constant-curvature solutions}.

\subsection{Regions in which the metric signature becomes unphysical.}

Last but not least, the existence of zeroes of functions $A$ and $B$ can also lead to changes in the metric signature. For example, if $A$ becomes negative for some values of $r$, then the metric becomes Euclidean (i.e.~all coordinates become space-like). If, on the contrary, $B$ becomes negative, then there are two time coordinates. Both situations are clearly unphysical and should be avoided. In particular, we can use again equation \eqref{Delta lambda} to know whether such regions with unphysical metric signatures can be reached by causal observers in finite affine parameter.

\section{Alternative derivation of the Class 1 line element.} 
\label{staticsol}

Let us now conclude this Appendix with an alternative derivation of the Class 1 line element \eqref{Class 1.1}, first appearing in Publication \cite{Casado-Turrion:2022xkl}. While the aforementioned article deals with stellar collapse in generic $f(\LCR)$ gravity models, we were able to obtain the (static) Class 1 metric by allowing the stellar surface to be static, i.e.~abandoning the assumption that the star is collapsing. Additionally, we found therein that Class 1 line elements can be matched to an interior Minkowski space-time, forming a vacuole-like solution.

Our starting point shall be the matching between the dust-star FLRW interior \eqref{FLRW} and a static exterior space-time of the form \eqref{MetricStaticAB} which will be a solution for some---still unspecified---$f(\LCR)$ gravity model. As seen in Section \ref{Proofs Results 1 2 Corollary}, junction condition \eqref{static beta dot} is satisfied provided that
\begin{equation}
\label{BNewSolution}
    B(r)=\dfrac{1}{\beta_0^2}=\const,
\end{equation}
although this constraint together with \eqref{1.1 f(R)} and \eqref{rp} would imply that the interior solution is also static:
\begin{equation}
    \dot{a}(\tau)=\dfrac{\dot{r}_*(\tau)}{\chi_*}=0.
\end{equation}
Using the standard normalisation $a(0)=1$ (which entails that $r_*=\chi_*=\const$), coordinate $\chi$ reduces to the areal radius $r$. The interior metric then becomes
\begin{equation} \label{staticint}
    \dif s_-^2=-\dif\tau^2+\dfrac{\dif r^2}{1-k r^2}+r^2\dif\Omega^2.
\end{equation}
As a straightforward consequence of this, both $f(\LCR^-)$ and the stellar density $\rho=\rho_0$ become constant in $\tau$ as well. The trace of the $f(\LCR)$ equations of motion for the interior space-time \eqref{staticint} reads
\begin{equation} 
\label{static trace}
    f'(\LCR^-)\LCR^- -2f(\LCR^-)=\kappa\stress^-=\kappa\rho_0.
\end{equation}
Assuming that $\LCR^-$ is a variable and not a value, expression \eqref{static trace} can be interpreted as a differential equation for $f(\LCR^-)$. This equation may be immediately integrated, revealing that \eqref{staticint} is a solution of
\begin{equation} \label{static theory}
    f(\LCR)=\alpha\LCR^2+\dfrac{\kappa\rho_0}{2},
\end{equation}
where $\alpha$ is an integration constant.\footnote{Notice that the statement that the correct theory is \eqref{static theory} is stronger than the assertion that Equation \eqref{static trace} holds for \eqref{staticint}, since \eqref{static theory} would accomplish \eqref{static trace} for every (interior) solution, not just \eqref{staticint}. Moreover, as we shall see later, the spurious dependence of $f(\LCR)$ on $\rho_0$ disappears when one takes into account the (vacuum) equations of motion for the exterior solution, which force $\rho_0=0$.}

On the other hand, should the interior solution have constant curvature scalar $\LCR^-$, the full set of equations of motion of $f(\LCR)$ gravity \eqref{f(R) EOM} become
\begin{equation} \label{EOM static theory}
    f'(\LCR^-)\LCR_{\mu\nu}^- -\dfrac{f(\LCR^-)}{2}g_{\mu\nu}^-=\kappa \stress_{\mu\nu}^-.
\end{equation}
Evaluating equations \eqref{EOM static theory} for theory \eqref{static theory} together with the static, constant-curvature interior metric \eqref{staticint}, one finds that equations \eqref{EOM static theory} only impose one additional constraint relating $k$ and $\rho_0$ as follows:
\begin{equation}
    k^2=\dfrac{\kappa\rho_0}{24\alpha}.
\end{equation}

We shall now see how the exterior solution is fixed by the remaining junction conditions, namely, the third and the fourth ones, as given by \eqref{3 f(R)} and \eqref{all t and r 4 f(R)}, respectively. Since $B(r)$ is given by \eqref{BNewSolution}, then
equation \eqref{all t and r 4 f(R)} is automatically satisfied, while equation \eqref{3 f(R)} becomes
\begin{equation}
    \LCR^+=\dfrac{6kr_*^2}{r^2}=\dfrac{6(1-\beta_0^2)}{r^2}.
\end{equation}
The left-hand side of this expression is obtained by substituting \eqref{BNewSolution} in the expression for $\LCR^+$:
\begin{equation}
    R^+(r)=\dfrac{2(1-\beta_0^2)}{r^2}-\beta_0^2\left[\dfrac{A''(r)}{A(r)}-\dfrac{A'^2(r)}{2A^2(r)}+\dfrac{2}{r}\dfrac{A'(r)}{A(r)}\right].
\end{equation}
Equating the two previous expressions, and solving the resulting ordinary differential equation for $A(r)$, one finds
\begin{equation}
    A(r)=D\,r^{-(\Delta+1)}\,(C+r^{\Delta})^2,
\end{equation}
where
\begin{equation}
    \Delta\equiv\sqrt{9-\dfrac{8}{\beta_0^2}},
\end{equation}
while $C$ and $D$ are integration constants. Notice that $D$ can be absorbed inside $\dif t^2$ by a redefinition of the time coordinate $Dt\rightarrow t$. In what follows, we may thus set, without loss of generality, $D=1$. Bearing all of this in mind, the exterior metric becomes
\begin{equation} \label{staticext}
    \dif s^2_+=-r^{-(\Delta+1)}\,(C+r^{\Delta})^2\,\dif t^2 +\dfrac{\dif r^2}{\beta_0^2} +r^2\,\dif\Omega^2.
\end{equation}
By construction, metrics \eqref{staticint} and \eqref{staticext} satisfy all the junction conditions at any given $r=r_*=\const$ Moreover, we have found that \eqref{staticint} solves the field equations of $f(\LCR)$ gravity provided that $f(R)=\alpha R^2+\kappa\rho_0/2$. Consequently, we must confirm whether the exterior space-time \eqref{staticext} is a vacuum solution for this class of $f(\LCR)$ models. By inserting the exterior line element \eqref{staticext} in the vacuum equations of motion of \eqref{static theory}, one finds that the latter are only satisfied provided that $\beta_0^2=1$, which, in turn, implies
\begin{equation} \label{Delta = 1 consequences}
    \Delta=1,\myhugeskip k=0\myskip\Rightarrow\myskip\rho_0=0.
\end{equation}
This entails that the $f(\LCR)$ models \eqref{static theory} reduce to 
\begin{equation} \label{R2}
    f(\LCR)=\alpha\LCR^2,
\end{equation}
i.e.~purely quadratic $f(\LCR)$ gravity.

Because of \eqref{Delta = 1 consequences}, the interior solution  reduces to Minkowski space-time. The exterior metric \eqref{staticext}, however, becomes
\begin{equation} \label{novelstatic}
    \dif s^2_+=-\left(1+\dfrac{C}{r}\right)^2\,\dif t^2 +\dif r^2+r^2\,\dif\Omega^2,
\end{equation}
which is the Class 1 line element discussed in Section \ref{Section: Class 1}. As discussed therein, Class 1 space-times turn out to be not only a solution of $f(\LCR)=\alpha\LCR^2$, but of \textit{any} theory satisfying $f(0)=0$ and $f'(0)=0$, such as the so-called `power-of-GR' models. Therefore, the following highly non-evident result can be established: the vacuole consisting of an interior Minkowski space-time smoothly matched to ab exterior Class 1 space-time \eqref{novelstatic} at any areal radius $r=r_*=\const$ is a properly matched solution of any $(\LCR_0=0)$-degenerate $f(\LCR)$ model, i.e.~it satisfies both the equations of motion and all of the junction conditions of those $f(\LCR)$ models.

\end{appendices}

\bibliographystyle{ieeetr}
\bibliography{bibliography}

\end{document}